\documentclass{SciPost}

\hypersetup{
	colorlinks,
	linkcolor={red!50!black},
	citecolor={blue!50!black},
	urlcolor={blue!80!black}
}

\usepackage[bitstream-charter]{mathdesign}
\DeclareSymbolFont{usualmathcal}{OMS}{cmsy}{m}{n}
\DeclareSymbolFontAlphabet{\mathcal}{usualmathcal}

\fancypagestyle{SPstyle}{
	\fancyhf{}
	
	\fancyfoot[C]{\textbf{\thepage}}
}

\usepackage{amsmath}    
\usepackage{graphicx}   
\usepackage{amscd}    
\usepackage{enumitem}
\usepackage{color}
\usepackage{mathrsfs}
\usepackage{bm}
\usepackage{bbm,wasysym}
\usepackage{multirow}
\usepackage{easyReview}
\usepackage{array}
\usepackage{empheq}
\usepackage{pifont}
\usepackage{soul}
\usepackage{svg}
\usepackage{float}

\renewcommand{\todo}[1]{\textcolor{red}{TODO: #1}\ }
\renewcommand{\todo}[1]{}								

\renewcommand{\(}{\left(}
\renewcommand{\)}{\right)}
\newcommand{\bb}[1]{\mathbb{#1}}

\newcommand{\fc}[1]{\mathcal{#1}}
\newcommand{\scr}[1]{\mathscr{#1}}
\newcommand{\dvec}[1]{\dot{\vec{#1}}}

\DeclareMathSymbol{\Alpha}{\mathalpha}{operators}{"41}
\DeclareMathSymbol{\Beta}{\mathalpha}{operators}{"42}
\DeclareMathSymbol{\Epsilon}{\mathalpha}{operators}{"45}
\DeclareMathSymbol{\Zeta}{\mathalpha}{operators}{"5A}
\DeclareMathSymbol{\Eta}{\mathalpha}{operators}{"48}
\DeclareMathSymbol{\Iota}{\mathalpha}{operators}{"49}
\DeclareMathSymbol{\Kappa}{\mathalpha}{operators}{"4B}
\DeclareMathSymbol{\Mu}{\mathalpha}{operators}{"4D}
\DeclareMathSymbol{\Nu}{\mathalpha}{operators}{"4E}
\DeclareMathSymbol{\Omicron}{\mathalpha}{operators}{"4F}
\DeclareMathSymbol{\Rho}{\mathalpha}{operators}{"50}
\DeclareMathSymbol{\Tau}{\mathalpha}{operators}{"54}
\DeclareMathSymbol{\Chi}{\mathalpha}{operators}{"58}
\DeclareMathSymbol{\omicron}{\mathord}{letters}{"6F}

\newcommand{\mc}[1]{\text{\fontencoding{U}\fontfamily{boondoxuprscr}\fontshape{n}\selectfont #1}}

\renewcommand{\l}{\mathcal{L}}

\newcommand{\h}{\mathcal{H}}

\renewcommand{\d}{\partial}

\newcommand{\8}{\infty}
\newcommand{\assert}{\overset{!}{=}}
\newcommand{\df}{\overset{\text{def}}{=}}

\newcommand{\p}{\hspace*{5ex}}
\newcommand{\ps}{\hspace*{2ex}}

\newcommand{\<}{\langle}
\renewcommand{\>}{\rangle}

\newcommand{\ec}{\mathtt{e}}

\newcommand{\OO}{\fc{O}}

\newcommand{\twid}[1]{\widetilde{#1}}
\newcommand{\eps}{\varepsilon}
\newcommand{\cmt}[1]{}

\newcommand{\ixt}{\text{int}}
\newcommand{\ee}{\text{E}}

\newcommand{\scri}{\mathscr{I}}

\newcommand{\mlb}{\bigl\<\!\!\bigl\<}
\newcommand{\mrb}{\bigr\>\!\!\bigr\>}

\makeatletter
\newcommand{\oset}[3][0ex]{%
	\mathrel{\mathop{#3}\limits^{
			\vbox to#1{\kern-.5\ex@
				\hbox{$\scriptstyle#2$}\vss}}}}
\makeatother

\makeatletter
\let\old@font@info\@font@info
\def\@font@info#1{%
	\expandafter\ifx\csname\detokenize{#1}\endcsname\relax
	\old@font@info{#1}%
	\fi
	\expandafter\xdef\csname\detokenize{#1}\endcsname{}%
}
\makeatother

\newcommand{\tens}[1]{
	\oset{\leftrightharpoonup}{#1}
}

\newcommand{\leftrightharpoonup}{%
	\mathrel{\mathpalette\lrhup\relax}%
}
\newcommand{\lrhup}[2]{%
	\ooalign{$#1\leftharpoonup$\cr$#1\rightharpoonup$\cr}%
}

\newcommand{\doot}[1]{\overset{\circ}{#1}}

\newcommand{\ghv}{\bar{\fc{G}}}
\newcommand{\ghz}{\hat{\fc{G}}}

\numberwithin{equation}{subsection}

\begin{document}
	
	\pagestyle{SPstyle}
	
	\begin{center}{\Large \textbf{\color{scipostdeepblue}{
					Variational Approach to the Mathisson-Papapetrou-Dixon Equations of Motion for Extended Objects in General Relativity\\
	}}}\end{center}
	
	\begin{center}\textbf{
			Trevor A. Scheopner\textsuperscript{1$\star$} and
			Felix A. Lichtner\textsuperscript{1$\dagger$}
	}\end{center}
	
	\begin{center}
		{\bf 1} Max Planck Institute for Gravitational Physics (Albert Einstein Institute), Am Mühlenberg 1, Potsdam 14476, Germany
		\\[\baselineskip]
		$\star$ \href{mailto:trevor.scheopner@aei.mpg.de}{\small trevor.scheopner@aei.mpg.de} \p
		$\dagger$ \href{mailto:felix.lichtner@aei.mpg.de}{\small felix.lichtner@aei.mpg.de}
	\end{center}

	\section*{\color{scipostdeepblue}{Abstract}}
	\textbf{\boldmath{%
			The exact motion of a finite size object in curved spacetime is captured by the Mathisson-Papapetrou-Dixon equations. They associate a unique worldline, momentum, and spin to that object and determine the evolution of such in terms of its fully independent multipole moments. We provide a thorough review of this theory and show how to build it directly via a manifestly gauge invariant variational principle well-suited for effective descriptions of relativistic compact objects. We discuss conditions under which the covariant spin supplementary condition can be enforced and identify the natural basis of curvature tensors conjugate to the object's multipole moments. 
	}}
	
	\vspace{\baselineskip}
	
	
	\vspace{10pt}
	\noindent\rule{\textwidth}{1pt}
	\tableofcontents
	\noindent\rule{\textwidth}{1pt}
	\vspace{10pt}
	
	\section{Introduction}
	
	\subsection{Background}
	
	The first observation of gravitational waves by the LIGO/Virgo collaboration~\cite{LIGOScientific:2016aoc, LIGOScientific:2017vwq} jump-started a new, exciting age of gravitational physics. In this modern era, we have unprecedented access to the experimental effects of strong gravitational fields on compact objects, such as black holes and neutron stars. Consequently, developing a precise theoretical understanding of the motion of such extended (and dynamically complicated) objects in full general relativity is extremely important for the continued advancement of gravitational wave science~\cite{Punturo:2010zz, LISA:2017pwj, Reitze:2019iox, Blanchet:2006zz, Porto:2016pyg, Levi:2018nxp, Buonanno:2022pgc}. Such analytical calculations are performed in several formalisms, typically in the post-Newtonian~\cite{Buonanno:2022pgc, Blanchet:2006zz, Brunello:2025gpf, Blanco:2026prj, Porto:2016pyg, Levi:2018nxp, Levi:2022rrq, Mandal:2022nty, Goldberger:2009qd, Goldberger:2004jt,Foffa:2012rn, Foffa:2019rdf, Blumlein:2020pog, Foffa:2019hrb, Blumlein:2021txe, Porto:2024cwd, Blumlein:2020znm, Porto:2005ac, Levi:2015msa, Kim:2021rfj, Ross:2012fc, Cho:2022syn, Siemonsen:2017yux, Porto:2006bt, Faye:2006gx, Damour:2007nc, Liu:2021zxr}, post-Minkowskian~\cite{Vines:2017hyw, Mogull:2020sak, Jakobsen:2023ndj, Jakobsen:2023pvx, Jakobsen:2022psy, Jakobsen:2023hig, Driesse:2024xad, Driesse:2024feo, Jakobsen:2021smu, Jakobsen:2021zvh, Cheung:2018wkq, Bern:2019nnu, Bern:2022kto, Cristofoli:2019neg, Bern:2021dqo, Bern:2021yeh, Georgoudis:2023lgf, Bohnenblust:2023qmy, Bohnenblust:2025gir, Bern:2020buy, Chiodaroli:2021eug, Aoude:2023vdk, Cangemi:2023ysz, Akpinar:2025bkt, FebresCordero:2022jts, Guevara:2019fsj, Guevara:2018wpp, Damgaard:2024fqj, Guevara:2020xjx, Alaverdian:2025jtw, Alaverdian:2024spu, Bern:2023ity, Porto:2026fsd, Dlapa:2026oyq, Amalberti:2023ohj, Amalberti:2024jaa, Mandal:2024iug, Dlapa:2021npj, Dlapa:2021vgp, Dlapa:2023hsl, Kalin:2020mvi, Kalin:2022hph, Dlapa:2022lmu, Dlapa:2024cje, Dlapa:2025biy,Bern:2026oqp, Driesse:2026qiz, Haddad:2025cmw, Mogull:2025cfn, Kim:2025gis, Scheopner:2023rzp, JanSteinhoff:2015ist}, or self-force expansions~\cite{Barack:2018yvs, Harte:2009yr, Harte:2008xq, Gralla:2008fg, Poisson:2011nh, Harte:2005yi, Blanco:2022mgd, Blanco:2023jxf, Blanco:2024fte, Harte:2025tmd, Blanco:2026wwi, Harte:2026ihj, Ramond:2024ozy, Ramond:2024sfp, Compere:2023alp, deFirmian:2026mln, Drummond:2022efc, Drummond:2022xej, Drummond:2023loz, Witzany:2022vck, Piovano:2024yks, Kerachian:2023oiw, Witzany:2024ttz, Skoupy:2024uan, Albertini:2024agg, Skoupy:2026ewu, Witzany:2026eqc, Bautista:2024agp,Harte:2014wya,Pound:2009sm,Pound:2021qin}. 
	
	The study of the motion of extended objects in Newtonian mechanics beyond the point-particle approximation dates back to Euler. The theory culminates in a system of equations of motion for the center of mass trajectory of the body, its total linear momentum, and its total spin angular momentum, all expressed in terms of the body's higher multipole moments. Those higher multipole moments do not satisfy universal equations of motion in the same way and are instead determined by whatever physical model is appropriate for the body of interest. While the general relativistic problem is vastly more complicated, these essential outcomes are exactly the same. 
	
	The motion of extended bodies in relativity was studied significantly but incompletely by several early relativists~\cite{Lense:1918zz, Papapetrou:1951pa, Mathisson:1937zz, madore, taub, Tulczyjew} with the most essential stepping stones provided by Mathisson and Papapetrou. Dixon eventually took up the problem and over the course of his original trilogy of papers~\cite{Dixon:1970mpd, dixon2, dixon3} thoroughly developed the core foundation of the theory. Those papers are over 50 years old, fully rigorous, and highly technical, but due to recent experimental and theoretical progress in gravitational wave physics, their contents have never been more relevant to modern research. The ultimate results of the theory are the so-called Mathisson-Papapetrou-Dixon (MPD) equations, which give \textit{exact} equations of motion for the body's total linear momentum, its total spin angular momentum, and (when supplemented by an appropriate definition of the body center) its worldline trajectory, provided the body and the spacetime around it satisfy mild regularity conditions.
	
	Since Dixon, many others~\cite{Beiglbock:1965fun, Ehlers:1977rud, BAILEY1980188, PhysRevD.95.044025, Steinhoff:2009tk, JanSteinhoff:2015ist, Steinhoff:2008zr, Steinhoff:2015ksa, Steinhoff:2016rfi, Levi:2014gsa, Harte:2014wya, Harte:2011ku, Witzany:2018ahb, Puetzfeld:2015cin, Ben-Shahar:2023djm, Saketh:2022wap, Saketh:2022xjb, Vines:2017hyw, Vines:2016unv, Guevara:2020xjx} have contributed significant additional insight to the theory. In the present, the MPD theory is the bedrock justification of the worldline methods used to study the general relativistic two-body problem analytically in perturbation theory beyond the point-particle approximation~\cite{Porto:2005ac, Siemonsen:2017yux, Porto:2006bt, Faye:2006gx, Damour:2007nc, Levi:2015msa, Liu:2021zxr, Jakobsen:2021zvh, Scheopner:2023rzp} and it is deeply useful for the analysis of gravitational self-force~\cite{Barack:2018yvs, Harte:2009yr, Harte:2008xq, Gralla:2008fg, Poisson:2011nh, Harte:2005yi, Pound:2009sm,Pound:2021qin}. The full-and-exact MPD equations contain all the physics of the local covariant conservation of energy-momentum for a compact object interacting with general relativity and electromagnetism. They can be systematically approximated in several ways (in for example the test-body limit or the weak-field expansion) and applied to multiple bodies to study many specific effects of interest such as adiabatic and dynamical tides~\cite{Steinhoff:2016rfi, Saketh:2023bul}, self-force~\cite{Harte:2014wya}, post-Minkowskian and post-Newtonian corrections~\cite{Levi:2018nxp, JanSteinhoff:2015ist, Levi:2015msa} in the two-body problem.
	
	This review aims to collect the many disparate pieces of the MPD theory of compact object motion into a coherent whole. Structurally, we will follow much of the same logic as Dixon's original papers. However, we will do so with modernized notation and style and we will provide further fleshed out motivations to guide our calculations. In preparing this review, we found several gaps in literature, such as in precisely stating the conditions under which the commonly used momentum direction time-slicing and covariant spin supplementary condition can be required to hold. We address these in Section \ref{sec:ps} and in greater detail in Appendices \ref{sec:uint} and \ref{sec:sscint}. We also found that the theory can be derived and developed more simply than in the original by using variational principles. Doing so allows some of the most important results in the MPD theory to take novel, very simple forms with clear conceptual meaning. This will be most strongly reflected in \eqref{eq:gravact}, which for the first time gives the simplified explicit formula for the energy-momentum tensor of a generic compact object in curved spacetime in terms of its reduced multipole moments. The variational approach is also optimal for performing approximations and making real calculations, as it guarantees the approximated equations of motion form a closed system at any order. 
	
	\subsection{Description of the Problem}
	
	We want to understand the motion of compact objects in general relativity (such as stars or black holes). A compact object (also called a body) is an object of finite spatial extent. Such a body\footnote{A black hole is a vacuum solution to Einstein's equations outside of its event horizons, but it is possible to construct an effective energy-momentum tensor that produces the same spacetime outside of the black hole. Such a source can be made to have support purely within the horizon. See, for example~\cite{Israel:1970kp, Levi:2015msa, Scheopner:2023rzp}. Using such effective sources, black holes can (at least in perturbation theory) be treated by the MPD formalism and the effective field theory methods established in Subsection \ref{sec:apps}.} has some energy-momentum tensor $T^{\alpha\beta}(x)$. Einstein's field equations guarantee that the energy-momentum tensor is symmetric:
	\begin{equation}
		T^{\alpha\beta} = T^{\beta\alpha} \label{eq:symt}
	\end{equation}
	and if the only forces the body interacts with are gravity and electromagnetism, they also guarantee that the body's energy-momentum tensor satisfies the energy-momentum transport equation:
	\begin{equation}
		\nabla_\beta T^{\alpha\beta} = \fc{F}^{\alpha\beta} J_\beta, \label{eq:tcons}
	\end{equation}
	where $\fc{F}_{\alpha\beta}$ is the electromagnetic field strength defined in terms of the vector potential $A_\alpha$ by:
	\begin{equation}
		\fc{F}_{\alpha\beta} = \nabla_\alpha A_\beta - \nabla_\beta A_\alpha \label{eq:fdef}
	\end{equation}
	and $J^\alpha$ is the body's electromagnetic charge current density which due to electromagnetic gauge invariance must satisfy the continuity equation:
	\begin{equation}
		\nabla_\alpha J^\alpha = 0. \label{eq:jcons}
	\end{equation}
	
	Equations \eqref{eq:symt}, \eqref{eq:tcons}, and  \eqref{eq:jcons} are exact and extremely general. However, they are not immediately helpful for understanding several basic questions (where is the body; how fast is it moving or rotating?). We seek to extract from the fundamental equations \eqref{eq:symt}, \eqref{eq:tcons}, and \eqref{eq:jcons} a reduced set of degrees of freedom which capture the relevant physics for applications concerned with the macroscopic behavior of the body. Specifically, we will look for reasonable definitions of the worldline curve $z^\mu(\lambda)$ (parameterized by the ``worldline time'' $\lambda$) traced out in time by the body's center of mass, the total linear momentum $p_\mu(\lambda)$ conjugate to translations of $z^\mu(\lambda)$ at each time, the total angular momentum $S^{\mu\nu}(\lambda)$ about the worldline at each time, and the total mass of the body. Then, we will develop exact equations of motion for how these quantities evolve with~$\lambda$. These total/average quantities often evolve on much longer time scales than the degrees of freedom of the body's internal dynamics, and so are the natural variables which are relevant for treating the effective interactions of the body with other bodies. Thus, many approximate forms of the full equations of motion are of fundamental import to the general relativistic two-body problem. 
	
	\subsection{Overview of the Main Results}
	
	Here we give a schematic description of the ultimate facts of compact object motion, with the precision of definitions and motivations developed over the rest of this review. All of the most important results are collected with proper details in Section \ref{sec:conc}. 
	
	There are unique definitions of the total linear momentum $p_\mu$ and total spin of the body $S^{\mu\nu}$ for any valid choice of body worldline $z^\mu$ and choice of how to define an equal timeslice with respect to an observer of interest. These definitions are unique in that they are the proper generators of translations and Lorentz transformations of that worldline. They are given by explicit integrals involving the energy-momentum tensor and the charge current density over a spatial slice at fixed worldline time. Their definitions and the equations that they satisfy are exact in full interacting general relativity (the body is not assumed point-like, the body is not assumed to be a test-body, the body is not assumed rigid, self-force is not neglected, the spin is not assumed small, nor any other approximation). It is possible to apply the MPD theory to the exact motion of multiple bodies, giving worldline descriptions for each body separately, whenever the energy-momentum tensors and current densities of the different bodies are always disjoint. 
	
	There is a unique set of multipole moments for the electromagnetic current density so that (I) the different multipole moments are fully reduced, meaning (beyond the monopole moment) they are completely unconstrained by the charge continuity equation and so that (II) the action for the electromagnetic interaction can be expressed as a series in these moments multiplying a natural curved spacetime generalization of the partial derivatives in spatial directions of the electromagnetic field strength. They are given by explicit integrals in terms of the current density over a spatial slice at fixed time along the worldline and may have arbitrary time evolution, determined only by the internal physics of the body. Using these moments, the electromagnetic continuity equation \eqref{eq:jcons} is completely equivalent to the conservation of the body's total charge $q$. 
	
	There is a unique set of multipole moments for the energy-momentum tensor so that (I) the different multipole moments are fully reduced, meaning (beyond the dipole moment) they are completely unconstrained by the covariant conservation of energy and momentum and so that (II) the action for the gravitational interaction can be expressed as a series in these moments multiplying a curved spacetime generalization of the spatial derivatives of the linearized Riemann tensor. They are given by explicit integrals in terms of the energy-momentum tensor and charge current density over a spatial slice at fixed worldline time and may have arbitrary time evolution, determined only by the internal physics of the body. Using these moments, the local covariant conservation of energy-momentum \eqref{eq:tcons} is completely equivalent to the Mathisson-Papapetrou-Dixon equations of motion for the momentum and spin: 
	\begin{equation}
		\doot p_\mu = - \frac{1}{2}R_{\mu\nu\rho\sigma}\doot z{}^\nu S^{\rho\sigma} + q\fc{F}_{\mu\nu} \doot z{}^\nu + \text{\textbf{F}}_\mu, \p \doot S{}^{\mu\nu} = p^\mu\doot z{}^\nu - \doot z{}^\mu p^\nu + \text{\textbf{T}}^{\mu\nu} \label{eq:mpd1}
	\end{equation}
	where the MPD force $\text{\textbf{F}}_\mu$ and torque $\text{\textbf{T}}^{\mu\nu}$ have expressions in terms of the charge and mass multipole moments of the body. 
	
	For physically reasonable bodies it is possible to choose the time direction on the worldline to coincide with the unit vector in the direction of the body's momentum and to choose the worldline $z^\mu$ so that the spin represents pure rotations, without boosts, in the center of mass frame. With these choices, the spin tensor of the body satisfies the covariant spin supplementary condition $S^{\mu\nu} p_\nu = 0$. The time derivative of this condition, using the MPD equations of motion, uniquely determines the worldline of the body.
	
	The entire MPD formalism can be packaged efficiently into a variational principle so that all of the main results of the theory are given by variations of that action. The unique definition of Dixon's moments is exactly born from attempting to write that action in a manifestly gauge invariant form.
	
	Very importantly, the exact equations of motion so described for $z$, $p$, and $S$ do not form a closed system of differential equations in general. The MPD force and torque depend on all higher multipole moments of the body, and those multipole moments have dynamics which are wholly unconstrained by the fundamental equations \eqref{eq:tcons} and \eqref{eq:jcons}. In order to form a closed system of equations, one needs to know enough about the physics of the body to additionally determine the time evolution of these multipole moments. In practice this is done using effective field theory and it is at this step that various approximations become useful, as described in Subsection \ref{sec:apps}.

	\subsection{Outline of Approach}
	
	The theory behind the MPD equations is quite towering when viewed all at once, but we will break it up into digestible pieces.
	We begin by first studying the motion of extended objects in Newtonian gravity in Section \ref{sec:newton}, which is important for motivating what we should be looking for in the relativistic theory.
	We then lay out our basic notations and physical assumptions for the relativistic problem in Section \ref{sec:prelim}. 
	We introduce some necessary theoretical tools in Section \ref{sec:bi} which will allow us to generalize our ideas from Newtonian mechanics to curved spacetime. 
	With these tools, in Section \ref{sec:ps} we find the general and exact definitions of the momentum, spin, and center of mass in general relativity. 
	In Section \ref{sec:em} we develop the theory of extended objects interacting with electromagnetism in flat spacetime as a warm-up. 
	This theory will be highly suggestive of the ultimate results in curved spacetime. 
	In Section \ref{sec:grfs} we then build the theory of an extended object interacting with gravity to first order in the metric perturbation about flat spacetime as a further step up in difficulty. 
	In Section \ref{sec:emgr} we establish the theory of the electromagnetic interaction in curved spacetime. 
	And then, in total glory, in Section \ref{sec:gr} we develop Dixon's theory for the coupling of an extended object to the metric, arriving ultimately at the MPD equations. 
	In doing so, we show that the multipole moments of the energy-momentum tensor are naturally conjugate to a set of tensors which we call the ``multipolar Riemann tensors''. The exact energy-momentum tensor of a generic compact object takes an especially simple form in terms of these tensors. 
	In Section \ref{sec:wlact}, we show how all of these results can be organized into a single simple variational principle and briefly describe several of its applications. 
	We take Section \ref{sec:conc} to summarize the entire theory and collect its most important results.
	
	\section{Newtonian Gravity} \label{sec:newton}
	
	Before studying the dynamics of compact objects in general relativity, we start with the much simpler problem of doing so in Newtonian gravity. This will help us establish a baseline of what sorts of structures we ought to look for when we level up to the full problem. 
	
	\subsection{Momentum, Spin, and the Center of Mass}
	
	Consider a body in Newtonian mechanics described by a mass density $\rho_m(t,\vec x)$, a local velocity field $\vec v(t,\vec x)$, and a symmetric stress tensor $\tens\sigma(t,\vec x)$ experiencing a force density $\vec f(t,\vec x)$. Such a body, in order to have local dynamics with Galilean symmetries, necessarily satisfies the nonrelativistic transport equations:
	\begin{equation}
		\frac{\d \rho_m}{\d t} +\nabla\cdot(\rho_m\vec v) = 0, \p \frac{\d}{\d t}(\rho_m\vec v) + \nabla\cdot(\rho_m\vec v\vec v-\tens\sigma) = \vec f. \label{eq:211}
	\end{equation}
	The first (the mass continuity equation) expresses local mass conservation while the second (the momentum transport equation) expresses local linear momentum conservation. Because $\rho_m(t,\vec x)$ is the mass density, there is a natural definition of the total mass $m$ of the body:
	\begin{equation}
		m = \int_{\bb{S}} \rho_m d\bb{S}
	\end{equation}
	where $\bb{S}$ is space at a fixed time and $d\bb{S}$ is the volume measure on that time-slice. The mass continuity equation guarantees that $m$ is constant. In the momentum transport equation, $\rho_m\vec v$ represents the linear momentum density of the body, so it is the generator density of translations of the body. Thus, there is a natural definition of its total linear momentum $\vec p(t)$:
	\begin{equation}
		\vec p(t) = \int_{\bb{S}} \rho_m\vec v d\bb{S}. \label{eq:newtpdef}
	\end{equation}
	Again thinking of $\rho_m\vec v$ as the generator density of body translations, the generator of rotations of the whole body about the origin (its angular momentum) $\vec J(t)$ and the generator of Galilean boosts of the whole body about the origin (called the dynamical mass moment) $\vec G(t)$ are given by:
	\begin{equation}
		\vec J(t) = \int_{\bb{S}} \rho_m \vec x\times\vec v d\bb{S}, \p \vec G(t) = \int_{\bb{S}} (\rho_m \vec v t - \rho_m \vec x) d\bb{S}.
	\end{equation}
	(This is because for a point mass at $\vec x$ the generators of rotations and Galilean boosts are $\vec x\times\vec p$ and  $ \vec p t - m \vec x$. Noether charges are extensive, so the generator of rotations/boosts of the whole body are the integrals of such for each point within the body.)
	
	We would like to define a natural trajectory $\vec z(t)$ giving the representative position for the body at each time $t$ and to find equations of motion for its evolution. If the body were a point-particle with the same mass and momentum, then it automatically would have angular momentum $\vec L(t)$ and dynamical mass moment $\vec N(t)$ (both about the origin):
	\begin{align}
		\vec L(t) = \vec z\times\vec p, \p \vec N(t) = t\vec p - m\vec z.
	\end{align}
	For a finite size body, we can define its intrinsic angular momentum $\vec S(t)$ (the spin) of the body and its intrinsic mass moment $\vec{\mc{K}}(t)$, both relative to the choice of trajectory, by the difference between the total angular momentum/mass moment and these hypothetical point-particle values:
	\begin{equation}
		\vec S(t) = \vec J - \vec L = \int_{\bb{S}} \rho_m (\vec x-\vec z)\times\vec v d\bb{S}, \p \vec {\mc{K}}(t) = \vec G - \vec N = -\int_{\bb{S}} \rho_m (\vec x-\vec z) d\bb{S}. \label{eq:newtsdef}
	\end{equation}
	The integral expression for $\vec {\mc{K}}$ shows that $-\vec {\mc{K}}$ is precisely the body's ``mass dipole moment''. It is suggestive to solve that expression for $\vec z(t)$:
	\begin{equation}
		\vec z(t) = \frac{1}{m}\int_{\bb{S}} \rho_m \vec x d\bb{S} + \frac{\vec {\mc{K}}(t)}{m}.
	\end{equation}
	From this we can see that choosing a $\vec z(t)$ is equivalent to choosing a $\vec {\mc{K}}(t)$. So far the trajectory $\vec z(t)$ has been arbitrary. But, if we choose $\vec z(t)$ to follow the center of mass of the body then we are  equivalently choosing $\vec {\mc{K}}(t)$ to be 0:
	\begin{equation}
		\vec {\mc{K}} = 0 \iff \vec z(t) = \frac{1}{m}\int_{\bb{S}} \rho_m \vec x d\bb{S}. \label{eq:newtkdef}
	\end{equation}
	It will be much easier to generalize the vanishing of the intrinsic mass moment to general relativity than to directly generalize the explicit formula for the center of mass. 
	
	Going forward in this section, we always take $\vec z(t)$ to be the center of mass trajectory. By differentiating \eqref{eq:newtkdef} using \eqref{eq:211}:
	\begin{equation}
		\vec p = m \dvec z. \label{eq:newtwl}
	\end{equation}
	So, the evolution of the trajectory is determined by the evolution of the total linear momentum. Differentiating the definitions of $\vec p(t)$ and $\vec S(t)$ and using \eqref{eq:211} gives:
	\begin{equation}
		\frac{d\vec p}{dt} = \vec{F}(t) = \int_{\bb{S}} \vec f d\bb{S}, \p \frac{d\vec S}{dt} = \vec{\fc{T}}(t) = \int_{\bb{S}} (\vec x-\vec z)\times\vec f d\bb{S}. \label{eq:newtfdef}
	\end{equation}
	The right sides above provide integral representations of the total force $\vec F$ and total torque $\vec{\fc{T}}$ on the body. If we had not chosen $\vec {\mc{K}} = 0$, the right hand side of the spin equation would have had an additional $\vec p\times\dvec z$ term. This provides another motivating reason for choosing the trajectory as we have; it is the unique choice for which the spin vector is conserved in the absence of any external forces. 
	
	\subsection{Mass Multipole Moments}
	
	Let us now assume that the only force acting on the body is Newtonian gravity, and so we have $\vec f = -\rho_m\nabla\phi$ where $\phi$ is the gravitational potential. Assuming that the potential is analytic in the region occupied by the body, we can write (for $\vec x$ a point in the body):
	\begin{equation}
		\phi(t,\vec x) = \sum_{n=0}^\8 \frac{1}{n!}(x-z(t))^{a_1}\dots(x-z(t))^{a_n} \d^n_{a_1\dots a_n}\phi(t,\vec z(t)). \label{eq:nphiexpand}
	\end{equation} 
	Inserting this into the first of \eqref{eq:newtfdef}, we find:
	\begin{equation}
		\frac{d p_a}{dt} = -\sum_{n=0}^\8 \frac{1}{n!} M_n^{b_1\dots b_n}(t) \d^{n+1}_{a b_1\dots b_n} \phi(t, \vec z), \label{eq:npeom}
	\end{equation}
	where we have defined the mass multipole moments $M_n^{b_1\dots b_n}$ by:
	\begin{equation}
		M_n^{a_1\dots a_n}(t) = \int_{\bb{S}} \rho_m(t,\vec x) (x-z(t))^{a_1}\dots(x-z(t))^{a_n} d\bb{S}. \label{eq:nmdef}
	\end{equation}
	Similarly, inserting $\vec f = -\rho_m\nabla\phi$ into the second of \eqref{eq:newtfdef} and performing the same manipulations yields:
	\begin{equation}
		\frac{d S_a}{dt} = -\sum_{n=0}^\8 \frac{1}{n!} \epsilon_{ab}{}^c M_{n+1}^{d_1\dots d_n b}(t) \d^{n+1}_{d_1\dots d_n c}\phi(t, \vec z). \label{eq:nseom}
	\end{equation}
	Equations \eqref{eq:newtwl}, \eqref{eq:npeom}, and \eqref{eq:nseom} provide a system of equations that determine the evolution of $\vec z$, $\vec p$, and $\vec S$ if the time dependence of the mass multipole moments is known. The time dependence of $M_0$ is easy, because it is just the mass:
	\begin{equation}
		M_0(t) = m. \label{eq:m00}
	\end{equation}
	Similarly, $M_1^a$ is 0 (when using the center of mass trajectory):
	\begin{equation}
		M_1^a(t) = \int_{\bb{S}} \rho x^a d\bb{S} - m z^a = -{\mc{K}}^a = 0. \label{eq:m10}
	\end{equation}
	For $n > 1$, the time dependence of $M_n^{\cdots}$ is not so easily determined. For different bodies there are generally different appropriate constitutive relations that determine the time dependence of its multipole moments. For example, for a rigid body, there exists a time dependent rotation matrix $R^a{}_b(t)$ so that if each index of the multipole moments is contracted with that rotation matrix, the resulting tensor is constant. It is not possible in general to find a closed system of equations for the time evolution of the mass multipole moments simply from the nonrelativistic transport equations. Mathematically this is because if one takes their definitions and differentiates them, the results do not close after any finite order of differentiation in terms of the previous quantities. Physically this is because different bodies have physically distinct internal interactions (characterized by $\tens{\sigma}$). 
	
	Using \eqref{eq:m00} and \eqref{eq:m10}, the equations of motion \eqref{eq:newtwl}, \eqref{eq:npeom}, and \eqref{eq:nseom} simplify to:
	\begin{align}
		\frac{d z^a}{dt} &= \frac{p^a}{m} \label{eq:ngz} \\
		\frac{d p_a}{dt} &= -m\d_a \phi(t,\vec z(t)) + \sum_{n=2}^{\8} \frac{1}{n!} M_n^{b_1\dots b_n}\d^{n+1}_{a b_1\dots b_n} \phi(t,\vec z(t)) \label{eq:ngp} \\
		\frac{d S_a}{dt} &= -\sum_{n=1}^\8 \frac{1}{n!} \epsilon_{ab}{}^c M_{n+1}^{d_1\dots d_n b} \d^{n+1}_{d_1\dots d_n c}\phi(t,\vec z(t)). \label{eq:ngs} 
	\end{align}
	These are the Newtonian equations of motion for a compact object in a gravitational field.
	
	\subsection{Forces from a Variational Principle}
	
	While variational methods were not present in Dixon's original analysis, we will find that their application to the theory is highly fruitful\footnote{A common misconception is that variational methods are only appropriate for conservative physics, but in fact if the correct phase space is used, they are fully applicable in general, even to systems with arbitrary dissipation. See~\cite{Galley:2012hx, Galley:2014wla}.}. The Newtonian gravitational potential $\phi$ satisfies the Poisson equation:
	\begin{equation}
		\nabla^2\phi = 4\pi G \rho_m,
	\end{equation}
	which is naturally produced as the $\phi$ equation of motion from the action:
	\begin{equation}
		\fc{S} = -\frac{1}{8\pi G}\int_{-\8}^\8\int_{\bb{S}}|\nabla\phi|^2 d\bb{S} dt + \fc{S}_m,
	\end{equation}
	where $\fc{S}_m$ is the action for the matter in the system when placed in the gravitational potential and the mass density $\rho_m$ is derived from it by:
	\begin{equation}
		\rho_m = -\frac{\delta \fc{S}_m}{\delta \phi}. \label{eq:rhomdef}
	\end{equation}
	In the case that the body's structure is modeled as fully independent of the gravitational field it experiences, the matter action takes the simple form:
	\begin{equation}
		\fc{S}_m = \fc{S}_0 + \fc{S}_\ixt, 
	\end{equation}
	where $\fc{S}_0$ only depends on the matter degrees of freedom and the interaction action is given by:
	\begin{equation}
		\fc{S}_\ixt = -\int_{-\8}^\8\int_{\bb{S}} \rho_m\phi d\bb{S} dt. \label{eq:nsint}
	\end{equation}
	Or, in the case that the body is modeled as having nontrivial gravitational susceptibility (tidal deformability), when $\phi$ is perturbed from some default potential $\phi_0$ (with associated mass density $\rho_m$) by an amount $\delta \phi$:
	\begin{equation}
		\fc{S}_m = \left.\fc{S}_m\right|_{\phi = \phi_0} - \int_{-\8}^\8\int_{\bb{S}} \rho_m \delta\phi d\bb{S} dt + \OO(\delta\phi^2). \label{eq:nsintexpand}
	\end{equation}
	In either case, $\fc{S}_m$ not only determines the effects of the mass density on the gravitational potential but necessarily also the effects of the gravitational potential on the mass density. Thus, we should be able to derive our force $\vec F$ and torque $\vec{\fc{T}}$ from this action. 
	
	We can express \eqref{eq:nsint} (by performing the same series expansion of $\phi$ as in \eqref{eq:nphiexpand}) nicely in terms of the multipole moments as:
	\begin{equation}
		\fc{S}_\ixt = \int_{-\8}^\8\(-m\phi(t,\vec z(t)) - \sum_{n=2}^\8 \frac{1}{n!} M_n^{a_1\dots a_n}(t) \d^n_{a_1\dots a_n}\phi(t,\vec z(t))\) dt. \label{eq:nsintseries}
	\end{equation}
	The fundamental definition of the generalized force on a coordinate $Q$ due to an interaction is $\frac{\delta \fc{S}_\ixt}{\delta Q}$. Using this, by varying $\fc{S}_\ixt$ with respect to the position and orientation of the body we can recover $\vec {F}$ and $\vec{\fc{T}}$. Under a translation of the entire body by $\vec a$ and a rotation about $\vec z(t)$ by a rotation matrix $\hat R$, the mass density of the body is transformed as:
	\begin{equation}
		\rho_m'(t,\vec x) = \rho_m(t,\vec z + \hat R^{-1}(\vec x-\vec z-\vec a)). 
	\end{equation}
	If we perform an infinitesimal translation by $\delta \vec z$ and an infinitesimal rotation by $\delta\theta^a$ (so that $R^a{}_b = \delta^a_b + \epsilon^a{}_{cb} \delta\theta^c$), then according to \eqref{eq:newtkdef} and \eqref{eq:nmdef}, the center of mass position and multipole moments of the body transform as:
	\begin{equation}
		\vec z' = \vec z + \delta\vec z, \ps M_n'^{a_1\dots a_n} = M_n^{a_1\dots a_n} + \epsilon^{a_1}{}_{bc}\delta\theta^b M_n^{ca_2\dots a_n} + \dots \epsilon^{a_n}{}_{bc} \delta\theta^b M_n^{a_1\dots a_{n-1} c} + \OO(\delta\theta^2). 
	\end{equation}
	Under these transformations, the variation of \eqref{eq:nsintseries} indeed becomes:
	\begin{equation}
		\delta \fc{S}_\ixt = \int_{-\8}^\8\(\vec F\cdot\delta\vec z(t) + \vec{\fc{T}}\cdot\delta\vec\theta(t)\) dt, \label{eq:nsintft}
	\end{equation}
	with the force and torque given exactly as in \eqref{eq:newtfdef}, as expected. It is important to note that in this calculation we never assume the body is rigid. The translations and rotations applied are of the entire body, but further variation of the multipole moments (corresponding to varying the internal structure of the body) are fully allowed. Including those variations in \eqref{eq:nsintft} would then provide a method to compute the generalized gravitational forces (such as tidal effects) on the corresponding internal body degrees of freedom. 
	
	Formally, \eqref{eq:nsintseries} provides a distributional expression for $\rho_m$ by following \eqref{eq:rhomdef}. In particular:
	\begin{equation}
		\rho_m(t,\vec x) = m \delta(\vec x-\vec z(t)) + \sum_{n=2}^\8 \frac{(-1)^n}{n!} M_n^{a_1\dots a_n}(t) \d^n_{a_1\dots a_n}\delta(\vec x-\vec z(t)). 
	\end{equation}
	
	\subsection{Moment Generating Function}
	
	By \eqref{eq:nsint}, the gravitational interaction action for a fixed mass density function is a linear functional of the gravitational potential. This linear functional structure will appear for us repeatedly, so it is useful to define the function inner product $\< , \>$ as usual on complex valued functions $\psi$ and $\varphi$ so that:
	\begin{equation}
		\<\psi, \varphi\> = \int_{-\8}^\8 \int_{\bb{S}} \psi^*(t,\vec x)\varphi(t,\vec x) d\bb{S} dt.
	\end{equation}
	We will also make regular use of the Fourier transform. The conventional Fourier transform is defined relative to the coordinate origin (one considers exponentials $e^{i\vec k\cdot\vec x}$). However, because our multipole moments are based around a particular curve $\vec z(t)$, it is convenient to introduce the Fourier transform as being about that curve. The spatial Fourier transform about the point $\vec z$ (for now not considered explicitly as a function of time) of a function $\psi$ is defined by:
	\begin{equation}
		\twid{\psi}(t, \vec z, \vec k) = \int e^{-i\vec k\cdot (\vec x-\vec z)} \psi(t,\vec x) d^3\vec x.
	\end{equation}
	By Parseval's theorem, for square-integrable $\varphi(t, \vec x)$:
	\begin{equation}
		\int \psi^*(t,\vec x) \varphi(t,\vec x) d^3\vec x = \int \twid{\psi}^*(t,\vec z, \vec k) \twid{\varphi}(t,\vec z, \vec k)  \frac{d^3\vec k}{(2\pi)^3}. \label{eq:parseval}
	\end{equation}
	This is true without placing any particular condition on $\vec z$. Due to this, we can immediately rewrite the interaction action as:
	\begin{equation}
		\fc{S}_\ixt = -\<\rho_m,\phi\> = -\int_{-\8}^\8\int M^*(t,\vec k) \twid{\phi}(t,\vec z(t), \vec k) \frac{d^3 \vec k}{(2\pi)^3} dt. \label{eq:nmomgen}
	\end{equation}
	where we have introduced the mass moment generating function:
	\begin{equation}
		M(t,\vec k) = \int e^{-i\vec k\cdot(\vec x-\vec z(t))}\rho_m(t,\vec x) d^3\vec x.
	\end{equation}
	In the Newtonian case, the moment generating function $M(t, \vec k)$ is just the Fourier transform of the associated source density $\rho_m$. In more complicated theories this will no longer be the case. By series expanding the exponential and using the definition of the mass multipole moments, we have:
	\begin{equation}
		M(t,\vec k) = \sum_{n=0}^\8 \frac{(-i)^n}{n!} M_n^{a_1\dots a_n}(t) k_{a_1}\dots k_{a_n} \\
		\implies M_n^{a_1\dots a_n}(t) = i^n \left.\frac{\d^n M(t, \vec k)}{\d k_{a_1}\dots\d k_{a_n}}\right|_{\vec k = 0}. \label{eq:nmomseries}
	\end{equation}
	Rather than trying to define the multipole moments directly in theories beyond Newton's, it is typically easier to find a reasonable definition of the moment generating function satisfying an appropriate generalization of \eqref{eq:nmomgen} and then define the moments as in \eqref{eq:nmomseries} from its series expansion. 
	
	\subsection{Full Variational Approach}
	
	While, through \eqref{eq:nsintft}, we produced the force and torque from a variational principle, we can actually go further and use a variational principle to produce the full system of equations of motion \eqref{eq:ngz}, \eqref{eq:ngp}, \eqref{eq:ngs} while allowing the body to be deformable by gravitational fields. Let $\vec z$ be defined by \eqref{eq:newtkdef} and $\hat R(t)$ be a rotation matrix tracking the overall orientation of the body. For each of the body's higher multipole moments ($M_n^{a_1\dots a_n}$ with $n \ge 2$) we can define a ``body frame multipole moment'' by applying to each of its indices a factor $\hat R^{-1}$, that is: $(\hat R^{-1}){}^{a_1}{}_{b_1}\dots (\hat R^{-1}){}^{a_n}{}_{b_n} M_n^{b_1\dots b_n}$. By construction, these body frame multipole moments are invariant under translations and rotations of the body. They may depend arbitrarily on the applied gravitational field, but given such an applied field they will then depend on \emph{some} internal body degrees of freedom $\scri^\Omega$ where $\Omega$ is an abstract index that runs over all the remaining body variables. This way, the dynamics of the body's mass density are captured fully by $\vec z, \hat R, \scri^\Omega$, and $\phi$ and the internal body degrees of freedom $\scri^\Omega$ are by construction invariant under overall translations and rotations of the body. For simple bodies several approximate reductions of these dependencies may be possible (no $\phi$ dependence for a non-gravitationally deformable body; no $\scri^\Omega$ dependence for a body with only spin induced and gravitationally induced deformations), but we will stay fully generic.
	
	If a small rotation $\delta^a_b + \epsilon^a{}_{cb}\delta\theta^c + \OO(\delta\theta)^2$ is applied to the body, then $\hat R$ is transformed as:
	\begin{equation}
		\delta R^a{}_b = \epsilon^a{}_{cd}\delta \theta^c R^d{}_b.
	\end{equation}
	Naturally then, the angular velocity $\vec\omega(t)$ of the body is the vector so that:
	\begin{equation}
		\frac{dR^a{}_b}{dt} = \epsilon^a{}_{cd}\omega^c R^d{}_b.
	\end{equation}
	Consequently, variations of the angular velocity are related to variations of $\hat R$ by:
	\begin{equation}
		\delta\vec\omega = \frac{d}{dt}\delta\vec\theta + \delta\vec\theta\times\vec\omega. 
	\end{equation}
	The spin vector $\vec S(t)$ is by definition the Legendre conjugate to $\vec\omega$. Let $\scr{P}_\Omega$ be the conjugate momenta to $\scri^\Omega$. Let $\h$ be the Hamiltonian of the matter in a gravitational field. Then, we seek the form of $\h$ so that the variation of the action:
	\begin{equation}
		\fc{S}_m = \int_{-\8}^\8\Bigl(\vec p\cdot\dvec z + \vec S\cdot\vec\omega + \scr{P}_\Omega\dot{\scri}^\Omega - \h(\vec p, \hat R, \vec S, \scri, \scr{P},\{\nabla^n\phi\}_n)\Bigr) dt \label{eq:nhamact}
	\end{equation}
	with respect to $\vec z, \vec p, \hat R, \vec S$ reproduces the equations of motion \eqref{eq:ngz}, \eqref{eq:ngp}, \eqref{eq:ngs} as well as the correct mass density \eqref{eq:rhomdef} when varied with respect to $\phi$. The notation $\{\nabla^n\phi\}_n$ indicates that the Hamiltonian is allowed to depend on $\phi$ evaluated on the worldline to arbitrary order of spatial differentiation.
	
	Directly varying \eqref{eq:nhamact} with respect to $\phi$ yields:
	\begin{equation}
		\delta\fc{S}_m = -\int_{-\8}^\8\(-\frac{\d\h}{\d\phi}\delta\phi - \sum_{n=1}^\8 \frac{\d\h}{\d\d^n_{a_1\dots a_n}\phi} \d^n_{a_1\dots a_n}\delta\phi\) dt.
	\end{equation}
	Requiring a match to \eqref{eq:nsintseries} then gives that the mass multipole moments are contained in the Hamiltonian by:
	\begin{equation}
		m = \frac{\d\h}{\d\phi}, \p \frac{\d\h}{\d\d_a\phi} = 0, \p M_n^{a_1\dots a_n} = n! \frac{\d\h}{\d \d^n_{a_1\dots a_n}\phi}. 
	\end{equation} 
	Varying \eqref{eq:nhamact} with respect to everything else yields the equations of motion:
	\begin{align}
		&\dvec z = \frac{\d\h}{\d\vec p},&	&\vec\omega = \frac{\d\h}{\d\vec S},&	&\dot{\scri}^\Omega = \frac{\d\h}{\d\scr{P}^\Omega},& \nonumber \\
		&\dvec p = -\frac{\d\h}{\d\vec x} + \vec F,& &\dvec S = -\vec{\scr{D}} \h + \vec{\fc{T}},& &\dot{\scr{P}}_\Omega = -\frac{\d\h}{\d\scri^\Omega}  \label{eq:257}
	\end{align}
	where the notation $\vec{\scr{D}}\h$ is defined so that under applying a rotation of the whole body about the worldline by $\delta^a_b + \epsilon^a{}_{cb}\delta\theta^c + \OO(\delta\theta^2)$, the variation of the Hamiltonian is $\delta\h = \delta\vec \theta\cdot\vec{\scr{D}}\h$. Explicitly, this overall rotation derivative is given by directly rotating everything in the Hamiltonian as:
	\begin{equation}
		\delta\theta^a \scr{D}_a\h = \frac{\d \h}{\d R^{a}{}_b}\delta R^a{}_b + \frac{\d \h}{\d S}\cdot\delta\vec{S} + \sum_{n=2}^\8 \frac{\d\h}{\d \d^n_{a_1\dots a_n}\phi}\delta(\d^n_{a_1\dots a_n}\phi). 
	\end{equation}
	By matching the Hamiltonian equations of motion against \eqref{eq:ngz}, \eqref{eq:ngp}, \eqref{eq:ngs}, we see that the Hamiltonian must take the form:
	\begin{equation}
		\h= \frac{|\vec p|^2}{2m} + m\phi(t,\vec z(t)) + U(\hat R, \vec S, \scri, \scr{P}, \{\nabla^n\phi\}_{n\ge 2}) \label{eq:259}
	\end{equation}
	where $U$ is an arbitrary function which is invariant under rotations of the body (and which represents the energy of the body in its center of mass frame beyond the point-particle gravitational potential energy $m\phi$). The mass multipole moments are then contained within the $\nabla^n\phi$ derivatives of $U$ as:
	\begin{equation}
		\vec{\scr{D}} U = 0, \p M_n^{a_1\dots a_n} = n! \frac{\d U}{\d \d^n_{a_1\dots a_n} \phi}. \label{eq:260}
	\end{equation}
	The specific function $U$ is determined by the internal physics of the body. With the Hamiltonian \eqref{eq:259}, \eqref{eq:nhamact} fully reproduces \eqref{eq:ngz}, \eqref{eq:ngp}, \eqref{eq:ngs}, and so contains the full dynamics of the body. In equations \eqref{eq:259} and \eqref{eq:260} we can already see the enormous power of using a variational principle to study compact object motion. Whereas \eqref{eq:ngz}, \eqref{eq:ngp}, and \eqref{eq:ngs} are not closed, due to depending on the unconstrained higher mass multipoles of the body, no matter what internal degrees of freedom we use for our body in the variational principle \eqref{eq:259}, the resulting equations of motion \eqref{eq:257} are guaranteed to be closed, and thus we can attempt to solve them (at least perturbatively). The entire specification of what a particular object of interest \textit{is} in this theory thus comes down to designating the correct set of internal degrees of freedom $\scri^\Omega$ and $\scr{P}_\Omega$ and to using the correct potential energy function $U$.
	
	Our goal for the general relativistic problem will be to find a definition of the worldline of the body analogous to \eqref{eq:newtkdef}, a definition of the multipole moments of the body analogous to \eqref{eq:nmdef} motivated through the moment generating function, and a system of equations of motion for the worldline, the total linear momentum of the body, and the spin angular momentum of the body, analogous to equations \eqref{eq:ngz}, \eqref{eq:ngp}, and \eqref{eq:ngs}. We will package those equations of motion into a variational principle analogous to \eqref{eq:nhamact}. 
	
	\section{Preliminaries} \label{sec:prelim}
	
	In this section we introduce the main notation and conventions necessary for working in general relativity, describe how to make a meaningful notion of equal time-slices in curved spacetime, and specify the physical assumptions we must place on the body for the theory to be well-defined. 
	
	\subsection{Basic Notation}
	
	Whenever we refer to a function as arbitrary, we mean an arbitrary smooth function of compact support. We call the spacetime manifold $\bb{T}$ and a point within it $x$. The tangent space of $\bb{T}$ at $x$ is $T_x(\bb{T})$ and it carries a pseudo-inner product denoted by $\cdot$ that provides a pseudo-metric structure which we take with mostly-plus signature. We define $\vec\iota_A(x)$, called a vierbein, as an orthonormal basis of $T_x(\bb{T})$:
	\begin{equation}
		\vec\iota_A\cdot\vec\iota_B = \eta_{AB} = \begin{pmatrix}-1 & 0 & 0 & 0 \\ 0 & 1 & 0 & 0 \\ 0 & 0 & 1 & 0 \\ 0 & 0 & 0 & 1\end{pmatrix}_{AB}. \label{eq:pseudometric}
	\end{equation}
	Capital Latin letter ($A, B, C, \dots $) indices always take the values $\{0,1,2,3\}$ and are always used to refer to these basis vectors or components of tensors expressed in this basis. Minuscule Latin letter ($a, b, c, \dots$) indices always take the values $\{1,2,3\}$ and are always used to refer to these vectors excluding $\vec\iota_0$ or the spacelike components of tensors in this basis. 
	
	We use coordinates $x^\alpha$ so that locally there is an invertible relationship between points in $\bb{T}$ and these coordinates in $\bb{R}^4$. The first five beginning alphabet Greek letter indices ($\alpha, \beta, \gamma, \delta, \epsilon$) are always used for reference to spacetime indices at a general location $x$ and take values $\{0,1,2,3\}$. All middle or later alphabet Greek letter indices ($\iota$, $\kappa, \lambda, \mu, \nu, \rho, \sigma, \tau, \omega$) are always used for reference to spacetime indices at a specific location $z$ of interest that will be specified by context. We use $\vec e_\alpha$ as the coordinate displacement basis vectors so that a generic displacement $dx \in T_x(\bb{T})$ can be written as:
	\begin{equation}
		dx = dx^\alpha \vec e_\alpha. \label{eq:coordinatebasis}
	\end{equation}
	The metric tensor $g_{\alpha\beta}$ and proper time element $d\tau$ are defined as:
	\begin{equation}
		g_{\alpha\beta} = \vec e_\alpha\cdot\vec e_\beta, \p d\tau^2 = -dx\cdot dx = -g_{\alpha\beta} dx^\alpha dx^\beta.
	\end{equation}
	The magnitude $|v|$ of any vector $v^\alpha$ and unit vector $\hat v^\alpha$ in the direction of that vector are defined by:
	\begin{equation}
		|v| = \sqrt{|g_{\alpha\beta} v^\alpha v^\beta|}, \p \hat v^\alpha = \frac{v^\alpha}{|v|}. 
	\end{equation}
	$\hat v^\alpha$ is not defined for lightlike vectors. Note that the magnitude is nonnegative for all vectors (in contrast to the self pseudo-inner product $v^\alpha v_\alpha$) and is 0 if and only if $v^\alpha$ is lightlike. 
	
	The covariant derivatives of the coordinate basis vectors define the connection coefficients, and we always use the Levi-Civita connection:
	\begin{equation}
		\nabla_\beta \vec e_\alpha = {\Gamma^\gamma}_{\alpha\beta} \vec e_\gamma, \p {\Gamma^\gamma}_{\alpha\beta} = \frac{1}{2}g^{\gamma\delta}\(\d_\alpha g_{\delta\beta} + \d_\beta g_{\delta\alpha} - \d_\delta g_{\alpha\beta}\) \label{eq:connectioncoefs}
	\end{equation}
	which is unique in being both metric compatible and torsionless: 
	\begin{equation}
		\nabla_\gamma g_{\alpha\beta} = 0, \p {\Gamma^\gamma}_{\alpha\beta} = {\Gamma^\gamma}_{\beta\alpha}.
	\end{equation}
	
	We will often have to consider tensors of high rank for which the particular details of their index structure is unimportant. When this is the case, we have two ways of condensing notation, depending on the necessary amount of resolution. We either use $\cdots$ on indices or a capital Greek letter to indicate a generic tensorial compound index. That is:
	\begin{equation}
		f^\Omega = f^{\alpha\cdots}{}_{\beta\cdots} = f^{\alpha_1\dots\alpha_m}{}_{\beta_1\dots\beta_n}.
	\end{equation}
	In this notation, we write that under a diffeomorphism $x' = x'(x)$, a tensor field $f^\Omega$ transforms as:
	\begin{equation}
		f'^{\alpha\cdots}{}_{\beta\cdots}(x') = \frac{\d x'^{\alpha}}{\d x^{\gamma}}\cdots\frac{\d x^{\delta}}{\d x'^{\beta}}\cdots f^{\gamma\cdots}{}_{\delta\cdots}(x).	\label{eq:coordchange}
	\end{equation}
	As well, the covariant derivative of a tensor is:
	\begin{equation}
		\nabla_\alpha f^\Omega = \d_\alpha f^\Omega + \Gamma^\gamma{}_{\alpha\beta} \fc{G}^{\beta}{}_\gamma f^{\Omega}
	\end{equation}
	where the DeWitt index shuffling operator $\fc{G}^\alpha{}_\beta$ is the linear operator on tensor fields defined so that:
	\begin{align}
		\fc{G}^\alpha{}_\beta f^{\gamma_1\dots\gamma_m}{}_{\delta_1\dots\delta_n} =& \delta^{\gamma_1}_{\beta} f^{\alpha \gamma_2\dots\gamma_m}{}_{\delta_1\dots\delta_n} + \dots + \delta^{\gamma_m}_{\beta} f^{\gamma_1\dots\gamma_{m-1}\alpha}{}_{\delta_1\dots\delta_n} \nonumber \\
		&\ -\delta^{\alpha}_{\delta_1} f^{\gamma_1\dots\gamma_m}{}_{\beta\delta_2\dots\delta_n} - \dots -\delta^{\alpha}_{\delta_n} f^{\gamma_1\dots\gamma_m}{}_{\delta_1\dots\delta_{n-1}\beta}.
	\end{align}
	
	The extent to which covariant derivatives fail to commute defines the Riemann curvature tensor ${R^\gamma}_{\alpha\delta\beta}$:
	\begin{equation}
		[\nabla_\delta,\nabla_\beta] f^\gamma = {R^\gamma}_{\alpha\delta\beta} f^\alpha
	\end{equation}
	for any vector with components $f^\alpha$. Consequently, in components the Riemann tensor is given by:
	\begin{equation}
		{R^\gamma}_{\alpha\delta\beta} = \d_\delta{\Gamma^\gamma}_{\alpha\beta} - \d_\beta{\Gamma^\gamma}_{\alpha\delta} + {\Gamma^\gamma}_{\epsilon\delta}{\Gamma^\epsilon}_{\alpha\beta} - {\Gamma^\gamma}_{\epsilon\beta}{\Gamma^\epsilon}_{\alpha\delta}. \label{eq:3112}
	\end{equation}
	It is also convenient to define the Ricci tensor $R_{\alpha\beta}$ and Ricci scalar $R$ by:
	\begin{equation}
		R_{\alpha\beta} = {R^\gamma}_{\alpha\gamma\beta}, \p R = R_{\alpha\beta} g^{\alpha\beta}.
	\end{equation}
	
	We use parentheses $(\, )$ for index symmetrization and brackets $[\, ]$ for index antisymmetrization. Symmetry factors are always included in (anti)symmetrization:
	\begin{equation}
		M^{(\alpha\beta)} = \frac{M^{\alpha\beta} + M^{\beta\alpha}}{2}, \p M^{[\alpha\beta]} = \frac{M^{\alpha\beta} - M^{\beta\alpha}}{2}.
	\end{equation}
	(Anti)symmetrization of $n$ indices is defined by $\frac{1}{n!}$ times the sum over all $n!$ permutations of those indices, with a minus sign included for odd permutations in an antisymmetrization. Indices caught in a(n) (anti)symmetrization but not intended to be part of it are blocked off by vertical bars $|$:
	\begin{equation}
		M^{(\alpha|\beta|\gamma)} = \frac{M^{\alpha\beta\gamma} + M^{\gamma\beta\alpha}}{2}, \p M^{[\alpha|\beta|\gamma]} = \frac{M^{\alpha\beta\gamma} - M^{\gamma\beta\alpha}}{2}.
	\end{equation}
	Sometimes it is useful to have symmetrization and antisymmetrization or two separate sets of antisymmetrization simultaneously active. When we do so, the different (anti)symmetrizations simply ignore each other. In the case of two antisymmetrizations, we call one of them $[\, ]$ and the other $['\, ]'$. For example:
	\begin{align}
		M^{(\alpha[\beta \gamma)\delta]} &= \frac{1}{4}(M^{\alpha\beta\gamma\delta} + M^{\gamma\beta\alpha\delta} - M^{\alpha\delta\gamma\beta} - M^{\gamma\delta\alpha\beta}) \\
		M^{[\alpha['\beta\gamma]\delta]'} &= \frac{1}{4}(M^{\alpha\beta\gamma\delta}- M^{\gamma\beta\alpha\delta} - M^{\alpha\delta\gamma\beta} + M^{\gamma\delta\alpha\beta}).
	\end{align}
	
	With our index symmetry notation, from its definition the Riemann tensor automatically has the index symmetries:
	\begin{equation}
		R_{\gamma\alpha\delta\beta} = R_{[\gamma\alpha][\delta\beta]} = R_{\delta\beta\gamma\alpha}.
	\end{equation}
	It also satisfies the algebraic and differential Bianchi identities:
	\begin{equation}
		R_{\gamma\alpha\delta\beta} + R_{\gamma\delta\beta\alpha} + R_{\gamma\beta\alpha\delta} = 0, \p \nabla_\epsilon R_{\gamma\alpha\delta\beta} + \nabla_\gamma R_{\alpha\epsilon\delta\beta} + \nabla_\alpha R_{\epsilon\gamma\delta\beta} = 0. 
	\end{equation}
	The electromagnetic field strength is just antisymmetric and satisfies the Bianchi identity for electromagnetism:
	\begin{equation}
		\fc{F}_{\alpha\beta} = \fc{F}_{[\alpha\beta]}, \p \nabla_\gamma \fc{F}_{\alpha\beta} + \nabla_\alpha \fc{F}_{\beta\gamma} + \nabla_\beta \fc{F}_{\gamma\alpha} = 0. 
	\end{equation}
	
	We use the notation $d\bb{T}$ or $D^4x$ to refer to the invariant spacetime volume measure:
	\begin{equation}
		d\bb{T} = D^4 x = \sqrt{-\det g} d^4 x.
	\end{equation}
	When we consider tensors, such as the Riemann tensor, contracted with some vectors of interest it is useful for keeping equations short to write the contracted vectors as subscripts or superscripts on the tensor. For example:
	\begin{equation}
		R_{uSkv} = R_{\alpha\gamma\beta\delta}u^\alpha S^\gamma k^\beta v^\delta. \label{eq:condensedices}
	\end{equation}
	By convention we define the Levi-Civita symbol $[\alpha\beta\gamma\delta]$ to be fully numerical with $[0123] = 1$ and antisymmetric under a swap of any two indices. Consequently, $[\alpha\beta\gamma\delta]$ does not transform properly as a tensor. Instead, the Levi-Civita pseudotensor ${\epsilon}_{\alpha\beta\gamma\delta}$ is defined by:
	\begin{equation}
		{\epsilon}_{\alpha\beta\gamma\delta} = [\alpha\beta\gamma\delta]\sqrt{-\det g} \implies {\epsilon}{}^{\alpha\beta\gamma\delta} = -\frac{[\alpha\beta\gamma\delta]}{\sqrt{-\det g}}
	\end{equation} 
	which does transform correctly as a tensor under orientation preserving diffeomorphisms. 
	
	The Lie derivative measures the extent to which a tensor field changes as a field when a small diffeomorphism is applied. For a real parameter $\eps$ and smooth vector field $\xi^\alpha(x)$, consider the diffeomorphism:
	\begin{equation}
		x'^\alpha = x^\alpha + \eps \xi^\alpha(x).	\label{eq:infdiff}
	\end{equation}
	The Lie derivative $L_\xi$ of a tensor field $\phi$ under this diffeomorphism is defined by:
	\begin{equation}
		L_\xi \phi^{\alpha\cdots}{}_{\beta\cdots} \df \lim_{\eps\to 0} \frac{\frac{\d x^\alpha}{\d x'^\gamma}\cdots\frac{\d x'^\delta}{\d x^\beta}\cdots \phi^{\gamma\cdots}{}_{\delta\cdots}(x') - \phi^{\alpha\cdots}{}_{\beta\cdots}(x)}{\eps}.
	\end{equation}
	Explicitly expanding \eqref{eq:coordchange} under \eqref{eq:infdiff} for small $\eps$ gives:
	\begin{equation}
		L_\xi \phi^\Omega = \xi^\alpha\d_\alpha \phi^\Omega - \d_\alpha \xi^\beta \fc{G}^\alpha{}_\beta \phi^{\Omega} = \xi^\alpha\nabla_\alpha \phi^\Omega - \nabla_\alpha \xi^\beta \fc{G}^\alpha{}_\beta \phi^{\Omega}	\label{eq:lieder}
	\end{equation}
	where the necessary terms to replace the partial derivatives with covariant derivatives in this exactly cancel between the two pieces. If the Lie derivative of a tensor field vanishes with respect to a particular vector field $\xi^\mu$, that vector field is a symmetry of the tensor field (in the sense that the diffeomorphism generated by that vector field leaves the tensor field unchanged). 
	
	Often we will need to perform surface integrals. Suppose we have a surface $\Sigma$ defined by a constraint $C(x) = 0$ for a smooth scalar field $C$. We perform a diffeomorphism to coordinates $y$ so that $y^0 = C(x)$ and $y^1, y^2, y^3$ provide coordinates on $\Sigma$, where we always order our coordinates so that this diffeomorphism is orientation preserving. Define $\eps$ to be $-1$ if the normal vector to $\Sigma$ is timelike and $+1$ if the normal vector to $\Sigma$ is spacelike. For our purposes, lightlike surfaces will not be necessary. Define the outward unit normal vector $\mc{n}_\alpha(x)$ to $\Sigma$, the directed measure $d\Sigma_\alpha$ on $\Sigma$, and the undirected measure $d\Sigma$ by:
	\begin{equation}
		\mc{n}_\alpha = \eps \frac{\nabla_\alpha C}{|\nabla C|}, \p d\Sigma_\alpha = \epsilon_{\alpha\beta\gamma\delta}\frac{\d x^\beta}{\d y^1}\frac{\d x^\gamma}{\d y^2}\frac{\d x^\delta}{\d y^3} dy^1 dy^2 dy^3 = \eps \mc{n}_\alpha d\Sigma.  \label{eq:volmeas}
	\end{equation}
	Then for a region $\bb{M}$ and vector field $f^\alpha$, Stokes' theorem is:
	\begin{equation}
		\int_{\bb{M}} \nabla_\alpha f^\alpha d\bb{T} = \int_{\Sigma} f^\alpha d\Sigma_\alpha
	\end{equation}
	where $\Sigma = \d\bb{M}$. 
	
	\subsection{Worldline Clock}
	
	In the Newtonian theory, the representative position of the body is given by $\vec z(t)$. In relativity, we package this more covariantly as a smooth future-oriented timelike curve $z^\mu(\lambda): \bb{R} \to \bb{T}$, called the worldline, identifying the representative position of the body at each worldline time $\lambda$. The Newtonian theory also depends heavily on the ability to take integrals over all space at a fixed time. The relativistic theory will need to generalize this concept, and so we need a covariant understanding of an equal time-slice. That is, we are going to separate spacetime $\bb{T}$ into a 1 parameter family of surfaces $\Sigma(\lambda)$ so that each such surface is spacelike and is intended to represent space at fixed time $\lambda$ according to some observer. At each time $\lambda$, at the worldline $z^\mu(\lambda)$, there will be some future-oriented timelike unit vector $u^\mu(\lambda)$ that the observer of interest identifies as the increasing time direction. Thus, for a point which is near the worldline to also lie in $\Sigma$, its displacement from the worldline ought to be orthogonal to $u$ (so that it is displaced from the worldline purely spatially). Continuing along that spatial displacement geodesically then gives a spacelike curve of points which also would naturally be identified as being at the ``same time'' according to the observer, due to the initial purely spatial displacement. Therefore, we assume going forward that we have a pair $(z, u)$ where $z^\mu(\lambda)$ is our timelike worldline and $u^\mu(\lambda)$ is at each time a future-oriented timelike unit vector in $T_{z(\lambda)}(\bb{T})$ and from this pair we define the time-slice $\Sigma(\lambda)$ as:
	\begin{equation}
		\Sigma(\lambda) = \{ x \in \bb{T}\ :\ \text{tangent of geodesic from }z(\lambda)\text{ to }x\text{ is orthogonal to }u(\lambda)\text{ at }z(\lambda)\}.\label{eq:sigdef}
	\end{equation}
	The overwhelming majority of our formalism will be essentially agnostic to what $(z,u)$ we use, but we will eventually specify the natural choices for them in Section \ref{sec:ps}. This shouldn't be too surprising; the Newtonian theory of compact objects is also almost entirely agnostic to which trajectory $\vec z(t)$ is used. Unless otherwise specified, we keep $(z, u)$ generic. 
	
	Each event $x$ in spacetime occurs at precisely one time in any healthy time-slicing. So, there exists a scalar function $\tau(x)$ called the worldline clock that gives the worldline time $\lambda$ at which $x$ occurs according to our observer:
	\begin{equation}
		\tau(x) = \lambda \iff x \in \Sigma(\lambda). 
	\end{equation}
	The worldline clock $\tau(x)$ has an unfortunate slight notational overlap with the proper time element $d\tau$, however we will never encounter any ambiguity because the proper time element will not appear any further in this work. $\tau(x)$ may not be a globally defined function; for our purposes we only actually need it to be defined in the neighborhood of our compact object, which will always be true due to the energy conditions we place on that object.
	
	The worldline time $\lambda$ is just an arbitrary parameter labeling which point along the worldline we are considering. Because of this, we can freely reparameterize the worldline from $\lambda$ to a new parameter $\lambda'(\lambda)$ satisfying $\frac{d\lambda'}{d\lambda} > 0$ (so that they are invertibly related and future-orientedness is preserved). The reparameterized worldline $z'^\mu(\lambda')$ is defined by:
	\begin{equation}
		z'^\mu(\lambda') = z^\mu(\lambda), 
	\end{equation}
	so that $z^\mu(\lambda)$ and $z'^\mu(\lambda')$ trace out the same curve. Consequently, the worldline velocity is not reparameterization invariant. Instead:
	\begin{equation}
		\dot z'^\mu = \dot z^\mu \frac{d\lambda}{d\lambda'}. 
	\end{equation}
	Because of this noninvariance, it is useful to introduce the einbein $\ec(\lambda)$, defined by:
	\begin{equation}
		\ec(\lambda) = -u\cdot \dot z \implies \ec'(\lambda') = \ec(\lambda) \frac{d\lambda}{d\lambda'}. \label{eq:ecdef}
	\end{equation}
	Since $u^\mu$ represents the unit increasing time direction according to our observer, $\ec$ gives the rate at which time passes per unit $\lambda$ according to that observer. We introduce the reparameterization invariant $\lambda$ derivative by:
	\begin{equation}
		\doot z{}^\mu = \frac{\dot z^\mu}{\ec} = \frac{\dot z'^\mu}{\ec'} = \doot z{}'^\mu. \label{eq:dootder}
	\end{equation}
	This is therefore the time derivative of the worldline position according to our observer. It is not the derivative with respect to proper time along the worldline, because $u^\mu$ and $\dot z^\mu$ are not the same. For a general tensor field $f^\Omega$ evaluated at $z(\lambda)$, the covariant $\lambda$ derivative and covariant reparameterization invariant $\lambda$ derivative are defined by:
	\begin{equation}
		\frac{D f^{\Omega}}{D\lambda} = \frac{d f^{\Omega}}{d\lambda} + \Gamma^\nu{}_{\mu\rho} \frac{dz^\rho}{d\lambda} \fc{G}^\mu{}_\nu f^{\Omega}, \p \doot f{}^\Omega = \frac{1}{\ec} \frac{D f{}^\Omega}{D\lambda}. 
	\end{equation}
	
	Using the worldline clock, we can turn an integral of a scalar field $f(x)$ over spacetime into an integral over each time-slice and then over all time. Consider an orientation preserving diffeomorphism to coordinates $y^0 = \tau(x)$, $y^1$, $y^2$, $y^3$ from generic coordinates $x^\alpha$. Let $w^\alpha(x)$ be the vector field:
	\begin{equation}
		w^\alpha(x) = \frac{1}{\ec(\tau(x))} \frac{\d x^\alpha}{\d y^0}.
	\end{equation}
	The exact $w^\alpha$ depends on how the full $y^\alpha$ coordinate system is constructed, but we will only need one property of it which is the same in all choices. In particular, by construction:
	\begin{equation}
		\ec w^\alpha\nabla_\alpha\tau = 1.
	\end{equation}
	That is, if an arbitrary point in $\Sigma(\lambda)$ is displaced by $w\ec\delta\lambda$, then the result is a point in $\Sigma(\lambda+\delta\lambda)$ for sufficiently small $\delta\lambda$. Contracting $\ec w^\alpha dy^0$ into $d\Sigma_\alpha$ in \eqref{eq:volmeas} gives:
	\begin{equation}
		D^4 x = d\Sigma_w D\lambda, \label{eq:volmeasdecomp}
	\end{equation}
	where (as defined in \eqref{eq:condensedices}) $d\Sigma_w = w^\alpha d\Sigma_\alpha$ and we have introduced the reparameterization invariant $\lambda$ integration measure $D\lambda$:
	\begin{equation}
		D\lambda = \ec(\lambda) d\lambda \implies D\lambda' = D\lambda. \label{eq:3213}
	\end{equation}
	Using the definition of $\mc{n}^\alpha$ in \eqref{eq:volmeas}, it also follows that:
	\begin{equation}
		\nabla_\alpha \tau = \frac{\mc{n}_\alpha}{\ec w\cdot \mc{n}}, \p d\Sigma_\alpha = \ec \nabla_\alpha \tau d\Sigma_w. \label{eq:volmeasder}
	\end{equation}
	Because of \eqref{eq:volmeasdecomp}, spacetime volume integrals can be written as:
	\begin{equation}
		\int_\bb{T} f(x) d\bb{T} = \int_{-\8}^\8 \int_{\Sigma(\lambda)} f(x) d\Sigma_w D\lambda, \label{eq:volintdecomp}
	\end{equation}
	explicitly establishing how to do an integral over a time-slice first and then over all time, covariantly. This can be made an integral over a finite interval of time by taking $f(x)$ to vanish outside of that interval. 
	
	We will also need to take time derivatives of spatial integrals. Consider a vector field $f^\alpha$ which is compactly supported in spatial directions. Let $\fc{C}(\lambda_1,\lambda_0)$ be the region in spacetime which is topologically a cylinder and which is infinite in the spatial directions with end caps $\Sigma(\lambda_1)$ and $\Sigma(\lambda_0)$ (where $\lambda_1 > \lambda_0$). Using Stokes' theorem and \eqref{eq:volintdecomp}:
	\begin{align}
		\frac{1}{\ec} \frac{d}{d\lambda}\int_{\Sigma(\lambda)} f^\alpha d\Sigma_\alpha &= \lim_{\delta\lambda\to 0}\frac{1}{\ec\delta\lambda}\int_{\d\fc{C}(\lambda+\delta\lambda,\lambda)} f^\alpha d\d\fc{C}_\alpha = \lim_{\delta\lambda\to 0} \frac{1}{\ec\delta\lambda}\int_{\fc{C}(\lambda+\delta\lambda,\lambda)} \nabla_\alpha f^\alpha d\bb{T}  \\
		&= \lim_{\delta\lambda\to 0}\frac{1}{\ec\delta\lambda}\int_\lambda^{\lambda+\delta\lambda} \int_{\Sigma(\lambda)} \nabla_\alpha f^\alpha d\Sigma_w D\lambda.
	\end{align}
	Completing the limit yields:
	\begin{equation}
		\frac{1}{\ec} \frac{d}{d\lambda}\int_{\Sigma(\lambda)} f^\alpha d\Sigma_\alpha = \int_{\Sigma(\lambda)} \nabla_\alpha f^\alpha d\Sigma_w. \label{eq:volintder}
	\end{equation}
	This formula will be critical as it allows us to relate continuity-type equations to equations of motion for worldline quantities. 
	
	\subsection{Basic Physical Assumptions} \label{sec:bpa}
	
	Let $V^\alpha$ be an arbitrary timelike vector representing the time direction in the frame of an observer at a particular point (the observer considers $V^\alpha$ to point purely in the direction of increasing time, at fixed spatial location). Then, the momentum density 4-vector $\wp^\alpha(V)$ of the body at the observer in the observer's frame and the physical mass density $\rho(V)$ measured by that observer are:
	\begin{equation}
		\wp^\alpha(V) = -T^{\alpha\beta} V_\beta, \p \rho(V) = \sqrt{-\wp_\alpha \wp^\alpha}.
	\end{equation}
	In order for this physical mass density to be real, $\wp_\alpha$ must be timelike or null. We require that the mass density of the body be nonzero anywhere that $\wp^\alpha$ is nonzero, so that the body's constituents are always moving slower than light. Thus:
	\begin{equation}
		T^{\alpha\beta} V_\beta\text{ is timelike wherever }T^{\alpha\beta} \neq 0\text{ for all }V^\alpha\text{ which are timelike.} \label{eq:timelike}
	\end{equation}
	The energy density of the body according to the observer is $T^{\alpha\beta} V_{\alpha} V_{\beta}$. We require that the body nowhere has negative energy density and at least somewhere has positive energy density. Therefore, we require that:
	\begin{equation}
		\text{For all }V^\alpha\text{ which are timelike, }T^{\alpha\beta} V_{\alpha} V_{\beta} \ge 0 \text{ and at least somewhere this is positive.} \label{eq:posenergy}
	\end{equation}
	
	Let the convex hull of the region in spacetime in which $T^{\alpha\beta}$ or $J^\alpha$ is nonzero be called the worldtube $\bb{W}$. In order for $T^{\alpha\beta}$, $J^\alpha$ to be said to meaningfully describe a body, we require that $\bb{W}$ is spatially compact (the intersection of $\bb{W}$ with any spacelike surface is compact), i.e. at any fixed time the spatial extents of $T^{\alpha\beta}$ and $J^\alpha$ are finite. Further, we require that $\bb{W}$ is spatially equibounded, meaning that there exists a constant $\ell$ such that the geodesic distance between any two spacelike separated points in $\bb{W}$ is always less than or equal to $\ell$. Conceptually this just requires that at any fixed time the distance between any points within the body is bounded and that there is a bound which is valid for all times; there is a size beyond which the body never expands. While the worldtube is required to be spatially finite, it will necessarily be infinite in extent in the time direction (due to the global conservation of charge and momentum). We require $T^{\alpha\beta}$ and $J^\alpha$ to be smooth (infinitely continuously differentiable). 
	
	There are further conditions we must place on how the worldtube relates to the spacetime in which the body moves. We assume that the spacetime is topologically like Minkowski space. Let $\Sigma$ be a spacelike surface and consider the intersection $\Sigma \cap \bb{W}$. We require that the size and mass of the body are small enough so that there is always precisely one geodesic connecting any two points within $\Sigma \cap \bb{W}$. This means that we require the spacetime to not be too curved on the scale of the body (i.e. the gravitational field to not be overwhelmingly strong). If this is satisfied, the ``self-force'' effects of the field produced by the body back on the body itself can be treated within the MPD formalism. We similarly require the electromagnetic field to not be overwhelmingly strong. In particular, we will show that the electromagnetic field makes contributions to the body's total linear momentum, and we require that these contributions keep the total linear momentum timelike.
	
	Because the worldtube at any single time in any observer's frame will generally have nonzero spatial size, there is no unique choice for the position of the body without providing some additional specifying condition. Similarly, there is no a priori unique choice for the instantaneous time direction of the body. Almost all of our formalism will be agnostic to our choices for these, so long as the position is within the worldtube. However, ultimately we will develop a unique method of selecting a point at each time to act, in an appropriate sense, as a best representative for the body's position analogous to the Newtonian concept of a center of mass and a unique definition of the instantaneous time direction to provide a definition for integrals over a spatial slice. In order for these to be well-defined and unique, we will have to place some specific technical conditions on the electromagnetic and gravitational fields which again amount to requiring them to not be overwhelming. These specific demands are made at the ends of Appendices~\ref{sec:uint} and \ref{sec:sscint}. 
	
	\section{Theory of Bitensors} \label{sec:bi}
	
	In building the Newtonian theory of compact objects, we used several mathematical tools which are not immediately valid in curved spacetime. While we have now established how to covariantly define equal time-slices and perform spatial integrals, in the Newtonian theory we also had to take the difference of finitely separated coordinates and integrate tensor fields over different points. These constructions (naively carried over) are geometrically meaningless in a curved spacetime. In order to generalize them, we first need to introduce the basics of the theory of bitensors. A bitensor is a mathematical object that is a function of two different spacetime points (call them $x$ and $z$) and may carry tensor indices for each point. Bitensors are critical to the development of the theory because they will allow us to convert tensor indices at a generic point to tensor indices on the worldline. Once all indices refer to the same tangent space, it becomes covariant to take differences or do integrals. The theory of bitensors has many applications beyond the MPD theory, see~\cite{Poisson:2011nh}. 
	
	\subsection{Synge's Worldfunction}
	
	Synge's worldfunction is the key to the entire theory of bitensors. Let $\zeta_1^\alpha(s)$ be a smooth curve in $\bb{T}$ with $\zeta^\mu_1(0) = z^\mu$ and $\zeta^\alpha_1(1) = x^\alpha$ and consider the functional $\sigma_1[\zeta_1]$:
	\begin{equation}
		\sigma_1[\zeta_1] = \int_0^1 \frac{1}{2} g_{\alpha\beta}(\zeta_1(s)) \frac{d\zeta_1^\alpha}{ds} \frac{d\zeta_1^\beta}{ds} ds.
	\end{equation}
	Under a variation of the path $\zeta^\alpha_1$ at fixed metric, we find:
	\begin{equation}
		\delta\sigma_1 = \frac{d\zeta_{1\alpha}}{ds}(1)\delta x^\alpha - \frac{d\zeta_{1\mu}}{ds}(0)\delta z^\mu - \int_0^1 \delta \zeta_{1\alpha}(s) \frac{D}{Ds}\(\frac{d\zeta_1^\alpha}{ds}\) ds. \label{eq:deltasig1}
	\end{equation}
	Let $\zeta^\alpha(s)$ be the specific choice of $\zeta_1^\alpha$ which is a geodesic with $s$ an affine parameter. That is, $\zeta^\alpha(s)$ satisfies the geodesic equation:
	\begin{equation}
		\frac{D}{Ds}\(\frac{d \zeta^\gamma}{ds}\) = \frac{d^2 \zeta^\gamma}{ds^2} + \Gamma^\gamma{}_{\alpha\beta} \frac{d\zeta^\alpha}{ds}\frac{d\zeta^\beta}{ds} = 0
	\end{equation}
	with the same boundary conditions as $\zeta_1$. We will use this \textit{standard geodesic} $\zeta^\alpha$ from $z^\mu$ to $x^\alpha$ repeatedly. Define Synge's worldfunction as the function of the end points resulting from evaluating $\sigma_1$ on the geodesic path $\zeta^\alpha(s)$:
	\begin{equation}
		\sigma(x, z) = \int_0^1 \frac{1}{2} g_{\alpha\beta}(\zeta) \frac{d\zeta^\alpha}{ds} \frac{d\zeta^\beta}{ds} ds. \label{eq:syngdef}
	\end{equation}
	Automatically by equation \eqref{eq:deltasig1} under variations in $x$ and $z$ (at fixed metric):
	\begin{equation}
		\delta\sigma = \frac{d\zeta_\alpha}{ds} \delta x^\alpha - \frac{d\zeta_\mu}{ds} \delta z^\mu.
	\end{equation}
	We place indices on $\sigma$ to indicate covariant differentiation. Beginning Greek alphabet indices indicate covariant differentiation with respect to $x$ while middle Greek alphabet indices indicate covariant differentiation with respect to $z$ so that:
	\begin{equation}
		\sigma_\alpha = \frac{\d \sigma}{\d x^\alpha} = \frac{d\zeta_\alpha}{ds}, \p \sigma_\mu = \frac{\d \sigma}{\d z^\mu} = -\frac{d\zeta_\mu}{ds}. \label{eq:syngder}
	\end{equation}
	Because covariant derivatives at the same point do not commute, the order in which they are placed matters if we have more than 2 indices of the same type. Indices are intended in reverse order so that:
	\begin{equation}
		\sigma_{\alpha\beta\gamma} = \nabla_\gamma\nabla_\beta\nabla_\alpha \sigma.
	\end{equation}
	The function $\sigma(x,z)$ is a biscalar and its covariant derivatives are bitensors. The indices on bitensors must be raised and lowered using the metric at the appropriate point corresponding to that index. We will often need the matrix inverse of $\sigma^\alpha{}_\mu$ which we write by putting a $-1$ or $-$ sign as a superscript:
	\begin{equation}
		\sigma^{-\mu}{}_\alpha \sigma^{\alpha}{}_\nu = \delta^\mu_\nu
	\end{equation}
	
	It follows from the geodesic equation that:
	\begin{equation}
		\frac{d}{ds}\(g_{\alpha\beta}(\zeta) \frac{d\zeta^\alpha}{ds} \frac{d\zeta^\beta}{ds}\) = 0. \label{eq:arcconst}
	\end{equation}
	Consider the distance element $\sqrt{|g_{\alpha\beta} d\zeta^\alpha d\zeta^\beta|}$ along the geodesic. By \eqref{eq:arcconst} that distance element is given by $S ds$ for some constant $S$. Integrating over the total path from $s=0$ to $s=1$ gives that $S$ is in fact the magnitude integrated geodesic distance from $z$ to $x$. Because the integrand of \eqref{eq:syngdef} is constant by \eqref{eq:arcconst}, we immediately have:
	\begin{equation}
		\eps S^2 = 2\sigma = \sigma^\alpha\sigma_\alpha = \sigma^\mu \sigma_\mu \label{eq:causalchar}
	\end{equation}
	where $\eps = 1$ if the geodesic between $x$ and $z$ is spacelike, $\eps = -1$ if it is timelike , and $\eps = 0$ if it is lightlike. Thus, the causal character (spacelikeness, timelikeness, or lightlikeness) of the geodesic connecting $x$ and $z$ is matched by the sign of $\sigma$ and by the causal characters of both $\sigma_\mu$ and $\sigma_\alpha$. Taking $x$ and/or $z$ derivatives of \eqref{eq:causalchar} gives useful identities which allow for common combinations involving the worldfunction to be simplified. For example, $\sigma^\mu{}_\alpha\sigma^\alpha = \sigma^\mu$. 
	
	\subsection{The Exponential Map}
	
	Let $\zeta(s, z, v)$ be a geodesic with affine parameter $s$ so that $\zeta(0, z, v) = z$ and $\frac{d\zeta}{ds}(0, z, v) = v$. Then, the exponential map is defined by the solution to the geodesic equation with these initial conditions:
	\begin{equation}
		\exp_z(v) = \zeta(1, z, v).
	\end{equation}
	$v$ is a tangent vector to the spacetime manifold at $z$, and so specifies a point in the tangent space $T_z(\bb{T})$. The ordered pair $(z,v)$ of a point $z$ in $\bb{T}$ together with a point in the corresponding tangent space $T_z(\bb{T})$ is just a generic point in the tangent bundle $T(\bb{T})$ of the spacetime. The exponential map is thus a map from the tangent bundle into the spacetime manifold. The ability to evaluate the exponential map $\exp_z(v)$ for a given $(z,v)$ relies on the existence and uniqueness of a geodesic $\zeta(s,z,v)$ as described, and so can only be ensured for sufficiently small $v$ in a curved spacetime. 
	
	The domain of the exponential map is not the full tangent bundle but a subset of it. Let $\fc{V}$ be an open set in the tangent bundle on which the exponential map is defined and which contains all points of the form $(z,0)$. We will assume we have chosen this $\fc{V}$ to be as large as needed, so that whenever we consider $\exp_z(v)$ going forward, it will always be with $(z, v) \in \fc{V}$. Let $\fc{V}_z$ be defined by:
	\begin{equation}
		\fc{V}_z = \fc{V}\cap T_z(\bb{T}).
	\end{equation}
	Because $\fc{V} \subset T(\bb{T})$ is the set of $(z,v)$ on which the exponential map is defined, this set $\fc{V}_z$ is the set of $(z,v)$ on which the exponential map is defined \textit{for this particular value of }$z$. Let:
	\begin{equation}
		\fc{U}_z = \exp(\fc{V}_z). 
	\end{equation}
	That is, $\fc{U}_z$ is the set of $x \in \bb{T}$ for a given $z \in \bb{T}$ so that there exists a $v \in T_z(\bb{T})$ satisfying $x = \exp_z(v)$. $\fc{U}_z$ is the set of points in spacetime reachable by the exponential map from $z$. For $x \in \fc{U}_z$ we define the function $v^\mu(x,z)$ as the tangent vector at $z^\mu$ of the geodesic which connects $z^\mu$ to $x^\mu$ in unit affine time. In other words, $v(x,z)$ satisfies:
	\begin{equation}
		\exp_z(v(x,z)) = x.
	\end{equation}
	By \eqref{eq:syngder} this implies:
	\begin{equation}
		v^\mu = -\sigma^\mu(x, z). \label{eq:vdef}
	\end{equation}
	In this way, the derivative of Synge's worldfunction provides the logarithm of the exponential map. 
	
	The exponential map allows us to view the final point $x$ as a function of $z$ and $v$, and so it is natural to compute how $x$ depends on each of these via differentiation. The result will be repeatedly useful.
	In order to present the result covariantly, we first need to introduce the idea of a covariant variation. Because $z^\mu$ is given by spacetime coordinates, the variation $\delta z^\mu$ is a proper vector at $z$. When we have a tensor field $f^\Omega$ at $z$ (such as $v^\mu$) and both the field and the base point are varied, one has to be careful in defining the variation to actually have the result be a tensor. The proper tensor $f$ (expressed using the coordinate basis vectors $\vec e_\alpha(x)$ from \eqref{eq:coordinatebasis}) is $f^{\mu\cdots}{}_{\nu\cdots} \vec e_\mu(z)\cdots \otimes \vec e^\nu(z) \cdots$. If the base point is taken to a new one $z'^\mu = z^\mu + \eps \delta z^\mu$ and the tensorial components to new components $f'^\Omega(z') = f^\Omega(z) + \eps \delta (f^\Omega(z))$, then $\delta (f^\Omega(z))$ is not an actual tensor. We need to parallel transport the varied $f'$ back to the original evaluation point $z$ in order to form a geometrically meaningful difference quotient between objects formed from the same tangent space. So, the covariant variation $\Delta f^\Omega$ is defined as:
	\begin{equation}
		\Delta f^{\mu\cdots}{}_{\nu\cdots}\vec e_\mu\cdots\otimes\vec e^\nu\cdots = \lim_{\eps\to 0} \frac{\text{PT}_{z'\to z}(f'^{\mu\cdots}{}_{\nu\cdots} \vec e_\mu(z')\cdots\otimes\vec e^\nu(z')\cdots) - f^{\mu\cdots}{}_{\nu\cdots}\vec e_\mu(z)\cdots\otimes\vec e^\nu(z)\cdots}{\eps} \label{eq:covaridef}
	\end{equation}
	where $\text{PT}_{z'\to z}(f')$ parallel transports the tensor $f'$ from $z'$ to $z$. Expanding out this definition (using the original definition of the connection coefficients in \eqref{eq:connectioncoefs}) gives that the covariant variation of a tensor when its base point is changed is explicitly:
	\begin{equation}
		\Delta f^\Omega = \delta (f^\Omega(z)) + \Gamma^\mu{}_{\nu\rho} \delta z^\rho \fc{G}^\nu{}_{\mu} f^\Omega. 
	\end{equation}
	Using this as we vary \eqref{eq:vdef} (acknowledging that $v^\mu$ is a vector at $z$) and solving that variation for $\delta x^\alpha$, we find: 
	\begin{equation}
		\delta x^\alpha = K^\alpha{}_\mu \delta z^\mu + H^\alpha{}_\mu \Delta v^\mu \label{eq:desicvari}
	\end{equation}
	where we have introduced the two bitensor propagators:
	\begin{equation}
		K^\alpha{}_\mu(x,z) = -\sigma^{-\alpha}{}_\nu \sigma^\nu{}_\mu, \p H^\alpha{}_\mu(x,z) = -\sigma^{-\alpha}{}_\mu. 
	\end{equation}
	$K^\alpha{}_\mu$ is the derivative of $x^\alpha$ with respect to $z^\mu$ at fixed $v$ while $H^\alpha{}_\mu$ is the derivative of $x^\alpha$ with respect to $v^\mu$ at fixed $z$. These bitensor propagators will play a central role in the theory. 
	
	\subsection{Geodesic Deviation}
	
	The bitensor propagators $K^\alpha{}_\mu$ and $H^\alpha{}_\mu$ have clear interpretations in terms of a familiar equation from general relativity, the geodesic deviation equation. Consider a one parameter family of geodesic curves $\zeta^\alpha(s, r)$ where $s$ is an affine parameter labeling points on a single geodesic and $r$ labels which geodesic is considered. Let $V^\alpha(x)$ and $W^\alpha(x)$ be smooth vector fields so that:
	\begin{equation}
		V^\alpha(\zeta(s,r)) = \frac{\d \zeta^\alpha}{\d s}(s,r) = \dot{\zeta}^\alpha, \p W^\alpha(\zeta(s, r)) = \frac{\d \zeta^\alpha}{\d r}(s, r). 
	\end{equation}
	Here we have introduced the notation that when considering geodesics, a dot indicates differentiation with respect to the affine parameter. The vector field $W^\alpha(x)$ is called a geodesic deviation field because it measures the displacement between points of nearby geodesics. Automatically by the equality of mixed partials:
	\begin{equation}
		\frac{D W^\alpha}{Ds} = \frac{D V^\alpha}{D r}. \label{eq:mixedpartial}
	\end{equation}
	Using this and taking the second covariant $s$-derivative of $W^\alpha$, we find after a short calculation that $W^\alpha$ satisfies the geodesic deviation equation:
	\begin{equation}
		\frac{D^2 W^\gamma}{Ds^2} + R^\gamma{}_{\alpha\delta\beta}\dot \zeta^\alpha \dot\zeta^\beta W^\delta = 0.
	\end{equation}
	
	We can solve the geodesic deviation equation using the worldfunction. Consider a one parameter of family of geodesics $\zeta^\alpha(s, r)$ so that $\zeta^\mu(0, r) = z^\mu(r)$ and $\zeta^\alpha(1,r) = x^\alpha(r)$. From these we can define $v(r)$ by:
	\begin{equation}
		v^\mu(r) = -\sigma^\mu(x(r), z(r)). 
	\end{equation}
	We can then invert this using the exponential map to alternatively think of $x$ as a function of $z,v$. If we take the covariant $r$ derivative of $x^\alpha(r)$ viewing it as a function of $z,v$ then we have by \eqref{eq:desicvari}:
	\begin{equation}
		\frac{dx^\alpha}{dr} = K^\alpha{}_\mu \frac{dz^\mu}{dr} + H^\alpha{}_\mu \frac{D v^\mu}{Dr}. 
	\end{equation}
	Rewritten in terms of $W^\alpha$ and using \eqref{eq:mixedpartial}, we have that the explicit solution to the geodesic deviation equation is:
	\begin{equation}
		W^\alpha(\zeta(1)) = K^\alpha{}_\mu W^\mu(\zeta(0)) + H^\alpha{}_\mu \frac{DW^\mu}{Ds}(\zeta(0)).  \label{eq:desicdevi}
	\end{equation}
	Thus, $K^\alpha{}_\mu$ measures the rate at which the end point of the geodesic is moved when its initial point is changed while keeping its initial tangent fixed and $H^\alpha{}_\mu$ measures the rate at which the end point is moved when its initial tangent is varied while its initial point is held fixed.
	
	\subsection{Coincidence Results}
	
	For various calculations, we will need the behavior of the worldfunction and its derivatives in the ``coincidence'' limit, meaning the limit in which $x \to z$. For a bitensor $\phi^\Omega(x,z)$, we denote the coincidence limit by the double angle brackets $\mlb \mrb$:
	\begin{equation}
		\mlb \phi^\Omega \mrb = \lim_{x\to z} \phi^\Omega(x,z). 
	\end{equation}
	We compute coincidence results by perturbatively solving the geodesic equation for $\zeta^\alpha(s)$ in powers of the the affine time $s$. The result through order $s^3$ is:
	\begin{equation}
		\zeta^\alpha(s) = z^\alpha + s v^\alpha(0) - \frac{s^2}{2}\Gamma^\alpha{}_{\beta\gamma}(z)v^\beta v^\gamma + \frac{s^3}{6}(2\Gamma^\alpha{}_{\beta\epsilon}\Gamma^\epsilon{}_{\gamma\delta} - \d_\delta \Gamma^\alpha{}_{\beta\gamma})(z) v^\beta v^\gamma v^\delta + \OO(s^4 v^4). \\
	\end{equation}
	$x^\alpha = \zeta^\alpha(1)$ then gives a perturbative relation between $z, x, v$. This perturbative relation can be inverted order by order to give $v$ (equivalently, $-\sigma^\mu$) as a function of $x$ and $z$ up to corrections that are $\OO(x-z)^4$. From this $\sigma^\mu$ and from $\sigma^\alpha = \dot \zeta^\alpha(1)$, each expressed in terms of $x$ and $z$, we can differentiate and take coincidence limits to yield:
	\begin{align}
		&\mlb \sigma \mrb = 0,&		 &\mlb \sigma^\alpha \mrb = 0,& 	&\mlb \sigma^\mu \mrb = 0,&	\nonumber \\
		&\mlb \sigma^\alpha{}_\beta \mrb = \delta^\alpha_\beta,&		&\mlb \sigma^\alpha{}_\mu \mrb = -\delta^\alpha_\mu,&		
		&\mlb \sigma^\mu{}_\nu \mrb = \delta^\mu_\nu,& 	 \nonumber \\
		&\mlb \sigma^\alpha{}_{\beta\gamma}\mrb = 0,& 		&\mlb \sigma^\alpha{}_{\beta\mu} \mrb = 0,&		&\mlb \sigma^\alpha{}_{\mu\nu} \mrb = 0,& \nonumber \\
		&\mlb \sigma^\rho{}_{\mu\nu} \mrb = 0&			&\mlb \sigma_{\alpha\beta\gamma\delta}\mrb = \frac{2}{3} R_{\alpha(\gamma\delta)\beta},&		&\mlb \sigma_{\mu\alpha\beta\gamma}\mrb = -\frac{2}{3} R_{\mu(\alpha\beta)\gamma},&		\nonumber \\
		&\mlb \sigma_{\mu\nu\alpha\beta}\mrb = \frac{2}{3} R_{\mu(\alpha\beta)\nu},&		&\mlb \sigma_{\mu\nu\rho\alpha}\mrb = -\frac{2}{3} R_{\alpha(\mu\nu)\rho},&			&\mlb \sigma_{\mu\nu\rho\sigma}\mrb = \frac{2}{3} R_{\mu(\rho\sigma)\nu}. \label{eq:coincide}&
	\end{align}
	
	\subsection{From Bitensors to Tangent Tensors and back}
	
	There is a natural way to turn bitensors at $(x,z)$ into tensors solely at $z$ that depend on $(z,v)$. At fixed $z$, the relationship between $z,v,x$ can be viewed as defining either $v$ as a function of $x$ (through $-\sigma^\mu$) or $x$ as a function of $v$ (through the exponential map), and so is a diffeomorphism between $x$ and $v$ as coordinate systems. Viewed as a diffeomorphism, the $v$ coordinates are exactly ``Riemann normal coordinates'' based at $z$. The coordinate transformation matrix is:
	\begin{equation}
		\frac{\d v^\mu}{\d x^\alpha} = -\sigma^\mu{}_\alpha = H^{-\mu}{}_\alpha(x,z), \p \frac{\d x^\alpha}{\d v^\mu} = -\sigma^{-\alpha}{}_{\mu} = H^\alpha{}_\mu(x,z). \label{eq:vxjacobi}
	\end{equation}
	
	Let $\phi^{\alpha\cdots}{}_{\beta\cdots}{}^\Omega(x,z)$ be a bitensor with $\Omega$ representing generic indices at $z$. We can ``change coordinates'' to the $v$ system from the $x$ system in the usual way:
	\begin{equation}
		\bar{\phi}^{\mu\cdots}{}_{\nu\cdots}{}^\Omega(z,v) = H^{-\mu}{}_{\alpha}\cdots H^{\beta}{}_{\nu}\cdots \phi^{\alpha\cdots}{}_{\beta\cdots}{}^\Omega(x,z) \label{eq:tangentize}
	\end{equation}
	where everywhere in this equation $v$ and $x$ are related by $v^\mu = -\sigma^\mu(x,z)$ or equivalently $x = \exp_z(v)$. There are two very different ways we can think about the quantity $\bar{\phi}^{\cdots}$. The first is to think of $\bar{\phi}^{\cdots}$ still just as a bitensor but now expressed in the $v$ coordinate system (rather than the $x$ system). We call this the Normal picture (since the $v$ coordinates are Riemann normal coordinates based at $z$). If you are thinking this way completely, then $\bar{\phi}^{\cdots}$ still carries two different species of indices ($z$-type and $v$-type). The $z$-type indices would then be properly raised and lowered using the metric at $z$, $g_{\mu\nu}(z)$, while the $v$-type indices would similarly be properly raised and lowered using the metric at $v$ expressed in the $v$-coordinate system, which is given by:
	\begin{equation}
		G_{\mu\nu}(z,v) = H^{\alpha}{}_\mu H^\beta{}_\nu g_{\alpha\beta}(x) \label{eq:tangentizemet}
	\end{equation}
	using the standard change of coordinate rule. In the Normal picture, all of the challenges of differentiating and integrating tensors fields at different points remain unchanged. So, we will almost entirely avoid this way of thinking.
	
	Instead, we will use a second perspective on $\bar{\phi}^{\cdots}$ that we call the Tangent picture. From $v^\mu = -\sigma^\mu$, we see that the $v$-type index is really \textit{also} a $z$-type index (with $v^\mu$ viewed as a vector in the tangent space at $z$), and so $\bar{\phi}^{\cdots}$ can be thought of as now only carrying tensor indices at $z$. $\bar{\phi}^{\cdots}$ is still not a tensor field, because it is a function on the tangent bundle $T(\bb{T})$ (i.e. of $(z,v)$). Because of this, we call $\bar{\phi}^{\cdots}$ a \textit{tangent tensor}. Specifically we say $\bar{\phi}^{\cdots}$ is the tangent tensor associated to $\phi^{\cdots}$ or the \textit{tangentization} of $\phi^{\cdots}$. We will in general write a tangent tensor associated to a bitensor using the same symbol as the bitensor with a bar over it. We do not use the notation $\bar{g}_{\mu\nu}$ as a special case solely for the metric to avoid it looking too similar to the actual metric at $z$, $g_{\mu\nu}(z)$. (For the tangentized metric, we use $G_{\mu\nu}$, as in \eqref{eq:tangentizemet}.)
	
	The subtlety of the Tangent picture is that raising and lowering the $x$ indices does not commute with tangentization, because now that we think of all of the indices as $z$-type, they must all be raised and lowered using $g_{\mu\nu}(z)$, not $G_{\mu\nu}(z,v)$. Thus, for any given quantity that becomes tangentized we will always explicitly write where its indices are considered to be naturally placed. While this is a little cumbersome, the advantages of working this way greatly outweigh the costs. In particular, because $\bar{\phi}^{\cdots}$ now has all indices based at $z$ regardless of the value of $v$, partial derivatives with respect to $v$ as well as integrals with respect to $v$ are covariantly meaningful. 
	
	We will stick to the Tangent picture as much as possible. However, in a very small number of places we will have to pay some care to whether a tangent tensor index originated from a $z$-type bitensor index (and so is unchanged by tangentization) or from an $x$-type bitensor index (and so is transformed by tangentization). The most basic case where this is important is in trying to invert back to a bitensor from a tangent tensor. In particular:
	\begin{equation}
		\phi^{\alpha\cdots}{}_{\beta\cdots}{}^\Omega(x,z) = H^{\alpha}{}_{\mu}\cdots H^{-\nu}{}_{\beta}\cdots  \bar{\phi}^{\mu\cdots}{}_{\nu\cdots}{}^\Omega(z,v)
	\end{equation}
	where only the natively $x$-type indices are transformed.
	
	\subsection{Differentiation of Bitensors}
	
	There are two related pairs of notions of differentiation on bitensors and tangent tensors. Because bitensors are naturally functions of $x$ and $z$ carrying indices at each, the natural derivatives on bitensors are covariant differentiation $\nabla_\alpha$ with respect to $x$ at fixed $z$ and covariant differentiation $\nabla_\mu$ with respect to $z$ at fixed $x$. Alternatively, tangent tensors are naturally functions of $z$ and $v$ and in the Tangent picture carry only $z$-type indices. $v$ just specifies a point in the tangent space $T_z(\bb{T})$ at fixed $z$, which is just a flat vector space. So, it is natural to define partial differentiation $\bar{\d}_\mu$ with respect to $v$ at fixed $z$ (because on vector spaces partial derivatives are geometrically meaningful) and still to define covariant differentiation $\bar{\nabla}_\mu$ with respect to $z$ at fixed $v$. 
	
	By using the exponential map to perform the coordinate transformation between $x$ and $v$, we can directly relate the derivatives of bitensors and derivatives of tangent tensors. By \eqref{eq:desicvari}, we have:
	\begin{equation}
		\bar{\d}_\rho\bar{\phi}^{\mu\cdots}{}_{\nu\cdots}{}^\Omega = H^\gamma{}_\rho \nabla_\gamma (H^{-\mu}{}_{\alpha}\cdots H^{\beta}{}_{\nu}\cdots \phi^{\alpha\cdots}{}_{\beta\cdots}{}^\Omega).
	\end{equation}
	This equation works because in the Tangent picture, $\bar{\phi}^{\mu\cdots}{}_{\nu\cdots}{}^\Omega$ is a scalar field with respect to the $v$ coordinates (since all of its indices are at $z$) and so its ``covariant'' $v$ derivative (if one were thinking in the Normal picture) is just its partial derivative. When converted to the $x$-covariant derivative by a standard change of coordinates, we have the right-hand side. Similarly, by using the chain rule and \eqref{eq:desicvari}:
	\begin{equation}
		\bar{\nabla}_\rho \bar{\phi}^{\mu\cdots}{}_{\nu\cdots}{}^\Omega = \nabla_\rho (H^{-\mu}{}_{\alpha}\cdots H^{\beta}{}_{\nu}\cdots \phi^{\alpha\cdots}{}_{\beta\cdots}{}^\Omega) + K^\gamma{}_\rho\nabla_\gamma(H^{-\mu}{}_{\alpha}\cdots H^{\beta}{}_{\nu}\cdots \phi^{\alpha\cdots}{}_{\beta\cdots}{}^\Omega).
	\end{equation}
	
	The Lie derivative can be extended directly to bitensors in a simple way. Under a diffeomorphism, both $x$ and $z$ are transformed to $x'$ and $z'$ and so a bitensor field $\phi^{\alpha\cdots}{}_{\beta\cdots}{}^{\mu\cdots}{}_{\nu\cdots}(x,z)$ transforms as:
	\begin{equation}
		\phi'^{\alpha\cdots}{}_{\beta\cdots}{}^{\mu\cdots}{}_{\nu\cdots}(x',z') = \frac{\d x'^\alpha}{\d x^{\gamma}}\cdots \frac{\d x^\delta}{\d x'^\beta}\cdots\frac{\d z'^\mu}{\d z^\rho}\cdots\frac{\d z^\sigma}{\d z'^\nu}\cdots \phi^{\gamma\cdots}{}_{\delta\cdots}{}^{\rho\cdots}{}_{\sigma\cdots}(x,z).
	\end{equation}
	As always, the Lie derivative measures the extent to which a bitensor field changes as a field when a small diffeomorphism is applied. So, under the diffeomorphism \eqref{eq:infdiff}, the Lie derivative $L_\xi$ of a bitensor field is defined by:
	\begin{equation}
		L_\xi \phi^{\alpha\cdots}{}_{\beta\cdots}{}^{\mu\cdots}{}_{\nu\cdots} \df \lim_{\eps\to 0} \frac{\frac{\d x^\alpha}{\d x'^{\gamma}}\cdots \frac{\d x'^\delta}{\d x^\beta}\cdots\frac{\d z^\mu}{\d z'^\rho}\cdots\frac{\d z'^\sigma}{\d z^\nu}\cdots\phi^{\gamma\cdots}{}_{\delta\cdots}{}^{\rho\cdots}{}_{\sigma\cdots}(x',z') -  \phi^{\alpha\cdots}{}_{\beta\cdots}{}^{\mu\cdots}{}_{\nu\cdots}(x,z)}{\eps}. 
	\end{equation}
	Explicitly computing, the Lie derivative is:
	\begin{equation}
		L_\xi \phi^\Omega = \xi^\alpha\nabla_\alpha \phi^\Omega - \nabla_\alpha \xi^\beta \fc{G}^\alpha{}_\beta \phi^{\Omega} + \xi^\mu \nabla_\mu \phi^\Omega - \nabla_\mu \xi^\nu \fc{G}^\mu{}_\nu \phi^{\Omega}  \label{eq:bilieder}
	\end{equation}	
	where, when acting on bitensors, DeWitt's index shuffling operator with beginning Greek indices is defined to only shuffle indices at $x$ and with middle Greek indices is defined to only shuffle indices at $z$. Because the Lie derivative of a bitensor is a bitensor of the same rank, the natural definition of how the Lie derivative translates to tangent tensors is to simply transform its indices as usual. That is, by definition:
	\begin{equation}
		L_{\bar{\xi}} \bar{\phi}^{\rho\cdots}{}_{\sigma\cdots}{}^{\mu\cdots}{}_{\nu\cdots} = H^{-\rho}{}_{\alpha}\cdots H^{\beta}{}_{\sigma}\cdots L_\xi \phi^{\alpha\cdots}{}_{\beta\cdots}{}^{\mu\cdots}{}_{\nu\cdots}. \label{eq:tanlieder}
	\end{equation}
	With some labor, this can be turned into:
	\begin{equation}
		L_{\bar{\xi}}\bar{\phi}^\Omega = \bar{\xi}^\mu\bar{\d}_\mu \bar{\phi}^\Omega - \bar{\d}_\mu \bar{\xi}^\nu \bar{\fc{G}}^\mu{}_\nu\bar{\phi}^\Omega + \xi^\mu \bar{\nabla}_\mu \bar{\phi}^\Omega - \sigma^\mu{}_\nu\xi^\nu \bar{\d}_\mu\bar{\phi}^\Omega + \xi^\rho\bar{\d}_\mu \sigma^\nu{}_\rho \bar{\fc{G}}^\mu{}_\nu \bar{\phi}^\Omega - \nabla_\mu \xi^\nu \hat{\fc{G}}^\mu{}_\nu \bar{\phi}^\Omega. \label{eq:evillieder}
	\end{equation}
	where $\ghv{}^\mu{}_\nu$ performs index shuffling only on indices that were tangentized from $x$ indices and $\ghz{}^\mu{}_\nu$ performs index shuffling only on native $z$ indices. This result is presently quite uninsightful, but we will find that for a particular class of $\xi$ of interest it will simplify significantly into a more useful form.
	
	\subsection{Tensor Extensions}
	
	Tangent tensors allow for a generalization of the Taylor series expansion to curved spacetime. We can think of a tensor field as the special case of a bitensor which is just a constant scalar field with respect to $z$. From a tensor field $\phi^\Omega(x)$ we can define the tangent tensor $\bar{\phi}^\Omega(z,v)$ according to \eqref{eq:tangentize}. It is not covariant to series expand $\phi^\Omega(x)$ because coordinate differences at separated points are not covariant and because at each different $x$, the tensor field lives in a different vector space. However, for all $v$, $\bar{\phi}^{\Omega}(z,v)$ is valued in the same vector space and because the tangent space is flat, repeated partial differentiation with respect to $v$ is meaningful. So, we can write (if the functions involved are analytic):
	\begin{equation}
		\bar{\phi}^\Omega(z,v) = \sum_{n=0}^\8 \frac{1}{n!} v^{\kappa_1}\dots v^{\kappa_n} \left.\frac{\d^n \bar{\phi}^\Omega(z,v')}{\d v'^{\kappa_1}\dots\d v'^{\kappa_n}}\right|_{v'= 0} = \sum_{n=0}^\8 \frac{1}{n!} v^{\kappa_1}\dots v^{\kappa_n} {\ee}_{\kappa_1\dots\kappa_n} \phi^\Omega(z) \label{eq:vtaylor}
	\end{equation}
	and every piece of this equation transforms properly as a tensor. In the last step we defined the order $n$ \textit{tensor extension} of $\phi^{\Omega}$ at $z$ as:
	\begin{equation}
		{\textup{E}}_{\kappa_1\dots\kappa_n} \phi^\Omega(z) = \left.\frac{\d^n \bar\phi^\Omega}{\d v^{\kappa_1}\dots\d v^{\kappa_n}}\right|_{v = 0}. \label{eq:tensexdef}
	\end{equation}
	This defines a tensor at the point $z$. Applying this construction at each point $z \in \bb{T}$ fully defines the tensor extension of a general tensor field as its own proper tensor field everywhere.
	
	Because normal coordinates are the closest approximate to the usual linear coordinates of flat spacetime possible in curved spacetime, repeated normal coordinate partial derivatives (tensor extensions) provide the best curved spacetime approximate concept to the idea of multiple partial derivatives. The first tensor extension is just the covariant derivative:
	\begin{equation}
		{\ee}_\mu \phi^\Omega(z) = \nabla_\mu \phi^\Omega(z).
	\end{equation}
	This can be seen by simply directly evaluating the definition and applying the coincidence properties of the worldfunction. The $n^\text{th}$ tensor extension is very much not simply the $n^\text{th}$ partial derivative, nor the $n^\text{th}$ covariant derivative, nor the symmetrization of the $n^\text{th}$ covariant derivative. One can develop explicit formulas relating tensor extensions to the symmetrization of the covariant derivative at higher orders, but with each order there is also an increasingly complex dependence on the Riemann tensor and its derivatives. The $2^\text{nd}$ tensor extension, for example, is given by:
	\begin{equation}
		\ee_{\alpha\beta} \phi^\Omega = \nabla^2_{(\alpha\beta)} \phi^\Omega - \frac{1}{3} R^\delta{}_{(\alpha\beta)\gamma} \fc{G}^\gamma{}_\delta \phi^\Omega. 
	\end{equation}
	This formula is in particular interesting for the metric, which due to metric compatibility yields:
	\begin{equation}
		\ee_{\alpha\beta} g_{\gamma\delta} = \frac{2}{3}R_{\gamma(\alpha\beta)\delta}.
	\end{equation}
	Using the index symmetries of the Riemann tensor, this formula can be ``inverted'' to give the Riemann tensor in terms of the second extension of the metric:
	\begin{equation}
		R_{\gamma\delta\alpha\beta} = \frac{1}{2}(\ee_{\delta\alpha}g_{\gamma\beta} - \ee_{\delta\beta}g_{\gamma\alpha} - \ee_{\gamma\alpha}g_{\delta\beta} + \ee_{\gamma\beta}g_{\delta\alpha})
	\end{equation}
	which is the exact curved spacetime generalization of the typical formula for the linearized Riemann tensor in terms of a metric perturbation about Minkowski space (equation \eqref{eq:linriem}). In Appendix \ref{sec:tensext} we show how to compute tensor extensions in terms of covariant derivatives at arbitrary order, but for our main developments we will not need explicit results beyond the $2^\text{nd}$ extension. 
	
	\subsection{Fourier Transform in Curved Spacetime}
	
	We can generalize the Fourier transform to curved spacetime very directly by using the theory of tangent tensors. The usual exponent $k_\mu(x-z)^\mu$ of the flat spacetime Fourier transform is not covariant in curved spacetime. However, we have already developed $v(x,z)$ as the natural generalization of $x-z$. Similarly, it does not make sense to integrate a tensor field $\phi^\Omega(x)$ over multiple values of $x$ because adding tensors at different spacetime points is not covariant. But, this can be solved by Fourier transforming $\bar{\phi}^\Omega(z,v)$ instead, as at each $v$ in the tangent picture it is a proper tensor only at $z$. Thus, in curved spacetime the Fourier transform acts on tangent tensors. Each $v$ is a vector in the tangent space $T_z(\bb{T})$ and so the natural integration measure is:
	\begin{equation}
		D^4 v = \sqrt{-\det g(z)} d^4 v.
	\end{equation}
	With this, the Fourier transform is now defined by:
	\begin{equation}
		\twid{\phi}^\Omega(z,k) = \int e^{-ik\cdot v} \bar{\phi}^\Omega(z,v) D^4 v. \label{eq:ftdef}
	\end{equation}
	Because $v^\nu$ is a vector in the tangent space $T_z(\bb{T})$ and $k\cdot v = k_\mu v^\mu$ produces an invariant number, $k_\mu$ is a vector in the cotangent space $T^*_z(\bb{T})$. So, the Fourier transform takes a function on the tangent bundle $T(\bb{T})$ and produces a function on the cotangent bundle $T^*(\bb{T})$. We in general write the Fourier transform of a tangent tensor by using the same symbol as the tangent tensor with a tilde over it. The inverse Fourier transform is:
	\begin{equation}
		\bar{\phi}^\Omega(z,v) = \int e^{ik\cdot v} \twid{\phi}^\Omega(z,k) \frac{D^4 k}{(2\pi)^4}
	\end{equation}
	where the invariant integration measure on cotangent space is:
	\begin{equation}
		D^4 k = \frac{d^4 k}{\sqrt{-\det g(z)}}.
	\end{equation}
	A careful reader may object that the exponential map is only well defined on $\fc{V}$ and so cannot be used across the whole tangent space to define the Fourier transform. This is true formally, but irrelevant because we will only ever consider tangent tensors with support contained in $\fc{V}$. For such functions there is no issue. With these ingredients, we have now generalized all the basic mathematical tools from the Newtonian case to curved spacetime.
	
	\section{Defining Momentum, Spin, and the Worldline} \label{sec:ps}
	
	The goal of this section is to generalize the definitions of the worldline, momentum, and spin from the Newtonian case in \eqref{eq:newtpdef}, \eqref{eq:newtsdef}, and \eqref{eq:newtkdef} to general relativity. In all contexts in physics the fundamental definitions of linear momentum and spin are unchanged. A system's linear momentum is its generator of overall translations. A system's angular momentum about a point is its generator of Lorentz transformations about that point. Due to Noether's theorem, we can equivalently define linear momentum and angular momentum as the extensive quantities which are conserved specifically in the presence of a translational or Lorentz symmetry. Both translations and Lorentz transformations are diffeomorphisms (spacetime transformations) and so to find these conserved quantities in general we simply need to understand the structure of diffeomorphisms as possible spacetime symmetries. For a spacetime transformation to be a physical symmetry of a gravitational and electromagnetic system it must be a Killing vector field of both the metric tensor and of the electromagnetic field strength. 
	
	\subsection{Symmetries of Minkowski Space}
	
	To understand how translations and Lorentz transformations relate to Killing vectors, let us begin by studying Killing vectors in Minkowski space. By equation \eqref{eq:lieder}, a diffeomorphism $x'^\alpha = x^\alpha + \eps\xi^\alpha(x) + \OO(\eps^2)$ is a symmetry of Minkowski space if and only if:
	\begin{equation}
		L_\xi \eta_{\alpha\beta} = \d_\alpha \xi_\beta + \d_\beta \xi_\alpha \assert 0. \label{eq:minklie}
	\end{equation}
	The partial derivative of this equation then gives (after several self reapplications with indices renamed):
	\begin{equation}
		\d^2_{\alpha\beta}\xi_\gamma = -\d^2_{\alpha\gamma}\xi_\beta = \d^2_{\beta\gamma}\xi_\alpha = -\d^2_{\alpha\beta}\xi_\gamma \implies \d^2_{\alpha\beta}\xi_\gamma = 0.
	\end{equation}
	Consequently, for some constants $a^\alpha$, $\theta_{\alpha\beta}$, and $z^\alpha$ ($z^\alpha$ is just a point and not presently intended to refer to any worldline):
	\begin{equation}
		\xi^\alpha(x) = a^\alpha - \theta^\alpha{}_\beta (x^\beta-z^\beta)
	\end{equation}
	and by using \eqref{eq:minklie}:
	\begin{equation}
		\theta_{\alpha\beta} = -\theta_{\beta\alpha}.
	\end{equation}
	The constant $z^\alpha$ can be shuffled into the definition of $a^\alpha$, however it is conceptually convenient to leave it arbitrary. The coordinate transformation $x'^\alpha = x^\alpha + \eps\xi^\alpha + \OO(\eps^2)$ to first order in $\eps$ is a Lorentz transformation with Lorentz transformation parameters $\eps \theta_{\alpha\beta}$ about the point $z$ followed by a translation by $\eps a^\alpha$. Therefore, the symmetries of Minkowski space are exactly the familiar Poincaré transformations. Moreover, we can identify the amount of translation and of Lorentz transformation by:
	\begin{equation}
		a^\mu = \xi^\mu(z), \p \theta_{\mu\nu} = \nabla_{[\mu}\xi_{\nu]}(z). \label{eq:liedata}
	\end{equation}
	(Because we are in Minkowski space and using globally Minkowskian coordinates, the partial and covariant derivatives coincide.)
	
	\subsection{Symmetries of Curved Spacetime}
	
	Let us now study Killing vectors in curved spacetime. By equation \eqref{eq:lieder}, a diffeomorphism $x'^\alpha = x^\alpha + \eps \xi^\alpha(x) + \OO(\eps^2)$ is a symmetry of the metric if and only if:
	\begin{equation}
		L_\xi g_{\alpha\beta} = \nabla_\alpha \xi_\beta + \nabla_\beta\xi_\alpha \assert 0. \label{eq:metlie}
	\end{equation}
	
	In Minkowski space any Killing vector field is determined by the value of that vector field (amount of translation) and its first derivative (amount of Lorentz transformation) at a single point. We can accomplish something similar in curved spacetime. Consider our standard geodesic $\zeta^\alpha(s)$ from $\zeta^\mu(0) = z^\mu$ to $\zeta^\alpha(1) = x^\alpha$. We want to see how $\xi^\alpha(\zeta(s))$ changes as we move along the geodesic. Automatically, if $\xi^\alpha$ is a Killing vector:
	\begin{equation}
		\frac{D \xi_\alpha}{Ds} = \dot \zeta^\beta\nabla_\beta \xi_\alpha = \dot\zeta^\beta \nabla_{[\beta}\xi_{\alpha]}. \label{eq:devider}
	\end{equation}
	Taking now a second covariant derivative of $\xi^\alpha$ with respect to $s$ and simplifying gives that $\xi^\alpha$ satisfies:
	\begin{equation}
		\frac{D^2 \xi^\gamma}{Ds^2} + R^\gamma{}_{\alpha\delta\beta} \dot\zeta^\alpha \dot\zeta^\beta \xi^\delta = 0. \label{eq:geodevi}
	\end{equation}
	This is just the geodesic deviation equation! Therefore, we can immediately recycle equation \eqref{eq:desicdevi} as:
	\begin{equation}
		\xi^\alpha(x) = K^\alpha{}_\mu(x,z) \xi^\mu(z) + H^\alpha{}_\mu(x,z) \frac{D\xi^\mu}{Ds}(0). 
	\end{equation}
	Because locally any Einsteinian manifold looks like Minkowski space, at the point $z$, by \eqref{eq:liedata}, $\xi^\mu(z)$ corresponds to a local translation and $\nabla_{[\alpha}\xi_{\beta]}(z)$ corresponds to a local Lorentz transformation in a neighborhood of $z$. Therefore, adopting that notation and using \eqref{eq:devider}, we have: 
	\begin{equation}
		\xi^\alpha(x) = K^\alpha{}_\mu(x,z) a^\mu - {H}^{\alpha\mu}(x,z) \theta_{\mu\nu} v^\nu(x,z). \label{eq:gkf}
	\end{equation}
	This tells us that, \textit{if} our spacetime has a Killing vector, then for an arbitrary point $z$ and some parameters $a^\mu$ and $\theta_{\mu\nu}$, that Killing vector can be put in the form of \eqref{eq:gkf}. Vector fields of the form \eqref{eq:gkf} (for some fixed parameters $z^\mu, a^\mu, \theta^{\mu\nu}$), regardless of whether they are actually Killing vectors, are called Generalized Killing Fields~\cite{Harte:2008xq}.
	
	At each worldline time $\lambda$, we will want to take the point $z$ in \eqref{eq:gkf} to be the representative worldline position $z^\mu(\lambda)$ of our body, so that $a^\mu$ corresponds to a translation of the worldline and $\theta_{\mu\nu}$ corresponds to a Lorentz transformation about the worldline. So, we will actually be interested in bivector fields of the form:
	\begin{equation}
		\xi^\alpha(x, z) = K^\alpha{}_\mu(x,z) a^\mu(z) - {H}^{\alpha[\mu}(x,z)v^{\nu]}(x,z) \theta_{\mu\nu}(z)  \label{eq:pkt}
	\end{equation}
	where now we allow this to properly depend on $z$ as a function so that we can evaluate it at different points along the worldline. Any bivector field of the form of \eqref{eq:pkt} we call a \textit{Poincaré-Killing transformation} (PKT). As promised, under the diffeomorphism provided by such a bitensor field, the Lie derivative on tangent tensors simplifies significantly. In particular after inserting \eqref{eq:pkt} into \eqref{eq:evillieder}, we find:
	\begin{equation}
		L_{\bar{\xi}} \bar{\phi}^\Omega = a^\mu \bar{\nabla}_\mu \bar{\phi}^\Omega + \theta^{\mu\nu} v_\mu \bar{\d}_\nu \bar{\phi}^\Omega + \theta^\mu{}_\nu \fc{G}^\nu{}_\mu\bar{\phi}^\Omega. \label{eq:goodlieder}
	\end{equation}
	The index shuffling appearing here shuffles over all of the indices of $\bar{\phi}^\Omega$, regardless of whether they were natively $z$ indices or tangentized from $x$ indices. Therefore, Lie differentiation by PKTs acts properly on tangent tensors with no regard to their bitensor origins. Each term on the right side of \eqref{eq:goodlieder} has a simple meaning. The first represents a change in the tensor due precisely to translation (thus contributes to linear momentum). The second represents the change in the tensor due to Lorentz transformations affecting its evaluation point. The third represents the intrinsic change in its components due to Lorentz transformations, not changing its evaluation point. (Thus, these second and third terms contribute to angular momentum.)
	
	\subsection{Symmetries of the Electromagnetic Field}
	
	Even if an electromagnetic field strength has a symmetry, the vector potential may not automatically manifest that symmetry in the wrong gauge. We will now show that it is always possible to go to a gauge in which the symmetry is manifest (at least in a coordinate patch). Suppose $\xi^\alpha$ is a symmetry of the electromagnetic field strength:
	\begin{equation}
		L_\xi \fc{F}_{\alpha\beta} = \xi^\gamma \nabla_\gamma \fc{F}_{\alpha\beta} + \nabla_\alpha \xi^\gamma \fc{F}_{\gamma\beta} + \nabla_\beta \xi^\gamma \fc{F}_{\alpha\gamma} \assert 0 
	\end{equation}
	Let $A^{(0)}_\alpha$ be a vector potential for this field strength (meaning $\fc{F}_{\alpha\beta} = \nabla_\alpha A^{(0)}_\beta - \nabla_\beta A^{(0)}_\alpha$). It automatically follows from the definition of the Lie derivative that:
	\begin{equation}
		\nabla_\alpha(L_\xi A_\beta^{(0)}) - \nabla_\beta(L_\xi A_\alpha^{(0)}) = L_\xi \fc{F}_{\alpha\beta} = 0.
	\end{equation}
	Thus by the Poincaré lemma, there exists a scalar field $\varphi^{(0)}(x)$ so that:
	\begin{equation}
		L_\xi A_\alpha^{(0)} = \nabla_\alpha \varphi^{(0)}. 
	\end{equation}
	
	Let us gauge transform, via a scalar function $\varphi^{(1)}$, to a different vector potential $A_\alpha$ for the same field strength:
	\begin{equation}
		A_\alpha = A^{(0)}_\alpha + \nabla_\alpha \varphi^{(1)} \\
		\implies L_\xi A_\alpha = \nabla_\alpha(\varphi^{(0)} + L_\xi \varphi^{(1)}). 
	\end{equation}
	Consider a hypersurface $\Xi$ such that nowhere on that surface does $\xi^\alpha$ lie in its tangent space and let $F(x)$ be an arbitrary scalar function on that surface. Then, the differential equation $\frac{dx^\alpha}{d\eps} = \xi^\alpha(x)$ has solutions which are unique in some neighborhood of $\Xi$ when the initial condition is taken to be a point on the $\Xi$. Let $x^\alpha(\eps, x_0)$ be this solution with initial condition $x^\alpha(0, x_0) = x_0^\alpha$. Choose $\varphi^{(1)}$ to be given by:
	\begin{equation}
		\varphi^{(1)}(x(\eps, x_0)) = F(x_0) - \int_0^\eps \varphi^{(0)}(x(\eps', x_0)) d\eps'
	\end{equation}
	With this gauge choice, the vector potential now does fully manifest the symmetry: 
	\begin{equation}
		L_\xi A_\alpha = 0. \label{eq:symvectorpot}
	\end{equation}
	This gauge is not unique ($F$ is arbitrary). We can modify $A_\alpha$ by an additional gauge transformation through $\varphi^{(2)}$ satisfying $\nabla_\alpha(L_\xi\varphi^{(2)}) = 0$ without breaking \eqref{eq:symvectorpot}. This allows us to choose $L_\xi\varphi^{(2)}= C$ for any constant $C$. As a result, any two vector potentials $A_\alpha$ and $A'_\alpha$ which manifest the symmetry will have $\xi^\alpha A'_\alpha = \xi^\alpha A_\alpha + C$ with $C$ some constant. 
	
	\subsection{Generators of Spacetime Transformations}
	
	For now, we assume that the worldline $z^\mu(\lambda)$ and family of time directions $u^\mu(\lambda)$ are generic and given. We define $\Sigma(\lambda)$ by \eqref{eq:sigdef}, which using the worldfunction we can express as:
	\begin{equation}
		\Sigma(\lambda) = \{ x\ :\ u^\mu(\lambda) \sigma_\mu(x, z(\lambda)) = 0 \}. 
	\end{equation}
	By applying \eqref{eq:volintder} to the continuity equation for electric charge \eqref{eq:jcons}, we have:
	\begin{equation}
		\frac{d}{d\lambda} \int_{\Sigma(\lambda)} J^\alpha d\Sigma_\alpha = 0. 
	\end{equation}
	Therefore, the conserved electric charge in curved spacetime is given by:
	\begin{equation}
		q = \int_{\Sigma(\lambda)} J^\alpha d\Sigma_\alpha.  \label{eq:qdef}
	\end{equation}
	
	Similarly, by Noether's theorem, when equation \eqref{eq:symt} has a spacetime symmetry there must be an associated extensive conserved quantity which is then the generator of that transformation. Consider $\xi^\alpha$ a vector field which is not necessarily a symmetry of the metric or electromagnetic field. By contracting it onto \eqref{eq:symt} and doing some minor manipulation:
	\begin{equation}
		 \nabla_\beta(\xi_\alpha T^{\alpha\beta} + J^\beta A_\alpha\xi^\alpha) = \frac{1}{2}T^{\alpha\beta}L_\xi g_{\alpha\beta} + J^\alpha L_\xi A_\alpha. 
	\end{equation}
	
	Let us now assume that $\xi^\alpha$ is simultaneously a symmetry of the metric and the electromagnetic field and that we use a gauge for the vector potential which manifests that symmetry. Then, the right hand side above vanishes. In such a case, by  \eqref{eq:volintder} the quantity:
	\begin{equation}
		\fc{Q}(\lambda) = \int_{\Sigma(\lambda)}(\xi_\alpha T^{\alpha\beta} + J^\beta A_\alpha\xi^\alpha) d\Sigma_\beta \label{eq:bigqdef}
	\end{equation}
	is extensive and conserved, and so by Noether's theorem is the generator of the transformation. Because $\xi^\alpha$ is a symmetry of the metric, by \eqref{eq:gkf} and \eqref{eq:pkt}, it must take the form, for some functions $a^\mu(\lambda)$ and $\theta_{\mu\nu}(\lambda) = \theta_{[\mu\nu]}$:
	\begin{equation}
		\xi^\alpha(x) = K^\alpha{}_\mu(x, z(\tau(x))) a^\mu(\tau(x)) - H^{\alpha\mu}(x, z(\tau(x))) v^\nu(x, z(\tau(x))) \theta_{\mu\nu}(\tau(x)).  \label{eq:546}
	\end{equation}
	Under a Poincaré-Killing transformation, the conjugate to the amount of translation $a^\mu$ is necessarily the conjugate momentum 4-vector. The conjugate momentum 4-vector of a point-particle is not gauge invariant (due to the $+q A_\mu \dot z^\mu$ in the point-particle Lagrangian). Instead, the conjugate momentum 4-vector can be written as $p_\mu + q A_\mu(z)$ where $p_\mu$ is the gauge invariant linear momentum of the particle. The conjugate to the amount of Lorentz transformation $\theta^{\mu\nu}$ about a point is the total angular momentum tensor about that point. If the point of transformation is taken to be the representative worldline location of the body, the conjugate angular momentum tensor is the spin $S^{\mu\nu}$ of the body. Therefore, in terms of $a^\mu$ and $\theta_{\mu\nu}$, whatever ultimate generator $\fc{Q}$ we use must be related to the (gauge-invariant) linear momentum and spin of the body by:
	\begin{equation}
		\fc{Q}(\lambda) = a^\mu(\lambda) (p_\mu(\lambda)+q A_\mu(z(\lambda))) + \frac{1}{2}\theta_{\mu\nu}(\lambda) S^{\mu\nu}(\lambda). \label{eq:bigqdecomp}
	\end{equation}
	The factor of $\frac{1}{2}$ is due to the normalization used for defining the Lorentz transformation parameters $\theta_{\mu\nu}$. We seek to express \eqref{eq:bigqdef} explicitly in terms of $a^\mu$ and $\theta_{\mu\nu}$ in a way so that the implied $p_\mu$ and $S^{\mu\nu}$ from comparing to \eqref{eq:bigqdecomp} are gauge invariant. 
	
	The appearance of $A^\alpha$ in the integrand of \eqref{eq:bigqdef} makes it unobvious how to separate $\fc{Q}$ into $p_\mu$ and $S^{\mu\nu}$ along the lines of \eqref{eq:bigqdecomp} in a gauge invariant way. However, we can disentangle this complexity by considering how $\xi^\alpha A_\alpha$ changes along our standard geodesic $\zeta$. Specifically:
	\begin{equation}
		\frac{d}{ds}(\xi^\alpha(\zeta) A_\alpha(\zeta)) = \fc{F}_{\alpha\beta}(\zeta) \dot \zeta^\alpha \xi^\beta(\zeta) + \dot\zeta^\alpha L_\xi A_\alpha = \fc{F}_{\alpha\beta}(\zeta) \dot \zeta^\alpha \xi^\beta(\zeta)
	\end{equation}
	where the final step uses that $\xi^\alpha$ is a symmetry of the vector potential. Integrating from $0$ to $1$ yields:
	\begin{equation}
		\xi^\alpha(x) A_\alpha(x) = \xi^\mu(z) A_\mu(z) + \int_0^1 \fc{F}_{\alpha\beta}(\zeta) \dot\zeta^\alpha \xi^\beta(\zeta) ds. \label{eq:Aalonggeo}
	\end{equation}
	Consider the geodesic $\zeta'^\alpha(s') = \zeta^\alpha(s s')$ which begins at $\zeta'^\mu(0) = z^\mu$ and reaches $\zeta'^\alpha(1) = \zeta^\alpha(s)$. Then:
	\begin{equation}
		\dot\zeta'^\alpha(1) = s \dot\zeta^\alpha(s), \ps \dot \zeta'^\mu(0) = s \dot\zeta^\mu(0) \ \ \implies\ \ \dot\zeta^\alpha(s) = \frac{1}{s} \sigma^\alpha(\zeta(s), z), \ps \dot\zeta'^\mu = s v^\mu(x,z). \label{eq:stdgeo}
	\end{equation}
	With these, using \eqref{eq:546}, \eqref{eq:Aalonggeo} becomes:
	\begin{equation}
		\xi^\alpha(x)A_\alpha(x) = a^\mu A_\mu(z) + a^\mu\Phi_{\mu}(x,z) + \theta_{\mu\nu} v^\mu(x,z) \Psi^\nu(x,z) \label{eq:Ageodecomp}
	\end{equation}
	where we have defined the bitensors $\Phi_{\mu}(x,z)$ and $\Psi_{\mu}(x,z)$ by:
	\begin{equation}
		\Phi_\mu(x,z) = \int_0^1 \fc{F}_{\alpha\beta}(\zeta) \sigma^\alpha(\zeta, z) K^\beta{}_\mu(\zeta, z) \frac{ds}{s}, \ps \Psi_\mu(x,z) = \int_0^1 \fc{F}_{\alpha\beta}(\zeta)\sigma^\alpha(\zeta,z) H^\beta{}_\mu(\zeta,z) ds. \label{eq:phipsidef}
	\end{equation}
	Both $\Psi_\mu$ and $\Phi_\mu$ are manifestly gauge invariant. Thus, the only non-gauge invariant piece of $\xi^\alpha A_\alpha$ is $a^\mu A_\mu(z)$. 
	
	Returning \eqref{eq:Ageodecomp} to \eqref{eq:bigqdef} and comparing to \eqref{eq:bigqdecomp} determines Dixon's ultimate formulas for the momentum and spin~\cite{Dixon:1970mpd}:
	\begin{align}
		p_\mu(\lambda) = \int_{\Sigma(\lambda)}\( T^{\alpha\beta} K_{\beta\mu} + \Phi_\mu J^\alpha\)d\Sigma_\alpha, \ps S^{\mu\nu}(\lambda) = 2\int_{\Sigma(\lambda)}\( v^{[\mu} H_\beta{}^{\nu]} T^{\alpha\beta} + v^{[\mu} \Psi^{\nu]} J^\alpha\) d\Sigma_\alpha. \label{eq:psdef}
	\end{align}
	Given a choice of worldline and time-slicing pair $(z,u)$, equation \eqref{eq:psdef} provides the fundamental formulas defining the linear momentum and spin of a body in curved spacetime interacting with gravity and electromagnetism; they are the analogs of \eqref{eq:newtpdef} and \eqref{eq:newtsdef} from the Newtonian case. Even though we assumed the presence of a spacetime and electromagnetic symmetry to derive these formulas, because the generator of a transformation is the same quantity regardless of whether the transformation is actually a symmetry of the system, equation \eqref{eq:psdef} remains the correct definition of linear momentum and spin even in a completely generic physical scenario. In Section \ref{sec:gr} we will find that these definitions of momentum and spin are indeed uniquely forced upon us.  

	\subsection{Selection of the Worldline}
	
	Almost all of the formalism we will develop will use no special properties of $(z, u)$, and unless we explicitly state otherwise, we never assume particular choices for them. But, in this subsection we discuss the \textit{natural} choice for the worldline given that you have already picked a time-slicing. Using a different worldline will still give physically correct results under appropriate mapping of quantities from one description to the other. As $u^\mu$ represents the time direction of the observer performing the time-slicing, we can decompose the spin angular momentum tensor $S^{\mu\nu}$ into components generating rotations about the worldline in the observer's frame and components generating Lorentz boosts about the worldline in the observer's frame. The vector $S^\mu$ of rotation generators (and thus the true spin angular momentum vector) is packaged into $S^{\mu\nu}$ so that in coordinates in which $u^\mu = \delta^\mu_0$, we have: $S^0 = 0$, $S^1 = S^{23}$, $S^2 = S^{31}$, $S^3 = S^{12}$. Similarly, the vector ${\mc{K}}^\mu$ of boost generators is packaged into $S^{\mu\nu}$ so that in those same coordinates, ${\mc{K}}^0 = 0$, ${\mc{K}}^1 = S^{01}$, ${\mc{K}}^2 = S^{02}$, ${\mc{K}}^3 = S^{03}$. We can express these both covariantly as:
	\begin{equation}
		S^\mu = -\frac{1}{2}{\epsilon}^{\mu\nu\rho\sigma} u_\nu S_{\rho\sigma}, \p {\mc{K}}^\mu = S^{\mu\nu} u_\nu.  \label{eq:skdef}
	\end{equation}
	The spin tensor can be recovered by using the double-Levi-Civita identity as:
	\begin{equation}
		S_{\mu\nu} = u_{\mu} {\mc{K}}_\nu - u_\nu {\mc{K}}_\mu + {\epsilon}_{\mu\nu\rho\sigma} u^\rho S^\sigma.
	\end{equation}
	In the Newtonian case, the natural center of mass worldline is selected by requiring the intrinsic boost generator about the worldline to vanish. Taking this as motivation, the center of mass worldline in general relativity is the worldline for which:
	\begin{equation}
		{\mc{K}}^\mu = S^{\mu\nu} u_\nu = 0.  \label{eq:ssc}
	\end{equation}
	For a given choice of $u^\mu$, the condition $S^{\mu\nu} u_\nu = 0$ is called the \textit{spin supplementary condition} (SSC) corresponding to $u^\mu$. Conceptually, the SSC means we are defining the spin of the body in the same Lorentz frame that we are doing the time-slicing. 
	
	To understand how the SSC works, it is useful to first look at the case of a free body in flat spacetime. Under such conditions, the spin part of \eqref{eq:psdef} becomes:
	\begin{equation}
		S^{\mu\nu}(\lambda) = \int_{\Sigma(\lambda)}( (x-z)^\mu T^{\nu\rho} - (x-z)^\nu T^{\mu\rho}) d\Sigma_\rho.
	\end{equation}
	Computing ${\mc{K}}^\mu$ from \eqref{eq:skdef} and solving for the worldline then gives:
	\begin{equation}
		z^\mu = \frac{\int_\Sigma x^\mu T^{\alpha\beta} u_{\alpha} u_\beta d\Sigma + {\mc{K}}^\mu}{\int_\Sigma T^{\alpha\beta} u_{\alpha} u_\beta d\Sigma}. \label{eq:555}
	\end{equation}
	So, under these simple conditions there is plainly a unique existent choice of $z^\mu$ so that ${\mc{K}}^\mu$ vanishes, corresponding to choosing the ``center of energy'' position of the body, meaning the weighted average position of the body using the energy density $T^{\alpha\beta}u_{\alpha} u_{\beta}$ as the weighting function. This is the special relativistic generalization of the concept of center of mass. We can rewrite this equation as:
	\begin{equation}
		z^\mu = z^\mu_{\text{com}} + \frac{{\mc{K}}^\mu}{\ee}
	\end{equation}
	where $\ee$ is the total energy of the body (given by the denominator in \eqref{eq:555}).
	
	Let's now generalize this analysis to curved spacetime~\cite{Beiglbock:1965fun}. If $u^\mu$ represents the time direction on the worldline, then $-u\cdot p$ is the corresponding total energy of the body on the worldline:
	\begin{equation}
		\ee = -u\cdot p = -\int_{\Sigma}\(T^{\alpha\beta} K_{\beta \mu} u^\mu + u^\mu \Phi_\mu J^\alpha\)d\Sigma_\alpha. 
	\end{equation} 
	The intrinsic boost generator is given by:
	\begin{equation}
		{\mc{K}}^\mu = S^{\mu\nu} u_\nu = \int_\Sigma v^\mu\(H_{\beta}{}^\nu u_\nu T^{\alpha\beta} + \Psi^\nu u_\nu J^\alpha\) d\Sigma_\alpha \label{eq:558}
	\end{equation}
	which is an integral of spatial displacement from the worldline times a density on the worldline, and so ${\mc{K}}^\mu$ is by construction purely spatial (orthogonal to $u^\mu$). It is not geometrically meaningful in a curved spacetime to try to add $-\frac{{\mc{K}}^\mu}{\ee}$ to the coordinates of $z^\mu$ to try to arrive at a center of mass worldline. Instead, at a given worldline time, from an initial candidate choice $z_0^\mu$ for the worldline position at that time, we can define a one-parameter family of such choices $z^\mu(r)$ by:
	\begin{gather}
		\frac{dz^\mu(r)}{dr} = -{\mc{K}}^\mu(z(r)) \label{eq:kpush}
	\end{gather}
	so that $z^\mu(0) = z_0^\mu$ (and where we use that by its definition ${\mc{K}}^\mu$ depends on the worldline point it is built around). In the case of flat spacetime, this differential equation is easily solved to give the flow in a straight line from $z_0^\mu$ to $z^\mu_{\text{com}}$ as $r\to \8$. In curved spacetime, $z^\mu_{\text{com}}$ is instead defined as the solution to this differential as $r\to \8$. We prove in Appendix \ref{sec:sscint} that, under reasonable regularity assumptions, regardless of the initial condition, all candidate points within the worldtube at a given worldline time flow toward the same fixed point, and plainly by the differential equation, that fixed point occurs at ${\mc{K}}^\mu = 0$, thus satisfying the SSC \eqref{eq:ssc}. This can be done at each worldline time to define the full worldline as a curve. 

	While the above process is how the worldline satisfying the SSC is formally constructed, in practical computations it is determined via an equation of motion. Though we haven't yet justified the MPD equations \eqref{eq:mpd1}, let us assume them for now to see how this works. Contracting $u^\nu$ with the spin equation in \eqref{eq:mpd1} and using the SSC and the definitions of the einbein and the body's energy, we find:
	\begin{equation}
		p^\mu = \ee \doot z{}^\mu + \text{\textbf{T}}^{\mu\nu} u_\nu + S^{\mu\nu}\doot u_\nu.  \label{eq:pvelrel}
	\end{equation}
	For a given choice of time direction function $u^\mu(\lambda)$, this equation determines $\doot z{}^\mu$, providing the curved spacetime interacting compact object generalization of \eqref{eq:ngz}. 

	\subsection{Selection of the Body Frame}
	
	In this subsection we discuss some natural choices for $u^\mu(\lambda)$ and assume that we always use the worldline satisfying the SSC given this time-slicing. Instantaneously, $u^\mu(\lambda)$ represents the increasing time direction at the worldline. There are (at least) four common choices for $u^\mu$. Typically the first or the fourth in our listing are optimal for most calculations. Again these are choices taken for the purpose of finding the simplest description, but they are physically equivalent under appropriate mappings.
	
	(I) \textit{Newton-Wigner}. If there is a particular observer of interest, we can let $t(x)$ be the function giving the time assigned by that observer to each point in spacetime. From $t(x)$ we can define $u^\mu$ along the worldline analogously to \eqref{eq:volmeas} as:
	\begin{equation}
		u_\mu(\lambda) = -\left.\frac{\nabla_\mu t}{|\nabla t|}\right|_{x = z(\lambda)}. \label{eq:wlclockdef}
	\end{equation}
	This is very useful for studying the motion of a body in a spacetime that has a timelike Killing vector, where $t$ is taken to be the time flowed by parallel transporting along that Killing vector, or when one wants to use the same time parameter for multiple different bodies, as is desirable in a Hamiltonian formulation. In this time-slicing, \eqref{eq:ssc} is called the \textit{canonical spin supplementary condition}. 
	
	(II) \textit{Worldline Rest-Frame}. You can take $u^\mu$ to point in the same direction as $\doot z{}^\mu$, so that the time-slicing is performed in the worldline's rest frame. Importantly, the worldline's rest frame is not the body's rest frame because no such frame exists in general; the different parts of a finite size body move at different velocities and there is no frame in which all are stationary. The worldline corresponding to this choice is never unique and exhibits helical motion even for free bodies~\cite{Dixon:1970mpd, Beiglbock:1965fun, Mathisson_1940, Costa:2014nta, Costa:2011zn}. While these helical worldlines are ``physical'', in the sense of being true solutions to the equations of motion, their unnecessary complexity is quite undesirable.
	
	Let's understand where these helical worldlines come from (assuming for now the MPD equations). For a free body, the MPD equations \eqref{eq:mpd1} become:	
	\begin{equation}
		\doot p_\mu = 0, \p \doot S{}^{\mu\nu} = p^\mu \doot z{}^\nu - p^\nu \doot z{}^\mu. 
	\end{equation}
	With $u^\mu \propto \doot z{}^\mu$, the einbein defined so that $u\cdot \doot z = -1$, and $u^\mu$ a unit vector, we have that actually $u^\mu = \doot z{}^\mu$. Now changing to the spin vector as defined in \eqref{eq:skdef}, we find that:
	\begin{equation}
		\doot S{}^\mu = u^\mu (\doot u\cdot S), \p \doot S{}^{\mu\nu} = -\epsilon^{\mu\nu\rho\sigma} S_\rho \doot u_\sigma. \label{eq:redsder}
	\end{equation}
	Contracting $u_\nu$ onto the second of these gives:
	\begin{equation}
		p^\mu = \ee u^\mu + \epsilon^{\mu\nu\rho\sigma} u_\nu S_\rho \doot u_\sigma.
	\end{equation}
	Contracting $\doot u_\mu$ and $S_\mu$ with this give $p\cdot \doot u = 0$ and  $p\cdot S = 0$. Using these it immediately follows that $\doot \ee = 0$ and $\doot S{}^\mu = 0$. Then, contracting $p_\nu$ onto the second of \eqref{eq:redsder} gives:
	\begin{equation}
		\fc{M}^2 u^\mu - \ee p^\mu + \epsilon^{\mu\nu\rho\sigma} p_\nu S_\rho \doot u_\sigma = 0 \label{eq:reduder}
	\end{equation}
	where we have defined the mass of the body:
	\begin{equation}
		\fc{M} = |p|. \label{eq:massdefi}
	\end{equation}
	Equation \eqref{eq:reduder} has a nice interpretation. Because $p_\mu$ is constant for a free body, we can go to the inertial frame in which it points purely in the time direction and define spatial vectors orthogonal to it. Let $\vec u_\perp$ be the spatial vector part of $u^\mu$ (so that $p^\mu - u^\mu_\perp$ points purely along $p^\mu$ and $\vec u_\perp$ is orthogonal to $p^\mu$). Similarly so for $\vec S$ from $S^\mu$. Then, \eqref{eq:reduder} becomes:
	\begin{equation}
		\vec S\times \doot{\vec u}{}_\perp + \fc{M} \vec u_\perp = 0. 
	\end{equation}
	Because $\vec S$ and $\fc{M}$ are constant for this body, this is the equation of motion of a helix of arbitrary radius. Thus, the worldline satisfying this spin supplementary condition is not unique. 
	
	(III) \textit{Ohashi-Kyrian-Semerák}. You can take $u^\mu$ to have some arbitrary initial condition at one point on the worldline and then to be parallel transported along the worldline under time evolution. This ``parallel slicing'' is unique up to the initial condition choice and leads to the simplest equation of motion for the worldline trajectory as in this choice the $\doot u{}^\mu$ term in \eqref{eq:pvelrel} vanishes. The difficulty of this approach is that the parallel transport of $u^\mu$ is yet another equation of motion to solve. As well, the body's energy $\ee$ (through $p^\mu$) then appears heavily in the equations of motion, which is usually a complicated function of momenta involving square roots akin to the free point-particle energy $\ee = \sqrt{|\vec p|^2 + m^2}$ in special relativity. This tends to make calculations challenging. 
	
	(IV) \textit{Tulczyjew-Dixon}. In generic relativistic calculations, this is typically the optimal time-slicing. You can take $u^\mu$ to point in the same direction as the body's momentum $p^\mu$, so that the time-slicing is performed in the body's center of momentum frame:
	\begin{equation}
		u^\mu = \hat p^\mu = \frac{p^\mu}{|p|}. \label{eq:udef}
	\end{equation}
	It is not at all obvious that this is generically possible. $p$ is dependent on $u$ in \eqref{eq:psdef} through the definition of $\Sigma$, so mathematically this requires there to be a solution to the implicit equation $p^\mu= |p| u^\mu$. At the very least, this requires $p^\mu$ to be timelike and future oriented. By equations \eqref{eq:timelike} and \eqref{eq:posenergy} and because $d\Sigma^\beta$ is timelike and future oriented, $T^{\alpha\beta} d\Sigma_{\beta}$ is timelike and future oriented. At coincidence $x\to z$, $K^\alpha{}_\mu \to \delta^\alpha_\mu$. Because we assume we choose the worldline so that there is only one geodesic from any point within $\Sigma(\lambda)$ to $z^\mu(\lambda)$, $K^\alpha{}_\mu$ must be invertible within the worldtube. Thus, it can never change signature within the worldtube, making $K^\alpha{}_\mu T^{\alpha\beta} d\Sigma_{\beta}$ also timelike and future oriented. The sum of future oriented timelike vectors is a future oriented timelike vector. Therefore, we have that its integral is timelike and future oriented. As explained in Subsection \ref{sec:bpa}, we require that the electromagnetic field experienced by the body is of manageable strength, so that the $\Phi_\mu$ contribution to \eqref{eq:psdef} does not break the future-orientedness or timelikeness of $p_\mu$. 
	
	Let $\scr{H}$ be the set of future oriented timelike unit vectors in $T_z(\bb{T})$. Let $V_1$ and $V_2$ be any two such vectors in $\scr{H}$ and let:
	\begin{equation}
		d(V_1, V_2) = (V_1-V_2)^2.
	\end{equation}
	One can show that $d \ge 0$ for all $V_1$ and $V_2$ in $\scr{H}$ and is 0 if and only if $V_1 = V_2$ (to do so, express $V_2$ a basis in which $V_1^\mu = (1,0,0,0)$). As well, plainly:
	\begin{equation}
		d(V_2,V_1) = d(V_1,V_2).
	\end{equation}
	Another short calculation shows that $d$ satisfies the triangle inequality:
	\begin{equation}
		d(V_1, V_2) \le d(V_1, V_3) + d(V_2, V_3)
	\end{equation}
	for any third vector $V_3 \in \scr{H}$. Therefore, $\scr{H}$ together with the distance function $d$ is in fact a metric space. For a given element of $V \in \scr{H}$, we can compute the resulting unit vector in the direction of the momentum implied by using $V^\mu$ to define our time-slicing. Call the map from $V^\mu$ to $\hat p^\mu$ as $F(V)$ (all done at a fixed $\lambda$ for now):
	\begin{equation}
		F^\mu(V) = \left.\hat p^\mu\right|_{u = V} \label{eq:5612}
	\end{equation}
	where $p^\mu$ is computed by \eqref{eq:psdef}. Because $p^\mu$ is timelike and future oriented, $F(V)$ acts as a map from $\scr{H}$ to $\scr{H}$. It is shown in Appendix \ref{sec:uint} (via methods motivated by~\cite{Beiglbock:1965fun, Costa:2014nta}) that for a body satisfying our basic physical assumptions in Subsection \ref{sec:bpa}:
	\begin{equation}
		d(F(V_1), F(V_2)) \le \eps_u d(V_1, V_2) \label{eq:banachdist}
	\end{equation}
	for some constant $\eps_u \in [0,1)$. Because of this, by the Banach fixed point theorem, the sequence $\{ V , F(V), F(F(V)), \dots\}$ limits to a vector $u^\mu$ which is the unique fixed point of $F(u)$. Such a $u^\mu$ thus necessarily satisfies:
	\begin{equation}
		u^\mu = F^\mu(u) = \left.\hat p^\mu\right|_u
	\end{equation}
	Therefore, for a given choice of base point $z$ there exists a unique choice of $u^\mu$ for time-slicing so that the resulting $\hat p^\mu = u^\mu$ in curved spacetime. 
	
	The sequence $\{ V , F(V), F(F(V)), \dots\}$ converging to $u^\mu = \hat p^\mu$ has a reasonable physical picture. If $V$ is used to time-slice, then $F(V)$ is the direction in which the momentum points in that slicing. An observer in the $V$ frame would to lowest order think that the ``body is in the $F(V)$ frame'' (in the sense of center of momentum). The reason this thinking is only valid at lowest order is that, since the body has finite size, the different points in it are moved by different amounts under a boost. But, this lowest order logic motivates why $F(V)$ would be closer to the body's center of momentum frame than $V$ is. Defining a new slicing by $F(V)$ then boosts to that new frame. Repeating this logic motivates why $\{ V , F(V), F(F(V)), \dots\}$ may get progressively closer to the body's actual center of momentum frame. The challenge is in rigorously proving that this intuition does what it promises (i.e. that the $\eps_u$ in \eqref{eq:banachdist} exists), as done in Appendix~\ref{sec:uint}. 
	
	Under the Tulczyjew-Dixon slicing, the SSC \eqref{eq:ssc} becomes:
	\begin{equation}
		S^{\mu\nu}(\lambda)p_\nu(\lambda) = 0 \label{eq:cssc}.
	\end{equation}
	This is known as the \textit{covariant spin supplementary condition} (CSSC). If we now assume the MPD equations \eqref{eq:mpd1} and subject them to \eqref{eq:udef} and \eqref{eq:cssc}, then \eqref{eq:pvelrel} becomes:
	\begin{equation}
		\fc{M}\doot z{}^\mu = p^\mu - \text{\textbf{T}}^{\mu\nu} u_\nu - S^{\mu\nu} \doot u_\nu. \label{eq:csscpvelrel}
	\end{equation}
	This form of the equation of motion still has not solved for the worldline trajectory explicitly, because $\doot u_\mu$ is determined by $\doot p_\mu$, the equation of motion of which depends on $\doot z{}^\mu$.
	
	To disentangle \eqref{eq:csscpvelrel} into an explicit equation for $\doot z{}^\mu$, we need to use that we are in 4 spacetime dimensions. For an antisymmetric tensor, such as the electromagnetic field strength, define the dual tensor by:
	\begin{equation}
		{}^\star \fc{F}_{\mu\nu} = \frac{1}{2}\epsilon_{\mu\nu\rho\sigma}\fc{F}^{\rho\sigma} \implies \fc{F}_{\mu\nu} = -\frac{1}{2}\epsilon_{\mu\nu\rho\sigma} {}^\star \fc{F}^{\rho\sigma}. 
	\end{equation}
	Similarly, for the Riemann tensor we define the left dual, right dual, and the double dual:
	\begin{equation}
		{}^\star R_{\rho\mu\sigma\nu} = \frac{1}{2}\epsilon_{\rho\mu}{}^{\iota\kappa} R_{\iota\kappa\sigma\nu}, \p R^\star_{\rho\mu\sigma\nu} = \frac{1}{2}\epsilon_{\sigma\nu}{}^{\iota\kappa} R_{\rho\mu\iota\kappa}, \p {}^\star R_{\rho\mu\sigma\nu}^\star = \frac{1}{2}\epsilon_{\rho\mu}{}^{\iota\kappa} R_{\iota\kappa\sigma\nu}^\star.
	\end{equation}
	In terms of the spin vector introduced in \eqref{eq:skdef}, the MPD equations \eqref{eq:mpd1} become:
	\begin{align}
		\doot p{}_\mu = -R^\star_{\mu\nu\rho\sigma} \doot z{}^\nu u^\rho S^\sigma + q \fc{F}_{\mu\nu} \doot z{}^\nu + \text{\textbf{F}}_\mu, \p \doot S{}^\mu = u^\mu \doot u{}_\nu S^\nu - {}^\star \text{\textbf{T}}^{\mu\nu} u_\nu.  \label{eq:redmpd}
	\end{align}
	Using this form of the equations of motion in the trajectory equation \eqref{eq:csscpvelrel}, we find:
	\begin{equation}
		-\fc{M} p^\mu + \fc{M}^2 \doot z{}^\mu + \text{\textbf{T}}^{\mu\nu} p_\nu + \frac{1}{2}S^{\mu\nu}\epsilon_{\nu\kappa\rho\sigma} {}^\star R^{\star\rho\sigma}{}_{uS} - \frac{q}{2}S^{\mu\nu}\epsilon_{\nu\rho\iota\kappa}{}^\star \fc{F}^{\iota\kappa} + \epsilon^{\mu\nu\rho\sigma} \text{\textbf{F}}_\nu u_\rho S_\sigma = 0. 
	\end{equation}
	With a little double-Levi-Civita manipulation, this can be rewritten as:
	\begin{align}
		(\fc{M}^2 + {}^\star R^\star_{uSuS} - q{}^\star \fc{F}_{uS})\doot z{}^\mu &+ (S\cdot \doot z)({}^\star R^{\star\mu}{}_{uuS} - q{}^\star \fc{F}^\mu{}_S) \nonumber \\
		&- \fc{M} p^\mu + {}^\star R^{\star\mu}{}_{SuS} - q{}^\star \fc{F}^\mu{}_S + \text{\textbf{T}}^{\mu\nu} p_\nu + \epsilon^{\mu\nu\rho\sigma} \text{\textbf{F}}_\nu u_\rho S_\sigma = 0. \label{eq:redvel}
	\end{align}
	Contracting with $S^\mu$ and simplifying gives:
	\begin{equation}
		S\cdot \doot z = -\frac{1}{\fc{M}^2} \text{\textbf{T}}^{\mu\nu} S_\mu p_\nu.
	\end{equation}
	Using this back in \eqref{eq:redvel}, we arrive at the fully explicit general relativistic analog of \eqref{eq:ngz}:
	\begin{align}
		(\fc{M}^2 + {}^\star R^\star_{uSuS} - q{}^\star \fc{F}_{uS})\doot z{}^\mu = &\ \fc{M}^2 u^\mu - {}^\star R^{\star\mu}{}_{SuS} + q{}^\star \fc{F}^{\mu}{}_S \nonumber \\
		&- \epsilon^{\mu\nu\rho\sigma} u_\nu S_\rho \text{\textbf{F}}_\sigma - \fc{M} \text{\textbf{T}}^{\mu\nu} u_\nu + \frac{\text{\textbf{T}}^{\rho\sigma} S_\rho u_\sigma}{\fc{M}}\({}^\star R^{\star\mu}{}_{uuS} - q{}^\star \fc{F}^\mu{}_S\). \label{eq:comwl}
	\end{align}
	This is the explicit equation for the center of mass worldline (first achieved by \cite{Ehlers:1977rud}) and together with the MPD equations, it completes our equations of motion for the universal degrees of freedom of a compact object. 
	
	For doing actual calculations, the Tulczyjew-Dixon time-slicing \eqref{eq:udef} and CSSC worldline \eqref{eq:cssc} are most common in generic physical scenarios. But, almost all of the formalism to follow does not depend on which $(z, u)$ are chosen. So, we will assume neither \eqref{eq:udef} nor \eqref{eq:cssc} in the coming sections. 
	
	 Our next major goal is to derive the MPD equations of motion properly. In order to do so, we will first need to thoroughly understand the multipole moments of electromagnetism and gravity, first in flat spacetime, and then in full general relativity. 
	
	\section{Electromagnetism in Minkowski Space} \label{sec:em}
	
	Based on the Newtonian theory, we expect the time derivatives of the momentum and spin from \eqref{eq:psdef} to be expressible as series over increasingly many spatial derivatives of the electromagnetic and gravitational fields, with the coefficients of those series representing the multipole moments of the current density and energy-momentum tensor. To get a basic idea of how this works, it is easiest to first work with pure electromagnetism. The goal of this section is find a set of multipole moments that describe the current density of a generic body in flat spacetime. 
	
	In Minkowski space, geodesics are simply straight lines. Consequently, \eqref{eq:sigdef} becomes:
	\begin{equation}
		\Sigma(\lambda) = \{ x \in \bb{T} \ :\ u_\mu(\lambda)(x^\mu - z^\mu(\lambda)) = 0\}. \label{eq:bigsigflatdef}
	\end{equation}
	Electromagnetism in flat spacetime introduces an important challenge over Newtonian gravity: the local conservation of electric charge. Specifically, the current density satisfies the continuity equation \eqref{eq:jcons}, which in flat spacetime becomes:
	\begin{equation}
		\d_\alpha J^\alpha = 0.
	\end{equation}
	and due to this, the different components of the current density are not independent from each other. In order to define a proper set of multipole moments of the current density that are fully reduced (contain no interdependencies), we will have to take care to remove all pieces which are annihilated by the continuity equation. The local conservation of charge is Noether-dual to the gauge invariance of electromagnetism, and so this reduction problem is exactly mirrored by the problem of trying to write the action for the electromagnetic interaction in a manifestly gauge invariant form. We can express the continuity equation as a functional statement by integrating it over all spacetime against an arbitrary function $f(x)$:
	\begin{equation}
		\int_\bb{T} f \d_\alpha J^\alpha d^4 x= 0.
	\end{equation}
	Since we only consider $f(x)$ with compact support, we can integrate the derivative onto it by parts. Thus:
	\begin{equation}
		\int_\bb{T} J^\alpha \d_\alpha f d^4 x = 0. \label{eq:funcjcons}
	\end{equation}
	Alternatively, by integrating $\d_\alpha(f J^\alpha)$ over a time-slice only, according to \eqref{eq:volintder}, we have:
	\begin{equation}
		\frac{1}{\ec}\frac{d}{d\lambda}\int_\Sigma f J^\alpha d\Sigma_\alpha = \int_\Sigma J^\alpha \d_\alpha f d\Sigma_w.  \label{eq:surfacejcons}
	\end{equation}
	Equation \eqref{eq:surfacejcons} is equivalent to \eqref{eq:funcjcons}; it can be derived from \eqref{eq:surfacejcons} by taking an $f$ that is only supported on an interval in the limit that the interval is taken toward width 0 and it can derive \eqref{eq:surfacejcons} by integrating over all $\lambda$. Equation \eqref{eq:surfacejcons} will be the useful form of the continuity equation for studying the impact of charge conservation on multipole moments. 
	
	\subsection{Gauge Invariant form of the Action}

	The vector potential satisfies Maxwell's equations:
	\begin{equation}
		\d_\beta \fc{F}^{\alpha\beta} = J^\alpha
	\end{equation}
	which are naturally produced by the action:
	\begin{equation}
		\fc{S}_{\text{em}} = -\frac{1}{4}\int_{\bb{T}} \fc{F}^{\alpha\beta}\fc{F}_{\alpha\beta} d^4 x + \fc{S}_m
	\end{equation}
	where $\fc{S}_m$ is the action for the electromagnetic matter and the current density is derived by:
	\begin{equation}
		J^\alpha = \frac{\delta \fc{S}_m}{\delta A_\alpha}.
	\end{equation}
	$\fc{S}_m$ may depend on $A_\alpha$ in a complicated nonlinear way for a body with nontrivial electromagnetic susceptibility. When $A_\alpha$ is perturbed from some default potential $A^{(0)}_\alpha$ (with associated current density $J^\alpha$) by an amount $\delta A_\alpha$:
	\begin{equation}
		\fc{S}_m = \left.\fc{S}_m\right|_{A = A^{(0)}} + \<J^\alpha, \delta A_\alpha\> + \OO(\delta A^2). \label{eq:jactpert}
	\end{equation}
	Thus, the behavior of $J$ as a linear functional $\<J^\alpha, \cdot\>$ determines the variations of the action and thus the physics of the electromagnetic interaction. It is all we need to study!
	
	Let $A_\alpha$ be a generic smooth vector potential of compact support. Using \eqref{eq:volintdecomp}:
	\begin{equation}
		\<J^\alpha, A_\alpha\> = \int_{-\8}^\8 \int_{\Sigma(\lambda)} J^\alpha(x) A_\alpha(x) d\Sigma_w D\lambda. 
	\end{equation}
	This double integral should be thought of as first integrating over a time-slice, then integrating over all time. We can series expand $A_\alpha(x)$ about the worldline analogously to the Newtonian logic for the gravitational potential:
	\begin{equation}
		A_\alpha(x) = \sum_{n=0}^\8 \frac{1}{n!}(x-z)^{\rho_1}\dots(x-z)^{\rho_n} \d^n_{\rho_1\dots\rho_n} A_\alpha(z). 
	\end{equation}
	Returning this to the action:
	\begin{equation}
		\<J^\alpha, A_\alpha\> = \int_{-\8}^\8 \sum_{n=0}^\8\frac{1}{n!} \mc{j}^{\rho_1\dots\rho_n\mu}_n \d^n_{\rho_1\dots\rho_n} A_\mu(z) D\lambda \label{eq:jafunc}
	\end{equation}
	where we have defined the ``naive multipole moments'' of the current density: 
	\begin{equation}
		\mc{j}^{\rho_1\dots\rho_n\mu}_n(\lambda) = \int_{\Sigma(\lambda)}(x-z(\lambda))^{\rho_1}\dots(x-z(\lambda))^{\rho_n} J^\mu(x) d\Sigma_w \label{eq:naivejdef}
	\end{equation}
	These moments are highly interdependent (due to the continuity equation) and do not put the action into a manifestly gauge invariant form. They are just integrals of each component of the current density against some number of powers of spatial displacement on a time-slice. 
	
	Because $\mc{n}^\alpha$, the unit vector normal to $\Sigma$ at $x$, locally represents the time direction on the time-slice, the quantity $-\mc{n}^\alpha J_\alpha$ locally represents the charge density. So, we can analogously define naive moments of the charge density by:
	\begin{equation}
		\mc{r}_n^{\rho_1\dots\rho_n}(\lambda) = \int_{\Sigma(\lambda)} (x-z(\lambda))^{\rho_1}\dots(x-z(\lambda))^{\rho_n} J^\alpha d\Sigma_\alpha.  \label{eq:naiverdef}
	\end{equation}
	Because of the continuity equation, the time evolution of the charge density is determined by the other components of the current density. We can see this by using \eqref{eq:surfacejcons} with appropriate choices of $f$ to relate the $\mc{r}_n^{\cdots}$ and the $\mc{j}_n^{\cdots}$. 
	
	For $f = 1$ and recalling \eqref{eq:qdef}, \eqref{eq:surfacejcons} gives:
	\begin{equation}
		\mc{r}_0 = q = \int_\Sigma J^\alpha d\Sigma_\alpha , \p  \frac{dq}{d\lambda}= 0.
	\end{equation}
	This just reproduces the conservation of total charge from \eqref{eq:qdef}. 
	
	In deriving \eqref{eq:surfacejcons}, we assumed that $f(x)$ is a field on $\bb{T}$ and so has no explicit $\lambda$ dependence. We can make $(x-z(\lambda))^\mu$ in \eqref{eq:naiverdef} an example of such a field\footnote{We should also in principle have $f$ have compact support, which can be accomplished by using $(x-z)^\mu h(x)$ where $h(x) = 1$ in a region containing $\Sigma \cap \bb{W}$ and decays to $0$ smoothly at finite $x$. Because $\Sigma\cap \bb{W}$ is spatially compact, we suffer no contributions from boundary terms from dropping this $h(x)$ in \eqref{eq:surfacejcons}.} by simply replacing it with $(x-z(\tau(x)))^\mu$. This is legitimate because on the region of integration, $\tau(x) = \lambda$. Using \eqref{eq:volmeasder}:
	\begin{equation}
		\d_\alpha (x-z(\tau(x)))^\mu = \delta^\mu_\alpha - \frac{\doot z{}^\mu \mc{n}_\alpha}{w\cdot \mc{n}}.
	\end{equation}
	Now with $f = (x-z)^\mu$ (it is legitimate to take a tensor field for $f$ instead of a scalar field in flat spacetime because global parallelism is well-defined), \eqref{eq:surfacejcons} becomes:
	\begin{equation}
		\mc{j}^\mu_0 = q\doot z{}^\mu + \doot {\mc{r}}  {}^\mu_1.
	\end{equation}
	Or more generally, for $f = (x-z)^{\rho_1}\dots(x-z)^{\rho_n}(x-z)^\mu$ with $n\ge 0$, we have:
	\begin{equation}
		\mc{j}_n^{(\rho_1\dots\rho_n\mu)} = \mc{r}_n^{(\rho_1\dots\rho_n} \doot z{}^{\mu)} + \frac{1}{n+1} \doot{\mc{r}}  {}^{\rho_1\dots\rho_n\mu}_{n+1}.
	\end{equation}
	This is the interrelationship of the naive multipole moments mandated by the continuity equation.
	
	In order to get the action contributions in \eqref{eq:jafunc} into a manifestly gauge invariant form, we need an antisymmetric pair of indices to turn $\d_\alpha A_\beta$ into $\fc{F}_{\alpha\beta}$. We can do this with some manipulations on $\mc{j}_n^{\cdots}$:
	\begin{align}
		(n+1) \mc{j}_n^{(\rho_1\dots\rho_n\mu)} &= \mc{j}_n^{\rho_1\dots\rho_n\mu} + n \mc{j}_n^{\mu(\rho_1\dots\rho_n)} = \mc{j}_n^{\rho_1\dots\rho_n\mu} + n(\mc{j}_n^{\mu(\rho_1\dots\rho_n)} - \mc{j}_n^{(\rho_1,,,\rho_n)\mu}) + n \mc{j}_n^{\rho_1\dots\rho_n\mu} \\
		&= (n+1) \mc{j}_n^{\rho_1\dots\rho_n\mu} - 2n \mc{j}_n^{(\rho_1\dots\rho_{n-1}[\rho_n)\mu]}. \\
		\implies &\ \mc{j}^{\rho_1\dots\rho_n\mu}_n = \mc{r}_n^{(\rho_1\dots\rho_n} \doot z{}^{\mu)} + \frac{1}{n+1} \doot{\mc{r}}  {}^{\rho_1\dots\rho_n\mu}_{n+1} + \frac{2n}{n+1} \mc{j}_n^{(\rho_1\dots\rho_{n-1}[\rho_n)\mu]}.
	\end{align}
	Returning our findings to \eqref{eq:jafunc}, simplifying, and dropping total $\lambda$ derivatives (because the vector potential is compactly supported): 
	\begin{align}
		\<J^\alpha, A_\alpha\> = \int_{-\8}^\8 \biggl(q \doot z{}^\mu A_\mu + \sum_{n=1}^\8 &\frac{1}{n!} \(\doot z{}^{(\mu} \mc{r}_n^{\rho_1\dots\rho_n)} - \doot z{}^{\rho_1} \mc{r}_n^{\rho_2\dots\rho_n\mu}\) \d^n_{\rho_1\dots\rho_n} A_\mu \nonumber \\
		+ \sum_{n=0}^\8 &\frac{n+1}{(n+2)!} \mc{j}^{\rho_1\dots\rho_n\mu\nu}_{n+1} \d^n_{\rho_1\dots\rho_n} \fc{F}_{\mu\nu}\biggr) D\lambda. \label{eq:6117}
	\end{align}
	We can use index symmetries to simplify further. In particular:
	\begin{align}
		(n+1)\doot z{}^{(\mu}\mc{r}_n^{\rho_1\dots\rho_n)} &= \doot z{}^{\mu} \mc{r}_n^{\rho_1\dots\rho_n} + n \doot z{}^{(\rho_1} \mc{r}_n^{\rho_2\dots\rho_n)\mu} \\
		&= (n+1) \doot z{}^{(\rho_1} \mc{r}_n^{\rho_2\dots\rho_n)\mu} + \doot z{}^\mu \mc{r}_n^{\rho_1\dots\rho_n} - \doot z{}^{(\rho_1} \mc{r}_n^{\rho_2\dots\rho_n)\mu} \\
		\implies &\doot z{}^{(\mu} \mc{r}_n^{\rho_1\dots\rho_n)} - \doot z{}^{(\rho_1} \mc{r}_n^{\rho_2\dots\rho_n)\mu} = \frac{2}{n+1} \mc{r}_n^{(\rho_1\dots\rho_{n-1}[\rho_n)} \doot z{}^{\mu]}.
	\end{align}
	Returning these to \eqref{eq:6117} and simplifying gives finally:
	\begin{equation}
		\<J^\alpha, A_\alpha\> = \int_{-\8}^\8\( q\doot z{}^\mu A_\mu + \sum_{n=0}^\8 \frac{n+1}{(n+2)!}Q_n^{\rho_1\dots\rho_n\mu\nu} \d^n_{\rho_1\dots\rho_n} \fc{F}_{\mu\nu}\) D\lambda \label{eq:flatjexpand}
	\end{equation}
	where we have defined the ``reduced field multipole moments'' of the current density by:
	\begin{equation}
		Q_n^{\rho_1\dots\rho_n\mu\nu}(\lambda) = \mc{j}^{\rho_1\dots\rho_n[\mu\nu]}_{n+1} + \frac{1}{n+1} \mc{r}^{\rho_1\dots\rho_n[\mu}_{n+1} \doot z{}^{\nu]}. \label{eq:flatqdef}
	\end{equation}
	We have now successfully put the action into the form of the point-particle term plus a tower of gauge invariant terms, each involving progressively higher spatial derivatives of the field strength! By expanding out their definition, these moments satisfy:
	\begin{align}
		&Q_n^{\rho_1\dots\rho_n\mu\nu} = Q_n^{(\rho_1\dots\rho_n)[\mu\nu]}, \p u_{\rho_1} Q_n^{\rho_1\dots\rho_n\mu\nu} = 0, \nonumber \\
		&Q_n^{\rho_1\dots\rho_{n-1} \rho_n\mu\nu} + Q_n^{\rho_1\dots\rho_{n-1}\mu\nu\rho_n} + Q_n^{\rho_1\dots\rho_{n-1}\nu\rho_n\mu} = 0. \label{eq:qmomcond} 
	\end{align}
	The first of these ensures (through the symmetry on $\rho_1\dots\rho_n$) that these moments don't contain any data that would be obfuscated by contracting into the partial derivatives $\d^n_{\rho_1\dots\rho_n} \fc{F}_{\mu\nu}$ and (through the antisymmetry on $\mu\nu$) that they don't contain any data that would be obfuscated by contracting into the field strength indices. The second ensures that the contraction into partial derivatives results in only taking spatial derivatives of the field strength. The third ensures that these moments don't contain any pieces that are annihilated by the electromagnetic Bianchi identity. Thus, the $Q_n^{\cdots}$ moments contain that and only that which is needed to write the action coupling of our current to electromagnetism. 
	
	Equation \eqref{eq:flatjexpand} is exactly what we wanted to accomplish. It and the index symmetries \eqref{eq:qmomcond} are the main results for this section. The rest of this section will motivate structures that will be useful for generalizing these to curved spacetime. Note that in the integral definition following from \eqref{eq:naivejdef}, \eqref{eq:naiverdef}, and \eqref{eq:flatqdef}, the integrand of moment $Q_n^{\cdots}$ always has $n+1$ factors of displacement from the worldline and consequently the total $Q_n^{\cdots}$ enters \eqref{eq:flatjexpand} coupled to the vector potential with $n+1$ derivatives. As such, in terms of the body size, $Q_n$ scales like $\ell^{n+1}$. $q \doot z{}^\mu$ is called the charge monopole moment, $Q_0$ the charge dipole moment (and can be decomposed into the familiar electric and magnetic dipoles), $Q_1$ the charge quadrupole moment, $Q_2$ the octopole, and so on. 
	
	\subsection{Reduced Multipole Moments}
	
	The $Q_n^{\cdots}$ moments provide a full description of the electromagnetic coupling (apart from the point-particle $q\doot z{}^\mu A_\mu$ term). We can expand out the field strength in the action \eqref{eq:flatjexpand} back into derivatives of the vector potential to achieve:
	\begin{equation}
		\<J^\alpha, A_\alpha\> = \int_{-\8}^\8\( q\doot z{}^\mu A_\mu + \sum_{n=1}^\8 \frac{2n}{(n+1)!}Q_{n-1}^{(\rho_1\dots\rho_n)\mu} \d^{n}_{\rho_1\dots\rho_{n}} A_\mu\) D\lambda.
	\end{equation}
	Then, if we set:
	\begin{equation}
		I_0^\mu = q\doot z{}^\mu, \p I_n^{\rho_1\dots\rho_n \mu} = \frac{2n}{n+1} Q_{n-1}^{(\rho_1\dots\rho_n)\mu} \label{eq:Imomdef}
	\end{equation}
	we automatically have that these satisfy:
	\begin{equation}
		I^{\rho_1\dots\rho_n\mu}_n = I^{(\rho_1\dots\rho_n)\mu}_n, \p I_n^{(\rho_1\dots\rho_n \mu)} = 0 \ps (n > 0), \p u_{\rho_1} I_n^{\rho_1\dots\rho_{n-1}[\rho_n\mu]} = 0 \ps (n > 1).  \label{eq:Imomprop}
	\end{equation}
	and that:
	\begin{equation}
		\<J^\alpha, A_\alpha\> = \int_{\-\8}^\8 \sum_{n=0}^\8 \frac{1}{n!} I^{\rho_1\dots\rho_n\mu}_n \d^n_{\rho_1\dots\rho_n} A_\mu(z) D\lambda \label{eq:flatjiexpand}
	\end{equation}
	for an arbitrary vector potential $A_\alpha(x)$. The $I_n^{\cdots}$ are called the ``reduced potential multipole moments'' of the current density. We refer to the $Q_n^{\cdots}$ as field moments because they naturally express $\<J^\alpha, A_\alpha\>$ in terms of the field strength in \eqref{eq:flatjexpand}, while we call the $I_n^{\cdots}$ potential moments because they do so in terms of the vector potential in \eqref{eq:flatjiexpand}. It is important to understand that both sets are multipole moments of the current density (not to be confused with literal multipole moments of the vector potential or the electromagnetic field strength, which we are not interested in in this work). Because of manifest gauge invariance, we tend to prefer expressions in terms of the field strength, and so the $Q_n^{\cdots}$ are more useful for calculations. The $Q_n^{\cdots}$ moments can be fully recovered from the potential moments by a little index manipulation. The result is simply:
	\begin{equation}
		Q_n^{\rho_1\dots\rho_n\mu\nu} = I_{n+1}^{\rho_1\dots\rho_n[\mu\nu]}. \label{eq:qfromi}
	\end{equation}
	
	We call our two sets of multipole moments reduced precisely because they have fully incorporated all of the constraints implied by the continuity equation. To see that this is the case, take \eqref{eq:funcjcons} and evaluate it using \eqref{eq:flatjiexpand}:
	\begin{equation}
		\<J^\alpha, \nabla_\alpha f\> = \int_{-\8}^\8\( I_0^\mu \d_\mu f + \sum_{n=1}^\8 \frac{1}{n!} I_n^{(\rho_1\dots\rho_n \mu)}\d^{n+1}_{\rho_1\dots\rho_n\mu} f\) D\lambda.
	\end{equation}
	This so far has not applied any of the properties of the multipole moments. If we now use \eqref{eq:Imomprop}, each term in the sum for $n \ge 1$ is identically 0 due to the symmetrization. The sole remaining term is $I_0^\mu\d_\mu f$ can be computed using the explicit form of $I_0^\mu$ in \eqref{eq:Imomdef} 
	\begin{equation}
		\<J^\alpha, \nabla_\alpha f\> = q \left. f(z(\lambda))\right|^{\lambda\to\8}_{\lambda\to-\8} = 0
	\end{equation}
	because $f$ is compactly supported. Therefore, the continuity equation is automatic from the reduced multipole moments and constrains them no further. 
	
	Equation \eqref{eq:flatjiexpand} formally provides a distributional expression for the current density by varying it with respect to $A_\mu$. In particular:
	\begin{equation}
		J^\mu(x) = \int_{-\8}^\8 \sum_{n=0}^\8 \frac{(-1)^n}{n!} I^{\nu_1\dots\nu_n\mu}_n(\lambda) \d^n_{\nu_1\dots\nu_n}\delta(x-z(\lambda)) D\lambda. 
	\end{equation}
	Unfortunately, formulas of this form do not generalize very well to curved spacetime, because Dirac deltas in curved spacetime are very poorly defined. 
	
	As a note on terminology, Dixon, in his original papers, only referred to the $I_n^{\cdots}$ as multipole moments and called them just the ``reduced multipole moments''. The $I_n^{\cdots}$ moments are natural in the sense that they uniquely satisfy \eqref{eq:flatjiexpand} while having no interdependencies. We extend the name multipole moments to the $Q_n^{\cdots}$, which are natural in the sense that they satisfy \eqref{eq:flatjexpand} and \eqref{eq:qmomcond}. The $Q_n^{\cdots}$ are more convenient in almost every calculation, due to coupling directly to the field strength, and contain the familiar charge multipoles in familiar form. In particular, the dipole moments $Q_0^{\mu\nu}$ can be decomposed into an electric part (which becomes the electric dipole moment) and a magnetic part (which becomes the magnetic dipole moment) via the same decomposition one would apply to separate a field strength $\fc{F}_{\mu\nu}$ into an electric field and a magnetic field.
	
	\subsection{Moment Generating Function}
	
	In the case of flat spacetime, the curved spacetime Fourier transform \eqref{eq:ftdef} simplifies to:
	\begin{equation}
		\twid{A}_\mu(z, k) = \int e^{-ik_\nu (x-z)^\nu} A_\mu(x) d^4 x. 
	\end{equation}
	This is unlike the Newtonian case in that we perform integration over all of spacetime rather than just a time-slice. This is very useful because it maintains covariance. If we had only integrated over $\Sigma$, then we would have also induced an unnecessarily complicated explicit $\lambda$ dependence. Because $\twid{A}_\mu(z, k)$ is now a function of 8 variables formed this way from a function of only~4, $\twid{A}_\mu(z, k)$ is not an arbitrary function; it is quite heavily restricted. In particular, the $\lambda$ dependence factors out:
	\begin{equation}
		\twid{A}_\mu(z,k) = e^{ik\cdot z} \twid{A}_\mu(0,k). 
	\end{equation}
	In analogy to the Newtonian case, we define the moment generating function $I^\mu(\lambda, k)$ for the current density by:
	\begin{equation}
		I^\mu(\lambda,k) = \sum_{k=0}^\8 \frac{(-i)^n}{n!} I_n^{\rho_1\dots\rho_n\mu}(\lambda) k_{\rho_1}\dots k_{\rho_n}.
	\end{equation}
	This is \textit{not} simply the Fourier transform of the current density. But, analogously to \eqref{eq:nmomgen}, it still follows from \eqref{eq:flatjiexpand} that:
	\begin{equation}
		\<J^\alpha, A_\alpha\> = I^\mu[A_\mu] = \int_{-\8}^\8\int I^{\mu*}(\lambda, k) \twid{A}_\mu(\lambda, k) \frac{d^4k}{(2\pi)^4} D\lambda \label{eq:imomgendef}
	\end{equation}
	where the last expression defines the moment functional $I^\mu[A_\mu]$. 
	
	\section{Gravity near Minkowski Space} \label{sec:grfs}
	
	We now take inspiration from electromagnetism and try to find an expression for the multipole moments of the energy-momentum tensor in flat spacetime by studying its interaction with linearized gravity. The local conservation of energy-momentum in flat spacetime is Noether-dual to the gauge invariance of linearized gravity, and much of the same structure relating the current density and the electromagnetic vector potential will repeat itself here. In this section we will neglect electromagnetism, meaning we set the current density $J^\alpha = 0$. Because we are still in flat spacetime, the surfaces $\Sigma(\lambda)$ retain their expression from~\eqref{eq:bigsigflatdef}. 
	
	In flat spacetime and in the absence of an electromagnetic current density, the energy-momentum transport equation \eqref{eq:tcons} becomes:
	\begin{equation}
		\d_\beta T^{\alpha\beta} = 0. \label{eq:flattcons}
	\end{equation}
	Thus, like the current density, the different components of the energy-momentum tensor are interdependent with each other. We can express this as a functional statement by integrating it against an arbitrary vector field $f_\alpha(x)$. Again, because of the compact support of $f_\alpha$, we can integrate the derivative by parts onto it to arrive at:
	\begin{equation}
		\int T^{\alpha\beta} \d_\beta f_\alpha d^4 x = 0.  \label{eq:flatfunct}
	\end{equation}
	Alternatively, we can also integrate $\d_\beta (T^{\alpha\beta} f_\alpha)$ over a time-slice and use \eqref{eq:volintder} to arrive at an equivalent condition that gives a nice formula for time derivatives:
	\begin{equation}
		\frac{1}{\ec}\frac{d}{d\lambda}\int_\Sigma T^{\alpha\beta} f_\alpha d\Sigma_\beta = \int_\Sigma T^{\alpha\beta}\d_\beta f_\alpha d\Sigma_w. \label{eq:spacetcons}
	\end{equation}
	Because we are in flat spacetime, we can replace $f_\alpha(x)$ by $c_\alpha f(x)$ where $c_\alpha$ is an arbitrary constant vector and $f(x)$ is an arbitrary scalar function. Then, because the vector $c_\alpha$ is arbitrary, we can differentiate with respect to it to yield:
	\begin{equation}
		\frac{1}{\ec}\frac{d}{d\lambda}\int_\Sigma f(x) T^{\alpha\beta} d\Sigma_\beta = \int_\Sigma T^{\alpha\beta}\d_\beta f(x) d\Sigma_w.  \label{eq:spacetcons2}
	\end{equation}
	Like \eqref{eq:surfacejcons}, this will be the useful form for studying the impact of energy-momentum conservation on multipole moments. 
	
	\subsection{Gauge Theory of Gravitational Perturbations}
	
	In full curved spacetime, the metric satisfies Einstein's equations:
	\begin{equation}
		R_{\alpha\beta} - \frac{1}{2}Rg_{\alpha\beta} = 8\pi G T_{\alpha\beta}
	\end{equation}
	which are naturally produced by the Einstein-Hilbert action:
	\begin{equation}
		\fc{S}_{\text{EH}} = \int \frac{R}{16\pi G} D^4 x + \fc{S}_m + \fc{S}_{\text{GHY}}
	\end{equation}
	where $\fc{S}_m$ is the action for the matter and $\fc{S}_{\text{GHY}}$ is the Gibbons-Hawking-York boundary term (which is independent of the matter fields, only depends on the metric and it's derivatives at infinity, and does not affect physics in bulk). The energy-momentum tensor is determined from the full action $\fc{S}_m$ of the body by:
	\begin{equation}
		T^{\alpha\beta}(x) = \frac{2}{\sqrt{-\det g}} \frac{\delta \fc{S}_m}{\delta g_{\alpha\beta}} \\
		\implies \delta \fc{S}_m = \frac{1}{2}\int T^{\alpha\beta} \delta g_{\alpha\beta} D^4 x \label{eq:stressdef}
	\end{equation}
	under a metric variation. 
	
	Now consider the metric ${g}_{\alpha\beta}$ to be perturbed by $\eps h_{\alpha\beta}$ from the Minkowski metric $\eta_{\alpha\beta}$ with $\eps$ a small parameter:
	\begin{equation}
		{g}_{\alpha\beta} = \eta_{\alpha\beta} + \eps h_{\alpha\beta}.
	\end{equation}
	The action $\fc{S}_m$ can then be series expanded in $\eps$:
	\begin{equation}
		\fc{S}_m = \fc{S}_0 + \eps \fc{S}_\ixt + \OO(\eps^2)
	\end{equation}
	where $\fc{S}_0$ is the action of the body in Minkowski space and $\fc{S}_\ixt$ gives the coupling of the body to the metric perturbation. Automatically, with $T^{\alpha\beta}$ meaning the energy-momentum tensor of the body in Minkowski space:
	\begin{equation}
		\fc{S}_{\ixt} = \frac{1}{2}\int T^{\alpha\beta}(x) h_{\alpha\beta}(x) d^4 x = \frac{1}{2}\<T^{\alpha\beta}, h_{\alpha\beta}\>. \label{eq:716}
	\end{equation}
	Therefore, the interactions of the body with linearized gravity are characterized by the behavior of the energy-momentum tensor as a functional on metric perturbations. 
	
	We will seek to express this coupling in a form which is (as much as possible) manifestly gauge invariant. The gauge transformations of general relativity are diffeomorphisms that are connected to the identity transformation. So, because we are working with gravitational perturbations and need to stay near the Minkowski metric, the gauge transformations are compactly supported infinitesimal diffeomorphisms. Let $x'^\alpha = x^\alpha + \eps \xi^\alpha(x)$ be a diffeomorphism with again $\eps$ small. The metric transforms as:
	\begin{equation}
		g'_{\alpha\beta}(x') = \frac{\d x^\gamma}{\d x'^\alpha} \frac{\d x^\delta}{\d x'^\beta} g_{\gamma\delta}(x).
	\end{equation}
	To linear order in $\eps$:
	\begin{equation}
		\frac{\d x^\gamma}{\d x'^\alpha} = \delta^\gamma_\alpha - \eps\d_\alpha \xi^\gamma + \OO(\eps^2).
	\end{equation}
	Thus, expanding ${g}'_{\alpha\beta}(x')$ in terms of the transformed metric perturbation $h'_{\alpha\beta}(x)$ as:
	\begin{equation}
		{g}'_{\alpha\beta}(x') = \eta_{\alpha\beta} +\eps h'_{\alpha\beta}(x') + \OO(\eps^2)
	\end{equation}
	we have:
	\begin{equation}
		h'_{\alpha\beta}(x) = h_{\alpha\beta}(x) - \d_\alpha \xi_\beta - \d_\beta \xi_\alpha
	\end{equation}
	where indices are raised and lowered using only the Minkowski metric. This is the gauge transformation property of metric perturbations.
	
	Clearly the value of $h_{\alpha\beta}$ at any one point is not gauge invariant. What about its first derivatives? The first derivatives of the metric are linearly related to the connection $\Gamma^\gamma{}_{\alpha\beta}$ and it is always possible to go to local coordinates so that the connection coefficients vanish at a given point (just use normal coordinates). Thus, the first derivatives of $h_{\alpha\beta}$ at a point also contain no gauge invariants. What about its second derivatives? It is always possible to go to local coordinates (again, normal coordinates) so that the second derivatives of the metric are fully determined by the Riemann tensor in those coordinates. Thus, if there are any gauge invariants of the second derivative, they must be contained in the Riemann tensor. Expanding the definition of the connection gives that its perturbation is:
	\begin{equation}
		{\Gamma}^\gamma{}_{\alpha\beta} = \frac{\eps}{2}(\d_\alpha h^\gamma{}_\beta + \d_\beta h^\gamma{}_\alpha - \d^\gamma h_{\alpha\beta}) + \OO(\eps^2).
	\end{equation}
	The variation of the definition of the Riemann tensor yields:
	\begin{equation}
		\delta R^\gamma{}_{\alpha\delta\beta} = \nabla_\delta \delta\Gamma^\gamma{}_{\alpha\beta} - \nabla_\beta\delta\Gamma^\gamma{}_{\alpha\delta}.
	\end{equation}
	Consequently, the linearized Riemann tensor is:
	\begin{equation}
		R_{\gamma\delta\alpha\beta} = \eps \delta R_{\gamma\delta\alpha\beta} + \OO(\eps^2).
	\end{equation}
	where: 
	\begin{equation}
		\delta R_{\gamma\delta\alpha\beta} = \frac{1}{2}(\d^2_{\alpha\delta} h_{\beta\gamma} - \d^2_{\beta\delta} h_{\alpha\gamma} -\d^2_{\alpha\gamma} h_{\beta\delta} + \d^2_{\beta\gamma} h_{\alpha\delta}). \label{eq:linriem}
	\end{equation}
	A short calculation reveals, in fact the entire linearized Riemann tensor $\delta R_{\gamma\delta\alpha\beta}$ is gauge invariant at leading order in $\eps$. Thus, it is the natural definition of the field strength for linearized gravity. 
	
	Due to the index symmetries of the Riemann tensor:
	\begin{equation}
		\delta R_{\gamma\delta\alpha\beta} = \delta R_{[\gamma\delta][\alpha\beta]} = \delta R_{\alpha\beta\gamma\delta}, \p \delta R_{\gamma\delta\alpha\beta} + \delta R_{\gamma\alpha\beta\delta} + \delta R_{\gamma\beta\delta\alpha} = 0
	\end{equation}
	and due to the differential Bianchi identity:
	\begin{equation}
		\d_{\epsilon}\delta R_{\gamma\delta\alpha\beta} + \d_\gamma\delta R_{\delta\epsilon\alpha\beta}+\d_\delta\delta R_{\epsilon\gamma\alpha\beta} = 0.
	\end{equation}
	Therefore, in analogy with the electromagnetic case, instead of expressing $\fc{S}_\ixt$ in terms of derivatives of $h$, we seek an expression in terms of derivatives of $\delta R$.
	
	\subsection{Free Body Linear Momentum and Spin}
	
	Due to the Poincaré symmetries of Minkowski space and our present neglect of electromagnetism, $T$ is the energy-momentum tensor of an isolated body. So, we expect its linear and angular momenta to be conserved (at least, if we choose $(z, u)$ appropriately). Let us see how these actually arise out of \eqref{eq:flattcons}. 
	
	We can series expand $h_{\alpha\beta}(x)$ about the worldline just like we did with $A_\alpha(x)$:
	\begin{equation}
		h_{\alpha\beta}(x) = \sum_{n=0}^\8 \frac{1}{n!}(x-z)^{\rho_1}\dots(x-z)^{\rho_n} \d^n_{\rho_1\dots\rho_n} h_{\alpha\beta}(z). 
	\end{equation}
	Returning this to the action \eqref{eq:716}:
	\begin{equation}
		\<T^{\alpha\beta}, h_{\alpha\beta}\> = \int_{-\8}^\8 \sum_{n=0}^\8\frac{1}{n!} \mc{t}^{\rho_1\dots\rho_n\mu\nu}_n \d^n_{\rho_1\dots\rho_n} h_{\mu\nu}(z) D\lambda  \label{eq:flathexpandt}
	\end{equation}
	where we have defined the ``naive multipole moments'' of the energy-momentum tensor:
	\begin{equation}
		\mc{t}^{\rho_1\dots\rho_n\mu\nu}_n(\lambda) = \int_{\Sigma(\lambda)}(x-z(\lambda))^{\rho_1}\dots(x-z(\lambda))^{\rho_n} T^{\mu\nu}(x) d\Sigma_w.  \label{eq:naivetdef}
	\end{equation}
	These moments are highly interdependent due to \eqref{eq:flattcons} and consequently do not put in the action contribution \eqref{eq:flathexpandt} into a manifestly gauge invariant form.
	
	Because $\mc{n}^\alpha$, the unit vector normal to $\Sigma$ at $x$, locally represents the time direction, $-T^{\alpha\beta} \mc{n}_\beta$ locally represents the linear momentum density. We can therefore define naive moments of the linear momentum density by:
	\begin{equation}
		\mc{p}_n^{\rho_1\dots\rho_n\mu}(\lambda) = \int_{\Sigma(\lambda)} (x-z)^{\rho_1}\dots(x-z)^{\rho_n} T^{\mu\nu} d\Sigma_\nu. \label{eq:naivepdef}
	\end{equation}
	Because of the energy-momentum transport equation, the time evolution of the linear momentum density is determined by the other components of the energy-momentum tensor. We can see this by using \eqref{eq:spacetcons2} to relate the $\mc{p}_n^{\cdots}$ and the $\mc{t}_n^{\cdots}$. For $f = 1$ and taking the Minkowski spacetime case of \eqref{eq:psdef}, \eqref{eq:spacetcons2} gives:
	\begin{equation}
		\mc{p}_0^\mu = p^\mu = \int_\Sigma T^{\mu\nu} d\Sigma_\nu, \p \frac{dp^\mu}{d\lambda} = 0. \label{eq:freepcons}
	\end{equation}
	Unsurprisingly, the total linear momentum of the body is conserved! Immediately, the mass $m$ of the body\footnote{We use $m$ for the constant mass of a free body in flat spacetime, as opposed to $\fc{M}$, the dynamical mass of a body genuinely interacting with electromagnetism and/or general relativity.} is constant:
	\begin{equation}
		m = |p| = \sqrt{-p_\mu p^\mu}.
	\end{equation}
	
	For $f = (x-z(\tau(x)))^\alpha$, \eqref{eq:spacetcons2} yields:
	\begin{equation}
		\doot{\mc{p}}{}^{\mu\nu}_1 = \mc{t}^{\mu\nu}_0 - \doot z{}^\mu p^\nu. \label{eq:p1dot}
	\end{equation}
	By the Minkowski spacetime case of \eqref{eq:psdef}, the spin of the body is given by:
	\begin{equation}
		S^{\mu\nu}(\lambda) = 2\mc{p}_1^{[\mu\nu]} = \int_{\Sigma(\lambda)}( (x-z)^\mu T^{\nu\rho} - (x-z)^\nu T^{\mu\rho}) d\Sigma_\rho. \label{eq:flatsdef}
	\end{equation}
	The antisymmetric part of \eqref{eq:p1dot} implies:
	\begin{equation}
		\doot S{}^{\mu\nu} = p^\mu \doot z{}^{\nu} - \doot z{}^\mu p^\nu. \label{eq:flatsdot}
	\end{equation}
	while the symmetric part gives:
	\begin{equation}
		\mc{t}^{\mu\nu}_0 = \doot z{}^{(\mu}p^{\nu)} + \doot{\mc{p}}{}^{(\mu\nu)}_1.
	\end{equation}
	It might be surprising that $S^{\mu\nu}$ is not automatically conserved. However, recall that we haven't yet put any conditions on the worldline $z^\mu$ or chosen what $u^\mu$ should be. In the Newtonian case in the absence of external forces, the spin vector and intrinsic boost generator are only conserved for a special choice of the trajectory (the center of mass). The same is true here. If we choose $(z, u)$ to satisfy \eqref{eq:udef} and \eqref{eq:cssc}, then it immediately follows from \eqref{eq:freepcons} and \eqref{eq:flatsdot} that $p^\mu = m\doot z{}^\mu$ and $\doot S{}^{\mu\nu} = 0$. 
	
	\subsection{Gauge Invariant form of the Action}
	
	Now let's finish finding all the relations between the $\mc{t}_n^{\cdots}$ and the $\mc{p}_n^{\cdots}$ and to using them to simplify $\<T^{\alpha\beta}, h_{\alpha\beta}\>$. For $f = (x-z)^{\rho_1}\dots(x-z)^{\rho_n}$, \eqref{eq:spacetcons2} becomes:
	\begin{gather}
		\doot{\mc{p}}{}_n^{\rho_1\dots\rho_n\mu} = n \mc{t}_{n-1}^{(\rho_1\dots\rho_n)\mu} - n \doot z{}^{(\rho_1} \mc{p}_{n-1}^{\rho_2\dots\rho_n)\mu}. \label{eq:pndot}
	\end{gather}
	For the special case of \eqref{eq:pndot} with $n=2$:
	\begin{equation}
		\mc{t}_1^{(\kappa\mu)\nu} = \doot z{}^{(\kappa} \mc{p}_1^{\mu)\nu} + \frac{1}{2}\doot{\mc{p}}{}_2^{\kappa\mu\nu}.
	\end{equation}
	Because $\mc{t}_1^{\kappa\mu\nu} = \mc{t}_1^{\kappa\nu\mu}$, it follows that:
	\begin{align}
		\mc{t}_1^{\kappa\mu\nu} = \mc{t}_1^{(\kappa\mu)\nu} + \mc{t}_1^{(\kappa\nu)\mu} - \mc{t}_1^{(\mu\nu)\kappa} = \doot z{}^\kappa \mc{p}_1^{(\mu\nu)} + S^{\kappa(\mu} \doot z{}^{\nu)} + \doot{\mc{p}}  {}_2^{\kappa(\mu\nu)} - \frac{1}{2}\doot {\mc{p}}  {}_2^{\mu\nu\kappa}.
	\end{align}
	Returning these low order results to \eqref{eq:flathexpandt} and dropping total time derivatives (because the metric perturbation is compactly supported):
	\begin{align}
		\<T^{\alpha\beta}, h_{\alpha\beta}\> = \int_{-\8}^\8&\biggl(\doot z{}^\mu p^\nu h_{\mu\nu} + S^{\kappa\mu}\doot z{}^{\nu} \d_\kappa h_{\mu\nu}  - \mc{p}_2^{\rho\mu\nu} \doot z{}^{\sigma} \d^2_{\rho\sigma}h_{\mu\nu} + \frac{1}{2}\mc{p}_2^{\mu\nu\rho} \doot z{}^\sigma \d^2_{\rho\sigma} h_{\mu\nu}  \nonumber \\
		&\ \ \  + \sum_{n=2}^\8 \frac{1}{n!} \mc{t}_n^{\kappa_1\dots\kappa_n \mu\nu} \d^n_{\kappa_1\dots\kappa_n} h_{\mu\nu}\biggr) D\lambda. \label{eq:hexpandpartsimp}
	\end{align}
	We've now successfully separated out all terms with fewer than two derivatives on $h$, and so next we will try to organize all of the second derivative and higher terms into gauge invariant form purely in terms of the linearized Riemann tensor and its derivatives. 
	
	In order to make the contributions from the moments to the action  gauge invariant, we need two pairs of antisymmetric indices so that when contracted with $\d^2_{\rho\sigma}h_{\mu\nu}$ they produce $\delta R_{\rho\mu\sigma\nu}$. A term with $n+2$ derivatives on $h$ will become a term with $n$ derivatives on $\delta R$. We can accomplish our desired reorganization with some manipulations on $\mc{t}_n^{\cdots}$. In order to keep the equations somewhat organized, we introduce two new sets of coefficients. Define:
	\begin{equation}
		\mc{x}_{n}^{\kappa_1\dots\kappa_n\rho\mu\sigma\nu} = \mc{t}_{n+2}^{(\kappa_1\dots\kappa_n\rho\mu\sigma)\nu}, \p \mc{y}_{n}^{\kappa_1\dots\kappa_n\rho\mu\sigma\nu} = \mc{t}_{n+2}^{\kappa_1\dots\kappa_n[\rho['\sigma\mu] \nu]'}. \label{eq:735}
	\end{equation}
	The $\mc{x}_n^{\cdots}$ coefficients are determined by \eqref{eq:pndot}:
	\begin{equation}
		\mc{x}_{n}^{\kappa_1\dots\kappa_n\rho\mu\sigma\nu} =  \doot z{}^{(\mu} \mc{p}_{n+2}^{\kappa_1\dots\kappa_n\rho\sigma)\nu} + \frac{1}{n+3} \doot{\mc{p}}{}_{n+3}^{\kappa_1\dots\kappa_n\rho\sigma\mu\nu}.
	\end{equation}
	The $\mc{y}_n^{\cdots}$ have the desired double-antisymmetry to help form the linearized Riemann tensor. So, we want to decompose $\mc{t}_{n+2}^{\cdots}$ into a linear combination of $\mc{x}_n^{\cdots}$ and $\mc{y}_n^{\cdots}$. 
	
	We can try to recover $\mc{t}_{n+2}^{\cdots}$ from the definition of $\mc{y}_n^{\cdots}$ by symmetrizing sufficiently. By definition:
	\begin{equation}
		4\mc{y}_n^{\kappa_1\dots\kappa_n\rho\mu\sigma\nu} = \mc{t}_{n+2}^{\kappa_1\dots\kappa_n\rho\sigma\mu\nu} - \mc{t}_{n+2}^{\kappa_1\dots\kappa_n\mu\sigma\rho\nu} - \mc{t}_{n+2}^{\kappa_1\dots\kappa_n\rho\nu\mu\sigma} + \mc{t}_{n+2}^{\kappa_1\dots\kappa_n\mu\nu\rho\sigma}.
	\end{equation}
	When contracted with $\d^{n+2}_{\kappa_1\dots\kappa_n\rho\sigma}h_{\mu\nu}$, the above will only contribute when $\rho\sigma$ are symmetrized together with the $\kappa_1\dots\kappa_n$, so we compute $\mc{y}_n^{(\kappa_1\dots\kappa_n\rho|\mu|\sigma)\nu}$. Applied to the above and using the index symmetries of $t_{n+2}^{\cdots}$:
	\begin{equation}
		4\mc{y}_n^{(\kappa_1\dots\kappa_n\rho|\mu|\sigma)\nu} = \mc{t}_{n+2}^{\kappa_1\dots\kappa_n\rho\sigma\mu\nu} - \mc{t}_{n+2}^{\mu(\kappa_1\dots\kappa_n\sigma\rho)\nu} - \mc{t}_{n+2}^{\nu(\kappa_1\dots\kappa_n\rho\sigma)\mu} + \mc{t}_{n+2}^{\mu\nu(\kappa_1\dots\kappa_n\rho\sigma)}. \label{eq:ysymexpand}
	\end{equation}
	The first term on the right hand side is already $\mc{t}_{n+2}^{\cdots}$, which is precisely what we want. The second and third terms match in form, and we will compute them next. The fourth term is more complicated.
	
	Beginning with the second/third terms in \eqref{eq:ysymexpand}, consider $\mc{t}_{n+2}^{(\kappa_1\dots\kappa_n\rho\sigma\mu)\nu}$. This expression contains $(n+3)!$ terms which can be organized into $(n+2)!$ terms in which $\mu$ appears in the second to last slot and $(n+2)(n+2)!$ terms in which $\mu$ does not appear in the second to last slot. Therefore:
	\begin{align}
		&(n+3)\mc{t}_{n+2}^{(\mu\kappa_1\dots\kappa_n\rho\sigma)\nu} = \mc{t}_{n+2}^{\kappa_1\dots\kappa_n\rho\sigma\mu\nu} + (n+2)\mc{t}_{n+2}^{\mu(\kappa_1\dots\kappa_n\rho\sigma)\nu} \\
		\implies &\mc{t}_{n+2}^{\mu(\kappa_1\dots\kappa_n\rho\sigma)\nu} = \frac{n+3}{n+2}\mc{x}_n^{\kappa_1\dots\kappa_n\rho\mu\sigma\nu} - \frac{1}{n+2}\mc{t}_{n+2}^{\kappa_1\dots\kappa_n\rho\sigma\mu\nu}. \label{eq:7310}
	\end{align}

	Now let us work on the fourth term in \eqref{eq:ysymexpand}. First consider $\mc{t}_{n+2}^{(\kappa_1\dots\kappa_n\rho\sigma\mu\nu)}$. Because this is fully symmetrized, we can trivially symmetrize any indices within it first applying the full symmetrization. Therefore: 
	\begin{equation}
		\mc{t}_{n+2}^{(\kappa_1\dots\kappa_{n}\rho\mu\sigma\nu)} = \mc{t}_{n+2}^{((\kappa_1\dots\kappa_{n}\rho\mu\sigma)\nu)} = \mc{x}_{n}^{(\kappa_1\dots\kappa_n\rho\sigma\mu\nu)}.
	\end{equation}
	Alternatively, $\mc{t}_{n+2}^{(\kappa_1\dots\kappa_n\rho\sigma\mu\nu)}$ can be decomposed into $(n+2)(n+3)!$ terms in which $\mu$ appears in neither of the last two slots and $2(n+3)!$ terms in which $\mu$ does appear in one of the last two slots. Thus:
	\begin{align}
		&(n+4)\mc{t}_{n+2}^{(\kappa_1\dots\kappa_n\rho\sigma\mu\nu)} = (n+2)\mc{t}_{n+2}^{\mu(\kappa_1\dots\kappa_n\rho\sigma\nu)} + 2\mc{t}_{n+2}^{(\kappa_1\dots\kappa_n\rho\sigma\nu)\mu} \\
		\implies &\mc{t}^{\mu(\kappa_1\dots\kappa_n\rho\sigma\nu)}_{n+2} = \frac{n+4}{n+2} \mc{x}_{n+2}^{(\kappa_1\dots\kappa_n\rho\sigma\mu\nu)} - \frac{2}{n+2} \mc{x}_{n+2}^{\kappa_1\dots\kappa_n\rho\sigma\nu\mu}. \label{eq:7313}
	\end{align}
	The left hand side of the above as a sum contains $(n+3)!$ terms which can be divided into $(n+1)(n+2)!$ terms in which $\nu$ appears in neither of the last two slots and $2(n+2)!$ terms in which $\nu$ does appear in either of the last two slots. Thus, using \eqref{eq:7310} and \eqref{eq:7313}:
	\begin{align}
		(n+3)\mc{t}^{\mu(\kappa_1\dots\kappa_n\rho\sigma\nu)}_{n+2} =&\ (n+1)\mc{t}^{\mu\nu(\kappa_1\dots\kappa_n\rho\sigma)} + 2\mc{t}^{\mu(\kappa_1\dots\kappa_n\rho\sigma)\nu} \\
		\implies \mc{t}^{\mu\nu(\kappa_1\dots\kappa_n\rho\sigma)}_{n+2} =&\ \frac{(n+3)(n+4)}{(n+1)(n+2)}\mc{x}_{n}^{(\kappa_1\dots\kappa_n\rho\sigma\mu\nu)} - \frac{4(n+3)}{(n+1)(n+2)}\mc{x}_n^{\kappa_1\dots\kappa_n\rho\sigma(\mu\nu)}\nonumber\\
		&\ + \frac{2}{(n+1)(n+2)}\mc{t}_{n+2}^{\kappa_1\dots\kappa_n\rho\sigma\mu\nu}. \label{eq:7315}
	\end{align}
	
	Using \eqref{eq:7310} and \eqref{eq:7315}, we can simplify \eqref{eq:ysymexpand} to give:
	\begin{equation}
		\mc{t}_{n+2}^{\kappa_1\dots\kappa_n\rho\sigma\mu\nu} = 4 \frac{n+1}{n+3} \mc{y}_n^{(\kappa_1\dots\kappa_n\rho|\mu|\sigma)\nu} + 2\frac{n+3}{n+2} \mc{x}_n^{\kappa_1\dots\kappa_n\rho\sigma(\mu\nu)} - \frac{n+4}{n+2}\mc{x}_n^{(\kappa_1\dots\kappa_n\rho\sigma\mu\nu)}. \label{eq:txysimp}
	\end{equation}
	Returning this to \eqref{eq:hexpandpartsimp} and simplifying:
	\begin{align}
		\<T^{\alpha\beta}, h_{\alpha\beta}\> =& \int_{-\8}^\8\left(\doot z{}^\mu p^\nu h_{\mu\nu} + S^{\kappa\mu} \doot z{}^\nu \d_\kappa h_{\mu\nu} - \mc{p}_2^{\rho\mu\nu}\doot z{}^\sigma \d^2_{\rho\sigma}h_{\mu\nu} + \frac{1}{2}\mc{p}_2^{\mu\nu\rho} \doot z{}^\sigma \d^2_{\rho\sigma}h_{\mu\nu} \right. \nonumber \\
		&\ps \ps \ps \ps \  -\sum_{n=0}^\8 2 \frac{n+1}{(n+3)!} \mc{t}_{n+2}^{\kappa_1\dots\kappa_n[\rho['\sigma \mu]\nu]'}\d^n_{\kappa_1\dots\kappa_n} \delta R_{\rho\mu\sigma\nu} \nonumber \\
		&\ps \ps \ps \ps \ +\sum_{n=2}^\8 \frac{2(n+1)}{n!n}\(\frac{\doot{\mc{p}}  {}^{\kappa_1\dots\kappa_n\mu\nu}_{n+1}}{n+1} + \doot z{}^{(\mu} \mc{p}_n^{\kappa_1\dots\kappa_n)\nu}\) \d^n_{\kappa_1\dots\kappa_n} h_{\mu\nu} \nonumber \\
		&\ps \ps \ps \ps \left. \vphantom{\int} - \sum_{n=2}^\8 \frac{n+2}{n! n} \(\frac{ \doot{\mc{p}}  {}^{(\kappa_1\dots\kappa_n\mu\nu)}_{n+1}}{n+1} + \doot z{}^{(\mu} \mc{p}_n^{\kappa_1\dots\kappa_n\nu)}\) \d^n_{\kappa_1\dots\kappa_n} h_{\mu\nu}\) D\lambda.
	\end{align}
	The time derivatives on the $\mc{p}_{n+1}^{\cdots}$ pieces can be integrated by parts onto the metric perturbation, dropping total time derivatives for the same reason as before. Doing so and simplifying yields:
	\begin{align}
		\<T^{\alpha\beta}, h_{\alpha\beta}\> =& \int_{-\8}^\8\left(\doot z{}^\mu p^\nu h_{\mu\nu} + S^{\kappa\mu} \doot z{}^\nu \d_\kappa h_{\mu\nu} -\sum_{n=0}^\8 2 \frac{n+1}{(n+3)!} \mc{t}_{n+2}^{\kappa_1\dots\kappa_n[\rho['\sigma \mu]\nu]'}\d^n_{\kappa_1\dots\kappa_n} \delta R_{\rho\mu\sigma\nu} \right. \nonumber \\
		&\ps \ps \ps\ps +\sum_{n=2}^\8 \frac{1}{n!}\left(2\frac{n+1}{n}\doot z{}^{(\mu}\mc{p}_n^{\kappa_1\dots\kappa_n)\nu} - \frac{n+2}{n}\doot z{}^{(\mu}\mc{p}_n^{\kappa_1\dots\kappa_n\nu)} \right. \nonumber \\
		&\ps \ps \ps \ps \ps \ps \ps \ps \ps  \left. \vphantom{\int} \left. - \frac{2n}{n-1}\doot z{}^{\kappa_1}\mc{p}_n^{\kappa_2\dots\kappa_n\mu\nu} + \frac{n+1}{n-1}\doot z{}^{\kappa_1}\mc{p}_n^{(\kappa_2\dots\kappa_n\mu\nu)}\)\d^n_{\kappa_1\dots\kappa_n}h_{\mu\nu}\right) D\lambda.
	\end{align}
	We can now take the above and expand out the remaining symmetrizations in the second and third lines. With a significant amount of algebra, the result of doing so is:
	\begin{equation}
		\<T^{\alpha\beta}, h_{\alpha\beta}\> = \int_{-\8}^\8\left(\doot z{}^\mu p^\nu h_{\mu\nu} + S^{\kappa\mu} \doot z{}^\nu \d_\kappa h_{\mu\nu} -2\sum_{n=0}^\8 \frac{n+1}{(n+3)!}\fc{J}^{\kappa_1\dots\kappa_n\rho\mu\sigma\nu}_n \d^n_{\kappa_1\dots\kappa_n}\delta R_{\rho\mu\sigma\nu} \right)D\lambda \label{eq:flatthexpand}
	\end{equation}
	where we have defined the $\fc{J}_n^{\cdots}$ ``reduced field multipole moments'' of the energy-momentum tensor by:
	\begin{equation}
		\fc{J}_n^{\kappa_1\dots\kappa_n\rho\mu\sigma\nu} = \mc{t}_{n+2}^{\kappa_1\dots\kappa_n[\rho['\sigma\mu]\nu]'} + \frac{1}{n+1}\(\mc{p}_{n+2}^{\kappa_1\dots\kappa_n[\rho['\sigma\mu]}\doot z{}^{\nu]'} + \mc{p}_{n+2}^{\kappa_1\dots\kappa_n[\rho['\sigma\nu]'}\doot z{}^{\mu]}\).\label{eq:flatjdef}
	\end{equation}
	Equation \eqref{eq:flatthexpand} is the desired form of the interaction action (up to a factor of $\frac{1}{2}$). Only the first two terms are not obviously gauge invariant, but we will show that their gauge invariance is implied by the equations of motion for the momentum and spin. 
	
	By expanding out the definition of the $\fc{J}_n^{\cdots}$ moments, we see that they satisfy all of the requisite index symmetries from the symmetries of the Riemann tensor:
	\begin{align}
		&\fc{J}_n^{\kappa_1\dots\kappa_n\rho\mu\sigma\nu} = \fc{J}_n^{(\kappa_1\dots\kappa_n)[\rho\mu][\sigma\nu]} = \fc{J}_n^{\kappa_1\dots\kappa_n\sigma\nu\rho\mu}, \p u_{\kappa_1} \fc{J}_n^{\kappa_1\dots\kappa_n\rho\mu\sigma\nu} = 0, \nonumber \\
		&\fc{J}_n^{\kappa_1\dots\kappa_n\rho\mu\sigma\nu} + \fc{J}_n^{\kappa_1\dots\kappa_n\rho\sigma\nu\mu} + \fc{J}_n^{\kappa_1\dots\kappa_n\rho\nu\mu\sigma} = 0,\nonumber \\
		&\fc{J}_n^{\kappa_1\dots\kappa_n\rho\mu\sigma\nu} + \fc{J}_n^{\kappa_1\dots\kappa_{n-1}\rho\mu\kappa_n\sigma\nu} + \fc{J}_n^{\kappa_1\dots\kappa_{n-1}\mu\kappa_n\rho\sigma\nu} = 0.  \label{eq:jmomprops}
	\end{align}
	The first of these ensures (through the symmetry on $\kappa_1\dots\kappa_n$) that these moments don't contain any data that would be obfuscated by contracting into the partial derivatives $\d^n_{\kappa_1\dots\kappa_n} \delta R_{\rho\mu\sigma\nu}$ and (through the antisymmetry on $\rho\mu, \sigma\nu$ and dual interchange symmetry between these pairs) that they don't contain any data that would be obfuscated by contracting into the linearized Riemann indices. The second ensures that the contraction into partial derivatives results in only taking spatial derivatives of the Riemann tensor. The third ensures that these moments don't contain any pieces that are annihilated by the algebraic Bianchi identity and the fourth ensures that they don't contain any pieces that are annihilated by the differential Bianchi identity. Thus, the $\fc{J}_n^{\cdots}$ moments contain that and only that which is needed to write the action coupling of the energy-momentum tensor to gravity. 
	
	Equation \eqref{eq:flatthexpand} is exactly what we wanted to accomplish. It and the index symmetries \eqref{eq:jmomprops} are the main results for this section. The rest of this section will motivate structures that will be useful for generalizing these to curved spacetime. 	Equation \eqref{eq:flatjexpand} is exactly what we wanted to accomplish. Note that in the integral definition following from \eqref{eq:naivetdef}, \eqref{eq:naivepdef}, and \eqref{eq:flatjdef}, the integrand of moment $\fc{J}_n^{\cdots}$ always has $n+2$ factors of displacement from the worldline and consequently the total $\fc{J}_n^{\cdots}$ enters \eqref{eq:flatthexpand} coupled to the metric with $n+2$ derivatives. As such, in terms of the body size, $\fc{J}_n$ scales like $\ell^{n+2}$. $p^{(\mu} \doot z{}^{\nu)}$ is called the mass monopole moment, $S^{\kappa(\mu} \doot z{}^{\nu)}$ the mass dipole moment, $\fc{J}_0$ the quadrupole moment, $\fc{J}_1$ the octopole moment, and so on. 
	
	\subsection{Reduced Multipole Moments}
	
	The $\fc{J}_n^{\cdots}$ moments provide a full description (apart from the first two terms in the series) of the gravitational coupling. We can expand out the linearized Riemann tensor back into derivatives of the metric perturbation to achieve:
	\begin{equation}
		\<T^{\alpha\beta}, h_{\alpha\beta}\> = \int_{-\8}^\8\(\doot z{}^\mu p^\nu h_{\mu\nu} + S^{\kappa\mu}\doot z{}^\nu \d_\kappa h_{\mu\nu} + \sum_{n=2}^\8 4 \frac{n-1}{(n+1)!} \fc{J}_{n-2}^{(\kappa_1\dots\kappa_{n-1}|\mu|\kappa_n)\nu}\d^n_{\kappa_1\dots\kappa_n}h_{\mu\nu}\)D\lambda. \label{eq:741}
	\end{equation}
	The last expression is automatically $\mu\nu$ symmetric due to the index symmetries of the $\fc{J}_n^{\cdots}$. We now set:
	\begin{equation}
		M_0^{\mu\nu} = \doot z{}^{(\mu} p^{\nu)}, \p M_1^{\kappa\mu\nu} = S^{\kappa(\mu}\doot z{}^{\nu)}, \p M_n^{\kappa_1\dots\kappa_n\mu\nu} = 4\frac{n-1}{n+1} \fc{J}_{n-2}^{(\kappa_1\dots\kappa_{n-1}|\mu|\kappa_n)\nu}. \label{eq:mfromj}
	\end{equation}
	Due to \eqref{eq:jmomprops}, these automatically satisfy:
	\begin{align}
		&M_n^{\kappa_1\dots\kappa_n\mu\nu} = M_n^{(\kappa_1\dots\kappa_n)(\mu\nu)}, \p M_1^{(\kappa\mu\nu)} = 0, \nonumber \\
		&M_n^{(\kappa_1\dots\kappa_n\mu)\nu} = 0 \ps (n > 1), \p u_{\kappa_1} M_n^{\kappa_1\dots\kappa_{n-2}[\kappa_{n-1}['\kappa_n \mu] \nu]'} = 0 \ps (n > 2) \label{eq:mmomprops}
	\end{align}
	and \eqref{eq:741} becomes:
	\begin{equation}
		\<T^{\alpha\beta}, h_{\alpha\beta}\> = \int_{\-\8}^\8 \sum_{n=0}^\8 \frac{1}{n!} M^{\kappa_1\dots\kappa_n\mu\nu}_n \d^n_{\kappa_1\dots\kappa_n} h_{\mu\nu}(z) D\lambda \label{eq:tmexpand}
	\end{equation}
	for arbitrary metric perturbations $h_{\alpha\beta}(x)$. The $M_n^{\cdots}$ moments are thus the ``reduced potential multipole moments'' for the energy-momentum tensor. The $\fc{J}_n^{\cdots}$ are the gravitational analogs of the electromagnetic $Q_n^{\cdots}$ and the $M_n^{\cdots}$ are the analogs of the $I_n^{\cdots}$. The $\fc{J}_n^{\cdots}$ are more convenient for calculations because they put \eqref{eq:flatthexpand} into manifestly gauge invariant form in terms of the linearized gravitational field strength while the $M_n^{\cdots}$ make \eqref{eq:tmexpand} look much more akin to the Newtonian formula \eqref{eq:nsintseries}. 
	
	The $\fc{J}_n^{\cdots}$ moments can be recovered from the potential multipole moments by following all of the index manipulations that related the $\mc{t}_n^{\cdots}$ and $\mc{y}_n^{\cdots}$ (equations \eqref{eq:735} to \eqref{eq:txysimp}) but replacing the $\mc{t}_n^{\cdots}$ with the $M_n^{\cdots}$, the $\mc{y}_n^{\cdots}$ with the $\fc{J}_n^{\cdots}$, and noting that in this analogy the $\mc{x}_n^{\cdots}$ are mapped to $0$ due to \eqref{eq:mmomprops}. Doing so yields:
	\begin{equation}
		\fc{J}_n^{\kappa_1\dots\kappa_n\rho\mu\sigma\nu} = M_{n+2}^{\kappa_1\dots\kappa_n[\rho['\sigma \mu] \nu]'}. \label{eq:jfromm}
	\end{equation}
	The potential multipole moments are reduced in the sense that they have fully incorporated all of the constraints implied by the conservation of the energy-momentum tensor. To see this, evaluate \eqref{eq:spacetcons} using \eqref{eq:tmexpand}:
	\begin{equation}
		\<T^{\alpha\beta}, \nabla_{(\alpha} f_{\beta)}\> = \int_{-\8}^\8 \sum_{n=0}^\8 \frac{1}{n!}M_n^{(\kappa_1\dots\kappa_n\mu)\nu} \d^{n+1}_{\kappa_1\dots\kappa_n \mu} f_\nu D\lambda.
	\end{equation}
	This follows without using any of the specific properties of the potential multipole moments. If we now use \eqref{eq:mmomprops}, each term in the series beyond the first two are identically 0 due to the symmetrization. The two remaining terms can then be explicitly computed to give:
	\begin{equation}
		\<T^{\alpha\beta}, \nabla_{(\alpha} f_{\beta)}\> = \left. p^\mu f_\mu \right|^{\8}_{-\8} + \left. S^{\mu\nu}\d_{[\mu} f_{\nu]}\right|^{\8}_{-\8} - \int_{-\8}^\8\(\doot p{}^\mu f_\mu + (\doot S{}^{\mu\nu} - p^\mu \doot z{}^\nu + p^\nu \doot z{}^\mu) \d_{[\mu} f_{\nu]}\)D\lambda.
	\end{equation}
	The first two terms vanish by the compactness of $f$ and the integrands vanish using the equations of motion for the momentum \eqref{eq:freepcons} and spin \eqref{eq:flatsdot}. Thus:
	\begin{equation}
		\<T^{\alpha\beta}, \nabla_{(\alpha} f_{\beta)}\> = 0
	\end{equation}
	as required by \eqref{eq:flatfunct}, without putting any further conditions on the $M_n^{\cdots}$ moments. 
	
	Equation \eqref{eq:tmexpand} formally provides a distribution expression for $T^{\alpha\beta}$ by varying it with respect to $h_{\alpha\beta}$ as:
	\begin{equation}
		T^{\mu\nu}(x) = \int_{-\8}^\8 \sum_{n=0}^\8 \frac{(-1)^n}{n!} M^{\kappa_1\dots\kappa_n\mu\nu}_n(\lambda) \d^n_{\kappa_1\dots\kappa_n}\delta(x-z(\lambda)) D\lambda.
	\end{equation}
	
	\subsection{Moment Generating Function}
	
	In flat spacetime, \eqref{eq:ftdef} gives that the Fourier transform of the metric perturbation about a point $z$ is:
	\begin{equation}
		\twid{h}_{\mu\nu}(z, k) = \int e^{-ik_\rho (x-z)^\rho} h_{\mu\nu}(x) d^4 x. 
	\end{equation}
	Like in the electromagnetic case, while $\twid{h}_{\mu\nu}(\lambda,k)$ is a function of 8 variables, it depends on the 4 corresponding to the point $z$ in a very restricted way:
	\begin{equation}
		\twid{h}_{\mu\nu}(z,k) = e^{ik\cdot z} \twid{h}_{\mu\nu}(0,k). 
	\end{equation}
	We also define the moment generating function $M^{\mu\nu}(\lambda,k)$ for the energy-momentum tensor by:
	\begin{equation}
		M^{\mu\nu}(\lambda,k) = \sum_{k=0}^\8 \frac{(-i)^n}{n!} M_n^{\kappa_1\dots\kappa_n\mu\nu}(\lambda) k_{\kappa_1}\dots k_{\kappa_n}.
	\end{equation}
	This is again not the Fourier transform of the energy-momentum tensor, but it follows from \eqref{eq:tmexpand} that:
	\begin{equation}
		\<T^{\alpha\beta}, h_{\alpha\beta}\> = M^{\mu\nu}[\bar{h}_{\mu\nu}] = \int_{-\8}^\8\int M^{\mu\nu*}(\lambda, k) \twid{h}_{\mu\nu}(\lambda, k) \frac{d^4k}{(2\pi)^4} D\lambda \label{eq:mfuncdef}
	\end{equation}
	where the final expression defines the moment functional $M^{\mu\nu}[\bar{h}_{\mu\nu}]$. 
	
	\subsection{What about Electromagnetism?}
	
	In this section we did not include the electromagnetic interaction. If we had, \eqref{eq:tcons} would not reduce to \eqref{eq:flattcons}. Instead, it would reduce to:
	\begin{equation}
		\d_\beta T^{\alpha\beta} = \fc{F}^{\alpha\beta}J_\beta. 
	\end{equation}
	The presence of the electromagnetic interaction changes the functional statement of energy-momentum conservation. In particular, now:
	\begin{equation}
		\<T^{\alpha\beta}, \nabla_{(\alpha}f_{\beta)}\> = \int T^{\alpha\beta}\d_\beta f_\alpha d^4 x = \int \fc{F}^{\alpha\beta} J_\alpha f_\beta d^4 x = \<J^\alpha, \fc{F}_{\alpha\beta}f^\beta\>. 
	\end{equation}
	Because of this, \eqref{eq:mfuncdef} and consequently the definition of the potential multipole moments need to be modified in the presence of the electromagnetic interaction in order to keep the multipole moments reduced (without interdependencies). Because the right hand side here depends on the current density and electromagnetic field, the proper definition of the reduced multipole moments of the energy-momentum tensor will necessarily also depend on them. We will see this explicitly in Section \ref{sec:gr}. 
	
	\section{Electromagnetism in curved spacetime} \label{sec:emgr}
	
	We will now find the appropriate definition of the reduced multipole moments of the current density in curved spacetime~\cite{dixon2}, with the ultimate goal of generalizing \eqref{eq:flatjexpand}. We learned several things in flat spacetime that will motivate how we define these multipole moments. We want to be able to express $\<J^\alpha, A_\alpha\>$, defined by:
	\begin{equation}
		\<J^\alpha, A_\alpha\> = \int J^\alpha A_\alpha D^4x = \int_{-\8}^\8 \int_\Sigma J^\alpha A_\alpha d\Sigma_w D\lambda \label{eq:jacurv}
	\end{equation}
	(using \eqref{eq:volintdecomp}) in terms of the multipole moments so that the only term which is not manifestly gauge invariant is the monopole (point-particle) term and so that we develop a series in ``spatial derivatives of the electromagnetic field strength'' along the worldline in some appropriate sense. The current density must always satisfy the continuity equation \eqref{eq:jcons}. In curved spacetime, the functional statement of the continuity equation is still achieved by simply integrating \eqref{eq:jcons} against an arbitrary scalar field $f(x)$:
	\begin{equation}
		\<J^\alpha, \nabla_\alpha f\> = -\int f \nabla_\alpha J^\alpha D^4 x = 0. \label{eq:jdivcurv}
	\end{equation}
	
	Because the multipole moments will get contracted into the vector potential and its derivatives evaluated on the worldline, the multipole moments are always purely tensors at $z$. We learned from \eqref{eq:qmomcond} that in order for the reduced field moments to properly couple to the something like the spatial derivatives of the electromagnetic field strength, they must satisfy:
	\begin{align}
		&Q_n^{\rho_1\dots\rho_n\mu\nu} = Q_n^{(\rho_1\dots\rho_n)[\mu\nu]}, \p u_{\rho_1} Q_n^{\rho_1\dots\rho_n\mu\nu} = 0, \nonumber \\
		&Q_n^{\rho_1\dots\rho_{n-1} \rho_n\mu\nu} + Q_n^{\rho_1\dots\rho_{n-1}\mu\nu\rho_n} + Q_n^{\rho_1\dots\rho_{n-1}\nu\rho_n\mu} = 0. \label{eq:qmomcond2} 
	\end{align}
	We learned from \eqref{eq:Imomdef} that it is convenient to define the reduced potential moments as:
	\begin{equation}
	I_0^\mu = q\doot z{}^\mu, \p	I_n^{\rho_1\dots\rho_n\mu} = \frac{2n}{n+1}Q_{n-1}^{(\rho_1\dots\rho_n)\mu} \label{eq:ifromq2}
	\end{equation}
	and that these then satisfy:
	\begin{gather}
		I_n^{(\rho_1\dots\rho_n)\mu} = I_n^{\rho_1\dots\rho_n\mu}, \ \ps \ I_n^{(\rho_1\dots\rho_n \mu)} = 0 \ps (n > 0), \ \ps \ u_{\rho_1} I_n^{\rho_1\dots\rho_{n-1}[\rho_n\mu]} = 0 \ps (n > 1) \label{eq:imomprops2}
	\end{gather}
	from which the field moments can be recovered by:
	\begin{equation}
		Q_n^{\rho_1\dots\rho_n\mu\nu} = I_{n+1}^{\rho_1\dots\rho_n[\mu\nu]}. \label{eq:804}
	\end{equation}
	
	From the potential multipole moments, we define the moment generating function:
	\begin{equation}
		I^\mu(\lambda, k) = \sum_{n=0}^\8 \frac{(-i)^n}{n!} I^{\rho_1\dots\rho_n\mu}_n k_{\rho_1}\dots k_{\rho_n}. \label{eq:igendef}
	\end{equation}
	We can integrate the moment generating function against an arbitrary Fourier transform of a bitensor vector field on the tangent bundle. So, for a generic bitensor $A_\alpha(x, z)$, not necessarily an ordinary vector field, with tangentization $\bar{A}_\mu(z, v)$, we define the moment functional $I^\mu[\bar {A}_\mu]$:
	\begin{equation}
		I^\mu[\bar {A}_\mu] = \int_{-\8}^\8 \int I^{\mu*}(\lambda,k)\twid{A}_\mu(z(\lambda),k) \frac{D^4k}{(2\pi)^4} D\lambda. 
	\end{equation}
	Because all of the moments beyond $I_0^\mu$ can be written in terms of the $Q_n^{\cdots}$ and we expect these moments to couple directly to the gauge invariant field strength, it is useful to also define the $Q$ function and $Q$ functional by subtracting off the contribution from $I_0^\mu$:
	\begin{align}
		Q^\mu(\lambda, k) &= \sum_{n=1}^\8 \frac{(-i)^n}{n!} I^{\rho_1\dots\rho_n\mu}_n k_{\rho_1}\dots k_{\rho_n} \label{eq:qserdef} \\
		Q^\mu[\bar{A}_\mu] &= \int_{-\8}^\8 \int Q^{\mu*}(\lambda,k)\twid{A}_\mu(z(\lambda),k) \frac{D^4k}{(2\pi)^4} D\lambda =  I^\mu[\bar{A}_\mu] - \int_{-\8}^\8 q \doot z{}^\mu \bar{A}_\mu(z, 0)D\lambda  \label{eq:qgendef}
	\end{align}
	Our goal will now be to find integral expressions for the $I_n^{\cdots}$ and $Q_n^{\cdots}$ moments in terms of $J^\alpha$ so that they satisfy all of these expectations.
	
	\subsection{Decomposition of the Vector Potential}

	In order to evaluate $\<J^\alpha, A_\alpha\>$ into a form that generalizes \eqref{eq:flatjexpand} to curved spacetime, we are going to separate the vector potential into three pieces: the point-particle contribution (the worldline integral of $+q\doot z{}^\mu A_\mu)$, a gradient (which will give no overall contribution due to \eqref{eq:jdivcurv}), and a manifestly gauge invariant remainder (like all of the remaining terms in \eqref{eq:flatjexpand}). We expect, based on that equation, that the contribution from the remainder will turn into a series in the $Q_n^{\cdots}$. It follows straight from \eqref{eq:igendef} and \eqref{eq:804} that the $Q_n^{\cdots}$ moments are contained in the moment generating function as:
	\begin{equation}
		\frac{i}{2}\(\frac{\d I^\nu}{\d k_\mu} - \frac{\d I^\mu}{\d k_\nu}\) = \sum_{n=0}^\8 \frac{(-i)^n}{n!} Q_n^{\rho_1\dots\rho_n\mu\nu} k_{\rho_1}\dots k_{\rho_n}. \label{eq:qfromider}
	\end{equation}
	Let $\mc{F}_{\alpha\beta}(x,z)$ be an arbitrary bitensor which is antisymmetric and which satisfies the electromagnetic Bianchi identity. Since we want the $Q_n^{\cdots}$ to couple to derivatives of the field strength, let us take the Fourier transform of $\mc{F}_{\alpha\beta}(x,z)$ and integrate the result against \eqref{eq:qfromider}. Each order of $k$ in the above series expansion under the Fourier transform becomes an order of differentiation with respect to $v$, and because $u$ contracted into any of the first $n$ indices of $Q_n^{\cdots}$ yields 0, all of those derivatives will be in spatial directions. A short calculation gives:
	\begin{equation}
		\int_{-\8}^\8\int \(\frac{i}{2}\(\frac{\d I^\nu}{\d k_\mu} - \frac{\d I^\mu}{\d k_\nu}\)\)^* \twid{\mc{F}}_{\mu\nu}(z(\lambda), k) \frac{D^4k}{(2\pi)^4} D\lambda = I^\mu[\bar{\mc{A}}_\mu]  \label{eq:redaq}
	\end{equation}
	where the ``bitensor vector potential'' is given by:
	\begin{equation}
		\mc{A}_\alpha(x,z) = -\mc{F}_{\alpha\beta}(x,z) \sigma^\beta(x,z).  \label{eq:bitensa}
	\end{equation}
	Therefore, if we compute $I^\mu[\bar{\mc{A}}_\mu]$ using an $\mc{A}_\alpha$ of this form, the result will only depend on the $Q_n^{\cdots}$ moments and form a series of spatial derivatives on $\mc{F}$. Plainly, any $\mc{A}_\alpha$ of the form \eqref{eq:bitensa} satisfies: 
	\begin{equation}
		\sigma^\alpha \mc{A}_\alpha = 0.  \label{eq:bitensacond}
	\end{equation}
	The converse is also true. That is, if a bitensor vector potential satisfies \eqref{eq:bitensacond} then for some $\mc{F}_{\alpha\beta}$ (which is antisymmetric and which satisfies the electromagnetic Bianchi identity) it can be put in the form of \eqref{eq:bitensa}. A quick proof of this fact is given in Appendix \ref{sec:poinca}.
	
	We ultimately want to reduce \eqref{eq:jacurv} to a series in terms of multipole moments when it is evaluated using a generic vector potential $A_\alpha(x)$. Based on our flat spacetime analysis, we expect that the result will be closely related to $I^\mu[\bar{A}_\mu]$. Because of \eqref{eq:redaq}, $I^\mu[\bar{\mc{A}}_\mu]$ takes a manifestly gauge invariant form when using a bitensor $\mc{A}_\alpha$ satisfying \eqref{eq:bitensacond}. We also know by \eqref{eq:jdivcurv} that the integral of $J$ against gradients of scalar functions vanishes. So, in order to evaluate \eqref{eq:jacurv} in terms of easier pieces, we seek a decomposition of a generic vector potential into such a bitensor and such a gradient of a biscalar function $\varphi(x,z)$ as:
	\begin{equation}
		A_\alpha(x) = \mc{A}_\alpha(x,z) + \nabla_\alpha\varphi(x,z). \label{eq:bitensagauge}
	\end{equation}
	That is, we want to take an arbitrary $A_\alpha(x)$ and show that it is possible to choose $\varphi(x,z)$ so that $\mc{A}_\alpha$ related to them by \eqref{eq:bitensagauge} satisfies \eqref{eq:bitensacond}. In order for \eqref{eq:bitensacond} to be true, we have (by simply contracting with $\sigma^\alpha$):
	\begin{equation}
		\sigma^\alpha\nabla_\alpha\varphi = \sigma^\alpha A_\alpha.  \label{eq:bitensagauge2}
	\end{equation}
	Taking $\zeta^\alpha(s)$ as our standard geodesic, this is equivalently:
	\begin{equation}
		\frac{d\varphi(\zeta,z)}{ds} = \frac{1}{s}\sigma^\alpha(\zeta, z)A_\alpha(\zeta).
	\end{equation}
	We can solve this by integrating from 0 to 1. Hence a desired $\varphi(x,z)$ does in fact exist, and in fact has $\varphi(z,z)$ arbitrary. We choose:
	\begin{equation}
		\varphi(z,z) = 0 \label{eq:varphiic}
	\end{equation}
	for simplicity, and so have:
	\begin{equation}
		\varphi(x,z) = \int_0^1 \sigma^\alpha(\zeta,z) A_\alpha(\zeta) \frac{ds}{s}. \label{eq:varphisol}
	\end{equation}
	With this, \eqref{eq:bitensacond} follows from \eqref{eq:bitensagauge}.
	
	It follows by a short calculation from \eqref{eq:bitensagauge} using \eqref{eq:bitensacond} that along the same geodesic:
	\begin{equation}
		\frac{d}{ds}(s H^\alpha{}_\mu \mc{A}_\alpha) = \fc{F}_{\alpha\beta}(\zeta) \sigma^\alpha(\zeta,z) H^\beta{}_\mu(\zeta, z).
	\end{equation}
	Therefore, again integrating from 0 to 1:
	\begin{equation}
		\bar{\mc{A}}_\mu = H^\alpha{}_\mu \mc{A}_\alpha = \int_0^1 \fc{F}_{\alpha\beta}\sigma^\alpha H^\beta{}_\mu ds = \Psi_\mu 
	\end{equation}
	where we have recalled the definition of the bitensor $\Psi_\mu$ from \eqref{eq:phipsidef}. This makes it manifest that indeed $\mc{A}_\alpha$ is gauge invariant.
	
	We want to use our decomposition \eqref{eq:bitensagauge} to evaluate $\<J^\alpha, A_\alpha\>$. We can't insert that decomposition directly because the integration in $\<J^\alpha, A_\alpha\>$ is over $x$, and the right hand side of \eqref{eq:bitensagauge} is a function of both $x$ and $z$. For the $\mc{A}_\alpha(x,z)$ term this is easily solved because we can evaluate its $z$ dependence at $z(\tau(x))$. For the $\nabla_\alpha\varphi(x,z)$ term there is a slight subtlety, because the differentiation with respect to $x$ and the evaluation of $z = z(\tau(x))$ do not commute. Instead, by the chain rule:
	\begin{equation}
		\nabla_\alpha(\varphi(x,z(\tau(x)))) = (\nabla_\alpha\varphi)(x,z(\tau(x))) + \nabla_\alpha\tau \left.\frac{\d\varphi(x,z(\lambda))}{\d\lambda}\right|_{\lambda = \tau(x)}. \label{eq:8112}
	\end{equation}
	And so, to express the right hand side of \eqref{eq:bitensagauge} purely in terms of $x$ we have:
	\begin{equation}
		A_\alpha(x) = \mc{A}_\alpha(x,z(\tau(x))) + \nabla_\alpha(\varphi(x,z(\tau))) - \nabla_\alpha \tau \left.\frac{\d\varphi}{\d\lambda}\right|_{\lambda = \tau}.  \label{eq:abisol}
	\end{equation}
	
	To finish simplifying \eqref{eq:abisol} we need to compute $\frac{\d\varphi}{\d\lambda}$. Since we want to maximally maintain manifest gauge invariance, we want to do so in terms of $\mc{A}_\alpha$ rather than in terms of $A_\alpha$. Because the left hand side of \eqref{eq:bitensagauge} does not depend on $z$ (and so does not depend on $\lambda$), by differentiating both sides with respect to $\lambda$ we find:
	\begin{equation}
		0 = \frac{\d\mc{A}_\alpha}{\d\lambda} + \nabla_\alpha \frac{\d\varphi}{\d\lambda}.
	\end{equation}
	Now contracting with $\sigma^\alpha$:
	\begin{equation}
		\sigma^\alpha\nabla_\alpha \frac{\d\varphi}{\d\lambda} = -\sigma^\alpha \frac{\d\mc{A}_\alpha}{\d\lambda}. 
	\end{equation}
	Expressed along the same geodesic as usual:
	\begin{equation}
		\frac{d}{ds}\(\frac{\d\varphi}{\d\lambda}\) = -\frac{1}{s}\sigma^\alpha(\zeta,z(\lambda))\frac{\d\mc{A}_\alpha(\zeta,z(\lambda))}{\d\lambda}.
	\end{equation}
	Integrating from $s= 0$ to $1$:
	\begin{equation}
		\frac{\d\varphi}{\d\lambda}(x,z(\tau)) = \left.\frac{\d\varphi(x,z(\lambda))}{\d\lambda}\right|_{x\to z}  - \ec(\tau(x))\psi(x) \label{eq:varphider}
	\end{equation}
	where we have defined the function $\psi(x)$ by:
	\begin{equation}
		\psi(x) = \frac{1}{\ec(\tau(x))}\int_0^1\sigma^\alpha(\zeta,z(\tau))\frac{\d\mc{A}_\alpha}{\d\lambda}(\zeta,z(\tau))\frac{ds}{s}. \label{eq:psisol}
	\end{equation}
	We can simplify $\psi$ slightly by taking the $\lambda$ derivative of \eqref{eq:bitensacond} when evaluating $z = z(\lambda)$:
	\begin{gather}
		\sigma^\alpha \frac{\d\mc{A}_\alpha}{\d\lambda} + \sigma^\alpha{}_\mu \dot z^\mu \mc{A}_\alpha = 0 \\
		\implies  \psi(x) = -\doot z{}^\mu(\tau)\int_0^1 \sigma^\alpha{}_\mu(\zeta, z(\tau)) \mc{A}_\alpha(\zeta, z(\tau)) \frac{ds}{s}.  \label{eq:psisol2}
	\end{gather}
	We can compute the initial condition in \eqref{eq:varphider} by taking the $\lambda$ derivative of \eqref{eq:varphiic}. Doing so and simplifying, we find:
	\begin{equation}
		\left.\frac{\d\varphi(x,z(\lambda))}{\d\lambda}\right|_{x\to z} = -\dot z^\mu A_\mu(z)
	\end{equation}
	and so returning to \eqref{eq:abisol} we have:
	\begin{equation}
		A_\alpha(x) = \mc{A}_\alpha(x, z(\tau)) + \nabla_\alpha(\varphi(x,z(\tau))) + \dot z^\mu(\tau) A_\mu(z(\tau)) \nabla_\alpha \tau + \ec(\tau)\psi(x)\nabla_\alpha \tau. \label{eq:abisol2}
	\end{equation}
	
	Equation \eqref{eq:abisol2} is precisely the decomposition of the vector potential we promised at the start of the section. We have now shown that given an arbitrary vector potential $A_\alpha(x)$, it is possible to define the function $\varphi(x,z)$ by \eqref{eq:varphisol} so that $\mc{A}_\alpha(x,z)$ defined by \eqref{eq:bitensagauge} satisfies \eqref{eq:bitensacond}. We can recover the original vector potential from these ingredients by \eqref{eq:abisol2}. The only pieces in \eqref{eq:abisol2} which are not manifestly gauge invariant are $\nabla_\alpha(\varphi(x,z(\tau)))$, which gives 0 when integrated against $J^\alpha$, and $\dot z^\mu(\tau) A_\mu(z(\tau)) \nabla_\alpha \tau$, which when integrated against $J^\alpha$ reproduces the electromagnetic coupling of a point-particle. In particular, if we now use \eqref{eq:abisol2} to evaluate \eqref{eq:jacurv} while using \eqref{eq:jdivcurv} and recalling \eqref{eq:volmeasder} and \eqref{eq:qdef}:
	\begin{equation}
		\<J^\alpha, A_\alpha\> = \int_{-\8}^\8 q \doot z{}^\mu A_\mu D\lambda + \int_{-\8}^\8\int_\Sigma\Bigl(J^\beta(x) \mc{A}_\beta(x,z(\lambda)) w^\alpha(x) + \psi(x) J^\alpha(x)\Bigr) d\Sigma_\alpha D\lambda \label{eq:jareduced}
	\end{equation}
	The right hand side is the monopole (i.e. point-particle) contribution plus pieces which only depend on $\mc{A}$ and therefore are manifestly gauge invariant. 
	
	\subsection{Determination of the Multipole Moments}
	
	We can now use \eqref{eq:jareduced} to determine the multipole moments. At each worldline time $\lambda$ in the integrand of \eqref{eq:jareduced}, all of the terms beyond the monopole are already gauge invariant. So, they have already incorporated all of the reductions due to the continuity equation (as that is the Noether-dual of gauge invariance). Thus, in analogy with \eqref{eq:imomgendef}, we can require at each worldline time that the moments satisfy:
	\begin{equation}
		\int I^{\mu*}(\lambda,k) \twid{\mc{A}}_\mu(z(\lambda),k) \frac{D^4k}{(2\pi)^4} = \int_\Sigma(J^\beta(x) \mc{A}_\beta(x,z(\lambda)) w^\alpha(x) + \psi(x) J^\alpha(x)) d\Sigma_\alpha. \label{eq:imomgencond}
	\end{equation}
	The left hand side, by \eqref{eq:qfromider}, is given by:
	\begin{equation}
		\int I^{\mu*}(\lambda,k) \twid{\mc{A}}_\mu(z(\lambda),k) \frac{D^4k}{(2\pi)^4} =  \int \sum_{n=0}^\8 \frac{i^n}{n!}Q_n^{\rho_1\dots\rho_n\mu\nu} k_{\rho_1}\dots k_{\rho_n} \twid{\mc{F}}_{\mu\nu}\frac{D^4k}{(2\pi)^4}. \label{eq:imomgenl}
	\end{equation}
	It will be our goal next to turn the right hand side of \eqref{eq:imomgencond} similarly into an integral of some series in $k$ against $\twid{\mc{F}}_{\mu\nu}$. 
	
	Explicitly expanding the first term on the right hand side of \eqref{eq:imomgencond} in terms of $\twid{\mc{F}}_{\mu\nu}$ using \eqref{eq:bitensa} gives:
	\begin{equation}
		\int_\Sigma J^\alpha \mc{A}_\alpha d\Sigma_w = \int \twid{\mc{F}}_{\mu\nu} \int e^{ik\cdot v} v^\mu H^{-\nu}{}_\alpha J^\alpha d\Sigma_w \frac{D^4k}{(2\pi)^4}.  \label{eq:823}
	\end{equation}
	We now define the curved spacetime generalization of the naive moments of the current density, in close analogy to \eqref{eq:naivejdef}:
	\begin{equation}
		\mc{j}_n^{\rho_1\dots\rho_n\mu} = \int_\Sigma v^{\rho_1}\dots v^{\rho_n} H^{-\mu}{}_\alpha J^\alpha d\Sigma_w.
	\end{equation}
	With these, after series expanding \eqref{eq:823} in $k$, we have:
	\begin{equation}
		\int J^\alpha \mc{A}_\alpha d\Sigma_w = \int \twid{\mc{F}}_{\mu\nu}\sum_{n=0}^\8 \frac{i^n}{n!} \mc{j}_{n+1}^{\rho_1\dots\rho_n[\mu\nu]} k_{\rho_1}\dots k_{\rho_n} \frac{D^4k}{(2\pi)^4}. \label{eq:imomgenr1}
	\end{equation}
	
	It takes a little more effort to evaluate the second term on the right hand side of \eqref{eq:imomgencond} because $\psi(x)$ is more complicated. Using \eqref{eq:bitensa} in \eqref{eq:psisol2}, we find:
	\begin{equation}
		\psi(x) = \int_0^1 \sigma^\alpha{}_\mu \doot z{}^\mu \mc{F}_{\alpha\beta} \sigma^\beta \frac{ds}{s}.
	\end{equation}
	Now rewrite $\mc{F}_{\alpha\beta}$ in terms of its tangentization and use that $\sigma^\rho(\zeta(s),z) = s \sigma^\rho(x,z)$:
	\begin{equation}
		\psi(x) = -\doot z{}_\mu(\tau) v^\rho(x,z(\tau)) \int_0^1 \sigma^\mu{}_\alpha(\zeta,z(\tau))\sigma^{\alpha\nu}(\zeta,z(\tau)) \bar{\mc{F}}_{\nu\rho}(z, sv) ds. 
	\end{equation}
	In terms of the Fourier transform:
	\begin{equation}
		\psi(x) = -\doot z{}_\mu v^\rho \int \twid{\mc{F}}_{\nu\rho} \sum_{n=0}^\8 \frac{i^n}{n!} (k\cdot v)^n \int_0^1 \sigma^\mu{}_\alpha(\zeta, z) \sigma^{\alpha\nu}(\zeta, z) s^n ds \frac{D^4k}{(2\pi)^4}.
	\end{equation}
	Now define the $\Theta^{\mu\nu}_n(x,z)$ bitensors by:
	\begin{equation}
		\Theta^{\mu\nu}_n(x,z) = (n+1)\int_0^1 \sigma^\mu{}_\alpha(\zeta, z) \sigma^{\alpha\nu} (\zeta, z) s^n ds. \label{eq:thetabidef}
	\end{equation}
	The factor of $n+1$ is included so that in flat spacetime $\Theta^{\mu\nu}_n = \eta^{\mu\nu}$ independent of $n$. With these, $\psi(x)$ becomes:
	\begin{equation}
		\psi(x) = -\doot z{}_\mu v^\rho \int \twid{\mc{F}}_{\nu\rho} \sum_{n=0}^\8 \frac{i^n}{(n+1)!} (k\cdot v)^n \Theta^{\mu\nu}_n \frac{D^4k}{(2\pi)^4}. 
	\end{equation}
	We can now evaluate the second part of the integral on the right hand side of \eqref{eq:imomgencond}:
	\begin{equation}
		\int \psi(x) J^\alpha d\Sigma_\alpha = \int \twid{\mc{F}}_{\mu\nu} \sum_{n=0}^\8 \frac{i^n}{(n+1)!} \mc{r}_{n+1}^{\rho_1\dots\rho_n[\mu\nu] \sigma} \doot z{}_\sigma  k_{\rho_1}\dots k_{\rho_n}\frac{D^4k}{(2\pi)^4} \label{eq:imomgenr2}
	\end{equation}
	where we have defined, similarly to \eqref{eq:naiverdef}, the curved spacetime generalization of the naive moments of the charge density:
	\begin{equation}
		\mc{r}_n^{\rho_1\dots\rho_n\mu\nu} = \int_\Sigma v^{\rho_1}\dots v^{\rho_n} \Theta^{\mu\nu}_{n-1} J^\alpha d\Sigma_\alpha.
	\end{equation}
	Notice that due to the additional complexity of curved spacetime, these coefficients carry more indices than their flat spacetime analogs. 
	
	Collecting \eqref{eq:imomgenl}, \eqref{eq:imomgenr1}, and \eqref{eq:imomgenr2} back into \eqref{eq:imomgencond} we have:
	\begin{equation}
		\int \twid{\mc{F}}_{\mu\nu}\sum_{n=0}^\8 \frac{i^n}{n!}k_{\rho_1}\dots k_{\rho_n}\(Q_n^{\rho_1\dots\rho_n\mu\nu} - \mc{j}^{\rho_1\dots\rho_n[\mu\nu]}_{n+1} - \frac{1}{n+1} \mc{r}_{n+1}^{\rho_1\dots\rho_n[\mu\nu]\sigma}\doot z{}_\sigma\) \frac{D^4k}{(2\pi)^4} = 0. 
	\end{equation}
	In order for this to hold for all possible $\twid{\mc{F}}_{\mu\nu}$ we must have:
	\begin{equation}
		Q_n^{\rho_1\dots\rho_n\mu\nu} = \mc{j}^{\rho_1\dots\rho_n[\mu\nu]}_{n+1} + \frac{1}{n+1} \mc{r}_{n+1}^{\rho_1\dots\rho_n[\mu\nu]\sigma}\doot z{}_\sigma.
	\end{equation}
	Thus, this defines the $Q_n^{\cdots}$ moments completely and in turn through \eqref{eq:ifromq2} determines the potential multipole moments $I_n^{\cdots}$. From this definition it is straightforward to check that all of the properties in \eqref{eq:qmomcond2} are satisfied.
	
	\subsection{Final Form of the Action}
	
	We can now relate the moment generating functional directly to $\<J^\alpha, A_\alpha\>$. By feeding \eqref{eq:imomgencond} into \eqref{eq:jareduced} we have:
	\begin{equation}
		\<J^\alpha, A_\alpha\> = \int_{-\8}^\8 q \doot z{}^\mu A_\mu D\lambda + I^\mu[\bar{\mc{A}}_\mu].
	\end{equation}
	Using \eqref{eq:bitensagauge}:
	\begin{equation}
		\<J^\alpha, A_\alpha\> = \int_{-\8}^\8 q \doot z{}^\mu A_\mu D\lambda + I^\mu[\bar{A}_\mu] - I^\mu[\overline{\nabla\varphi}_\mu] \label{eq:jared2}
	\end{equation}
	where here $\nabla_\alpha\varphi$ literally means $\nabla_\alpha\varphi(x, z(\lambda))$ (as opposed to the same expression with $\lambda$ replaced by $\tau(x)$). We can compute $I^\mu[\overline{\nabla\varphi}_\mu]$ directly from the definition of $I$ to get:
	\begin{equation}
		I^\mu[\overline{\nabla\varphi}_\mu] = \int_{-\8}^\8\int \int i I^{\mu*}(\lambda,k) k_\mu \varphi(\exp_z v, z) e^{-ik \cdot v} D^4 v \frac{D^4k}{(2\pi)^4} D\lambda. \label{eq:inabphi}
	\end{equation} 
	Using \eqref{eq:igendef} and \eqref{eq:imomprops2}:
	\begin{equation}
		I^\mu k_\mu = q \doot z{}^\mu k_\mu.  \label{eq:ikq}
	\end{equation}
	With this, \eqref{eq:inabphi} becomes:
	\begin{equation}
		I^\mu[\overline{\nabla\varphi}_\mu] = \int_{-\8}^\8 q \doot z{}^\mu \delta^\alpha_\mu \left.\nabla_\alpha\varphi(x,z)\right|_{x\to z} D\lambda. \label{eq:inabphi2}
	\end{equation}
	We can evaluate the integrand using \eqref{eq:bitensagauge2} in the limit $x\to z$ where we take the initial displacement between $x$ and $z$ to be colinear with $\dot z$. Then:
	\begin{equation}
		\doot z{}^\mu \delta^\alpha_\mu \left.\nabla_\alpha\varphi(x,z)\right|_{x\to z} = \doot z{}^\mu A_\mu(z) \\
		\implies I^\mu[\overline{\nabla\varphi}_\mu] = \int_{-\8}^\8 q \doot z{}^\mu A_\mu D\lambda.
	\end{equation}
	This is again just the point-particle piece. Returning to \eqref{eq:jared2} we have:
	\begin{equation}
		\<J^\alpha, A_\alpha\> = I^\mu[\bar{A}_\mu].\label{eq:ijarelate}
	\end{equation}
	Thus the moment generating functional is effectively the same as the functional $\<J^\alpha, \cdot\>$ (up to tangentization) when acting on genuine vector fields, just like in the flat spacetime case \eqref{eq:imomgendef}. 
	
	Using \eqref{eq:ijarelate}, we can verify that the potential multipole moments are in fact reduced. The reduction requires the satisfaction of the functional continuity equation \eqref{eq:jdivcurv}. Enforced on the multipole moments this requires:
	\begin{equation}
		I^\mu[\overline{\nabla f}_\mu] = \<J^\alpha, \nabla_\alpha f\> = 0.
	\end{equation}
	Explicitly computing $I^\mu[\overline{\nabla f}_\mu]$ gives:
	\begin{equation}
		I^\mu[\overline{\nabla f}_\mu] = \int_{-\8}^\8 q \doot z{}^\mu \nabla_\mu f(z) D\lambda = q \left.f(z)\right|^{\lambda\to\8}_{\lambda\to-\8} = 0
	\end{equation}
	for any compactly supported $f$. Thus, there are no further conditions required of the multipole moments. 
	
	From \eqref{eq:ijarelate} we can now express $\<J^\alpha, A_\alpha\>$ explicitly as a series in the derivatives of the vector potential. Putting in the Fourier transform and series expanding the tangentized vector potential, we find:
	\begin{equation}
		\<J^\alpha, A_\alpha\> = \int_{-\8}^\8 \sum_{n=0}^\8 \frac{1}{n!} I_n^{\rho_1\dots\rho_n \mu} {\ee}_{\rho_1\dots\rho_n} A_\mu (z) D\lambda. \label{eq:8310}
	\end{equation}
	However, we can form an equation much more like \eqref{eq:flatjexpand} by rewriting in terms of the $Q_n^{\cdots}$. Momentarily thinking in the Normal picture, it follows quickly that:
	\begin{equation}
		\frac{\d \bar A_\nu}{\d v^\mu} - \frac{\d \bar A_\mu}{\d v^\nu} = \bar{\fc{F}}_{\mu\nu}.
	\end{equation}
	Then, using \eqref{eq:ifromq2}, we have from \eqref{eq:8310} that:
	\begin{equation}
		\<J^\alpha, A_\alpha\> = \int_{-\8}^\8\( q\doot z{}^\mu A_\mu(z) + \sum_{n=0}^\8 \frac{n+1}{(n+2)!} Q_n^{\rho_1\dots\rho_n\mu\nu} \left.\frac{\d^n}{\d v^{\rho_1}\dots\d v^{\rho_n}}\(\frac{\d \bar A_\nu}{\d v^\mu} - \frac{\d \bar A_\mu}{\d v^\nu}\)\right|_{v = 0}\) D\lambda.
	\end{equation}
	Finally then, in terms of tensor extensions, we have that the electromagnetic interaction action for an electromagnetic compact object in curved spacetime is:
	\begin{equation}
		\<J^\alpha, A_\alpha\> = \int_{-\8}^\8\( q\doot z{}^\mu A_\mu(z) + \sum_{n=0}^\8 \frac{n+1}{(n+2)!} Q_n^{\rho_1\dots\rho_n\mu\nu} {\ee}_{\rho_1\dots\rho_n} \fc{F}_{\mu\nu}\) D\lambda. \label{eq:jacurvexpand}
	\end{equation}
	The only difference between this and \eqref{eq:flatjexpand} is the replacement of partial derivatives by tensor extensions! Equation \eqref{eq:jacurvexpand} is the main result for electromagnetism. $\<J^\alpha, A_\alpha\>$ represents the interaction term in the electromagnetic action by \eqref{eq:jactpert}. It characterizes both the effects of a charge current on the electromagnetic field and the effects of the electromagnetic field on the charge current. Using \eqref{eq:jacurvexpand} we can uniquely rewrite it as a series in terms of multipole moments of our current distribution which are fully reduced (i.e. containing no interdependencies) coupled with the spatial derivatives of the field strength on the worldline.
	
	\subsection{General Evaluation of the Moment Functional}
	
	When we study the multipole moments of the energy-momentum tensor in curved spacetime, it happens that we need to be able to evaluate $I^\mu[\bar{\fc{B}}_\mu]$, for $\fc{B}_\alpha(x,z)$ an arbitrary bitensor vector field (depending on both $x$ and $z$), in terms of the current density $J^\alpha(x)$. We already accomplished this for a genuine vector potential $A_\alpha(x)$ (depending only on $x$) in~\eqref{eq:ijarelate}. The $z$ dependence in $\fc{B}_\alpha(x,z)$ will induce an extra term. To approach this, all we have to do is slightly generalize our calculation for an arbitrary vector potential from equation \eqref{eq:bitensagauge} to \eqref{eq:jareduced}. Given arbitrary $\fc{B}_\alpha(x,z)$, we search for a bitensor vector field $\mc{A}_\alpha(x,z)$ and biscalar field $\fc{C}(x,z)$ related by:
	\begin{equation}
		\fc{B}_\alpha(x,z) = \mc{A}_\alpha(x,z) + \nabla_\alpha\fc{C}(x,z) \label{eq:badef}
	\end{equation}
	so that:
	\begin{equation}
		\mc{A}_\alpha\sigma^\alpha = 0. \label{eq:bacond}
	\end{equation}
	If we can find these, then by \eqref{eq:imomgencond} and \eqref{eq:ikq} we have:
	\begin{equation}
		I^\mu[\bar{\fc{B}}_\mu] = \int_{-\8}^\8 \int_{\Sigma} (J^\beta \mc{A}_\beta w^\alpha + \psi(x) J^\alpha) d\Sigma_\alpha D\lambda + \int_{-\8}^\8 q \doot z{}^\mu \delta^\alpha_\mu \left.\nabla_\alpha \fc{C}\right|_{x\to z} D\lambda \label{eq:ibsimp}
	\end{equation}
	with $\psi(x)$ defined by \eqref{eq:psisol}. 
	
	We can determine the necessary $\fc{C}$ by contracting \eqref{eq:badef} with $\sigma^\alpha$ and integrating along a geodesic. In particular:
	\begin{equation}
		\fc{B}_\alpha \sigma^\alpha = \sigma^\alpha\nabla_\alpha \fc{C} = s\frac{D\fc{C}}{Ds} \\
		\implies \fc{C}(x,z) = \int_0^1 \sigma^\alpha \fc{B}_\alpha \frac{ds}{s}
	\end{equation}
	where we have chosen the initial condition:
	\begin{equation}
		\fc{C}(z,z) = 0. \label{eq:cic}
	\end{equation}
	Equation \eqref{eq:bacond} combined with \eqref{eq:badef} requires:
	\begin{equation}
		\left.\nabla_\alpha \fc{C}\right|_{x\to z} = \fc{B}_\alpha(z,z). \label{eq:nabcic}
	\end{equation}
	To finish evaluating $\fc{I}^\mu[\bar{\fc{B}}_\mu]$ in \eqref{eq:ibsimp}, we need to reexpress $\mc{A}_\alpha$ on the right hand side in terms of $\fc{B}_\alpha$ as simply as possible. We can do so using \eqref{eq:badef} where, like we did with $\varphi$, we turn the $\fc{C}$ term into a total gradient. Just like in \eqref{eq:8112}, we must be careful about differentiation before and after evaluation as:
	\begin{equation}
		(\nabla_\alpha\fc{C})(x,z(\tau)) = \nabla_\alpha(\fc{C}(x,z(\tau))) - \left.\frac{\d\fc{C}(x,z(\lambda))}{\d\lambda}\right|_{\lambda\to\tau} \nabla_\alpha \tau.
	\end{equation}
	We can compute $\frac{\d\fc{C}}{\d\lambda}$ by differentiating \eqref{eq:badef} and contracting with $\sigma^\alpha$:
	\begin{equation}
		\sigma^\alpha \frac{\d\fc{B}_\alpha}{\d\lambda} = \sigma^\alpha\frac{\d\mc{A}_\alpha}{\d\lambda} + \sigma^\alpha \nabla_\alpha\frac{\d\fc{C}}{\d\lambda} \\
		\implies s\frac{D}{Ds}\(\frac{\d\fc{C}}{\d\lambda}\) = \sigma^\alpha \frac{\d\fc{B}_\alpha}{\d\lambda} - \sigma^\alpha\frac{\d\mc{A}_\alpha}{\d\lambda}.
	\end{equation}
	Then, integrating along a geodesic we have:
	\begin{equation}
		\frac{\d\fc{C}}{\d\lambda}(x,z(\tau(x))) = \frac{\d\fc{C}}{\d\lambda}(z,z) + \ec(\tau)\Upsilon(x) - \ec(\tau)\psi(x)
	\end{equation}
	where we have recalled the definition of $\psi(x)$ and defined the function $\Upsilon(x)$, which is analogous to $\psi(x)$ but for $\fc{B}$ instead of $\mc{A}$:
	\begin{equation}
		\Upsilon(x) = \frac{1}{\ec(\tau(x))} \int_0^1 \sigma^\alpha(\zeta,z(\tau)) \left.\frac{\d \fc{B}_\alpha}{\d\lambda}(\zeta,z)\right|_{\lambda = \tau} \frac{ds}{s}. \label{eq:Upsilondef}
	\end{equation}
	Differentiating \eqref{eq:cic} with respect to $\lambda$ together with \eqref{eq:nabcic} implies:
	\begin{equation}
		\frac{\d\fc{C}}{\d\lambda}(z,z) = -\dot z^\mu \fc{B}_\mu(z,z). 
	\end{equation}
	Returning this result to \eqref{eq:badef} evaluated at $z = z(\tau(x))$, we have:
	\begin{equation}
		\mc{A}_\alpha(x,z(\tau)) = \fc{B}_\alpha(x,z(\tau)) - \nabla_\alpha(\fc{C}(x,z(\tau))) - \dot z^\mu \fc{B}_\mu(z,z) \nabla_\alpha\tau + \ec \Upsilon \nabla_\alpha\tau - e\psi\nabla_\alpha\tau. 
	\end{equation}
	We can now take this and use it to simplify \eqref{eq:ibsimp}. With only a little effort, we find:
	\begin{equation}
		I^\mu[\bar{\fc{B}}_\mu] = \int_{-\8}^\8\int_{\Sigma}(J^\beta \fc{B}_\beta w^\alpha + J^\alpha \Upsilon(x)) d\Sigma_\alpha D\lambda \label{eq:ibfinal}
	\end{equation}
	for an arbitrary bitensor vector field $\fc{B}_\alpha(x,z)$ where $\Upsilon(x)$ is defined by \eqref{eq:Upsilondef}. The first term on the right side is precisely $\<J^\alpha, \fc{B}_\alpha\>$ while the second is the contribution due to the change in $\fc{B}_\alpha$ with $z$ (look at the definition of $\Upsilon$ in \eqref{eq:Upsilondef}). 
	
	\section{Full General Relativity} \label{sec:gr}
	
	We will now work out how to define the reduced multipole moments of the energy-momentum tensor in curved spacetime~\cite{dixon3}, with the ultimate goal of generalizing \eqref{eq:flatthexpand}. We learned several things in flat spacetime that will motivate how we go about doing so. By \eqref{eq:stressdef}, the interaction term contributed to the action by a metric perturbation $h_{\alpha\beta}(x)$, even in curved spacetime, is $\frac{1}{2}\<T^{\alpha\beta}, h_{\alpha\beta}\>$ where $\<T^{\alpha\beta}, h_{\alpha\beta}\>$ is defined by: 
	\begin{equation}
		\<T^{\alpha\beta}, h_{\alpha\beta}\> = \int T^{\alpha\beta} h_{\alpha\beta} D^4x. \label{eq:tfuncdef}
	\end{equation}
	Like \eqref{eq:flatthexpand}, we want to express $\<T^{\alpha\beta}, h_{\alpha\beta}\>$ in terms of the multipole moments so that the only pieces which are not manifestly gauge invariant are the monopole and dipole terms and so that the rest is a series in something like the ``spatial derivatives of the gravitational field strength'' along the worldline. The energy-momentum tensor must always satisfy the continuity equation \eqref{eq:tcons}. In curved spacetime, the functional statement of the energy-momentum transport equation is still achieved by simply integrating \eqref{eq:tcons} against an arbitrary vector field $f_\alpha(x)$:
	\begin{equation}
		\<T^{\alpha\beta}, \nabla_{(\alpha} f_{\beta)}\> = -\int f_\alpha \nabla_\beta T^{\alpha\beta} D^4 x = \int J_\alpha \fc{F}^{\alpha\beta} f_\beta D^4 x = \< J^\alpha, \fc{F}_{\alpha\beta}f^\beta\>. \label{eq:tfuncnabf}
	\end{equation}
	
	Because the multipole moments will get contracted into the metric perturbation and its derivatives evaluated on the worldline, the multipole moments are always purely tensors at $z$. We learned from \eqref{eq:jmomprops} that in order for the reduced field moments to couple to something like the spatial derivatives of the gravitational field strength, they must satisfy:
	\begin{align}
		&\fc{J}_n^{\kappa_1\dots\kappa_n\rho\mu\sigma\nu} = \fc{J}_n^{(\kappa_1\dots\kappa_n)[\rho\mu][\sigma\nu]} = \fc{J}_n^{\kappa_1\dots\kappa_n\sigma\nu\rho\mu}, \p u_{\kappa_1} \fc{J}_n^{\kappa_1\dots\kappa_n\rho\mu\sigma\nu} = 0, \nonumber \\
		&\fc{J}_n^{\kappa_1\dots\kappa_n\rho\mu\sigma\nu} + \fc{J}_n^{\kappa_1\dots\kappa_n\rho\sigma\nu\mu} + \fc{J}_n^{\kappa_1\dots\kappa_n\rho\nu\mu\sigma} = 0,\nonumber \\
		&\fc{J}_n^{\kappa_1\dots\kappa_n\rho\mu\sigma\nu} + \fc{J}_n^{\kappa_1\dots\kappa_{n-1}\rho\mu\kappa_n\sigma\nu} + \fc{J}_n^{\kappa_1\dots\kappa_{n-1}\mu\kappa_n\rho\sigma\nu} = 0.  \label{eq:jmomprops2}
	\end{align}
	We learned from \eqref{eq:mfromj} that it is convenient to define the reduced potential moments as:
	\begin{gather}
		M_0^{\mu\nu} = \doot z{}^{(\mu} p^{\nu)}, \p M_1^{\kappa\mu\nu} = S^{\kappa(\mu}\doot z{}^{\nu)}, \p M_n^{\kappa_1\dots\kappa_n\mu\nu} = 4\frac{n-1}{n+1} \fc{J}_{n-2}^{(\kappa_1\dots\kappa_{n-1}|\mu|\kappa_n)\nu} \label{eq:mfromj2}
	\end{gather}
	and that these satisfy:
	\begin{align}
		&M_n^{\kappa_1\dots\kappa_n\mu\nu} = M_n^{(\kappa_1\dots\kappa_n)(\mu\nu)}, \p M_1^{(\kappa\mu\nu)} = 0, \nonumber \\
		&M_n^{(\kappa_1\dots\kappa_n\mu)\nu} = 0 \ps (n > 1), \p u_{\kappa_1} M_n^{\kappa_1\dots\kappa_{n-2}[\kappa_{n-1}['\kappa_n \mu] \nu]'} = 0 \ps (n > 2) \label{eq:mmomprops2}
	\end{align}
	from which the field moments can be recovered by:
	\begin{equation}
		\fc{J}_n^{\kappa_1\dots\kappa_n\rho\mu\sigma\nu} = M_{n+2}^{\kappa_1\dots\kappa_n[\rho['\sigma \mu] \nu]'}.  \label{eq:904}
	\end{equation}
	
	From the potential multipole moments, we define the moment generating function:
	\begin{equation}
		M^{\mu\nu}(\lambda, k) = \sum_{n=0}^\8 \frac{(-i)^n}{n!} M^{\rho_1\dots\rho_n\mu\nu}_n k_{\rho_1}\dots k_{\rho_n} \label{eq:mfuncdef2}
	\end{equation}
	and we define the moment functional $M^{\mu\nu}[\bar{h}_{\mu\nu}]$ acting on the tangentization of a generic bitensor $h_{\alpha\beta}(x, z)$ as:
	\begin{equation}
		M^{\mu\nu}[\bar{h}_{\mu\nu}] = \int_{-\8}^\8 \int M^{\mu\nu*}(\lambda,k)\twid{h}_{\mu\nu}(z(\lambda),k) \frac{D^4k}{(2\pi)^4} D\lambda. 
	\end{equation}
	Because all of the moments other than $M_0^{\mu\nu}$ and $M_1^{\rho\mu\nu}$ can be written in terms of the $\fc{J}_n^{\cdots}$ and we expect these moments to couple directly to something like the linearized Riemann tensor, it is useful to also define the $\fc{J}^{\mu\nu}$ function and $\fc{J}^{\mu\nu}$ functional by subtracting off the contributions from the first two moments:
	\begin{align}
		\fc{J}^{\mu\nu}(\lambda, k) &= \sum_{n=2}^\8 \frac{(-i)^n}{n!} M^{\rho_1\dots\rho_n\mu\nu}_n k_{\rho_1}\dots k_{\rho_n} \label{eq:jserdef} \\
		\fc{J}^{\mu\nu}[\bar{h}_{\mu\nu}] &= \int_{-\8}^\8 \int \fc{J}^{\mu\nu*}\twid{h}_{\mu\nu}\frac{D^4k}{(2\pi)^4} D\lambda \label{eq:jfuncdef0} \\
		&= M^{\mu\nu}[\bar{h}_{\mu\nu}] - \int_{-\8}^\8\(p^\mu \doot z{}^\nu \bar{h}_{\mu\nu}(z,0) + S^{\rho\mu} \doot z{}^\nu \bar{\d}_\rho \bar{h}_{\mu\nu}(z,0)\) D\lambda. \label{eq:jfuncdef}
	\end{align}
	
	We found that near Minkowski space, the linearized Riemann tensor is the gauge invariant field strength of the metric perturbation. If we now imagine perturbing about a generic metric:
	\begin{equation}
		g_{\alpha\beta}^\text{full} = g_{\alpha\beta} + \eps h_{\alpha\beta} + \OO(\eps^2)
	\end{equation}
	then the gauge transformations by infinitesimal diffeomorphisms $x'^\alpha = x^\alpha + \eps \xi^\alpha + \OO(\eps^2)$ become:
	\begin{equation}
		h'_{\alpha\beta}(x) = h_{\alpha\beta}(x) - \nabla_\alpha \xi_\beta - \nabla_\beta\xi_\alpha.
	\end{equation}
	Perturbing the Riemann tensor gives:
	\begin{equation}
		R^\text{full}_{\gamma\alpha\delta\beta} = R_{\gamma\alpha\delta\beta} + \eps \delta R_{\gamma\alpha\delta\beta} + \OO(\eps^2)
	\end{equation}
	where, quite nontrivially, the linearized Riemann tensor is given nicely in terms of the second tensor extensions of the metric perturbation as:
	\begin{equation}
		\delta R_{\gamma\alpha\delta\beta} = \frac{1}{2}(\ee_{\alpha\delta}h_{\gamma\beta} - \ee_{\alpha\beta}h_{\gamma\delta} - \ee_{\gamma\delta} h_{\alpha\beta} + \ee_{\gamma\beta} h_{\alpha\delta}). 
	\end{equation}
	This is the curved spacetime analog of \eqref{eq:linriem}. However, in a generic spacetime the linearized Riemann tensor is not gauge invariant. Instead:
	\begin{equation}
		\delta R'_{\gamma\alpha\delta\beta}(x) = \delta R_{\gamma\alpha\delta\beta}(x) - L_\xi R_{\gamma\alpha\delta\beta}.
	\end{equation}
	As well, while the linearized Riemann tensor has all of the correct algebraic index symmetries, it does not satisfy the differential Bianchi identity (which can be seen quickly by perturbing that identity). The origin of this difficulty is that metric perturbations affect the covariant derivative itself. Thus, unlike in flat spacetime, the linearized Riemann tensor does not provide a satisfactory definition for the perturbative gravitational field strength in curved spacetime. 
	
	We will show by the end of this section that at each multipole order there is a particular relevant combination of the Riemann tensor and its derivatives whose variations serve as the variations of the gravitational field strength in the desired sense. We dub them the ``multipolar Riemann tensors'':
	\begin{equation}
		R^{(n)}_{\kappa_1\dots\kappa_n\rho\mu\sigma\nu}(z) = -2\ee_{\kappa_1\dots\kappa_n [\rho['\sigma}g_{\mu]\nu]'}.  \label{eq:mrtdef}
	\end{equation}
	$R^{(0)}_{\rho\mu\sigma\nu}$ is literally the Riemann tensor, while the others all satisfy $R^{(n)}_{\kappa_1\dots\kappa_n\rho\mu\sigma\nu} = \nabla^n_{(\kappa_1\dots\kappa_n)} R_{\rho\mu\sigma\nu} +$ corrections in terms of Riemann tensors with fewer than $n$ derivatives. These multipolar Riemann tensors have all of the same index symmetries as the corresponding $\fc{J}_n^{\cdots}$ moment, and so naturally include the consequences of the differential Bianchi identity. The relevant tensors for perturbations of a given spacetime will be $\delta R^{(n)}_{\cdots}$, meaning the perturbation in the entire multipolar Riemann tensor under a metric perturbation. Following the calculation strategy outlined in appendix \ref{sec:tensext}, the first few multipolar Riemann tensors are explicitly given by:
	\begin{align}
		& \p \p R^{(0)}_{\rho\mu\sigma\nu} = R_{\rho\mu\sigma\nu}, \p R^{(1)}_{\kappa\rho\mu\sigma\nu} = \nabla_\kappa R_{\rho\mu\sigma\nu}, \nonumber \\
		R^{(2)}_{\kappa\lambda\rho\mu\sigma\nu} = &\ \nabla^2_{(\kappa\lambda)} R_{\rho\mu\sigma\nu} + \frac{8}{9} R^\tau{}_{(\rho\sigma)(\kappa} R_{\lambda)(\mu\nu) \tau} - \frac{8}{9} R^\tau{}_{(\mu\sigma)(\kappa}R_{\lambda)(\rho\nu)\tau} \nonumber \\
		& + \frac{1}{3}R^\tau{}_{\mu\sigma\nu} R_{\tau(\kappa\lambda)\rho}  - \frac{1}{3}R^\tau{}_{\rho\sigma\nu} R_{\tau(\kappa\lambda)\mu} + \frac{1}{3} R^\tau{}_{\nu\rho\mu} R_{\tau(\kappa\lambda)\sigma} - \frac{1}{3} R^\tau{}_{\sigma\rho\mu} R_{\tau(\kappa\lambda)\nu}. 
	\end{align}
	The multipolar Riemann tensors did not appear in Dixon's original papers or in follow up expositions on the MPD theory because, while Dixon considered the energy-momentum tensor as a functional $\<T^{\alpha\beta}, \cdot\>$, he did not frame quantities it was integrated against as metric perturbations. It is that framing that will allow us to make new simplifications of the results of the MPD theory and which will cause the multipolar Riemann tensors to appear automatically. 
	
	Because we will encounter metric perturbations of tensor extensions and those extensions are constructed using Synge's worldfunction, we will need to understand the variation of the worldfunction $\delta\sigma(x,z)$ at fixed arguments $(x, z)$ under changes in that metric. We compute this by varying \eqref{eq:syngdef}. In doing so, note that the geodesic path $\zeta^\alpha(s)$ between $z$ and $x$ is itself varied (so that the varied path satisfies the geodesic equation with the varied metric). It happens by carrying out the calculation though that the effect of the path variation drops out (precisely because the worldfunction is stationary with respect to the path when evaluated on the geodesic), so that the metric variation of the worldfunction is:
	\begin{equation}
		\delta\sigma(x,z) = \frac{1}{2}\int_0^1 h_{\alpha\beta}(\zeta)\dot\zeta^\alpha\dot\zeta^\beta ds.  \label{eq:deltasig}
	\end{equation}
	
	\subsection{Decomposition of the Metric Perturbation}
	
	The structure of our calculations in this section will be highly analogous to our treatment of electromagnetism in curved spacetime. In order to evaluate $\<T^{\alpha\beta}, h_{\alpha\beta}\>$ into a form that generalizes \eqref{eq:flatthexpand} to curved spacetime, we are going to separate the metric perturbation into three pieces: the monopole (momentum) and dipole (spin) contributions that lead \eqref{eq:flatthexpand}, a symmetrized gradient of a vector field (which will be turned into an electromagnetic contribution due to \eqref{eq:tfuncnabf}), and a manifestly gauge invariant remainder (like all of the remaining terms in \eqref{eq:flatjexpand}). We expect, based on that equation, that the contribution from the remainder will turn into a series in the $\fc{J}_n^{\cdots}$. Using \eqref{eq:904}, we can see that the $\fc{J}_n^{\cdots}$ are contained in the moment generating function \eqref{eq:mfuncdef2} as: 
	\begin{equation}
		-\frac{1}{4}\(\frac{\d^2 M^{\mu\nu}}{\d k_\rho \d k_\sigma} - \frac{\d^2 M^{\rho\nu}}{\d k_\mu \d k_\sigma} - \frac{\d^2 M^{\mu\sigma}}{\d k_\rho \d k_\nu} + \frac{\d^2 M^{\rho\sigma}}{\d k_\mu \d k_\nu}\) = \sum_{n=0}^\8 \frac{(-i)^n}{n!} \fc{J}_n^{\kappa_1\dots\kappa_n \rho\mu\sigma\nu} k_{\kappa_1}\dots k_{\kappa_n}.  \label{eq:jfrommder}
	\end{equation}
	We want the $\fc{J}_n^{\cdots}$ to couple to something like the spatial derivatives of the gravitational field strength. So, let $\mc{R}_{\gamma\alpha\delta\beta}(x,z)$ be an arbitrary bitensor with all of the same index symmetries that a Riemann tensor must have and let us take its Fourier transform and integrate the result against \eqref{eq:jfrommder}. Each order of $k$ in the above series expansion under the Fourier transform becomes an order of differentiation with respect to $v$, and because $u$ contracted into any of the first $n$ indices of $\fc{J}_n^{\cdots}$ yields 0, all of those derivatives will be in spatial directions. A short calculation gives:
	\begin{equation}
		\int_{-\8}^\8\! \int \!\frac{-1}{4}\!\(\!\frac{\d^2 M^{\mu\nu}}{\d k_\rho \d k_\sigma} \!-\! \frac{\d^2 M^{\rho\nu}}{\d k_\mu \d k_\sigma} \!-\! \frac{\d^2 M^{\mu\sigma}}{\d k_\rho \d k_\nu} \!+\! \frac{\d^2 M^{\rho\sigma}}{\d k_\mu \d k_\nu} \!\)^* \!\!\twid{\mc{R}}_{\rho\mu\sigma\nu}(z(\lambda), k) \frac{D^4k}{(2\pi)^4} D\lambda  \!=\! M^{\mu\nu}[{\bar{{\mc{H}}}}_{\mu\nu}] \label{eq:mbitensh}
	\end{equation}
	where the ``bitensor metric perturbation'' is given by:
	\begin{equation}
		{\mc{H}}_{\alpha\beta}(x,z) = \mc{R}_{\gamma\alpha\delta\beta}(x,z) \sigma^\gamma(x,z)\sigma^\delta(x,z).  \label{eq:bitensh}
	\end{equation}
	Therefore, if we compute $M^{\mu\nu}[\bar{{\mc{H}}}_{\mu\nu}]$ using a bitensor metric perturbation of this form, the result will only depend on the $\fc{J}_n^{\cdots}$ moments and form a series of spatial derivatives on $\mc{R}$. Plainly, any bitensor metric perturbation of this form satisfies:
	\begin{equation}
		{\mc{H}}_{\alpha\beta} \sigma^\beta = 0.  \label{eq:bitenshcond}
	\end{equation}
	The converse is also true. That is, if a bitensor metric perturbation satisfies \eqref{eq:bitenshcond} then for some $\mc{R}_{\gamma\alpha\delta\beta}$ (with the Riemann tensor's index symmetries and satisfying the Bianchi identities) it can be put in the form of \eqref{eq:bitensh}. A quick proof of this fact is given in Appendix \ref{sec:poincah}.
	
	We ultimately want to reduce \eqref{eq:tfuncdef} to a series in terms of multipole moments when it is evaluated using a generic metric perturbation $h_{\alpha\beta}(x)$. Based on our flat spacetime analysis, we expect that the result will be closely related to $M^{\mu\nu}[\bar{h}_{\mu\nu}]$. Because of \eqref{eq:mbitensh}, $M^{\mu\nu}[\bar{{\mc{H}}}_{\mu\nu}]$ takes a manifestly gauge invariant form when using a bitensor ${\mc{H}}_{\alpha\beta}$ satisfying \eqref{eq:bitenshcond}. We also know by \eqref{eq:tfuncnabf} that the integral of $T$ against gradients of vector fields reduces to an electromagnetic expression, which we already know how to evaluate using the results of the previous section. So, in order to evaluate \eqref{eq:tfuncdef} in terms of easier pieces, we seek a decomposition of a generic metric perturbation into such a bitensor and such a gradient of a bivector $\Omega^\alpha(x,z)$ as:
	\begin{equation}
		h_{\alpha\beta}(x) = {\mc{H}}_{\alpha\beta}(x,z) + \nabla_\alpha\Omega_\beta(x,z) + \nabla_\beta\Omega_\alpha(x,z). \label{eq:bitenshgauge}
	\end{equation}
	That is, we want to take an arbitrary $h_{\alpha\beta}(x)$ and show that it is possible to choose $\Omega^\alpha(x,z)$ so that ${\mc{H}}_{\alpha\beta}$ related to them by \eqref{eq:bitenshgauge} satisfies \eqref{eq:bitenshcond}. Note that we define $\Omega^\alpha$ to naturally have its index raised as opposed to lowered. This distinction will matter as soon as we do some tangentization, because tangentization does not commute with raising and lowering indices. Conceptually, this decomposition is like a gauge transformation of the metric perturbation and so $\Omega^\alpha$ is like the vector field generating a diffeomorphism, which explains why it is natural for its index to be raised by default. 
	
	The coincidence value ($x\to z$) $\Omega_\alpha(z,z)$ falls out of \eqref{eq:bitenshgauge} entirely, so we set by definition:
	\begin{equation}
		\Omega_\alpha(z,z) = 0. \label{eq:omegaic}
	\end{equation}
	The coincidence value of $\nabla_\alpha\Omega_\beta(z,z)$ is not arbitrary. By \eqref{eq:bitensh} and \eqref{eq:bitenshgauge}, $\nabla_{(\alpha}\Omega_{\beta)}$ is completely determined. However, $\nabla_{[\alpha}\Omega_{\beta]}$ is free, and so we choose it vanish:
	\begin{equation}
		\delta^\alpha_\mu \delta^\beta_\nu\left.\nabla_\alpha\Omega_\beta\right|_{x\to z} = \frac{1}{2}h_{\mu\nu}(z). \label{eq:deromegaic}
	\end{equation}
	Now let's actually compute $\Omega_\alpha(x,z)$. Because of \eqref{eq:bitenshcond}:
	\begin{equation}
		\sigma^\beta(\nabla_\alpha\Omega_\beta+\nabla_\beta\Omega_\alpha) = h_{\alpha\beta}\sigma^\beta.
	\end{equation}
	So, along our standard geodesic:
	\begin{equation}
		\frac{d}{ds}(\dot\zeta^\alpha\Omega_\alpha) = \frac{1}{2}h_{\alpha\beta}\dot\zeta^\alpha\dot\zeta^\beta. 
	\end{equation}
	Integrating with respect to $s$ from 0 to 1 and using both \eqref{eq:omegaic} and \eqref{eq:deltasig}:
	\begin{equation}
		\sigma^\alpha(x,z)\Omega_\alpha(x,z) =  \delta\sigma(x,z). \label{eq:omegadeltasig}
	\end{equation}
	Already we are starting to see the metric variation of the worldfunction! Equation \eqref{eq:omegadeltasig} still doesn't have $\Omega_\alpha$ explicitly solved for, but we can make progress by taking its $z$ derivative:
	\begin{equation}
		\nabla_\mu \delta\sigma = \sigma^\alpha{}_\mu\Omega_\alpha + \sigma^\alpha\nabla_\mu\Omega_\alpha. \label{eq:9111}
	\end{equation}
	We can compute $\sigma^\alpha\nabla_\mu\Omega_\alpha$ by taking the $z$ derivative of \eqref{eq:bitenshgauge}, which gives:
	\begin{equation}
		0 = \nabla_\mu {\mc{H}}_{\alpha\beta} + \nabla_\alpha\nabla_\mu\Omega_\beta + \nabla_\beta\nabla_\mu\Omega_\alpha. 
	\end{equation}
	Contracting with $\sigma^\alpha\sigma^\beta$ and simplifying using \eqref{eq:bitenshcond} brings us to:
	\begin{equation}
		\frac{D}{Ds}(\dot\zeta^\alpha\nabla_\mu\Omega_\alpha) = 0 \implies \sigma^\alpha\nabla_\mu\Omega_\alpha = v^\nu \delta^\alpha_\nu \left.\nabla_\mu\Omega_\alpha\right|_{x\to z}. 
	\end{equation}
	We can determine the $z$ derivative of $\Omega_\alpha$ at coincidence by taking the $z$ derivative of the boundary condition \eqref{eq:omegaic}:
	\begin{equation}
		\delta^\beta_\nu\left.\nabla_\beta\Omega_\alpha\right|_{x\to z} + \left.\nabla_\nu\Omega_\alpha\right|_{x\to z} = 0 \implies \delta^\alpha_\nu \left.\nabla_\mu\Omega_\alpha\right|_{x\to z} = -\frac{1}{2}h_{\mu\nu}(z) \label{eq:omegader}
	\end{equation}
	where we have applied the boundary condition \eqref{eq:deromegaic}. Therefore \eqref{eq:9111} becomes:
	\begin{equation}
		\bar{\Omega}^\mu = H^{-\mu}{}_\alpha\Omega^\alpha = -\nabla^\mu\delta\sigma - \frac{1}{2} h^{\mu\nu} v_\nu. \label{eq:omegabarsol}
	\end{equation}
	The right hand side here is now an explicit functional of $h_{\alpha\beta}$, and so given any metric perturbation we can build $\Omega^\alpha$ so that \eqref{eq:bitenshcond} and \eqref{eq:bitenshgauge} are true. 
	
	We want to use our decomposition \eqref{eq:bitenshgauge} to evaluate $\<T^{\alpha\beta}, h_{\alpha\beta}\>$. We can't insert that decomposition directly because the integration in $\<T^{\alpha\beta}, h_{\alpha\beta}\>$ is over $x$, and the right hand side of \eqref{eq:bitenshgauge} is a function of both $x$ and $z$. For the ${\mc{H}}_{\alpha\beta}(x,z)$ term this is easily solved because we can evaluate its $z$ dependence at $z(\tau(x))$. For the $\nabla_\alpha\Omega_\beta(x,z)$ term there is a slight subtlety, because the differentiation with respect to $x$ and the evaluation $z = z(\tau(x))$ do not commute. Instead, by the chain rule:
	\begin{equation}
		\nabla_\alpha(\Omega_\beta(x,z(\tau(x)))) = (\nabla_\alpha\Omega_\beta)(x,z(\tau(x))) + \nabla_\alpha\tau \left.\frac{\d\Omega_\beta(x,z(\lambda))}{\d\lambda}\right|_{\lambda = \tau(x)}. 
	\end{equation}
	And so, to express the right hand side of \eqref{eq:bitensagauge} purely in terms of $x$ we have:
	\begin{align}
		h_{\alpha\beta}(x) =&\ {\mc{H}}_{\alpha\beta}(x,z(\tau(x))) + \nabla_\alpha(\Omega_\beta(x,z(\tau))) + \nabla_\beta(\Omega_\alpha(x,z(\tau))) \nonumber \\
		&\ - \nabla_\alpha \tau \left.\frac{\d\Omega_\beta}{\d\lambda}\right|_{\lambda = \tau} - \nabla_\beta\tau \left.\frac{\d\Omega_\alpha}{\d\lambda}\right|_{\lambda=\tau}.  \label{eq:hbisol}
	\end{align}
	
	To finish simplifying \eqref{eq:hbisol} we need to compute $\frac{\d\Omega_\alpha}{\d\lambda}$. Since we want to maximally maintain manifest gauge invariance, we want to do so in terms of ${\mc{H}}_{\alpha\beta}$ rather than in terms of $h_{\alpha\beta}$. Because the left hand side of \eqref{eq:bitenshgauge} does not depend on $z$ (and so does not depend on $\lambda$), by differentiating both sides with respect to $\lambda$ we find:
	\begin{equation}
		\nabla_\alpha\frac{\d\Omega_\beta}{\d\lambda} + \nabla_\beta\frac{\d\Omega_\alpha}{\d\lambda} = -\frac{\d{\mc{H}}_{\alpha\beta}}{\d\lambda}.\label{eq:omegalambdader}
	\end{equation}

	In order to integrate \eqref{eq:omegalambdader} to determine $\frac{\d\Omega_\alpha}{\d\lambda}$, we will need the values of $\frac{\d\Omega_\alpha}{\d\lambda}$ and its first $x$ derivatives at coincidence, just like we needed the corresponding coincidence to determine $\Omega_\alpha$. We can calculate those by taking the $\lambda$ derivatives of the coincidence values \eqref{eq:omegaic} and \eqref{eq:deromegaic} (and using \eqref{eq:omegader}). Doing so for \eqref{eq:omegaic}, we find:
	\begin{equation}
		\delta^\alpha_\mu \left.\frac{\d\Omega_\alpha}{\d\lambda}(x,z)\right|_{x\to z} = -\frac{1}{2}h_{\mu\nu}\dot z^\nu.  \label{eq:omegadotic} 
	\end{equation}
	The $\lambda$ derivative of \eqref{eq:deromegaic} is more complicated; it gives:
	\begin{equation}
		\delta^\alpha_\mu \delta^\beta_\nu \delta^\gamma_\rho \dot z^\rho \left.\nabla_\gamma \nabla_\alpha \Omega_\beta\right|_{x\to z} + \delta^\alpha_\mu \delta^\beta_\nu \left.\nabla_\alpha \frac{\d\Omega_\beta}{\d\lambda}\right|_{x\to z} = \frac{1}{2}\dot z^\rho \nabla_\rho h_{\mu\nu}.  \label{eq:omegadotic2}
	\end{equation}
	By \eqref{eq:bitensh}:
	\begin{equation}
		\left. \nabla_\gamma {\mc{H}}_{\alpha\beta}\right|_{x\to z} = 0 \label{eq:hderic}
	\end{equation}
	With this and \eqref{eq:bitenshgauge}, the symmetric part of \eqref{eq:omegadotic2} can be simplified significantly. After only a little algebra:
	\begin{equation}
		\delta^\alpha_{(\mu}\delta^\beta_{\nu)} \nabla_\alpha\left.\frac{\d\Omega_\beta}{\d\lambda}\right|_{x\to z} = 0.
	\end{equation}
	To simplify the antisymmetric part of \eqref{eq:omegadotic2}, we use \eqref{eq:bitenshgauge} and that the Riemann tensor is the commutator of covariant derivatives. After a bit more algebra:
	\begin{align}
		2\delta^\alpha_{[\mu}\delta^\beta_{\nu]} \nabla_\alpha\left.\frac{\d\Omega_\beta}{\d\lambda}\right|_{x\to z} =&\ -2R^\delta{}_{\rho\mu\nu}(z)\dot z^\rho \Omega_\delta(z,z) - \delta^\alpha_\mu \delta^\beta_\nu\delta^\gamma_\rho \dot z^\rho(\nabla_\alpha h_{\gamma\beta} - \nabla_\beta h_{\gamma\alpha}) \nonumber \\
		&\ + \delta^\alpha_\mu \delta^\beta_\nu\delta^\gamma_\rho \dot z^\rho(\nabla_\alpha {\mc{H}}_{\gamma\beta}-\nabla_\beta {\mc{H}}_{\gamma\alpha}). 
	\end{align}
	Then, with \eqref{eq:omegaic} and \eqref{eq:hderic}, our initial condition analogous to \eqref{eq:deromegaic} is ultimately:
	\begin{equation}
		\delta^\alpha_{\mu}\delta^\beta_{\nu} \nabla_\alpha\left.\frac{\d\Omega_\beta}{\d\lambda}\right|_{x\to z} = -\dot z^\rho\nabla_{[\mu} h_{\nu]\rho}.  \label{eq:hderic2}
	\end{equation}
	
	Now that we have the initial conditions we need, let us see how to integrate \eqref{eq:omegalambdader} to solve for $\frac{\d\Omega_\alpha}{\d\lambda}$. By contracting $\sigma^\beta$ onto \eqref{eq:omegalambdader} and simplifying, we find:
	\begin{equation}
		s\frac{D}{Ds}\frac{\d\Omega_\alpha}{\d\lambda} - \sigma^\beta{}_\alpha \frac{\d\Omega_\beta}{\d\lambda} + \nabla_\alpha\(\sigma^\beta\frac{\d\Omega_\beta}{\d\lambda}\) = -\frac{\d{\mc{H}}_{\alpha\beta}}{\d\lambda}\sigma^\beta.\label{eq:omegadiffeq}
	\end{equation}
	Let us simplify the $\sigma^\beta\frac{\d\Omega_\beta}{\d\lambda}$ term. The $\lambda$ derivative of \eqref{eq:bitenshcond} gives:
	\begin{equation}
		\frac{\d\mc{H}_{\alpha\beta}}{\d\lambda}\sigma^\beta + {\mc{H}}_{\alpha\beta}\sigma^\beta{}_\mu \dot z^\mu = 0 \label{eq:diffhcond}
	\end{equation}
	and so:
	\begin{equation}
		\frac{\d{\mc{H}}_{\alpha\beta}}{\d\lambda}\sigma^\alpha \sigma^\beta = 0. \label{eq:hsigsigder}
	\end{equation}
	Contracting $\sigma^\alpha$ into \eqref{eq:omegadiffeq} and using \eqref{eq:hsigsigder} and \eqref{eq:omegadotic}, we find:
	\begin{equation}
		\frac{d}{ds}\(\dot\zeta^\alpha\frac{\d\Omega_\alpha}{\d\lambda}\) = 0 \implies \sigma^\alpha\frac{\d\Omega_\alpha}{\d\lambda} = -\frac{1}{2}h_{\mu\nu} v^\mu \dot z^\nu.
	\end{equation}
	Returning this and \eqref{eq:diffhcond} to \eqref{eq:omegadiffeq}, we have:
	\begin{equation}
		s\frac{D}{Ds}\frac{\d\Omega_\alpha}{\d\lambda} - \sigma^\beta{}_\alpha\frac{\d\Omega_\beta}{\d\lambda} + \frac{1}{2} h_{\mu\nu}\sigma^\mu{}_\alpha\dot z^\nu = {\mc{H}}_{\alpha\beta}\sigma^\beta{}_\mu \dot z^\mu. 
	\end{equation}
	This can be reorganized into a form more suitable for integration as:
	\begin{equation}
		\frac{d}{ds}\(\frac{1}{s}\(\sigma^{\mu\alpha}\frac{\d\Omega_\alpha}{\d\lambda} - \frac{1}{2}\sigma^{\mu\nu}h_{\nu\rho} \dot z^\rho\)\) = \frac{1}{s^2} \sigma^{\mu\alpha}\sigma^\beta{}_\nu {\mc{H}}_{\alpha\beta}\dot z^\nu. \label{eq:omegadiffeq2}
	\end{equation}
	\eqref{eq:bitensh} guarantees that for small $s$, ${\mc{H}}_{\alpha\beta}$ is of order $s^2$ the right-hand side here is finite as $s\to 0$. Similarly, the expression inside the derivative on the left-hand side looks poorly behaved for small $s$, but by using the coincidence behavior of the worldfunction and \eqref{eq:omegadotic}, the combination multiplying the $\frac{1}{s}$ vanishes as $s\to 0$. Thus, the the expression inside the derivative for small $s$ becomes a difference quotient:
	\begin{equation}
		\lim_{s\to 0}\frac{1}{s}\(\sigma^{\mu\alpha}\frac{\d\Omega_\alpha}{\d\lambda} - \frac{1}{2}\sigma^{\mu\nu}h_{\nu\rho} \dot z^\rho\) =\ \left.\frac{d}{ds}\(\sigma^{\mu\alpha}\frac{\d\Omega_\alpha}{\d\lambda} - \frac{1}{2}\sigma^{\mu\nu}h_{\nu\rho} \dot z^\rho\)\right|_{s=0}= -g^{\mu\rho} v^\sigma \dot z^\nu \nabla_{[\rho}h_{\sigma]\nu}
	\end{equation}
	by using \eqref{eq:hderic2} and the coincidence properties of the worldfunction. We can now integrate \eqref{eq:omegadiffeq2} from $s=0$ to $s=1$ to find:
	\begin{equation}
		\frac{\d\Omega_\alpha}{\d\lambda}(x,z(\tau(x))) = -\frac{1}{2}K_\alpha{}^\mu h_{\mu\nu}\dot z^\nu + H_\alpha{}^{\mu} v^\nu \dot z^\rho \nabla_{[\mu}h_{\nu]\rho} - \ec(\tau)\phi_\alpha(x) \label{eq:dotomegasol}
	\end{equation}
	where we have defined the function $\phi_\alpha(x)$ by:
	\begin{equation}
		\phi_\alpha(x) = \doot z{}^\nu(\tau) H_\alpha{}^\mu(x,z(\tau)) \int_0^1 \sigma^\beta{}_\mu(\zeta,z(\tau)) \sigma^\gamma{}_\nu(\zeta,z(\tau)) {\mc{H}}_{\beta\gamma}(\zeta,z(\tau)) \frac{ds}{s^2}.  \label{eq:9133}
	\end{equation}
	
	We now have everything we need to return to \eqref{eq:hbisol} explicitly and to feed it directly into $\<T^{\alpha\beta}, h_{\alpha\beta}\>$. Doing so using \eqref{eq:tfuncnabf} while recalling \eqref{eq:volmeasder} and \eqref{eq:qdef}:
	\begin{align}
		\<T^{\alpha\beta}, h_{\alpha\beta}\> = &\int_{-\8}^\8 \int_\Sigma(T^{\alpha\beta}{\mc{H}}_{\alpha\beta}w^\gamma + 2T^{\alpha\gamma}\phi_\alpha)d\Sigma_\gamma D\lambda + 2\int_{-\8}^\8\int_\Sigma J^\alpha \fc{F}_{\alpha\beta}\Omega^\beta d\Sigma_w D\lambda \nonumber \\
		&+ \int_{-\8}^\8 \int_\Sigma T^{\alpha\beta}\(K_\beta{}^\mu \doot z{}^\nu h_{\mu\nu} + 2 v^{[\mu} H_\beta{}^{\nu]} \doot z{}^\rho \nabla_\mu h_{\nu\rho}\) d\Sigma_\alpha D\lambda. \label{eq:tevalred}
	\end{align}
	Using \eqref{eq:ibfinal}, we can evaluate all of the electromagnetic contributions to \eqref{eq:tevalred}. Let $\fc{B}_\alpha$ be defined by:
	\begin{equation}
		\fc{B}_\alpha(x,z) = \fc{F}_{\alpha\beta}(x)\Omega^\beta(x,z) \label{eq:9135}
	\end{equation}
	and let $\Upsilon(x)$ be defined by \eqref{eq:Upsilondef} in terms of this $\fc{B}_\alpha$. Using \eqref{eq:ibfinal} we have:
	\begin{equation}
		\int_{-\8}^\8\int_\Sigma J^\alpha \fc{F}_{\alpha\beta}\Omega^\beta d\Sigma_w D\lambda = I[\fc{B}] - \int_{-\8}^\8 \Upsilon J^\alpha d\Sigma_\alpha D\lambda.
	\end{equation}
	The function $\Upsilon$ can be computed directly using \eqref{eq:9135} and \eqref{eq:dotomegasol}:
	\begin{equation}
		\Upsilon(x) = -\frac{1}{2}h_{\mu\nu} \doot z{}^\nu \Phi^\mu + v^\nu \doot z{}^\rho\nabla_{[\mu} h_{\nu]\rho} \Psi^\mu - \chi(x)
	\end{equation}
	where we have recalled the definitions of the bitensors in \eqref{eq:phipsidef} and also defined the scalar function $\chi(x)$ by:
	\begin{equation}
		\chi(x) = \int_0^1 \sigma^\alpha(\zeta,z(\tau))\fc{F}_{\alpha\beta}(\zeta)\phi^\beta(\zeta) \frac{ds}{s}. \label{eq:9138}
	\end{equation}
	With this, \eqref{eq:tevalred} becomes:
	\begin{align}
		\<T^{\alpha\beta}, h_{\alpha\beta}\> =&\ \int_{-\8}^\8\(h_{\mu\nu} p^\mu \doot z{}^\nu + S^{\rho\mu}\doot z{}^\nu \nabla_\rho h_{\mu\nu}\)D\lambda + 2I^\mu[\bar{\fc{F}}_{\mu\nu}\bar{\Omega}^\nu] \nonumber \\
		&\  + \int_{-\8}^\8 \int_\Sigma \(T^{\alpha\beta}{\mc{H}}_{\alpha\beta}w^\gamma + 2 T^{\alpha\gamma}\phi_\alpha + 2\chi J^\gamma\)d\Sigma_\gamma D\lambda. \label{eq:tevalred2}
	\end{align}
	Highly nontrivially, the exact definitions of momentum and spin from \eqref{eq:psdef} have appeared automatically in this decomposition! Equation \eqref{eq:tevalred2} is the gravitational analog of \eqref{eq:jareduced}. In it, we have succeeded in manifestly separating out the monopole (momentum) and dipole (spin) terms. The remaining pieces either (in the first line) can already be evaluated using the charge multipole moments and depend on the metric perturbation only through $\Omega^\alpha$ or (in the second line) are genuinely gravitational and depend on the metric perturbation on through ${\mc{H}}_{\alpha\beta}$. 
	
	\subsection{Determination of the Multipole Moments}
	
	We can now use \eqref{eq:tevalred2} to determine the multipole moments. The first line of \eqref{eq:tevalred2} does not involve the energy-momentum tensor beyond its monopole and dipole moments. At each worldline time $\lambda$ the entire integrand of the second line of \eqref{eq:tevalred2} already only depends on the metric perturbation through ${\mc{H}}_{\alpha\beta}$. So, in analogy with \eqref{eq:imomgencond}, we can require that at each worldline time the moments satisfy:
	\begin{equation}
		\int M^{\mu\nu*}(\lambda,k) \twid{{\mc{H}}}_{\mu\nu}(z(\lambda),k) \frac{D^4k}{(2\pi)^4} = \int_\Sigma \(T^{\alpha\beta}{\mc{H}}_{\alpha\beta}w^\gamma + 2 T^{\alpha\gamma}\phi_\alpha + 2\chi J^\gamma\)d\Sigma_\gamma. \label{eq:mevalcond}
	\end{equation}
	The left hand side, by \eqref{eq:jfrommder}, is given by:
	\begin{equation}
		\int M^{\mu\nu*}(\lambda,k)\twid{{\mc{H}}}_{\mu\nu}(z(\lambda),k) \frac{D^4k}{(2\pi)^4} = \int \twid{\mc{R}}_{\rho\mu\sigma\nu}\sum_{n=0}^\8 \frac{i^n}{n!} \fc{J}_n^{\kappa_1\dots\kappa_n\rho\sigma\mu\nu} k_{\kappa_1}\dots k_{\kappa_n} \frac{D^4k}{(2\pi)^4}. \label{eq:mevall}
	\end{equation}
	It will be our goal next to evaluate the right hand side of \eqref{eq:mevalcond} similarly into an integral of some series of in $k$ against $\twid{\mc{R}}_{\rho\mu\sigma\nu}$. 
	
	Explicitly expanding the first term on the right hand side of \eqref{eq:mevalcond} in terms of $\twid{\mc{R}}_{\rho\mu\sigma\nu}$ gives:
	\begin{equation}
		\int_\Sigma T^{\alpha\beta}{\mc{H}}_{\alpha\beta} d\Sigma_w = \int \twid{\mc{R}}_{\rho\mu\sigma\nu} \int e^{ik\cdot v} v^\rho v^\sigma H^{-\mu}{}_\alpha H^{-\nu}{}_\beta T^{\alpha\beta} d\Sigma_w \frac{D^4k}{(2\pi)^4}.  \label{eq:923}
	\end{equation}
	We now define the curved spacetime generalization of the naive moments of the energy-momentum tensor, in close analogy to \eqref{eq:naivetdef}:
	\begin{equation}
		\mc{t}_n^{\kappa_1\dots\kappa_n\mu\nu} = \int_\Sigma v^{\kappa_1}\dots v^{\kappa_n} H^{-\mu}{}_\alpha H^{-\nu}{}_\beta T^{\alpha\beta} d\Sigma_w
	\end{equation}
	With these, after series expanding \eqref{eq:923} in $k$, we have:
	\begin{equation}
		\int_\Sigma T^{\alpha\beta}{\mc{H}}_{\alpha\beta} d\Sigma_w = \int \twid{\mc{R}}_{\rho\mu\sigma\nu}\sum_{n=0}^\8 \frac{i^n}{n!} \mc{t}_{n+2}^{\kappa_1\dots\kappa_n[\rho['\sigma \mu] \nu]'} k_{\kappa_1}\dots k_{\kappa_n} \frac{D^4k}{(2\pi)^4}. \label{eq:mevalr1}
	\end{equation}
	
	It takes a little more effort to evaluate the second and third terms on the right hand side of \eqref{eq:mevalcond} because $\phi_\alpha(x)$ and $\chi(x)$ are more complicated. Using \eqref{eq:bitensh} in \eqref{eq:9133}, we find: 
	\begin{equation}
		\phi_\alpha = H_{\alpha\mu} \doot z{}_\nu v^\tau v^\omega \int_0^1 \sigma^{\mu\beta}\sigma_\beta{}^\rho \sigma^{\nu\gamma}\sigma_\gamma {}^\sigma \bar{\mc{R}}_{\tau\rho\omega\sigma} ds. 
	\end{equation}
	In terms of the Fourier transform:
	\begin{equation}
		\phi_\alpha \!=\! H_{\alpha\mu}\doot z{}_\nu v^\tau v^\omega \!\!\int\!\! \twid{\mc{R}}_{\tau\rho\omega\sigma} \sum_{n=0}^\8 \frac{i^n}{n!}(k\cdot v)^n\int_0^1 \sigma^{\mu\beta}(\zeta,z) \sigma_\beta{}^\rho(\zeta,z) \sigma^{\nu\gamma}(\zeta,z)\sigma_\gamma{}^\sigma(\zeta,z) s^n ds \frac{D^4k}{(2\pi)^4}.
	\end{equation}
	Define the bitensors $\text{V}_n^{\rho\sigma\mu\nu}(x,z)$ by:
	\begin{equation}
		\text{V}_n^{\rho\sigma\mu\nu} = (n-1)\int_0^1 \sigma^\rho{}_\alpha(\zeta,z)\sigma^{\alpha(\mu}(\zeta,z) \sigma^{\nu)}{}_\beta(\zeta, z) \sigma^{\beta\sigma}(\zeta,z) s^{n-2} ds.
	\end{equation}
	The factor of $(n-1)$ is chosen so that in flat spacetime these bitensors all become $\eta^{\rho(\mu}\eta^{\nu)\sigma}$. With these, $\phi_\alpha$ becomes:
	\begin{equation}
		\phi_\alpha(x) = H_{\alpha\mu}\doot z{}_\nu v^\iota v^\kappa \int \twid{\mc{R}}_{\iota\rho\kappa\sigma} \sum_{n=0}^\8 \frac{i^n}{(n+1)!}(k\cdot v)^n \text{V}^{\rho\sigma\mu\nu}_{n+2} \frac{D^4k}{(2\pi)^4}. 
	\end{equation}
	We can now feed this directly into \eqref{eq:9138} to get:
	\begin{equation}
		\chi(x) = \int\twid{\mc{R}}_{\rho\mu\sigma\nu}\sum_{n=0}^\8 \frac{i^n}{(n+1)!} (k\cdot v)^n v^\tau \doot z{}_\kappa v^\rho v^\sigma \int_0^1 \fc{F}_{\alpha\beta} H^\alpha{}_\tau H^\beta{}_\iota s^{n+2} \text{V}^{\mu\nu\iota\kappa}_{n+2} ds \frac{D^4k}{(2\pi)^4}.
	\end{equation}
	Define the bitensors $\text{W}_n^{\rho\sigma\mu\nu}(x,z)$ by:
	\begin{equation}
		\text{W}_n^{\rho\sigma\mu\nu}(x,z) = (n+1)\int_0^1 \fc{F}_{\alpha\beta}(\zeta) H^{\alpha\rho}(\zeta,z)H^\beta{}_\kappa(\zeta,z) \text{V}^{\kappa\sigma\mu\nu}_n(\zeta,z) s^n ds
	\end{equation}
	The factor of $(n+1)$ is chosen so that in flat spacetime and for a uniform electromagnetic field strength these bitensors all become $\fc{F}{}^{\rho(\mu} \eta{}^{\nu)\sigma}$. With these, $\chi(x)$ simplifies to:
	\begin{equation}
		\chi(x) = v_\iota \doot z{}_\kappa v^\rho v^\sigma\int\twid{\mc{R}}_{\rho\mu\sigma\nu} \sum_{n=0}^\8 \frac{i^n}{(n+1)!(n+3)} (k\cdot v)^n \text{W}_{n+2}^{\iota\kappa\mu\nu} \frac{D^4k}{(2\pi)^4}.
	\end{equation}
	With our simplified $\phi_\alpha$ and $\chi$, we can now evaluate the second and third parts of the integral on the right hand side of \eqref{eq:mevalcond}:
	\begin{equation}
		\int(2 T^{\alpha\beta} \phi_\beta + 2\chi J^\alpha) d\Sigma_\alpha = \int \twid{\mc{R}}_{\rho\mu\sigma\nu}\sum_{n=0}^\8 \frac{i^n}{(n+1)!} k_{\kappa_1}\dots k_{\kappa_n} \mc{p}_{n+2}^{\kappa_1\dots\kappa_n[\rho['\sigma\mu]\nu]'\iota}\doot z{}_\iota \frac{D^4k}{(2\pi)^4} \label{eq:mevalr2}
	\end{equation}
	where we have defined, similarity to \eqref{eq:naivepdef}, the curved spacetime generalization of the naive moments of the momentum density:
	\begin{equation}
		\mc{p}_n^{\kappa_1\dots\kappa_n\mu\nu\sigma} = \int_{\Sigma} v^{\kappa_1}\dots v^{\kappa_n} \(T^{\alpha\beta} H_{\beta\rho} \text{V}_n^{\rho\sigma\mu\nu} + \frac{1}{n+1} J^\alpha v_\rho \text{W}_n^{\rho\sigma\mu\nu}\) d\Sigma_\alpha. 
	\end{equation}
	Notice that due to the additional complexity of curved spacetime, these coefficients carry more indices than their flat spacetime analogs. 
	
	Collecting \eqref{eq:mevall}, \eqref{eq:mevalr1}, and \eqref{eq:mevalr2} back into \eqref{eq:mevalcond} we have:
	\begin{equation}
		\int\!\! \twid{\mc{R}}_{\rho\mu\sigma\nu}\!\sum_{n=0}^\8 \frac{i^n}{n!} k_{\kappa_1}\dots k_{\kappa_n}\!\!\(\! \fc{J}_n^{\kappa_1\dots\kappa_n \rho\mu\sigma\nu} - \mc{t}_{n+2}^{\kappa_1\dots\kappa_n[\rho['\sigma\mu]\nu]'} \!- \frac{2}{n+1}\mc{p}_{n+2}^{\kappa_1\dots\kappa_n[\rho['\sigma\mu]\nu]'\iota}\doot z{}_\iota\!\! \) \frac{D^4k}{(2\pi)^4} = 0. 
	\end{equation}
	In order for this to hold for all possible $\twid{\mc{R}}_{\rho\mu\sigma\nu}$ we must have:
	\begin{equation}
		\fc{J}_n^{\kappa_1\dots\kappa_n \rho\mu\sigma\nu} = \mc{t}_{n+2}^{\kappa_1\dots\kappa_n[\rho['\sigma\mu]\nu]'} + \frac{2}{n+1}\mc{p}_{n+2}^{\kappa_1\dots\kappa_n[\rho['\sigma\mu]\nu]'\iota}\doot z{}_\iota.
	\end{equation}
	Thus, this defines the $\fc{J}_n^{\cdots}$ moments completely and in turn through \eqref{eq:mfromj2} determines the potential multipole moments $M_n^{\cdots}$. It is straightforward (though algebraically tedious) from this definition to check that all of the index symmetries from \eqref{eq:jmomprops2} are satisfied.
	
	\subsection{Final Form of the Action}
	
	We can now relate the moment generating functional directly to $\<T^{\alpha\beta}, h_{\alpha\beta}\>$. By feeding \eqref{eq:mevalcond} into \eqref{eq:tevalred2} we have:
	\begin{equation}
		\<T^{\alpha\beta}, h_{\alpha\beta}\> = \int_{-\8}^\8(p^\mu \doot z{}^\nu h_{\mu\nu} + S^{\rho\mu} \doot z{}^\nu \nabla_\rho h_{\mu\nu}) D\lambda + 2Q^\mu[\bar{\fc{F}}_{\mu\nu}\bar{\Omega}^\nu] + \fc{J}^{\mu\nu}[\bar{{\mc{H}}}_{\mu\nu}]. \label{eq:tact1}
	\end{equation}
	Here we have used that by \eqref{eq:mevall}, $M^{\mu\nu}[\bar{{\mc{H}}}_{\mu\nu}]$ does not depend on $M^{\mu\nu}_0$ or $M^{\rho\mu\nu}_1$ and so is actually equal to $\fc{J}^{\mu\nu}[\bar{{\mc{H}}}_{\mu\nu}]$ and that similarly by \eqref{eq:qgendef}, $I^\mu[\bar{\fc{F}}_{\mu\nu}\bar{\Omega}^\nu]$ does not depend on $I^\mu_0$ because $\bar{\Omega}^\nu(z, 0 ) = 0$ (due to \eqref{eq:omegaic}), and so is actually equal to $Q[\bar{\fc{F}}_{\mu\nu}\bar{\Omega}^\nu]$.

	We want to reexpress $\fc{J}^{\mu\nu}[\bar{{\mc{H}}}_{\mu\nu}]$ directly in terms of $h$ to develop the gravitational analog of \eqref{eq:ijarelate} and the curved spacetime generalization of \eqref{eq:mfuncdef}. Remember that by definition $\Omega^\alpha$ has an upper index naturally. With that, we can rewrite \eqref{eq:bitenshgauge} as:
	\begin{equation}
		{\mc{H}}_{\alpha\beta} = h_{\alpha\beta} - L_\Omega g_{\alpha\beta}.
	\end{equation}
	By equation \eqref{eq:tanlieder}:
	\begin{equation}
		\bar{{\mc{H}}}_{\mu\nu} = \bar{h}_{\mu\nu} - L_{\bar{\Omega}} G_{\mu\nu} \label{eq:hbarlie}
	\end{equation}
	where we have recalled the definition of the tangentized metric from \eqref{eq:tangentizemet}. $\bar{\fc{F}}_{\mu\nu}\bar{\Omega}^\nu$ can be simplified similarly. In particular, because of \eqref{eq:bitensagauge}:
	\begin{equation}
		\fc{F}_{\alpha\beta} = \nabla_\alpha \mc{A}_\beta - \nabla_\beta \mc{A}_\alpha. \label{eq:fcaf}
	\end{equation}
	Using this, we can reorganize $\fc{F}_{\alpha\beta}\Omega^\beta$ using \eqref{eq:bilieder} as:
	\begin{equation}
		\fc{F}_{\alpha\beta}\Omega^\beta = -L_\Omega \mc{A}_\alpha + \nabla_\alpha(\mc{A}_\beta\Omega^\beta) + \Omega^\mu(z,z) {\nabla}_\mu \mc{A}_\alpha \label{eq:935}
	\end{equation}
	where in principle it is important to remember when Lie differentiating that $\mc{A}_\alpha$ is a bitensor that is a vector at $x$ and a scalar at $z$ (recall \eqref{eq:bilieder}). In this specific case, the scalar term does not matter because $\Omega^\mu(z,z) = 0$. Under tangentization, \eqref{eq:935} becomes:
	\begin{equation}
		\bar{\fc{F}}_{\mu\nu}\bar{\Omega}^\nu = -L_{\bar{\Omega}} \bar{\mc{A}}_\mu + \bar{\d}_\mu (\bar{\mc{A}}_\nu \bar{\Omega}^\nu). \label{eq:fomega}
	\end{equation}
	Due to the index symmetries of the $Q^{\cdots}_n$ moments, the total derivative term does not actually contribute to $Q^\mu[\bar{\fc{F}}_{\mu\nu}\bar{\Omega}^\nu]$. Returning \eqref{eq:hbarlie} and \eqref{eq:fomega} to \eqref{eq:tact1}, we arrive at:
	\begin{equation}
		\<T^{\alpha\beta}, h_{\alpha\beta}\> = M^{\mu\nu}[\bar{h}_{\mu\nu}] - 2Q^\mu[L_{\bar\Omega} \bar{\mc{A}}_\mu] - \fc{J}^{\mu\nu}[L_{\bar\Omega}G_{\mu\nu}]. \label{eq:tact2}
	\end{equation}
	
	Equation \eqref{eq:tact2} is the gravitational analog of \eqref{eq:ijarelate} and the curved spacetime generalization of \eqref{eq:mfuncdef}. There is an important difference between \eqref{eq:tact2} and those previous results: the $\Omega^\alpha$ terms on the right-hand side. These terms prevent $\<T^{\alpha\beta}, h_{\alpha\beta}\>$ and $M^{\mu\nu}[\bar{h}_{\mu\nu}]$ from coinciding the way they did in flat spacetime (or the equivalent electromagnetic quantities did even in curved spacetime). There is a simple reason something like the $\Omega^\alpha$ terms have to be there. The time evolution of momentum and spin are determined by generalized forces (linear force and torque) which originate from Poincaré-Killing transformations (translations and Lorentz transformations in the neighborhood of the worldline). Such transformations are diffeomorphisms and under such a diffeomorphism, the metric perturbation becomes (up to a sign) the Lie derivative of the spacetime metric with respect to that diffeomorphism. Because the interaction action is proportional to $\<T^{\alpha\beta}, h_{\alpha\beta}\>$, that combination must contain the forces due to the diffeomorphism when viewed as a variational principle, and the $\Omega^\alpha$ terms in \eqref{eq:tact2} will turn out to be precisely the terms responsible for the force and torque in the equations of motion for the body's momentum and spin. 
	
	Next we will develop a curved spacetime generalization of \eqref{eq:flatthexpand}/a gravitational analog of \eqref{eq:jacurvexpand}. The $\Omega^\alpha$ terms will be crucial in getting us a simple result. The key to our approach is to reexpress $\bar{h}_{\mu\nu}$, which is the tangentization of the metric perturbation, in terms of $\delta G_{\mu\nu}$, meaning the perturbation of the tangentized metric. Because the act of tangentization itself depends on the metric, these two do not coincide. By varying $G_{\mu\nu}(z,v)$ at fixed $(z, v)$, we find by using the chain rule:
	\begin{equation}
		\delta G_{\mu\nu} = \delta H^\alpha{}_\mu H^\beta{}_\nu g_{\alpha\beta} + H^\alpha{}_\mu \delta H^\beta{}_\nu g_{\alpha\beta} + H^\alpha{}_\mu H^\beta{}_\nu \delta g_{\alpha\beta} + \delta x^\gamma\nabla_\gamma (H^\alpha{}_\mu H^\beta{}_\nu g_{\alpha\beta}) \label{eq:deltabigg}
	\end{equation}
	where the variations in $H^\alpha{}_\mu$ and $g_{\alpha\beta}$ are taken at fixed $(x,z)$ and $\delta x^\gamma$ is the variation in $x^\gamma$ under a change in the metric at fixed $(z,v)$. From $H^\alpha{}_\mu = -\sigma^{-\alpha}{}_\mu$, it immediately follows that:
	\begin{equation}
		\delta H^\alpha{}_\mu = H^\alpha{}_\nu \bar{\d}_\mu \delta\sigma^\nu
	\end{equation}
	where $\delta\sigma^\nu$ is the metric variation of $\sigma^\nu$. Importantly, $\delta \sigma^\mu \neq \nabla^\mu \delta\sigma$ as $\sigma^\mu = g^{\mu\nu}\nabla_\nu \sigma$. Instead:
	\begin{equation}
		\delta\sigma^\mu = \nabla^\mu\delta\sigma + h^{\mu\nu}v_\nu \label{eq:deltasignab}
	\end{equation}
	after using $v^\mu = -\sigma^\mu(x,z)$. By differentiating $v^\mu = -\sigma^\mu(x,z)$ it immediately follows that at fixed $(z,v)$:
	\begin{equation}
		\delta x^\alpha = H^\alpha{}_\mu \delta \sigma^\mu. 
	\end{equation}
	Returning these to \eqref{eq:deltabigg}, we find:
	\begin{equation}
		\delta G_{\mu\nu}(z,v) = \bar{h}_{\mu\nu} + \delta \sigma^\rho \bar{\d}_\rho G_{\mu\nu} + G_{\mu\rho}\bar{\d}_\nu\delta\sigma^\rho + G_{\rho\nu}\bar{\d}_\mu\delta\sigma^\rho.  \label{eq:9312}
	\end{equation}
	Using \eqref{eq:deltasignab} and \eqref{eq:omegabarsol}, this becomes:
	\begin{equation}
		\delta G_{\mu\nu}(z,v) = \bar{h}_{\mu\nu} - L_{\bar{\Omega}} G_{\mu\nu} + \frac{1}{2}h^{\rho\sigma} v_\rho \bar{\d}_\sigma G_{\mu\nu} + \frac{1}{2}h^\rho{}_\mu G_{\rho\nu} + \frac{1}{2}h^\rho{}_\nu G_{\rho\mu}. 
	\end{equation}
	So, by \eqref{eq:hbarlie}:
	\begin{equation}
		\bar{{\mc{H}}}_{\mu\nu} = \delta G_{\mu\nu} - \frac{1}{2}h^{\rho\sigma} v_\rho \bar{\d}_\sigma G_{\mu\nu} - \frac{1}{2}h^\rho{}_\mu G_{\rho\nu} - \frac{1}{2}h^\rho{}_\nu G_{\rho\mu}.  \label{eq:hbardeltag}
	\end{equation}
	
	We now want to feed \eqref{eq:hbardeltag} into $\fc{J}^{\mu\nu}[\bar{{\mc{H}}}_{\mu\nu}]$ in \eqref{eq:tact1}. The $\delta G_{\mu\nu}$ piece is straightforward to evaluate just by using \eqref{eq:jserdef}, \eqref{eq:jfuncdef0}, and \eqref{eq:mfromj2}:
	\begin{equation}
		\fc{J}^{\mu\nu}[\delta G_{\mu\nu}] = \int_{-\8}^\8 4\sum_{n=0}^\8 \frac{n+1}{(n+3)!} \fc{J}_n^{\kappa_1\dots\kappa_n\rho\mu\sigma\nu} \left.\bar{\d}^{n+2}_{\kappa_1\dots\kappa_n\rho\sigma} \delta G_{\mu\nu}\right|_{v=0} D\lambda. 
	\end{equation}
	Finally it is time for the multipolar Riemann tensors to appear. By \eqref{eq:mrtdef} and \eqref{eq:tensexdef}, the metric variations of the multipolar Riemann tensors are given by:
	\begin{equation}
		\delta R^{(n)}_{\kappa_1\dots\kappa_n\rho\mu\sigma\nu} = -2\left.\bar{\d}^{n+2}_{\kappa_1\dots\kappa_n[\rho['\sigma} \delta G_{\mu]\nu]'}\right|_{v=0}
	\end{equation}
	and so, using the index symmetries of the $\fc{J}^{\cdots}_n$ moments \eqref{eq:904}, we find:
	\begin{equation}
		\fc{J}^{\mu\nu}[\delta G_{\mu\nu}] = -2\int_{-\8}^\8 \sum_{n=0}^\8 \frac{n+1}{(n+3)!} \fc{J}_n^{\kappa_1\dots\kappa_n\rho\mu\sigma\nu} \delta R^{(n)}_{\kappa_1\dots\kappa_n\rho\mu\sigma\nu} D\lambda. \label{eq:jeval1}
	\end{equation}
	
	The remaining pieces of \eqref{eq:hbardeltag} can also be evaluated nicely in terms of the multipolar Riemann tensors with a bit of index manipulation. Computing directly gives:
	\begin{align}
		 \fc{J}^{\mu\nu}[h^{\rho\sigma} v_\rho &\bar{\d}_\sigma G_{\mu\nu} + h^\rho{}_\mu G_{\rho\nu} + h^\rho{}_\nu G_{\rho\mu}] \nonumber \\
		 &= -2\int_{-\8}^\8\sum_{n=0}^\8 \frac{n+1}{(n+3)!} \fc{J}_n^{\kappa_1\dots\kappa_n\rho\mu\sigma\nu}(n h^\tau{}_{(\kappa_1} R^{(n)}_{\kappa_2\dots\kappa_n)\tau\rho\mu\sigma\nu} + 4 h^\tau{}_\rho R^{(n)}_{\kappa_1\dots\kappa_n\tau\mu\sigma\nu}) D\lambda \\
		 &= 2\int_{-\8}^\8 \sum_{n=0}^\8\frac{n+1}{(n+3)!} \fc{J}_n^{\kappa_1\dots\kappa_n\rho\mu\sigma\nu} h_\tau{}^\omega \fc{G}^\tau{}_\omega R^{(n)}_{\kappa_1\dots\kappa_n\rho\mu\sigma\nu} D\lambda \label{eq:jeval2}
	\end{align}
	where the last step requires being a bit clever to recognize the appearance of index shuffling. Together equations \eqref{eq:hbardeltag}, \eqref{eq:jeval1}, and \eqref{eq:jeval2} give:
	\begin{equation}
		\fc{J}^{\mu\nu}[\bar{{\mc{H}}}_{\mu\nu}] = -2\int_{-\8}^\8\sum_{n=0}^\8 \frac{n+1}{(n+3)!} \fc{J}_n^{\kappa_1\dots\kappa_n\rho\mu\sigma\nu}\(\delta R^{(n)}_{\kappa_1\dots\kappa_n\rho\mu\sigma\nu} + \frac{1}{2} h_{\tau\omega} \fc{G}^{\tau\omega} R^{(n)}_{\kappa_1\dots\kappa_n\rho\mu\sigma\nu}\) D\lambda.  \label{eq:9320}
	\end{equation}
	
	By following a similar strategy, we can evaluate the $Q^\mu[\bar{\fc{F}}_{\mu\nu}\bar{\Omega}^\nu]$ in terms of the metric perturbations of tensor extensions of the electromagnetic field strength. Just following through the evaluation of $Q^\mu[\bar{\fc{F}}_{\mu\nu}\bar{\Omega}^\nu]$ directly via \eqref{eq:qserdef}, \eqref{eq:qgendef}, and \eqref{eq:ifromq2}, we find:
	\begin{equation}
		Q^\mu[\bar{\fc{F}}_{\mu\nu}\bar{\Omega}^\nu] = \int_{-\8}^\8\sum_{n=0}^\8 \frac{n+1}{(n+2)!}Q_n^{\rho_1\dots\rho_n\mu\nu}\left.\bar{\d}^n_{\rho_1\dots\rho_n}(\bar{\d}_\mu(\bar{\fc{F}}_{\nu\sigma}\bar{\Omega}^\sigma) - \bar{\d}_\nu(\bar{\fc{F}}_{\mu\sigma}\bar{\Omega}^\sigma))\right|_{v=0} D\lambda. 
	\end{equation}
	Under a metric variation, by \eqref{eq:tensexdef}:
	\begin{equation}
		\delta\ee_{\rho_1\dots\rho_n}\fc{F}_{\mu\nu} = \left.\bar{\d}^n_{\rho_1\dots\rho_n}\delta\bar{\fc{F}}_{\mu\nu}\right|_{v=0}. 
	\end{equation}
	Now let's compute $\delta\bar{\fc{F}}_{\mu\nu}$, varied with respect to the metric at fixed $(z,v)$. By the same reasoning as \eqref{eq:deltabigg}:
	\begin{equation}
		\delta\bar{\fc{F}}_{\mu\nu} = \delta H^\alpha{}_\mu H^\beta{}_\nu \fc{F}_{\alpha\beta} + H^\alpha{}_\mu H^\beta{}_\nu \fc{F}_{\alpha\beta} + H^\alpha{}_\mu H^\beta{}_\nu \delta\fc{F}_{\alpha\beta} + \delta x^\gamma \nabla_\gamma(H^\alpha{}_\mu H^\beta{}_\nu\fc{F}_{\alpha\beta}). 
	\end{equation}
	Upon simplification by the same steps leading to \eqref{eq:9312}, we have:
	\begin{equation}
		\delta\bar{\fc{F}}_{\mu\nu} = \delta\sigma^\rho\bar{\d}_\rho \bar{\fc{F}}_{\mu\nu} + \bar{\fc{F}}_{\mu\rho} \bar{\d}_\nu \delta\sigma^\rho + \bar{\fc{F}}_{\rho\nu} \bar{\d}_\mu \delta \sigma^\rho. 
	\end{equation}
	Using the Bianchi identity and rewriting $\delta\sigma^\mu$ in terms of $\bar{\Omega}^\mu$ using \eqref{eq:omegabarsol} brings us to:
	\begin{equation}
		\delta\bar{\fc{F}}_{\mu\nu} = \bar{\d}_\mu(\bar{\fc{F}}_{\nu\rho}\bar{\Omega}^\rho) - \bar{\d}_\nu(\bar{\fc{F}}_{\mu\rho}\bar{\Omega}^\rho) + \frac{1}{2}h^{\rho\sigma} v_\rho \bar{\d}_\sigma \bar{\fc{F}}_{\mu\nu} + \frac{1}{2}h^\rho{}_\mu \bar{\fc{F}}_{\rho\nu} + \frac{1}{2}h^\rho{}_\nu \bar{\fc{F}}_{\mu\rho}. 
	\end{equation}
	Differentiating $n$ times at $v=0$ gives:
	\begin{equation}
		\left.\bar{\d}^n_{\rho_1\dots\rho_n}(\bar{\d}_\mu(\bar{\fc{F}}_{\nu\sigma}\bar{\Omega}^\sigma) - \bar{\d}_\nu(\bar{\fc{F}}_{\mu\sigma}\bar{\Omega}^\sigma))\right|_{v=0} = \delta \ee_{\rho_1\dots\rho_n}\fc{F}_{\mu\nu}+ \frac{1}{2}h_{\tau\omega}\fc{G}^{\tau\omega} E_{\rho_1\dots\rho_n}\fc{F}_{\mu\nu}. \label{eq:9326}
	\end{equation}
	Now returning \eqref{eq:9320} and \eqref{eq:9326} back to \eqref{eq:tact1}, we have:
	\begin{align}
		\delta \fc{S}_m =&\ \frac{1}{2}\<T^{\alpha\beta}, h_{\alpha\beta}\> \nonumber \\
		=&\ \frac{1}{2}\int_{-\8}^\8\(p^\mu \doot z{}^\nu h_{\mu\nu} + S^{\rho\mu} \doot z{}^\nu \nabla_\rho h_{\mu\nu}\) D\lambda \nonumber \\
		&\ + \int_{-\8}^\8 \sum_{n=0}^\8 \frac{n+1}{(n+2)!} Q_n^{\rho_1\dots\rho_n\mu\nu}\(\delta \ee_{\rho_1\dots\rho_n}\fc{F}_{\mu\nu} + \frac{1}{2}h_{\tau\omega} \fc{G}^{\tau\omega} \ee_{\rho_1\dots\rho_n}\fc{F}_{\mu\nu}\) D\lambda \nonumber \\
		&\ - \int_{-\8}^\8 \sum_{n=0}^\8 \frac{n+1}{(n+3)!} \fc{J}_n^{\kappa_1\dots\kappa_n\rho\mu\sigma\nu}\(\delta R^{(n)}_{\kappa_1\dots\kappa_n\rho\mu\sigma\nu} + \frac{1}{2} h_{\tau\omega}\fc{G}^{\tau\omega} R^{(n)}_{\kappa_1\dots\kappa_n\rho\mu\sigma\nu}\) D\lambda. \label{eq:gravact}
	\end{align}
	This is the replacement of \eqref{eq:flatthexpand} in full general relativity and one of the most important equations in the entire MPD theory. It fully determines $T^{\alpha\beta}$ as a distribution, reducing it to only quantities valued on the worldline. Previous literature only considered $\<T^{\alpha\beta}, h_{\alpha\beta}\>$ with $h_{\alpha\beta}$ thought as a generic test function, and so while Dixon's original papers do have \eqref{eq:tact2}, equation \eqref{eq:gravact} has not appeared before. While we still treat $h_{\alpha\beta}$ as a generic test function, by interpreting it as a metric perturbation, we were uniquely able to motivate the multipolar Riemann tensors and to use them to reorganize the calculation so that the result takes this simple, exact form. 
	
	\subsection{MPD Equations of Motion}
	
	We now want to show that our multipole moments are indeed fully reduced, meaning that there are no restrictions on the dynamics of the $\fc{J}_n^{\cdots}$ provided by the energy-momentum transport equation. By equation \eqref{eq:tfuncnabf}, we need to require:
	\begin{equation}
		\<T^{\alpha\beta}, \nabla_{(\alpha} f_{\beta)}\> - \<J^\alpha, \fc{F}_{\alpha\beta} f^\beta\> = 0. \label{eq:tnabf}
	\end{equation}
	for arbitrary $f^\alpha(x)$. Consider the metric perturbation:
	\begin{equation}
		h_{\alpha\beta}(x) = \nabla_\alpha f_\beta(x) + \nabla_\beta f_\alpha(x) = L_f g_{\alpha\beta}. \label{eq:hlieg}
	\end{equation}
	For this special case, using \eqref{eq:tact1} and \eqref{eq:ijarelate} equation, \eqref{eq:tnabf} becomes:
	\begin{equation}
		\int_{-\8}^\8 \!\!\(\frac{1}{2}p^\mu \doot z{}^\nu h_{\mu\nu} + \frac{1}{2}S^{\rho\mu}\doot z{}^\nu \nabla_\rho h_{\mu\nu} + q \fc{F}_{\mu\nu} \doot z{}^\nu f^\mu(z)\!\!\)D\lambda + Q^\mu[\bar{\fc{F}}_{\mu\nu}(\bar{\Omega}^\nu-\bar{f}^\nu)] + \frac{1}{2} \fc{J}^{\mu\nu}[\bar{{\mc{H}}}_{\mu\nu}] = 0. \label{eq:tnabf2}
	\end{equation}
	The first two pieces in terms of the momentum and spin can be simplified significantly using \eqref{eq:hlieg}. With a little algebra and integration by parts:
	\begin{align}
		\frac{1}{2}\int_{-\8}^\8\(p^\mu \doot z{}^\nu h_{\mu\nu}\right. &+ \left. S^{\rho\mu}\doot z{}^\nu \nabla_\rho h_{\mu\nu}\)D\lambda \nonumber \\
		&= \int_{-\8}^\8\(f^\mu\(-\doot p{}_\mu - \frac{1}{2}R_{\mu\nu\rho\sigma} \doot z{}^\nu S^{\rho\sigma}\) + \frac{1}{2}\nabla_{[\mu}f_{\nu]}(-\doot S{}^{\mu\nu} + p^\mu\doot z{}^\nu - p^\nu\doot z{}^\mu)\)D\lambda. \label{eq:tnabf3}
	\end{align}
	These pieces will eventually give major contributions to the equations of motion for the momentum and spin when we enforce \eqref{eq:tnabf}. 
	
	For the special case \eqref{eq:hlieg}, $\Omega^\alpha$ can be simplified significantly. The integrand in \eqref{eq:omegadeltasig} becomes a total derivative, so that:
	\begin{equation}
		\delta\sigma(x,z) = \sigma^\alpha(x,z) f_\alpha(x) + \sigma^\mu(x,z) f_\mu(z).
	\end{equation}
	Returning this to \eqref{eq:omegabarsol} and simplifying then gives:
	\begin{equation}
		\Omega^\alpha(x,z) = f^\alpha(x) - \xi^\alpha(x,z)
	\end{equation}
	where we have defined the bitensor $\xi^\alpha(x,z)$ by:
	\begin{equation}
		\xi^\alpha(x,z) = K^\alpha{}_\mu(x,z) f^\mu(z) + v^\mu(x,z) H^{\alpha\nu}(x,z) \nabla_{[\mu}f_{\nu]}(z).  \label{eq:lilomegadef}
	\end{equation}
	$\xi^\alpha$ is exactly a Poincaré-Killing transformation! Its amount of translation and Lorentz transformation are precisely those given by using $f^\alpha(x)$ as a diffeomorphism in the neighborhood of the worldline. Returning this to \eqref{eq:bitenshgauge}, we have simply:
	\begin{equation}
		{\mc{H}}_{\alpha\beta} = L_\xi g_{\alpha\beta}.
	\end{equation}
	Or, tangentized:
	\begin{equation}
		\bar{{\mc{H}}}_{\mu\nu} = L_{\bar\xi} G_{\mu\nu}.  \label{eq:hbarsimp}
	\end{equation}
	
	The electromagnetic contributions to \eqref{eq:tnabf2} can be simplified similarly. The $Q^\mu[\cdots]$ contribution to \eqref{eq:tnabf2} is:
	\begin{equation}
		Q^\mu[\bar{\fc{F}}_{\mu\nu}(\bar{\Omega}^\nu - \bar{f}^\nu)] = -Q^\mu[\bar{\fc{F}}_{\mu\nu}\xi^\nu]. 
	\end{equation}
	Using \eqref{eq:fcaf} and \eqref{eq:bilieder}, we can reorganize $\fc{F}_{\alpha\beta}\xi^\beta$ as:
	\begin{equation}
		\fc{F}_{\alpha\beta}\xi^\beta = -L_\xi \mc{A}_\alpha + \nabla_\alpha(\mc{A}_\beta\xi^\beta) + f^\mu {\nabla}_\mu \mc{A}_\alpha
	\end{equation}
	where it is important to remember when Lie differentiating that $\mc{A}_\alpha$ is a bitensor that is specifically a vector at $x$ and a scalar at $z$. Under tangentization:
	\begin{equation}
		\bar{\fc{F}}_{\mu\nu}\bar{\xi}^\nu = -L_{\bar{\xi}} \bar{\mc{A}}_\mu + \bar{\d}_\mu (\bar{\mc{A}}_\nu \bar{\xi}^\nu - f^\nu(z){\nabla}_\nu \varphi)  \label{eq:fbaromega}
	\end{equation}
	where we have used that, due to \eqref{eq:bitensagauge} and because $A_\alpha(x)$ does not depend on $z$:
	\begin{equation}
		\nabla_\nu\mc{A}_\alpha = -\nabla_\alpha{\nabla}_\nu\varphi.
	\end{equation}
	Due to the index symmetries of the $Q^{\cdots}_n$ moments, the total derivative term does not actually contribute to $Q^\mu[\bar{F}_{\mu\nu}\bar{\xi}^\nu]$. Thus:
	\begin{equation}
		Q^\mu[\bar{\fc{F}}_{\mu\nu}(\bar{\Omega}^\nu - \bar{f}^\nu)] = Q^\mu[L_{\bar\xi} \bar{\mc{A}}_\mu]. \label{eq:9414}
	\end{equation}
	
	Returning \eqref{eq:tnabf3}, \eqref{eq:hbarsimp}, and \eqref{eq:9414} to \eqref{eq:tnabf2}, we find:
	\begin{align}
		 \int_{-\8}^\8\( f^\mu\(-\doot p_\mu - \frac{1}{2}R_{\mu\nu\rho\sigma}\doot z{}^\nu S^{\rho\sigma} + q\fc{F}_{\mu\nu} \doot z{}^\nu\) + \frac{1}{2}\nabla_{[\mu} f_{\nu]}\(-\doot S{}^{\mu\nu} + p^\mu\doot z{}^\nu - \doot z{}^\mu p^\nu\)\)D\lambda \nonumber \\
		 + Q^\mu[L_{\bar{\xi}} \bar{\mc{A}}_\mu] + \frac{1}{2}\fc{J}^{\mu\nu}[L_{\bar{\xi}} G_{\mu\nu}] = 0. \label{eq:tnabf4}
	\end{align}
	$\xi^\alpha$ depends on $f^\alpha(x)$ only linearly and only through $f^\mu(z)$ and $\nabla^{[\mu} f^{\nu]}(z)$ by \eqref{eq:lilomegadef}. Therefore, for some functions $\text{\textbf{F}}_\mu(\lambda)$ (the ``non-point force vector'') and $\text{\textbf{T}}_{\mu\nu}(\lambda)$ (the ``non-point torque tensor'' ) which are independent of $f^\alpha(x)$: 
	\begin{equation}
		Q^\mu[L_{\bar\xi}\bar{\mc{A}}_\mu] + \frac{1}{2}\fc{J}^{\mu\nu}[L_{\bar\xi} G_{\mu\nu}] = \int_{-\8}^\8\(f^\mu(z){\text{\textbf{F}}}_\mu + \frac{1}{2}\nabla_{[\mu}f_{\nu]}(z){\text{\textbf{T}}}^{\mu\nu}\)D\lambda. \label{eq:ftintro}
	\end{equation}
	Because $f^\alpha(x)$ is arbitrary, $f^\mu(z)$ and $\nabla^{[\mu} f^{\nu]}(z)$ can be independently specified at each worldline time. Returning \eqref{eq:ftintro} to \eqref{eq:tnabf4} and using that \eqref{eq:tnabf4} must hold for arbitrary $f^\alpha(x)$ then gives that the pieces multiplying $f^\mu(z)$ and $\nabla^{[\mu} f^{\nu]}(z)$ must each vanish at each worldline time. Therefore, \eqref{eq:tnabf} is equivalent to:
	\begin{equation}
	\doot p_\mu = - \frac{1}{2}R_{\mu\nu\rho\sigma}\doot z{}^\nu S^{\rho\sigma} + q\fc{F}_{\mu\nu} \doot z{}^\nu + \text{\textbf{F}}_\mu, \p \doot S{}^{\mu\nu} = p^\mu\doot z{}^\nu - \doot z{}^\mu p^\nu + \text{\textbf{T}}^{\mu\nu}. \label{eq:mpd}
	\end{equation}
	We did it! These are the \textbf{Mathisson--Papapetrou--Dixon (MPD)} equations of motion for the momentum and spin! They are the general relativistic analogs of \eqref{eq:newtfdef}. And, as we have just shown, they are the necessary and sufficient conditions for the energy-momentum transport equation \eqref{eq:tcons} to hold.
	
	Now that we have properly derived our equations of motion, we next move to simplify the force vector and torque tensor as much as possible. Because $\xi^\alpha$ is a PKT, using \eqref{eq:goodlieder}:
	\begin{align}
		L_{\bar{\xi}}\bar{\mc{A}}_\mu &= f^\rho(z)\bar{\nabla}_\rho \bar{\mc{A}}_\mu + \nabla^{[\rho} f^{\sigma]}(z)(v_\rho \bar{\d}_\sigma \bar{\mc{A}}_\mu + g_{\mu\rho} \bar{\mc{A}}_\sigma) \nonumber \\
		L_{\bar{\xi}}G_{\mu\nu} &= f^\rho(z)\bar{\nabla}_\rho G_{\mu\nu} + \nabla^{[\rho} f^{\sigma]}(z)(v_\rho \bar{\d}_\sigma G_{\mu\nu} + g_{\mu\rho} G_{\sigma\nu} + g_{\nu\rho} G_{\mu \sigma}). 
	\end{align}
	Let us use these to compute $Q^\mu[L_{\bar{\xi}} \bar{\mc{A}}_\mu]$ and $\fc{J}^{\mu\nu}[L_{\bar{\xi}} G_{\mu\nu}]$. The steps of calculation are mathematically very similar to those followed in deriving \eqref{eq:gravact}. Evaluating $Q^\mu[L_{\bar{\xi}} \bar{\mc{A}}_\mu]$ directly, we find:
	\begin{equation}
		Q^\mu[L_{\bar{\xi}} \bar{\mc{A}}_\mu] = \int_{-\8}^\8\sum_{n=0}^\8 \frac{n+1}{(n+2)!}Q_n^{\rho_1\dots\rho_n\mu\nu}\left.\bar{\d}^n_{\rho_1\dots\rho_n}(\bar{\d}_\mu(L_{\bar{\xi}}\bar{\mc{A}}_\nu) - \bar{\d}_\nu(L_{\bar{\xi}}\bar{\mc{A}}_\mu))\right|_{v=0} D\lambda. 
	\end{equation}
	Expanding the derivatives:
	\begin{equation}
		\bar{\d}_\mu(L_{\bar{\xi}}\bar{\mc{A}}_\nu) - \bar{\d}_\nu(L_{\bar{\xi}}\bar{\mc{A}}_\mu) = f^\rho\bar{\nabla}_\rho \bar{\fc{F}}_{\mu\nu} + \nabla^{[\rho}f^{\sigma]}(v_\rho \bar{\d}_\sigma \bar{\fc{F}}_{\mu\nu} + g_{\mu\rho} \bar{\fc{F}}_{\sigma\nu} + g_{\nu\rho}\bar{\fc{F}}_{\mu\sigma}). 
	\end{equation}
	Consequently, at $v=0$:
	\begin{equation}
		\left.\bar{\d}^n_{\kappa_1\dots\kappa_n}(\bar{\d}_\rho(L_{\bar{\xi}}\bar{\mc{A}}_\sigma) - \bar{\d}_\sigma(L_{\bar{\xi}}\bar{\mc{A}}_\rho))\right|_{v=0} = f^\mu \nabla_\mu \ee_{\kappa_1\dots\kappa_n} \fc{F}_{\rho\sigma} - \nabla^{[\mu}f^{\nu]} \fc{G}_{[\mu\nu]} \ee_{\kappa_1\dots\kappa_n}\fc{F}_{\rho\sigma}. 
	\end{equation}
	Thus, the total electromagnetic contribution to \eqref{eq:ftintro} is:
		\begin{equation}
		Q^\mu[L_{\bar{\xi}} \bar{\mc{A}}_\mu] = \int_{-\8}^\8\sum_{n=0}^\8 \frac{n+1}{(n+2)!}Q_n^{\kappa_1\dots\kappa_n\rho\sigma}(f^\mu \nabla_\mu \ee_{\kappa_1\dots\kappa_n} \fc{F}_{\rho\sigma} - \nabla^{[\mu}f^{\nu]} \fc{G}_{[\mu\nu]} \ee_{\kappa_1\dots\kappa_n}\fc{F}_{\rho\sigma}) D\lambda.  \label{eq:qlieeval}
	\end{equation}
	
	Similarly paralleling the steps for evaluating \eqref{eq:jeval1} and \eqref{eq:jeval2}, we directly find:
	\begin{equation}
		\fc{J}^{\mu\nu}[L_{\bar{\xi}}G_{\mu\nu}] = -2\int_{-\8}^\8\sum_{n=0}^\8\frac{n+1}{(n+3)!} \fc{J}^{\kappa_1\dots\kappa_n\rho\mu\sigma\nu}_n \left.\bar{\d}^n_{\kappa_1\dots\kappa_n}(-2 \bar{\d}^2_{[\rho['\sigma} L_{\bar{\xi}} G_{\mu]\nu']})\right|_{v=0} D\lambda. 
	\end{equation}
	Expanding the derivatives:
	\begin{align}
		-2 \bar{\d}^2_{[\rho['\sigma} L_{\bar{\xi}} G_{\mu]\nu']} =\ f^\kappa \bar{\nabla}_\kappa \fc{R}_{\rho\mu\sigma\nu} + \nabla^{[\tau}f^{\omega]}(& v_\tau \bar{\d}_\omega \fc{R}_{\rho\mu\sigma\nu} + g_{\tau\rho}\fc{R}_{\omega\mu\sigma\nu} + g_{\tau\mu} \fc{R}_{\rho\omega\sigma\nu} \nonumber \\
		&\ + g_{\tau\sigma} \fc{R}_{\rho\mu\omega\nu} + g_{\tau\nu} \fc{R}_{\rho\mu\sigma\omega})
	\end{align}
	where we have defined the tangent tensor:
	\begin{equation}
		\fc{R}_{\rho\mu\sigma\nu}(z,v) = -2 \bar{\d}^2_{[\rho['\sigma} G_{\mu]\nu']}.
	\end{equation}
	Consequently, at $v=0$:
	\begin{equation}
		\left.\bar{\d}^n_{\kappa_1\dots\kappa_n}(-2 \bar{\d}^2_{[\rho['\sigma} L_{\bar{\xi}} G_{\mu]\nu']})\right|_{v=0} = f^\tau\nabla_\tau R^{(n)}_{\kappa_1\dots\kappa_n\rho\mu\sigma\nu} - \nabla^{[\tau}f^{\omega]} \fc{G}_{[\tau\omega]} R^{(n)}_{\kappa_1\dots\kappa_n\rho\mu\sigma\nu}. 
	\end{equation}
	Thus, the total gravitational contribution to \eqref{eq:ftintro} is:
	\begin{align}
		\fc{J}^{\mu\nu}[L_{\bar{\xi}}G_{\mu\nu}] = -2\int_{-\8}^\8\sum_{n=0}^\8\frac{n+1}{(n+3)!} \fc{J}^{\kappa_1\dots\kappa_n\rho\mu\sigma\nu}_n \bigl(&f^\tau\nabla_\tau R^{(n)}_{\kappa_1\dots\kappa_n\rho\mu\sigma\nu} \nonumber \\
		&- \nabla^{[\tau}f^{\omega]} \fc{G}_{[\tau\omega]} R^{(n)}_{\kappa_1\dots\kappa_n\rho\mu\sigma\nu} \bigr) D\lambda. \label{eq:jlieeval}
	\end{align}
	
	By feeding \eqref{eq:qlieeval} and \eqref{eq:jlieeval} into \eqref{eq:ftintro}, we can read off the force and torque as:
	\begin{align}
		\text{\textbf{F}}_\mu =&\  - \sum_{n=0}^\8\frac{n+1}{(n+3)!} \fc{J}^{\kappa_1\dots\kappa_n\rho\tau\sigma\omega}_n \nabla_\mu R^{(n)}_{\kappa_1\dots\kappa_n\rho\tau\sigma\omega} + \sum_{n=0}^\8 \frac{n+1}{(n+2)!}Q_n^{\kappa_1\dots\kappa_n\rho\sigma}\nabla_\mu \ee_{\kappa_1\dots\kappa_n} \fc{F}_{\rho\sigma} \nonumber \\
		\text{\textbf{T}}_{\mu\nu} =&\ 2\sum_{n=0}^\8\frac{n+1}{(n+3)!} \fc{J}^{\kappa_1\dots\kappa_n\rho\tau\sigma\omega}_n\fc{G}_{[\mu\nu]} R^{(n)}_{\kappa_1\dots\kappa_n\rho\tau\sigma\omega} -2\sum_{n=0}^\8 \frac{n+1}{(n+2)!}Q_n^{\kappa_1\dots\kappa_n\rho\sigma}\fc{G}_{[\mu\nu]} \ee_{\kappa_1\dots\kappa_n}\fc{F}_{\rho\sigma}. \label{eq:ftexpr}
	\end{align}
	These are the ultimate expressions for the force and torque in terms of the multipole moments! They are the analogs of the right hand sides of \eqref{eq:ngp} and \eqref{eq:ngs} in Newtonian gravity. 

	We can now understand very plainly the sense in which the MPD equations are exact but do not form a closed system of differential equations. For a fixed choice of worldline and time-slicing $(z, u)$, if we are simply given the full time dependence of the charge $Q^{\cdots}_n$ and mass $\fc{J}^{\cdots}_n$ multipole moments of the body, then equations \eqref{eq:mpd} and \eqref{eq:ftexpr} give exact equations of motion for the body's momentum and spin. Because \eqref{eq:mpd} is exactly equivalent to \eqref{eq:tcons}, absolutely nothing whatsoever about the time evolution of those higher multipole moments can be inferred from the universal facts of compact object motion in general relativity. In order to form a closed system of equations that can actually be used to make calculations, on top of the MPD theory one needs a model specific to the body of interest that determines the time evolution of its higher multipole moments. We discuss how to do this in practice at the end of Section \ref{sec:wlact}.
	
	\section{MPD from a Variational Principle} \label{sec:wlact}
	
	Dixon's original theory consisted of the MPD equations \eqref{eq:mpd}, the expressions for the MPD force and torque \eqref{eq:ftexpr} (in a less simplified form), the integral definitions of the charge and mass multipole moments in terms of the current density and energy-momentum tensor, and their consequent behaviors when the current density and energy-momentum tensor are integrated against test functions (\eqref{eq:jacurvexpand} and \eqref{eq:gravact}, again in less simplified form). Using the center of momentum time-slicing \eqref{eq:udef} and the specific worldline which satisfies the covariant spin supplementary condition \eqref{eq:cssc}, the worldline trajectory is determined explicitly by \eqref{eq:comwl}. This set of equations contain all of the physics \eqref{eq:tcons} and \eqref{eq:jcons}. However, the various definitions of the full MPD theory expressed in this form are extremely complicated and do not form a closed system of equations of motion. In this section we will how the entire theory can be packaged much more simply into a worldline variational principle. That variational form is ultimately the most useful approach for doing real calculations as it manifestly closes the equations of motion when a body specific model is available.

	\subsection{Covariant Worldline Variational Principles}
		
		We seek to describe the evolution of a compact object in curved spacetime interacting with gravity and electromagnetism via a variational principle in terms of the location of its center of mass position and its internal degrees of freedom. Because we are now going at the problem with a different approach, we will reintroduce several established concepts (the worldline time and einbein, for example) based on how they appear in this different thinking. So, temporarily suspend our preexisting definitions until they are recalled. At each time $t$ (according to whoever is doing the time-slicing), the body's position is specified by three spatial coordinates $z^1, z^2, z^3$ which we package as $\vec z$. As well, at each $t$, the body has some internal degrees of freedom $\psi^\Omega(t)$ where $\Omega$ is an index running over all of those internal variables, whose specific structure we pay no mind to yet. The Lagrangian form of the action for the body in arbitrary background fields is then:
		\begin{equation}
			\fc{S}[\vec z, \psi] = \int_{-\8}^\8 L(t, \vec z, \vec z_t, \psi, \psi_t) dt \label{eq:sbase}
		\end{equation}
		where $\vec z_t$ and $\psi_t$ are the time derivatives defined as:
		\begin{equation}
			\vec z_t = \frac{d\vec z}{dt}, \p \psi^\Omega_t = \frac{d\psi^\Omega}{dt}.
		\end{equation}
		This action is allowed to depend on (presently unwritten) background fields such as the metric (and its derivatives) as well as the electromagnetic field (and its derivatives). Later when we need to think about those dependencies, we will write them as arguments explicitly. 
		
		The Euler-Lagrange equations of motion from the stationary variation of the action are exactly as usual:
		\begin{equation}
			\delta \fc{S} \assert 0 \implies \frac{\d L}{\d\vec z} = \frac{d}{dt} \frac{\d L}{\d\vec z_t}, \p \frac{\d L}{\d \psi^\Omega} = \frac{d}{dt} \frac{\d L}{\d\psi^\Omega_t}. 
		\end{equation}
		Already we can see the inherent advantage of a variational principle: the equations of motion are always closed in terms of the variables in the action. The conjugate momentum $\vec{p}$ to $\vec z$, conjugate momenta $\pi_\Omega$ to $\psi^\Omega$, and Hamiltonian $H$ are defined as usual:
		\begin{equation}
			\vec p = \frac{\d L}{\d \vec z_t}, \p \pi_\Omega = \frac{\d L}{\d \psi^\Omega_t}, \p H(t, \vec z, \vec p, \psi, \pi) = \vec p\cdot\vec z_t + \pi_\Omega \psi^\Omega_t - L(t, \vec z, \vec z_t, \psi, \psi_t). \label{eq:1014}
		\end{equation}
		It quickly follows from the Euler-Lagrange equations that the momenta and Hamiltonian have time evolution given by:
		\begin{equation}
			\frac{d\vec p}{dt} = -\frac{\d H}{\d \vec z}, \p \frac{d \pi_\Omega}{dt} = -\frac{\d H}{\d \psi^\Omega}, \p  \frac{dH}{dt} = \frac{\d H}{\d t}. \label{eq:hbase}
		\end{equation}
		
		The action as written in \eqref{eq:sbase} is not manifestly covariant due to the separation of the time and space coordinates. We covariantize the action by introducing $z^0 = t$ as well as the worldline time function $\lambda(t)$, which is an arbitrary smooth and increasing function of $t$. Since $\lambda(t)$ is invertible, we can define $z^\mu(\lambda)$ and $\psi^\Omega(\lambda)$ so that when $\lambda$ is evaluated at $\lambda(t)$ they agree with the previous meanings of these symbols. We define the dot-over-a-symbol derivative as the $\lambda$ time-derivative:
		\begin{equation}
			\dot z^\mu = \frac{dz^\mu}{d\lambda}, \p \dot\psi^\Omega = \frac{d\psi^\Omega}{d\lambda}.
		\end{equation}
		Now we can rewrite the action in \eqref{eq:sbase} as:
		\begin{equation}
			\fc{S}[z,\psi] = \int_{-\8}^\8 \l(z, \dot z, \psi, \dot\psi) d\lambda \label{eq:slbase}
		\end{equation}
		where the $\lambda$-Lagrangian $\l$ is given by:
		\begin{equation}
			\l(z,\dot z,\psi,\dot\psi) = \dot z^0 L\(z, \frac{\dvec z}{\dot z^0}, \psi, \frac{\dot\psi}{\dot z^0}\). \label{eq:lambdal}
		\end{equation}
		Applying the chain rule to \eqref{eq:lambdal} and using \eqref{eq:1014}, it follows that: 
		\begin{align}
			&\frac{\d\l}{\d \vec z} = \dot z^0 \frac{\d L}{\d\vec z},& &\frac{\d\l}{\d\psi^\Omega} = \dot z^0\frac{\d L}{\d\psi^\Omega},& &\frac{\d \l}{\d z^0} = \dot z^0\frac{\d L}{\d t}, \nonumber \\
			&\frac{\d\l}{\d \dvec z} = \vec p,& &\frac{\d \l}{\d\dot\psi^\Omega} = \pi_\Omega,& &\frac{\d\l}{\d\dot z^0} = L - \vec p\cdot\vec z_t - \pi_\Omega\psi^\Omega_t = -H. \label{eq:lambdaleom}
		\end{align}
		With these, the equations of motion from \eqref{eq:hbase} can be rewritten as:
		\begin{equation}
			\frac{\d\l}{\d z^\mu} = \frac{d}{d\lambda}\frac{\d\l}{\d\dot z^\mu}, \p \frac{\d\l}{\d\psi^\Omega} = \frac{d}{d\lambda}\frac{\d\l}{\d\dot\psi^\Omega}
		\end{equation}
		which are now manifestly covariant. Therefore, for arbitrary worldline time parameter $\lambda$, the action as written in \eqref{eq:slbase} is stationary with respect to variations of the covariant degrees of freedom $z^\mu(\lambda), \psi^\Omega(\lambda)$. As well, by identifying the the energy $H = -p_0$, we have:
		\begin{equation}
			p_\mu = \frac{\d\l}{\d\dot z^\mu}.  \label{eq:pmul}
		\end{equation}
		
		Not just any function $\l(z,\dot z,\psi, \dot\psi)$ can be used for the action in \eqref{eq:slbase}. Due to the form in \eqref{eq:lambdal}, $\l$ has a very particular kind of dependence on $\dot z^0$. This is reflected in the final condition of \eqref{eq:lambdaleom} which can be rewritten covariantly as:
		\begin{equation}
			\dot z^\mu \frac{\d\l}{\d \dot z^\mu} + \dot \psi^\Omega \frac{\d \l}{\d\dot\psi^\Omega } - \l = 0. \label{eq:reparam}
		\end{equation}
		This differential equation fully determines the $\dot z^0$ dependence of $\l$ and is solved identically and exclusively by a function of the form \eqref{eq:lambdal}. Now let $\ec$ be some real parameter and observe by use of the chain rule that \eqref{eq:reparam} is equivalent to:
		\begin{equation}
			\frac{\d}{\d\ec}\(\ec \l\(z, \frac{\dot z}{\ec}, \psi, \frac{\dot\psi}{\ec}\)\) = 0. 
		\end{equation}
		Integrating with respect to $\ec$ gives:
		\begin{equation}
			\ec \l\(z, \frac{\dot z}{\ec}, \psi, \frac{\dot\psi}{\ec}\) = \l\(z, \dot z, \psi, \dot\psi\). \label{eq:einbeinl}
		\end{equation}
		This condition really just expresses that the parameter $\lambda$ is arbitrary. Under a reparameterization to a new parameter $\lambda' = \lambda'(\lambda)$:
		\begin{gather}
		z'^\mu(\lambda') = z^\mu(\lambda), \ps \psi'^\Omega(\lambda') = \psi^\Omega(\lambda) \ps \implies \ps \dot z'^\mu = \frac{d\lambda}{d\lambda'}\dot z^\mu, \ps \dot \psi'^\Omega = \frac{d\lambda}{d\lambda'} \dot \psi^\Omega.
		\end{gather}
		Due to \eqref{eq:einbeinl} with $\ec$ replaced by $\frac{d\lambda'}{d\lambda}$:
		\begin{equation}
			\l(z',\dot z',\psi',\dot\psi') = \frac{d\lambda}{d\lambda'} \l(z,\dot z, \psi, \dot\psi). \label{eq:lreparam}
		\end{equation}
		Integrating this over $d\lambda'$, we have that the action is reparameterization invariant:
		\begin{equation}
			\fc{S}[z',\psi'] = \fc{S}[z,\psi].
		\end{equation}
		The reverse reasoning holds as well. Reparameterization invariance of the action for arbitrary smooth increasing $\lambda'(\lambda)$ implies \eqref{eq:lreparam} which in turn can be differentiated back into \eqref{eq:reparam}. 
		
		Instead of trying to build the function $\l$ to satisfy \eqref{eq:lreparam} manually, we can be a little clever and enforce it indirectly by introducing an auxiliary field (akin to a Lagrange multiplier). Let $\ec(\lambda)$, called the einbein, be an arbitrary, smooth, positive function of the worldline time. We let the einbein transform under reparameterization as in the result of \eqref{eq:ecdef} and recall the reparameterization invariant worldline time derivative from \eqref{eq:dootder} and the reparameterization invariant worldline time measure $D\lambda$ from \eqref{eq:3213}. Now consider the action:
		\begin{equation}
			\fc{S}[\ec, z, \psi] = \int_{-\8}^\8 \l(z,\doot z, \psi, \doot\psi) D\lambda \label{eq:seinbein2}
		\end{equation}
		where we allow $\l$ to not necessarily satisfy \eqref{eq:reparam}. If we require the vanishing variation of this action with respect to $\ec$ as a function (noting carefully that it appears in $\doot z$, $\doot \psi$, and $D\lambda$), we find that reparameterization invariance emerges as a constraint:
		\begin{equation}
			\frac{\delta\fc{S}}{\delta \ec} = \frac{\d}{\d\ec}\(\ec \l\(z, \frac{\dot z}{\ec}, \psi, \frac{\dot\psi}{\ec}\)\) \assert 0. \label{eq:deltaseinbein}
		\end{equation}	
		Along the actual solution to this equation, due to \eqref{eq:einbeinl}, the action \eqref{eq:seinbein2} just reduces back to \eqref{eq:slbase}. So, going forward, we can just take \eqref{eq:seinbein2} as the form of our action and in doing so we free ourselves of having to worry about choosing $\l$ to satisfy \eqref{eq:reparam}. 
		
		From \eqref{eq:seinbein2}, we define the reparameterization invariant momenta $p_\mu$ and $\pi_\Omega$ and covariant Hamiltonian $\h(z,p,\psi,\pi)$ by:
		\begin{equation}
			p_\mu = \frac{\d\l}{\d\doot z{}^\mu}, \p \pi_\Omega = \frac{\d\l}{\d\doot \psi{}^\Omega}, \p \h(z,p,\psi,\pi) = p_\mu\doot z{}^\mu + \pi_\Omega \doot \psi{}^\Omega - \l(z,\doot z, \psi, \doot \psi). 
		\end{equation}
		Thus, in Hamiltonian form our action \eqref{eq:seinbein2} is simply:
		\begin{equation}
			\fc{S}[\ec, z, p, \psi, \pi] = \int_{-\8}^\8\(p_\mu\dot z{}^\mu + \pi_\Omega\dot\psi^\Omega - \ec \h(z, p, \psi, \pi)\) d\lambda. \label{eq:shamact}
		\end{equation}
		The Legendre transformation drastically simplifies the einbein dependence of the action so that now the only condition required by reparameterization invariance is that on-shell the covariant Hamiltonian must vanish:
		\begin{equation}
			\frac{\delta\fc{S}}{\delta\ec} = -\h \assert 0. 
		\end{equation}
		This is known as the On-Shell Constraint (OSC). 
		
		Our next step is to simplify the on-shell constraint into maximally convenient form. Because all that is required is $\h = 0$ on-shell, there are many different choices of $\h$ which give the same physics. To motivate what form we coax $\h$ into, let us first look at the dynamics of a point-particle in a generic metric. The action is simply:
		\begin{equation}
			\fc{S}_{\text{geo}} = -m\int_{-\8}^\8|\dot z|d\lambda.
		\end{equation}
		This action gives the equation of motion:
		\begin{equation}
			\frac{\delta \fc{S}_{\text{geo}}}{\delta z^\mu} = -m g_{\mu\nu}\frac{D}{D\lambda}\left(\frac{\dot z^\nu}{|\dot z|}\right)  \label{eq:geod}
		\end{equation}
		which becomes the usual geodesic equation when set to 0. The linear momentum produced by this action is:
		\begin{equation}
			p_\mu = \frac{\d}{\d\dot z^\mu}(-m|\dot z|) = m g_{\mu\nu}\frac{\dot z^\nu}{|\dot z|}
		\end{equation}
		which plainly satisfies $|p| = m$ on-shell. Now consider the action $\fc{S}_m$ which enforces $|p| = m$ as a constraint:
		\begin{equation}
			\fc{S}_m = \int_{-\8}^\8(p_\mu\dot z^\mu + \mathtt{e}(|p| - m)) d\lambda \label{eq:actfree}.
		\end{equation}
		The einbein and momentum variations give:
		\begin{align}
			\frac{\delta \fc{S}_m}{\delta \mathtt{e}} \assert 0 \implies |p| = m, \p \frac{\delta \fc{S}_m}{\delta p_\mu} \assert 0\implies p^\mu = m\doot z{}^\mu. 
		\end{align}
		The resulting equation of motion for the trajectory with these imposed is:
		\begin{equation}
			\frac{\delta \fc{S}_m}{\delta z^\mu} = -\dot p_\mu + \mathtt{e} \d_\mu\sqrt{-g^{\rho\sigma} p_\rho p_\sigma} = -m g_{\mu\nu}\frac{D}{D\lambda}\left(\frac{\dot z^\nu}{|\dot z|}\right) 
		\end{equation}
		which exactly matches equation \ref{eq:geod}. Therefore, $\fc{S}_m$ and $\fc{S}_{\text{geo}}$ are physically equivalent. $\fc{S}_m$ is already in the form of \eqref{eq:shamact} and so a valid choice of covariant Hamiltonian for a free point-particle is:
		\begin{equation}
			\h_\text{free} = m-|p|. \label{eq:freeham}
		\end{equation}
		Such an expression is necessarily not unique. The covariant Hamiltonian can be modified to any new function $\h' = \h F$ where $F$ is an arbitrary function without zeros. Doing so will also rescale the einbein unless $F$ limits to $1$ when on-shell ($\h = 0$). Another very reasonable and common choice of covariant Hamiltonian is $\frac{p^2 + m^2}{2m}$. 
		
		For a generic (i.e. non-point) body which has internal degrees of freedom and which may not be free, its covariant Hamiltonian will not in general be given by \ref{eq:freeham}. Instead, define $\delta\h$ as the difference between its actual Hamiltonian and the would-be free point-particle Hamiltonian:
		\begin{equation}
			\delta\h \df \h - \h_\text{free} \implies \h(z, p, \psi, \pi) = -|p| + m + \delta\h(z, p, \psi, \pi).
		\end{equation}
		Then, the OSC gives an implicit equation for $|p|$:
		\begin{equation}
			\h \assert 0 \implies |p| = m + \delta\h(z, |p|u, \psi, \pi) \label{eq:onshellp}
		\end{equation}
		where we have defined the unit vector in the direction of momentum:
		\begin{equation}
			u^\mu = \frac{p^\mu}{|p|}.
		\end{equation}
		In order for the variational principle to give well-defined equations of motion, the implicit equation \eqref{eq:onshellp} must only have 1 solution for real non-negative $|p|$. Solving this equation for $|p|$ will give an expression for $|p|$ in terms of $z, u, \psi, \pi$:
		\begin{equation}
			\text{OSC}\implies |p| = \mathcal{M}(z, u, \psi, \pi)
		\end{equation}
		where $\mathcal{M}$, the solution to the OSC, is called the dynamical mass function~\cite{BAILEY1980188, JanSteinhoff:2015ist, Vines:2016unv}.
		
		Using the dynamical mass function, we can rewrite the covariant Hamiltonian as:
		\begin{equation}
			\h(z, p, \psi, \pi) = \frac{-|p|+m+\delta\h(z,p,\psi,\pi)}{-|p|+\fc{M}(z,u,\psi,\pi)}\Bigl(-|p|+\fc{M}(z,u,\psi,\pi)\Bigr).
		\end{equation}
		The prefactor fraction here remains finite in the limit that the constraint is applied. We can define the new covariant Hamiltonian $\h'$ by:
		\begin{equation}
			\h'(z, p, \psi, \pi) = -|p| + \fc{M}(z,u,\psi,\pi)
		\end{equation}
		and the new einbein $\ec'$ by:
		\begin{equation}
			\ec' = \ec \frac{-|p|+m+\delta\h(z,p,\psi,\pi)}{-|p|+\fc{M}(z,u,\psi,\pi)} \implies \ec \h = \ec'\h'. \label{eq:einrescale}
		\end{equation}
		Because $\ec$ is arbitrary, we can equivalently treat $\ec'$ as arbitrary and thus can treat $\h'$ equally well as the effective Hamiltonian in the variational principle. We assume going forward we have already taken them as such and so no longer write the primes. 
		With the preceding considerations, we have that the action for a body in curved spacetime with a worldline $z$ and arbitrary internal degrees of freedom can be put into the form:
		\begin{equation}
			\fc{S}[\ec, z, p, \psi, \pi] = \int_{-\8}^\8\(p_\mu \dot z{}^\mu + \pi_\Omega \dot\psi^\Omega + \ec(|p| - \fc{M}(z, u, \psi, \pi))\) d\lambda.  \label{eq:dmfact}
		\end{equation}
	
	\subsection{Spin Degrees of Freedom}
	
	Now we will decompose our previously generic internal degrees of freedom $\psi^\Omega$ into rotational degrees of freedom $\phi^a$ which control the overall orientation of the body and other degrees of freedom $\scri^\Omega$ which determine the configuration of the body when it is in a given orientation. We define these degrees of freedom so that under a rotation of the body, the $\phi^a$ change and the $\scri^\Omega$ do not. 
	
	Recall the definition of the vierbein as a local basis for the tangent space at $x$ from \eqref{eq:pseudometric}, which can be written in components as:
	\begin{equation}
		g_{\alpha\beta} \iota^\alpha{}_A\iota^\beta{}_B = \eta_{AB}. \label{eq:vierbein}
	\end{equation}
	We can express any tensor in terms of the vierbein, for example a vector field $f^\alpha$ with corresponding components $f^A$ as $f^\alpha = \iota^\alpha{}_A f^A$. We call components in terms of the vierbein the local Lorentz frame. The indices of such components are raised and lowered using the Minkowski metric.

	At each worldline time, the body may be oriented differently with respect to the local Lorentz frame (due to rotating and being boosted relative to that frame). We define the body frame as the instantaneous frame comoving with the body in which the orientation of the body is instantaneously constant, and we represent tensors expressed in this frame with middle alphabet capital Latin indices $I, J, K, L$. These indices are raised and lowered using the Minkowski metric. At each worldline time, there will be a set of four vectors (with $I$ indicating which vector is chosen) so that their components $\Lambda^\mu{}_I(\lambda)$, called the body-frame tetrad (or just tetrad for short), transform spacetime tensors to the body frame. That is, we can express any tensor on the worldline in terms of the these $\Lambda^\mu{}_I(\lambda)$, for example a vector $f^\mu$ with corresponding components $f^I$ as $f^\mu = \Lambda^\mu{}_I f^I$. Since these $\Lambda^\mu{}_I$ are a local Minkowski basis:
	\begin{equation}
		g_{\mu\nu} \Lambda^\mu{}_I \Lambda^\nu{}_J = \eta_{IJ}.  \label{eq:tetrad}
	\end{equation}
	We assume we have ordered and oriented our coordinates so that $\Lambda^0{}_0 > 0$ (we agree in both frames on the direction of time) and so that $\det \Lambda > 0$ (we agree on the parity orientation of space). Define the matrix $L^A{}_I$:
	\begin{equation}
		L^A{}_I(\lambda) = \iota^{\mu A}(z(\lambda)) \Lambda_{\mu I}(\lambda). \label{eq:lorentztrans}
	\end{equation}
	It follows from \eqref{eq:tetrad} that:
	\begin{equation}
		\Lambda^\mu{}_I = L^A{}_I \iota^\mu{}_A \label{eq:lorentztetrad}
	\end{equation}
	and from feeding this into \eqref{eq:tetrad} and simplifying using \eqref{eq:vierbein} that:
	\begin{equation}
		L^A{}_I L^B{}_J \eta_{AB} = \eta_{IJ}. \label{eq:islorentz}
	\end{equation}
	By \eqref{eq:islorentz} and \eqref{eq:lorentztetrad}, $L^A{}_I$ is the Lorentz transformation that takes quantities in the body frame to quantities in the Lorentz frame ($f^A = L^A{}_I f^I$).
	
	The Lorentz transformation $L^A{}_I$ can be divided up naturally into a rotation and a boost. In the body-frame, the vector with components $\delta^I_0$ by definition represents the increasing time-direction of the body. As a spacetime vector, $\delta^I_0$ becomes $\Lambda^\mu{}_0$, thus this vector gives the instantaneous future-oriented direction of the body in the coordinate frame. We can define a Lorentz transform $B^A{}_I$ that Lorentz boosts from the time direction of the body frame to that of the local Lorentz frame. That is, $B^A{}_I$ must satisfy:
	\begin{equation}
		\eta_{AB} B^A{}_I B^B{}_J = \eta_{IJ}, \p B^A{}_0 \iota^\mu{}_A = \Lambda^\mu{}_0.  \label{eq:boosttrans}
	\end{equation}
	We will never actually need the exact formula for $B^A{}_I$, but because we want it to be a pure boost independent of the body orientation, we require it is constructed from $\Lambda^\mu{}_A$ using only the time direction $\Lambda^\mu{}_0$. With this condition, it is possible to find a unique explicit formula:
	\begin{equation}
		B^A{}_I = \delta^A_I - \Lambda^\mu{}_0 \iota_{\mu}{}^{A} \delta^0_I + \frac{1}{1-\Lambda^\mu{}_0\iota_{\mu 0}}\(\(\delta^A_0 + \Lambda^\nu{}_0\iota_{\nu}{}^A\)\Lambda^\rho{}_0\iota_{\rho I} - \(\delta^A_0 +  \Lambda^\nu{}_0\iota_{\nu 0}  \Lambda^\rho{}_0\iota_{\rho}{}^A\)\delta^0_I\). 
	\end{equation}
	Next, define the rotation matrix $R^I{}_J(\lambda)$:
	\begin{equation}
		R^I{}_J = B_A{}^I L^A{}_J \implies L^A{}_I = B^A{}_J R^J{}_I. \label{eq:rottrans}
	\end{equation}
	It immediately follows that $R^I{}_J$ is genuinely a Lorentz transformation:
	\begin{equation}
		\eta_{KL}R^K{}_I R^L{}_J = \eta_{IJ}.
	\end{equation}
	Combining \eqref{eq:lorentztetrad}, \eqref{eq:boosttrans}, and \eqref{eq:rottrans} we see that $R^A{}_B$ must satisfy:
	\begin{equation}
		(1-R^0{}_0) \Lambda^\mu{}_0 + B^A{}_i R^i{}_0\iota^\mu{}_A = 0
	\end{equation}
	where middle letter minuscule Latin indices indicate spatial $\{1,2,3\}$ components in the body frame. Due to the linear independence of the $\iota^\mu{}_A$ as a list of four vectors indexed by $A$, this requires that the coefficient of each vanish separately. Thus $R^I{}_0 = \delta^I_0$, which is precisely what we mean when we say that $R^I{}_J$ is a rotation (it is a Lorentz transformation leaving the time components unchanged). In 4 spacetime dimensions, such $R^I{}_J$ can be written, for some parameters $\phi^1(\lambda)$, $\phi^2(\lambda)$, $\phi^3(\lambda)$ as:
	\begin{equation}
		R^I{}_J = \(e^{\phi^i \textbf{J}_i}\)^I{}_J \label{eq:10211}
	\end{equation}
	where the rotation generating matrices $\textbf{J}_i$ are defined by:
	\begin{equation}
		(\textbf{J}_i)^I{}_J = \eta^{Ij}\delta^k_J\epsilon_{ijk}
	\end{equation}
	and $\epsilon_{ijk}$ is the 3 dimensional Levi-Civita symbol with $\epsilon_{123} = 1$. The three parameters $\phi^i(\lambda)$ are thus a valid set of independent degrees of freedom characterizing the orientation of the body. 
	
	While we could use the angular degrees of freedom$\phi^i$ in our variational principle for the body, they are horribly inconvenient for actual calculations because there are caught up in matrix exponentials. We can formulate rotations much more easily by mapping them to a set of constrained degrees of freedom, just like in the Newtonian case. Under a pure rotation of the body, $\Lambda^\mu{}_I$ will change to a new tetrad $\Lambda'^\mu{}_I$ which also satisfies \eqref{eq:tetrad} and which has $\Lambda'^{\mu}{}_0 = \Lambda^\mu{}_0$. Thus, there exists some rotation matrix $D^I{}_J$ so that:
	\begin{equation}
		\Lambda'^\mu{}_I = D^J{}_I \Lambda^\mu{}_J, \p \eta_{KL}D^K{}_I D^L{}_J = \eta_{IJ}, \p D^I{}_0 = \delta^I_0. 
	\end{equation}
	Under a small rotation, $D^I{}_J = \delta^I_J - \eps \delta\Theta^I{}_J + \OO(\eps^2)$ with $\eps$ near 0, for some $\delta\Theta^I{}_J$ which satisfies: 
	\begin{equation}
		\delta\Lambda^\mu{}_I = -\delta\Theta^J{}_I\Lambda^\mu{}_J, \p \delta\Theta_{JI} = -\delta\Theta_{IJ}, \p \delta\Theta_{I0} = 0.  \label{eq:deltabigthetai}
	\end{equation}
	Because of the last two of these properties, under a small rotation there exist three parameters $\delta\theta^i$ so that:
	\begin{equation}
		\delta\Theta^I{}_J = \delta\theta^i (\textbf{J}_i)^I{}_J \implies \delta\theta^i = \frac{1}{2}\epsilon^{ijk} \delta\Theta_{jk}. \label{eq:10215}
	\end{equation}
	We see that these three parameters $\delta\theta^i$ represent the amount of infinitesimal rotation about the current orientation performed by a small application of the rotation in question in the body frame. The generator corresponding to such a rotation is precisely the definition of the spin of the body and so by Noether's theorem, for a particular rotation direction the corresponding generator is:
	\begin{equation}
		G = \varphi_i \delta\phi^i = S_i \delta\theta^i \label{eq:directedspin}
	\end{equation}
	where $\delta\phi^i$ represents the variation in the genuine $\phi^i$ coordinates under the rotation, $\varphi_i$ are the conjugate momenta to the $\phi^i$, and the far right side serves to define the components of the spin $S_i$ in the frame of the body. 
	
	The variations $\delta\phi^i$ and $\delta\theta^i$ completely determine each other. In particular, by contracting the first of \eqref{eq:deltabigthetai} with $\Lambda_\mu{}^I$ we have:
	\begin{equation}
		\delta\Theta_{IJ} = -\Lambda_{\mu I}\delta\Lambda^{\mu}{}_J
	\end{equation}
	and by using \eqref{eq:rottrans} we have:
	\begin{equation}
		\delta\Theta_{IJ} = -R_{KI}\delta R^K{}_J.
	\end{equation}
	The left hand side here determines $\delta\theta^i$ via \eqref{eq:10215} while the right has some linear expression in terms of $\delta\phi^i$ following from \eqref{eq:10211}. Simplifying both gives:
	\begin{equation}
		\delta\theta^i = M^i{}_j(\phi)\delta\phi^j
	\end{equation}
	for some matrix $M^i{}_j$ which is invertible and whose particular expression we will never actually use (we just need to know that it exists). Given this matrix, we immediately have from \eqref{eq:directedspin} that the $\varphi_i$ and $S_i$ are invertibly related by:
	\begin{equation}
		\varphi_i = M^j{}_i S_j. \label{eq:phims}
	\end{equation}
	
	We define the spin tensor in the frame of the body $S_{IJ}$ just like the relationship between $\delta\Theta_{IJ}$ and $\delta\theta^i$ in \eqref{eq:10215}:
	\begin{equation}
		S_{IJ} = S_i (\textbf{J}^i)_{IJ} \implies S^i = \frac{1}{2}\epsilon^{ijk}S_{jk}. \label{eq:svectens1}
	\end{equation}
	From this it follows immediately that:
	\begin{equation}
		S_{JI} = -S_{IJ}, \p S_{I0} = 0, \p \frac{1}{2}S_{IJ}\delta\Theta^{IJ} = S_i\delta\theta^i.  \label{eq:stensprops1}
	\end{equation}
	
	We define the Lorentz frame angular variation $\delta\Theta^{AB}$ and the Lorentz frame spin tensor $S_{AB}$ simply by transforming the components of each from the body frame to the Lorentz frame using $L^A{}_I$:
	\begin{equation}
		\delta\Theta^{AB} = L^A{}_IL^B{}_J\delta\Theta^{IJ}, \p S^{AB} = L^A{}_I L^B{}_J S^{IJ}.
	\end{equation}
	These quantities are more convenient for a variational principle because the body frame (obviously) depends on the state of the body and so gets varied under body variations, while the Lorentz frame is totally independent of the body. It follows quickly from \eqref{eq:deltabigthetai}  that (with $\delta L^A{}_I$ being the variation in $L^A{}_I$ under a rotation of the body):
	\begin{equation}
		\delta L^A{}_I = -\delta\Theta^A{}_B L^B{}_I, \p \delta\Theta_{BA} = -\delta\Theta_{AB}, \p \Lambda^\mu{}_0 \iota^B{}_\mu \delta\Theta_{AB} = 0, \p \delta\Theta^{AB} = L^{AI}\delta L^B{}_I \label{eq:lorentzvari}
	\end{equation}
	and from \eqref{eq:stensprops1} that the Lorentz frame spin tensor satisfies:
	\begin{equation}
		S_{BA} = -S_{AB}, \p \Lambda^\mu{}_0 \iota^B{}_\mu S_{AB} = 0, \p \frac{1}{2} S_{AB}\delta\Theta^{AB} = \varphi_i \delta\phi^i. \label{eq:stensprops2}
	\end{equation}
	The last equation of \eqref{eq:lorentzvari} gives the instantaneous rotational variation of the body in terms of the variation of the Lorentz transformation $L^A{}_I$. Thus, the instantaneous angular velocity tensor of the body in terms of the time derivative of that Lorentz transformation is:
	\begin{equation}
		\Omega^{AB} = \Omega^{[AB]} = L^{AI}\frac{dL^B{}_I}{d\lambda}. \label{eq:angveldef}
	\end{equation}
	It immediately follows from the last condition of \eqref{eq:stensprops2} that:
	\begin{equation}
		\varphi_i\dot\phi^i = \frac{1}{2}S_{AB}\Omega^{AB}. \label{eq:angkinetic}
	\end{equation}
	
	Now let us return to our variational principle \eqref{eq:dmfact}. We first separate $\psi^\Omega$ into the rotational degrees of freedom $\phi^i$ and the other internal body degrees of freedom $\scri^\Omega$. Then, $\pi_\Omega\dot\psi^\Omega = \varphi_i \dot\phi^i + \scr{P}_\Omega\dot\scri^\Omega$ where $\varphi_i$ and $\scr{P}_\Omega$ are the conjugate momenta to $\phi^i$ and $\scri^\Omega$. The $\phi^i$ are determined by $L^A{}_I$ by \eqref{eq:rottrans} which itself is determined by $\Lambda^\mu{}_I$ by \eqref{eq:lorentztrans}. Given that, the $\varphi_i$ are determined by $S_{AB}$ via \eqref{eq:phims}. Using \eqref{eq:angkinetic}, \eqref{eq:dmfact} becomes:
	\begin{equation}
		\fc{S}[\ec, z, \twid{p}, \Lambda, S, \scri, \scr{P}] = \int_{-\8}^\8\(\twid{p}_\mu\dot z^\mu + \frac{1}{2}S_{AB}\Omega^{AB} + \scr{P}_\Omega\dot\scri^\Omega +\ec(|\twid{p}| - \twid{\fc{M}}(z,\twid{u},\Lambda, S, \scri, \scr{P}))\)d\lambda \label{eq:tildeact}
	\end{equation}
	where variations of this action are subject to the constraints:
	\begin{equation}
		g_{\mu\nu}\Lambda^\mu{}_I \Lambda^\nu{}_J = \eta_{IJ}, \p \Lambda^\mu{}_0\iota^B{}_\mu S_{AB} = 0. \label{eq:10229}
	\end{equation}
	In \eqref{eq:tildeact} we have started writing $\twid{p}_\mu$ to mean the canonically conjugate momentum to $z^\mu$ instead of just the undecorated symbol $p_\mu$. We similarly use $\twid{u}^\mu$ to mean the unit vector in the direction of $\twid{p}^\mu$ and $\twid{\fc{M}}$ to be the dynamical mass function corresponding to $\twid{p}_\mu$. We have introduced this notation because we want to reserve the use of $p_\mu$ for Dixon's momentum, which includes the effects of interactions. We will establish how $p_\mu$ and $\twid{p}_\mu$ are related momentarily. Do not forget, as we established at the beginning of the section, that we define all of the remaining body degrees of freedom $\scri^\Omega$ and $\scr{P}_\Omega$ purely within in the body frame so that they are unchanged by rotations, boosts, and translations of the body. 
	
	The variational principle in \eqref{eq:tildeact} relies on a choice of vierbein $\iota^\mu{}_A$ in order to define the Lorentz frame spin tensor and angular velocity. We can simplify and fully eliminate any reference to Lorentz frame at the level of the action by defining the angular momentum and spin tensor in the coordinate frame. Define the angular velocity tensor $\Omega^{\mu\nu}$ analogously to \eqref{eq:angveldef}:
	\begin{equation}
		\Omega^{\mu\nu} = \Omega^{[\mu\nu]} = \Lambda^\mu{}_I \frac{D\Lambda^{\nu I}}{D\lambda}. \label{eq:omegaspacetime}
	\end{equation}
	Here it is important to use the covariant $\lambda$ derivative because $\Lambda^\mu{}_I$ is a vector field with respect to changes of spacetime coordinates. By \eqref{eq:lorentztetrad}:
	\begin{equation}
		\frac{D\Lambda^\mu{}_I}{D\lambda} = \frac{d\Lambda^\mu{}_I}{d\lambda} + \Gamma^\mu{}_{\nu\rho} \Lambda^\nu{}_I \dot z^\rho = \frac{d L^A{}_I}{d\lambda} \iota^\mu{}_A + L^A{}_I \frac{D\iota^\mu{}_A}{D\lambda}.
	\end{equation}
	Returning this to $\Omega^{\mu\nu}$, we have:
	\begin{equation}
		\Omega^{\mu\nu} = \iota^\mu{}_A \iota^\nu{}_B \Omega^{AB} + \dot z^\rho \iota^{\mu A}\nabla_\rho \iota^\nu{}_A \label{eq:10232}
	\end{equation}
	(where the covariant derivative acting on the vierbein is understood as thinking of it as a list of spacetime vectors). Now define the coordinate frame spin tensor by simply transforming the indices of the Lorentz frame spin tensor:
	\begin{equation}
		S_{\mu\nu} = \iota_\mu{}^A \iota_\nu{}^B S_{AB}.
	\end{equation}
	Immediately (using \eqref{eq:10229} and \eqref{eq:10232}):
	\begin{equation}
		S_{\mu\nu}\Lambda^\nu{}_0 = 0, \p \frac{1}{2}S_{AB}\Omega^{AB} = \frac{1}{2}S_{\mu\nu}\Omega^{\mu\nu} + \frac{1}{2}\dot z^\mu S^{AB} \iota^\nu{}_A \nabla_\mu \iota_{\nu B}. 
	\end{equation}
	With this, our variational principle \eqref{eq:tildeact} becomes:
	\begin{equation}
		\fc{S} = \int_{-\8}^\8\(\(\twid{p}_\mu+\frac{1}{2}S^{AB} \iota^\nu{}_A \nabla_\mu \iota_{\nu B}\)\dot z^\mu + \frac{1}{2}S_{\mu\nu}\Omega^{\mu\nu} + \scr{P}_\Omega\dot\scri^\Omega +\ec\(|\twid{p}| - \twid{\fc{M}}\,\)\)d\lambda
	\end{equation}
	where variations of this action are now subject to the constraints:
	\begin{equation}
		g_{\mu\nu}\Lambda^\mu{}_I \Lambda^\nu{}_J = \eta_{IJ}, \p S_{\mu\nu}\Lambda^\nu{}_0 = 0. \label{eq:actconstraints}
	\end{equation}
	
	Under the electromagnetic interaction, the monopole contribution to the $\lambda$ Lagrangian is $qA_\mu\dot z^\mu$. This term, unlike the higher multipole moments, directly shifts the expression for the conjugate momentum in a non-gauge-invariant way. Similarly, we see that under the gravitational interaction the linear in spin interaction shifts the conjugate momentum by a piece proportional to the spin connection. It is convenient to absorb these shifts by defining the gauge covariant momentum~\cite{Witzany:2018ahb}:
	\begin{equation}
		p_\mu = \twid{p}_\mu + \frac{1}{2}S^{AB}\iota^\nu{}_A\nabla_\mu \iota_{\nu B} - q A_\mu.
	\end{equation}
	In the case of a free particle, $p_\mu$ and $\twid{p}_\mu$ coincide and so all of the same arguments which led from the form of the action in terms of a generic Hamiltonian in \eqref{eq:shamact} into dynamical mass function form in \eqref{eq:dmfact} can be repeated using $p_\mu$. We will see shortly that the gauge covariant momentum $p_\mu$ exactly coincides with Dixon's momentum in \eqref{eq:psdef}. 
	
	The vector $\Lambda^\mu{}_0$ is the instantaneous future-time direction of the body. It is determined by how the body-frame is defined. So far we haven't actually specified what it means to be comoving with the body, simply how rotations are defined once you are in a comoving frame. Which direction is identified as the future-time direction of the body is a \textit{choice} of frame in which rotations are defined. There is a particularly natural choice of the future-time direction because it uniquely requires no extra structure for the body and gives a holonomic constraint. The only available expressions for $\Lambda^\mu{}_0$ in terms of quantities already in the problem are functions of $p^\mu$, $\dot z^\mu$, $S_{\mu\nu}$, and their derivatives. Any expression in terms of $\dot z$ or any other time derivatives gives a nonholonomic (path dependent) constraint on variations, which is never desirable in a variational principle. The spin is defined by rotations in the body frame, and so any definition of the body frame which is spin dependent gives that frame only a recursive definition. Thus the simplest choice is thus a function of $p^\mu$ only, in which case the only possible expression is:
	\begin{equation}
		\Lambda^\mu{}_0 = u^\mu = \frac{p^\mu}{|p|}. \label{eq:framedef}
	\end{equation}
	Going forward, we always take this as defining the body frame. Conceptually, this means we define the body's spin in the body's center of momentum frame. With this choice, among $u^\mu$ and $\Lambda^\mu{}_I$ there are only four independent vectors, and so we take by definition the dynamical mass function to depend on them through $u^\mu$ and $\Lambda^\mu{}_i$ (with $i$ taking values $\{1, 2, 3\}$). Altogether, in terms of the gauge covariant momentum in dynamical mass function form, the action becomes:
	\begin{equation}
		\fc{S} = \int_{-\8}^\8\(p_\mu\dot z^\mu + \frac{1}{2}S_{\mu\nu}\Omega^{\mu\nu} + q A_\mu \dot z^\mu + \scr{P}_\Omega\dot\scri^\Omega  +\ec(|{p}| - {\fc{M}}(z,{u},\Lambda, S, \scri, \scr{P}))\)d\lambda \label{eq:simpact}
	\end{equation}
	where variations of this action are subject to the constraints \eqref{eq:actconstraints} and \eqref{eq:framedef}. 
	
	\subsection{Variations of the Action}
	
	It is about time to actually use our action to find equations of motion. In order to do so, recall the covariant variation as defined in \eqref{eq:covaridef}. For a tensor field $f^\Omega$, let $\delta f^\Omega(z)$ mean the variation $\delta f^\Omega(x)$ in the components of $f^\Omega$ evaluated at a fixed $x$ and then the final result put on the worldline ($x = z$). By the chain rule, $\delta(f^\Omega(z)) = \delta f^\Omega(z) + \delta z^\rho \d_\rho f^\Omega(z)$ and so:
	\begin{equation}
		\Delta f^\Omega = \delta f^\Omega(z) + \delta z^\rho \nabla_\rho f^\Omega. \label{eq:1031}
	\end{equation}
	For tensors that live purely on the worldline, like the momentum vector $p_\mu$, we write $\delta p_\mu$ precisely to mean the non-tensorial quantity analogous to $\delta (f^\Omega(z))$. As in, suppose $P_\alpha(x)$ is a vector field so that on the worldline $P_\mu(z(\lambda)) = p_\mu(\lambda)$. Then, $\delta p_\mu = \delta(P_\mu(z))$. Similarly, $\Delta p_\mu(\lambda) = \Delta P_\mu(z(\lambda))$. 
	
	We will need to vary the action with respect to all of its dependencies, including the metric. As usual, we call the metric variation $\delta g_{\alpha\beta}(x) = h_{\alpha\beta}(x)$. The indices on $h_{\alpha\beta}$ are raised and lowered using the metric as normal. Due to the variation of the metric, $\Delta p^\mu$ is not the raising of $\Delta p_\mu$. Whenever we write $\Delta p^\mu$ we unambiguously mean $\Delta(p^\mu)$ so that any factors of the metric caught inside are varied as well.

	To vary the action \eqref{eq:simpact} properly we need to take careful consideration of the constraints in \eqref{eq:actconstraints} and \eqref{eq:framedef}. The constraint \eqref{eq:framedef} can be easily treated either by simply replacing $\Lambda^\mu{}_0$ everywhere in the action with $u^\mu$ or through the use of a Lagrange multiplier $\chi_\mu$ (so that the contribution to the integrand of the action is $+\chi_\mu(\Lambda^\mu{}_0 - u^\mu)$). The second constraint in \eqref{eq:actconstraints} is $S_{\mu\nu}\Lambda^\nu{}_0 = 0$, which under consideration of the constraint $\Lambda^\mu{}_0 = u^\mu$ is equivalent to $S_{\mu\nu} p^\nu = 0$. We can implement this constraint with a Lagrange multiplier $\beta_\mu$ (so that the contribution to the integrand of the action is $+\beta_\mu S^{\mu\nu} p_\nu$). 
	
	The first constraint in \eqref{eq:actconstraints} is more challenging. We define the covariant angular variation tensor $\Delta\Theta^{\mu\nu}$ under a variation of the worldline degrees of freedom analogously to the angular velocity: 
	\begin{equation}
		\Delta\Theta^{\mu\nu} = \eta^{IJ} \Lambda^{[\mu}{}_I \Delta\Lambda^{\nu]}{}_J.
	\end{equation}
	Then, the variation of the first constraint of \eqref{eq:actconstraints} just gives that the covariant variations of the tetrad are given by:
	\begin{equation}
		\Delta\Lambda^\mu{}_I = -\frac{1}{2}h^\mu{}_\nu \Lambda^\nu{}_I - \Delta\Theta^{\mu\nu}\Lambda_{\nu I}.\label{eq:deltatetrad}
	\end{equation}
	Hence all we need to require is that at each worldline time $\Delta\Theta^{\mu\nu}$ is an arbitrary antisymmetric tensor. The covariant variation of the angular velocity tensor can be directly calculated in terms of $\Delta\Theta^{\mu\nu}$ and our other perturbations by expanding out \eqref{eq:omegaspacetime}. Explicitly (and carefully appreciating that the connection coefficients get varied under a variation of the worldline and of the metric):
	\begin{align}
		\Delta\Omega^{\mu\nu} =&\ \frac{D}{D\lambda}\Delta\Theta^{\mu\nu} + \Omega^\mu{}_\rho \Delta\Theta^{\rho\nu} - \Omega^\nu{}_\rho \Delta\Theta^{\rho\mu} + R^{\mu\nu}{}_{\rho\sigma}\dot z^\rho \delta z^\sigma \nonumber \\
		&-\frac{1}{2}\Omega^\mu{}_\rho h^{\rho\nu} + \frac{1}{2}\Omega^\nu{}_\rho h^{\rho\mu} + \frac{1}{2}\dot z^\rho \nabla^\mu h^\nu{}_\rho - \frac{1}{2}\dot z^\rho \nabla^\nu h^\mu{}_\rho.   \label{eq:bigomegavari}
	\end{align}	
	We can now organize our variational principle cleanly into its absolute final form:
	\begin{equation}
		\fc{S} = \int_{-\8}^\8\(p_\mu\dot z^\mu + \frac{1}{2}S^{\mu\nu}\Omega_{\mu\nu} + q A_\mu \dot z^\mu + \scr{P}_\Omega\dot\scri^\Omega + \ec(|{p}| - \fc{M}) + \beta_{\mu} S^{\mu\nu} p_\nu + \chi_{\mu}(\Lambda^\mu{}_0 - u^\mu)\)d\lambda \label{eq:lagrangeact}
	\end{equation}
	where the variation of the tetrad is given by \eqref{eq:deltatetrad} and of the angular velocity by \eqref{eq:bigomegavari}.
	
	In order to vary the action \eqref{eq:lagrangeact} covariantly, we need to specify how precisely the dynamical mass function depends on the worldline position $z^\mu(\lambda)$. It may depend on the worldline through the evaluation of the metric at the worldline position $g_{\mu\nu}(z)$ and it may also depend on some other tensor fields $\Phi^\Omega(z)$ evaluated along the worldline. Any explicit real-valued function of the spacetime coordinates can be made manifestly covariant by treating it as a scalar field with that function expression in that particular coordinate system. Doing so, we can always introduce background fields $\Phi^\Omega(z)$ so that the dynamical mass function is a scalar function $\fc{M}(g, \Phi, u, \Lambda, S, \scr{I}, \scr{P})$ with no further $z$-dependence. Because we are now thinking about the explicit metric dependence of the dynamical mass function, it matters whether we consider it as a function of $u^\mu$ or of $u_\mu$ (and similarly so for the index placements on other tensors). By convention we take it to depend on $u^\mu, \Lambda^\mu{}_i$, $S^{\mu\nu}$ with all spacetime indices raised. Since $\fc{M}$ is now manifestly a scalar function, it does not change value under a change of coordinates: 
	\begin{equation}
		\fc{M}(g', \Phi', u', \Lambda', S', \scri', \scr{P}') = \fc{M}(g, \Phi, u, \Lambda, S, \scri, \scr{P}) \label{eq:1035}
	\end{equation}
	where all primed quantities are given by coordinate transforming their spacetime indices. Consider a coordinate transformation $x'^\alpha = x^\alpha + \eps\xi^\alpha(x)$. Then, the $\eps$ derivative of \eqref{eq:1035} at $\eps = 0$ gives (because it must hold for any $\xi^\alpha$)~\cite{JanSteinhoff:2015ist}:
	\begin{equation}
		2\frac{\d\fc{M}}{\d g_{\mu\nu}} = \frac{\d\fc{M}}{\d \Phi^\Omega} \fc{G}^{\mu\nu} \Phi^\Omega +  u^\mu \frac{\d \fc{M}}{\d u_\nu} + \Lambda^\mu{}_i \frac{\d\fc{M}}{\d \Lambda_{\nu i}} + 2S^\mu{}_\rho\frac{\d\fc{M}}{\d S_{\nu\rho}}. \label{eq:scalarity}
	\end{equation}
	The are no terms due to $\scri$ or $\scr{P}$ because they only possess body frame indices, thus $\scri'^\Omega = \scri^\Omega$ and $\scr{P}'_\Omega = \scr{P}_\Omega$. We can use \eqref{eq:scalarity} to eliminate the metric derivative of the dynamical mass function from any equation. Additionally, the left hand side is automatically symmetric, so the antisymmetrization of the right hand side must vanish. Define the suggestively named ``provisional torque'' $\text{\textbf{T}}^{\mu\nu}$ by:
	\begin{equation}
		\text{\textbf{T}}^{\mu\nu} = 2\frac{\d\fc{M}}{\d\Phi^\Omega} \fc{G}^{[\mu\nu]} \Phi^\Omega. 
	\end{equation}
	Then, the antisymmetrization of \eqref{eq:scalarity} gives:
	\begin{equation}
		\text{\textbf{T}}^{\mu\nu} = \frac{\d\fc{M}}{\d u_\mu} u^\nu -  \frac{\d\fc{M}}{\d u_\nu} u^\mu +  \frac{\d\fc{M}}{\d \Lambda_{\mu i}} \Lambda^\nu{}_i -  \frac{\d\fc{M}}{\d \Lambda_{\nu i} } \Lambda^\mu{}_i + 2\frac{\d\fc{M}}{\d S_{\mu\rho}}S^\nu{}_\rho - 2\frac{\d\fc{M}}{\d S_{\nu\rho}} S^\mu{}_\rho. 
	\end{equation}
	
	Now consider a generic variation of all of the dependencies of the dynamical mass function. By the chain rule:
	\begin{align}
		\delta\fc{M} =&\ \frac{\d\fc{M}}{\d g_{\mu\nu}}\delta(g_{\mu\nu}(z)) + \frac{\d\fc{M}}{\d \Phi^\Omega}\delta(\Phi^\Omega(z)) + \frac{\d\fc{M}}{\d u^\mu}\delta u^\mu \nonumber \\
		&\ + \frac{\d\fc{M}}{\d \Lambda^\mu{}_i}\delta \Lambda^\mu{}_i + \frac{\d\fc{M}}{\d S^{\mu\nu}}\delta S^{\mu\nu} + \frac{\d\fc{M}}{\d \scri^\Omega}\delta \scri^\Omega + \frac{\d\fc{M}}{\d\scr{P}_\Omega}\delta\scr{P}_\Omega. 
	\end{align}
	Using \eqref{eq:1031} and \eqref{eq:scalarity} we can rewrite this purely in terms of covariant variations as:
	\begin{align}
		\delta\fc{M} =&\ \frac{1}{2}\(\frac{\d\fc{M}}{\d \Phi^\Omega} \fc{G}^{\mu\nu} \Phi^\Omega +  u^\mu \frac{\d \fc{M}}{\d u_\nu} + \Lambda^\mu{}_i \frac{\d\fc{M}}{\d \Lambda_{\nu i}} + 2S^\mu{}_\rho\frac{\d\fc{M}}{\d S_{\nu\rho}}\) h_{\mu\nu} + \frac{\d\fc{M}}{\d\Phi^\Omega} \delta\Phi^\Omega  \nonumber \\
		&\ + \frac{\d\fc{M}}{\d\Phi^\Omega} \nabla_\mu \Phi^\Omega \delta z^\mu + \frac{\d\fc{M}}{\d u^\mu}\Delta u^\mu + \frac{\d\fc{M}}{\d \Lambda^\mu{}_i}\Delta \Lambda^\mu{}_i + \frac{\d\fc{M}}{\d S^{\mu\nu}}\Delta S^{\mu\nu} + \frac{\d\fc{M}}{\d \scri^\Omega}\delta \scri^\Omega + \frac{\d\fc{M}}{\d\scr{P}_\Omega}\delta\scr{P}_\Omega.
	\end{align}
	We now suggestively define the ``provisional force'' $\text{\textbf{F}}_\mu$ to capture only the position variations under this covariant organization:
	\begin{equation}
		\text{\textbf{F}}_\mu = -\frac{\d\fc{M}}{\d\Phi^\Omega} \nabla_\mu \Phi^\Omega. 
	\end{equation}
	
	Because we are interested in the interactions of our body with gravity and electromagnetism, we take the fields $\Phi^\Omega$ that body couples to to be the electromagnetic field strength, the gravitational Riemann tensor, and their derivatives. We can organize these derivative dependencies to be in particular be expressed in terms of the tensor extensions of the field strength and the multipolar Riemann tensors. Doing so, the provisional force and torque automatically become:
	\begin{align}
		\text{\textbf{F}}_\mu &= -\sum_{n=0}^\8\frac{\d\fc{M}}{\d \ee_{\kappa_1\dots\kappa_n}\fc{F}_{\rho\sigma}} \nabla_\mu \ee_{\kappa_1\dots\kappa_n}\fc{F}_{\rho\sigma} -\sum_{n=0}^\8\frac{\d\fc{M}}{\d R^{(n)}_{\kappa_1\dots\kappa_n\rho\tau\sigma\omega}} \nabla_\mu R^{(n)}_{\kappa_1\dots\kappa_n\rho\tau\sigma\omega} \nonumber \\
		\text{\textbf{T}}^{\mu\nu} &= 2\sum_{n=0}^\8\frac{\d\fc{M}}{\d \ee_{\kappa_1\dots\kappa_n}\fc{F}_{\rho\sigma}} \fc{G}^{[\mu\nu]} \ee_{\kappa_1\dots\kappa_n}\fc{F}_{\rho\sigma} + 2\sum_{n=0}^\8\frac{\d\fc{M}}{\d R^{(n)}_{\kappa_1\dots\kappa_n\rho\tau\sigma\omega}} \fc{G}^{[\mu\nu]} R^{(n)}_{\kappa_1\dots\kappa_n\rho\tau\sigma\omega} \label{eq:provft}
	\end{align}
	which are shaping up to a very similar form to \eqref{eq:ftexpr}. 
	
	Now let us vary the action \eqref{eq:lagrangeact} with respect to all of its dependencies. First we consider stationary variations with respect to all of the body degrees of freedom and Lagrange multipliers to find the body equations of motion. Afterward, we will vary with respect to the vector potential and metric. Requiring stationary variation with respect to the body, we find:
	\begin{align}
		\Delta p_\mu	&: \p					\dot z^\mu - \ec u^\mu - \frac{\ec}{|p|} \gamma^{\mu\nu} \frac{\d\fc{M}}{\d u^\nu} - S^{\mu\nu} \beta_\nu - \frac{1}{|p|}\gamma^{\mu\nu}\chi_\nu = 0 \nonumber \\
		\delta z^\mu	&: \p					-\dot p_\mu -\frac{1}{2}R_{\mu\nu\rho\sigma} \dot z^\nu S^{\rho\sigma} + q\fc{F}_{\mu\nu}\dot z^\nu + \ec \text{\textbf{F}}_\mu = 0 \nonumber \\
		\Delta S^{\mu\nu}	&: \p			\Omega_{\mu\nu} - 2\ec\frac{\d\fc{M}}{\d S^{\mu\nu}} + \beta_\mu p_\nu - \beta_\nu p_\mu = 0 \nonumber \\
		\Delta \Theta^{\mu\nu} &: \p	-\dot S_{\mu\nu} +\Omega^\rho{}_\mu S_{\rho\nu} - \Omega^\rho{}_\nu S_{\rho\mu} + \ec \frac{\d\fc{M}}{\d \Lambda^\mu{}_i} \Lambda_{\nu i} - \ec \frac{\d\fc{M}}{\d\Lambda^\nu{}_i} \Lambda_{\mu i}  + u_\mu \chi_\nu - u_\nu\chi_\mu = 0 \nonumber \\
		\delta \scr{P}_\Omega &: \p		\dot\scri^\Omega - \ec\frac{\d\fc{M}}{\d \scr{P}_\Omega} = 0 \nonumber \\
		\delta \scri^\Omega &: \p 			-\dot{\scr{P}}_\Omega - \ec\frac{\d\fc{M}}{\d\scri^\Omega} = 0 \nonumber \\
		\delta \ec &: \p							|p| - \fc{M} = 0 \nonumber \\
		\Delta\beta_\mu &: \p					S^{\mu\nu} p_\nu = 0 \nonumber \\
		\Delta\chi_\mu &: \p 					\Lambda^\mu{}_0 - u^\mu = 0.  \label{eq:longeom}
	\end{align}
	Here we have introduced the spatial metric on the worldline $\gamma_{\mu\nu}$ which projects out the time direction:
	\begin{equation}
		\gamma_{\mu\nu} = g_{\mu\nu} + u_\mu u_\nu \implies \gamma^\mu{}_\rho \gamma^\rho{}_\nu = \gamma^\mu{}_\nu, \p \gamma^\mu{}_\nu u^\nu = 0. 
	\end{equation}
	
	We now move to eliminate the Lagrange multipliers from our equations of motion in \eqref{eq:longeom} and to simplify the results. By the $\Delta\chi_\mu$ constraint and \eqref{eq:omegaspacetime}:
	\begin{equation}
		\dot u^\mu = -\Omega^{\mu\nu} u_\nu. 
	\end{equation}
	Using this while contracting $u^\mu$ onto the $\Delta S^{\mu\nu}$ equation of motion gives:
	\begin{equation}
		\fc{M}\gamma_{\mu\nu}\beta^\nu = -\dot u_\mu - 2\ec \frac{\d\fc{M}}{\d S^{\mu\nu}} u^\nu. \label{eq:10317}
	\end{equation}
	Returning this to the $\Delta S^{\mu\nu}$ equation of motion allows the elimination of $\beta_\mu$ from it. Instead contracting one index of the spin tensor onto \eqref{eq:10317} and using the $\delta\ec$ and $\Delta\beta_\mu$ constraints gives:
	\begin{equation}
		S^{\mu\nu}\beta_\nu = -\frac{1}{\fc{M}^2} S^{\mu\nu}\dot p_\nu - \frac{2\ec}{\fc{M}}S^{\mu\nu}\frac{\d\fc{M}}{\d S^{\nu\rho}} u^\rho. 
	\end{equation}
	Using this, $\beta_\mu$ can be eliminated from the $\Delta p_\mu$ equation of motion. We next eliminate $\chi_\mu$. By contracting $u^\nu$ onto the $\Delta\Theta^{\mu\nu}$ equation of motion, we find:
	\begin{equation}
		\gamma_{\mu\nu}\chi^\nu = \ec u^\nu \frac{\d\fc{M}}{\d \Lambda^\nu{}_i} \Lambda_{\mu i}. 
	\end{equation}
	With this, $\chi_\mu$ can be directly eliminated from both the $\Delta p_\mu$ and $\Delta\Theta^{\mu\nu}$ equations. After all of these removals and a little algebra, we arrive at the fully simplified equations of motion:
	\begin{align}
		&\fc{M}\doot z{}^\mu = p^\mu - \text{\textbf{T}}{}^{\mu\nu} u_\nu - S^{\mu\nu} \doot u_\nu, &
		&\doot p{}_\mu = -\frac{1}{2} R_{\mu\nu\rho\sigma} \doot z{}^\nu S^{\rho\sigma} + q\fc{F}_{\mu\nu}\doot z{}^\nu + \text{\textbf{F}}_\mu, &\nonumber \\
		&\doot \Lambda{}^\mu{}_i = \(u^\mu \doot u{}^\nu + 2\gamma^{\mu\rho}\gamma^{\nu\sigma}\frac{\d\fc{M}}{\d S^{\rho\sigma}}\) \Lambda_{\nu i}, &
		&\doot S{}^{\mu\nu} = p^\mu \doot z{}^\nu - p^\nu \doot z{}^\mu + \text{\textbf{T}}{}^{\mu\nu},&\nonumber \\
		&\doot{\scri}{}^\Omega = \frac{\d\fc{M}}{\d \scr{P}_{\Omega}},&
		&\doot{\scr{P}}{}_\Omega = -\frac{\d\fc{M}}{\d \scri^\Omega}. \label{eq:shorteom}
	\end{align}
	Wow! The $z, p, S$ equations of motion are almost identical to the MPD equations \eqref{eq:mpd} (with center of mass worldline \eqref{eq:csscpvelrel}). In order to finish proving that they are actually the same, we need to show that the provisional force and torque coincide with the MPD force and torque. Doing so will be our next task. The remaining equations of motion in \eqref{eq:shorteom} determine the evolution of the body frame and the internal body degrees of freedom. 
	
	We can see by comparing \eqref{eq:ftexpr} and \eqref{eq:provft} that in order to relate the provisional force and torque to the MPD force and torque, we first need to relate the reduced field moments to the derivatives of the dynamical mass function. To do this, we simply need to calculate the variation of our action under variations of the vector potential and metric and compare the results to \eqref{eq:jacurvexpand} and \eqref{eq:gravact}. The extensions of the electromagnetic field strength depend on both the vector potential and the metric, so we divide up their variations as fields into pieces due to the variation of each:
	\begin{equation}
		\delta\ee_{\kappa_1\dots\kappa_n}\fc{F}_{\mu\nu} = \delta_A \ee_{\kappa_1\dots\kappa_n}\fc{F}_{\mu\nu} + \delta_g \ee_{\kappa_1\dots\kappa_n}\fc{F}_{\mu\nu}
	\end{equation}
	where $\delta_A$ varies only the vector potential at fixed metric and $\delta_g$ varies only the metric at fixed vector potential. Under a vector potential variation of \eqref{eq:lagrangeact}, we find:
	\begin{equation}
		\delta_A\fc{S} = \int_{-\8}^\8 \(q \doot z{}^\mu \delta A_\mu - \sum_{n=0}^\8 \frac{\d\fc{M}}{\d \ee_{\kappa_1\dots\kappa_n} \fc{F}_{\rho\sigma}}\ee_{\kappa_1\dots\kappa_n}\delta_A \fc{F}_{\rho\sigma}\) D\lambda. \label{eq:avariact}
	\end{equation}
	The derivatives of the dynamical mass function with respect to the tensor extensions of the field strength have all of the same index symmetries as the $Q^{\cdots}_n$ moments, but those $Q_n^{\cdots}$ moments also satisfy the orthogonality condition $u_{\kappa_1} Q^{\kappa_1\dots\kappa_n \mu\nu} = 0$. By the chain of reasoning leading to \eqref{eq:jacurvexpand}, such moments are unique (as we were able to construct explicit formulas for them in terms of integrals over the current density using these conditions). Therefore, by comparing \eqref{eq:avariact} to \eqref{eq:jacurvexpand}, we see that the field moments for the current density are related to the dynamical mass function precisely by:
	\begin{equation}
		Q_n^{\rho_1\dots\rho_n\mu\nu} = -(n+2) n! \frac{\d\fc{M}}{\d \ee_{\rho_1\dots\rho_n}\fc{F}_{\mu\nu}} \label{eq:qfromdmf}
	\end{equation}
	where to make this deduction the dynamical mass function must satisfy the orthogonality relation:
	\begin{equation}
		u_{\rho_1} \frac{\d \fc{M}}{\d \ee_{\rho_1\dots\rho_n}\fc{F}_{\mu\nu}} = 0. \label{eq:dmfqortho}
	\end{equation}
	This orthogonality relation is not a condition on physics. Instead, it is the condition satisfied automatically when one has written the dynamical mass function so that it is fully reduced in terms of the independent multipole moments of the current density . Many distinct dynamical mass functions which differ in how they depend on derivatives in the time direction of the field strength may produce the same physics, because those higher time derivatives can be iteratively eliminated through use of the equations of motion and integration by parts at each order. Such different expressions will give different integrands at each $\lambda$ even though the integrated result is unchanged. Because any covariantly conserved current density can be put into the form of \eqref{eq:jacurvexpand} uniquely, we can use that form to uniquely fix how time derivatives of the field strength appear in the dynamical mass function, the result being that it must satisfy \eqref{eq:dmfqortho}. 
	
	Under a metric variation of \eqref{eq:lagrangeact}, we find (after using \eqref{eq:shorteom} and \eqref{eq:qfromdmf}):
	\begin{align}
		\delta_g \fc{S} = \int_{-\8}^\8\Bigl(&\ \frac{1}{2}h_{\mu\nu}\doot z{}^\mu p^\nu + \frac{1}{2}S^{\mu\nu} \doot z{}^\rho \nabla_\mu h_{\nu\rho} \nonumber \\
		&\ + \sum_{n=0}^\8 \frac{n+1}{(n+2)!} Q_n^{\rho_1\dots\rho_n\mu\nu}\bigl(\delta _g \ee_{\rho_1\dots\rho_n}\fc{F}_{\mu\nu} + \frac{1}{2} h_{\tau\omega} \fc{G}^{\tau\omega} \ee_{\rho_1\dots\rho_n} \fc{F}_{\mu\nu}\bigr) \nonumber \\
		&\ -\sum_{n=0}^\8 \frac{\d\fc{M}}{\d R^{(n)}_{\kappa_1\dots\kappa_n\rho\mu\sigma\nu}}\bigl(\delta R^{(n)}_{\kappa_1\dots\kappa_n\rho\mu\sigma\nu} + \frac{1}{2}h_{\tau\omega}\fc{G}^{\tau\omega} R^{(n)}_{\kappa_1\dots\kappa_n\rho\mu\sigma\nu}\bigr) \Bigr) D\lambda. 
	\end{align}
	By the exact same reasoning as in the electromagnetic case, the uniqueness of Dixon's moments allow us to compare this to \eqref{eq:gravact} and identify that the field moments for the energy-momentum tensor are related to the dynamical mass function precisely by:
	\begin{equation}
		\fc{J}_n^{\kappa_1\dots\kappa_n \rho\mu\sigma\nu} = (n+2)(n+3) n! \frac{\d\fc{M}}{\d R^{(n)}_{\kappa_1\dots\kappa_n\rho\mu\sigma\nu}} \label{eq:jfromdmf}
	\end{equation}
	where the fully reduced dynamical mass function satisfies the orthogonality relation:
	\begin{equation}
		u_{\kappa_1} \frac{\d\fc{M}}{\d R^{(n)}_{\kappa_1\dots\kappa_n\rho\mu\sigma\nu}} = 0. 
	\end{equation}
	The monopole and dipole terms of \eqref{eq:gravact} confirm that the momentum and spin from the variational principle exactly match Dixon's momentum and spin. Identifying \eqref{eq:qfromdmf} and \eqref{eq:jfromdmf} in \eqref{eq:provft}, the provisional force and torque become exactly the MPD force and torque \eqref{eq:ftexpr}. Thus, the variational principle \eqref{eq:lagrangeact} is perfectly and fully equivalent to the MPD theory with the correspondences \eqref{eq:qfromdmf} and \eqref{eq:jfromdmf}!
	
	There are several slightly different common forms in literature for the action \eqref{eq:lagrangeact}. The $\chi_\mu$ constraint can be completed removed by replacing $\Lambda^\mu{}_0$ in the angular velocity tensor by $u^\mu$ at the level of the action. The result is that the kinetic term becomes more complicated and the $\Delta\Theta^{\mu\nu}$ variations are not fully independent of $\Delta u^\mu$ variations, as they have to maintain the constraint. The $\beta_\mu$ constraint can be reduced or removed in several ways. The constraint can be mostly solved by use of the spin vector as in \eqref{eq:skdef} and replacing $S^{\mu\nu} = \epsilon^{\mu\nu\rho\sigma} u_\rho S_\sigma$ at the level of the action. However, $S^\mu u_\mu = 0$ and so a scalar Lagrange multiplier is still necessary. That scalar constraint can be solved as well by use of the body frame spin vector $S^i$ via $S^\mu = \Lambda^\mu{}_i S^i$. Each reduction of constraints this way makes the variational principle just a bit more complicated, as this introduces further tetrad dependence which needs to be varied. Alternatively still, the constraints can be simply not written and instead the phase space enlarged to include $\Lambda^\mu{}_0$ (independent of $u^\mu$) and $S^{\mu\nu}$ (unconstrained apart from anti-symmetry). If the constraints are imposed on them as initial conditions and the equations of motion computed, by requiring the constraints to be preserved dynamically there will be resulting differential equations the dynamical mass function must satisfy. These additional conditions turn out to be completely equivalent to having used the constraints via Lagrange multipliers in the first place. Another common reformulation of \eqref{eq:lagrangeact} is to write the einbein constraint as $-\frac{1}{2}\alpha(p^2 + \fc{M}^2)$ with Lagrange multiplier $\alpha$ instead of as $+\ec(|p|-\fc{M})$. This is really just a reparameterization as in the same spirit as \eqref{eq:einrescale}. At the level of the equations of motion they are mapped into each other by $\ec = \alpha \fc{M}$. This $\alpha$ constraint has the advantage of avoiding the square root in $|p|$ in favor of $p^2$. However, $u^\mu$ dependence in $\fc{M}$ still carries $|p|$ dependence, so this effect is marginal. 
	
	\subsection{Applications and Limitations of the MPD Equations} \label{sec:apps}
	
	The variational principle \eqref{eq:lagrangeact} captures the full MPD theory. The failure of the MPD equations to close is expressed by that action carrying as yet unspecified internal body degrees of freedom $\scri$ and $\scr{P}$ and a generic dynamical mass function $\fc{M}$. Here we will discuss how to specify these quantities so that \eqref{eq:lagrangeact} becomes useful for real applications. In this subsection we refer only to gravitational fields and mass multipole moments, but the extension to electromagnetism is immediate. 
	
	The most generic thing one could possibly do is to just take the $\scri$ to be all of the reduced $\fc{J}_n$ field multipole moments of the body expressed in the body frame. There are infinitely many ways to contract the indices of these degrees of freedom to form scalars. Since $\fc{M}$ is an arbitrary function of all of these infinite scalar contractions, we do not have enough structure to do anything useful. In order to get the problem under control, we need a \textit{separation of scales} so that we can use the methods of effective field theory. We will discuss a few of the approximations most commonly available to us. Usually multiple or even all of these are applicable. 
	
	\textit{Hierarchy of time-dependences.} When a test-body is placed in Minkowski space, its mass monopole (characterized by its momentum) and dipole (characterized by its spin) are conserved\footnote{This is assuming you use the Tulczyjew-Dixon time-slicing and the covariant spin supplementary condition.} but its higher multipole moments still have arbitrary dynamics. This limiting case is illustrative of the fact that even for a non-test-body in a curved spacetime, the body's momentum and spin evolve on the relevant length/time scale of the curvature of that spacetime, while the time scales of the dynamics of the body's higher multipole moments are typically determined by the body's internal physics. Because of this, it is very often a reasonable approximation that the higher multipole moments of a body evolve much more quickly than its momentum and spin, getting faster at each multipole order. Suppose this is so and that for some $N_t$, the $\fc{J}_n$ for $n > N_t$ evolve too quickly for their precise time dependence to be resolvable to us at our desired level of precision. Then, we can approximate each of these fast multipole moments by their time averages over our time scale of interest (say, that of the fastest relevant moment $\fc{J}_{N_t}$). In doing so, they become functions of the other degrees of freedom and fields in the problem and thus we reduce the \textit{independent} body degrees of freedom $\scri$ to just $\fc{J}_1$, ..., $\fc{J}_{N_t}$. The moments for $n > N_t$ are not set to 0 by this, they are just now determined in terms of our other quantities by \eqref{eq:jfromdmf}. In the most extreme case, one takes $N_t = 0$ and says all of the higher multipole moments are too fast. This leaves us with no $\scri$, and the only body variables become the so-called \textit{minimal degrees of freedom} $z, p, S, \Lambda$. By \eqref{eq:jfromdmf} then, the higher multipole moments will be functions of the body's spin (giving \textit{spin-induced tides}~\cite{Porto:2005ac,Porto:2006bt,Porto:2008jj,Steinhoff:2009tk,Levi:2015msa}) and of the multipolar Riemann tensors on the worldline (giving \textit{adiabatic tides}~\cite{Henry:2019xhg,Mandal:2024iug}). The next-to-minimal approximation $N_t=1$ retains explicit dynamical independence of the quadrupole moment $\fc{J}_0$, which then satisfies an equation of motion (giving \textit{dynamical tides}~\cite{Chakraborty:2026dox, Steinhoff:2016rfi, Mandal:2023lgy, Chakraborty:2026qru, Saketh:2026trm, Jarequi:2026cyp}) rather than having a simple algebraic expression. 
	
	\textit{Hierarchy of length-scales.} It is often the case that one is interested in the motion of a body experiencing a field that varies over a characteristic length scale $L$ that is much larger than the body size $\ell$. Each higher multipole moment scales with one higher power of the body size $\ell$ due to the factors of $v^\mu$ appearing in their integral representations and enters the action paired with a commensurately higher order of spatial differentiation (tensor extension) of the metric. Thus, each higher multipole moment contributes to the action at one higher order in $\frac{\ell}{L}$ and at a given level of precision only multipole moments and multipolar Riemann tensors up to some order $N_x$ need to be considered nonzero. This truncation leaves $\fc{M}$ a function of a finite number of variables (the minimal degrees of freedom, the $\fc{J}_n$ and their conjugate momenta for $n \le N_t \le N_x$, and the $R^{(n)}$ for $n \le N_x$). Only a finite number of scalars can be made out of this list, and because one only needs precision up to a given order in $\frac{\ell}{L}$, there are only a finite number possible terms in the resulting dynamical mass function. Each of these finite number of terms carries a Wilson coefficient and contributes a concrete effect to observables~\cite{Levi:2015msa}. These Wilson coefficients are distinct for different bodies of interest, and can be matched through observables to external data to fully determine the EFT for the body at hand to the given order. At the absolute lowest in the hierarchy of length scales, one takes $N_x = 0$ and keeps only terms through a single power in $\frac{\ell}{L}$. The resulting equations of motion are then \eqref{eq:shorteom} with the force and torque set to 0, and only keeping terms up through linear order in the spin. These are known as the \textit{pole-dipole order MPD equations} and contain no Wilson coefficients, i.e. they are universal. This universality is lost as soon as spin-squared effects or $N_x > 0$ are considered. 

	\textit{Self-Force Expansion.} Throughout this work we have not discussed at all the way in which a body deforms the spacetime in which it exists. Due to Einstein's equations, the presence of the body's energy-momentum tensor induces a curvature. The self-field produced by a body is generally not slowly varying over the size of that body, preventing the preceding hierarchy of length-scales arguments from applying. However, work in gravitational self-force~\cite{Harte:2025tmd, Harte:2014wya} suggests that it is possible to decompose the full field produced by the body into a fast-varying self-field that renormalizes the body parameters and a slow-varying remainder effective field, at least at low orders in the body size. This approximation scheme also requires scaling the total mass of the body with $\ell$. One way to motivate this scaling is that for a body of mass $m$, the characteristic length scale of the curvature it produces is the Schwarzschild radius~$2Gm$. At $0^\text{th}$ order in self-force, the body is treated as a test-body in that it moves purely in the background spacetime in which it is placed. At $1^\text{st}$ order, the $0^\text{th}$ order motion of the body deforms the background spacetime, which creates a correction to the body's motion. Practically speaking, calculations with the MPD equations are almost always done in the test-body limit, though it is possible by implementing these schemes to include the effects of self-force in the MPD dynamics. 
	
	\textit{Weak-Field Expansion.} Often it is reasonable to approximate the spacetime external to the body as nearly Minkowskian and so to expand it in powers of $G$ (with appropriate other factors so that a dimensionless parameter is formed, dependent on the scenario). For example, in the two-body problem, when the interbody separation $b$ is large compared to the would-be Schwarzschild radius $2GM$ corresponding to a black hole with the same total mass $M$ as the system itself, then the field produced by each body at the location of the other (and the resulting affects on motion from this field) can be calculated perturbatively in $\frac{GM}{b}$. This naturally leads to the post-Minkowskian and post-Newtonian expansions. In this case, terms in the dynamical mass function that contain too many powers of the Riemann tensor and its derivatives will not contribute to observables at the relevant order of approximation. 

	\section{Summary} \label{sec:conc}

	This section is a self-contained summary giving all of the important definitions and results of Dixon's theory.
	
	\subsection{Notation Reminders}
	
	Synge's worldfunction $\sigma(x, z)$ is the biscalar field defined as:
	\begin{equation}
		\sigma(x, z) = \frac{1}{2}\int_0^1 g_{\alpha\beta}(\zeta)\frac{d\zeta^\alpha}{ds}\frac{d\zeta^\beta}{ds} ds
	\end{equation}
	where $\zeta^\alpha(s)$ is the affinely-parameterized minimal geodesic connecting $x$ and $z$: $\zeta^\mu(0) = z^\mu$ and $\zeta^\alpha(1) = x^\alpha$. Synge's worldfunction only exists for $x, z$ close enough so that there is a unique such geodesic. Beginning Greek indices ($\alpha, \beta, \gamma, \delta, \epsilon $) on $\sigma$ indicate covariant $x^\alpha$ derivatives; middle Greek indices ($\iota$, $\kappa, \lambda, \mu, \nu, \rho, \sigma, \tau, \omega$) indicate covariant $z^\mu$ derivatives. Indices are raised and lowered using the metric at the appropriate point. Define:
	\begin{equation}
		v^\mu = -\sigma^\mu(x,z), \p H^\alpha{}_\mu(x,z) = \sigma^{-\alpha}{}_\mu, \p K^\alpha{}_\mu(x,z) = H^\alpha{}_\nu\sigma^\nu{}_\mu
	\end{equation}
	where $\sigma^{-\alpha}{}_\mu$ is the matrix inverse of $\sigma^\mu{}_\alpha$. $v^\mu = -\sigma^\mu$ at fixed $z$ provides a coordinate transformation from the $x^\alpha$ coordinate system to the $v^\mu$ Riemann normal coordinate system based at $z$. The partial derivatives in the $v^\mu$ coordinate system at $z$ define tensor extensions:
	\begin{equation}
		\ee_{\rho_1\dots\rho_n} \phi^{\mu\cdots}{}_{\nu\cdots}(z) = \left.\frac{\d^n}{\d v^\rho_1\dots\d v^\rho_n}\(H^{-\mu}{}_\alpha(x(z,v),z)\cdots H^\beta{}_\nu(x(z,v), z)\cdots \phi^{\alpha\cdots}{}_{\beta\cdots}(x(z,v))\)\right|_{v= 0}.
	\end{equation}
	DeWitt's index shuffling operator is defined to act on tensor fields as:
	\begin{align}
		\fc{G}^\alpha{}_\beta f^{\gamma_1\dots\gamma_m}{}_{\delta_1\dots\delta_n} =& \delta^{\gamma_1}_{\beta} f^{\alpha \gamma_2\dots\gamma_m}{}_{\delta_1\dots\delta_n} + \dots + \delta^{\gamma_m}_{\beta} f^{\gamma_1\dots\gamma_{m-1}\alpha}{}_{\delta_1\dots\delta_n} \nonumber \\
		&\ -\delta^{\alpha}_{\delta_1} f^{\gamma_1\dots\gamma_m}{}_{\beta\delta_2\dots\delta_n} - \dots -\delta^{\alpha}_{\delta_n} f^{\gamma_1\dots\gamma_m}{}_{\delta_1\dots\delta_{n-1}\beta}.
	\end{align}
	
	\subsection{Basic Assumptions on the Body}
	
	Suppose we have a compact object described by an energy-momentum tensor $T^{\alpha\beta}$ and electromagnetic charge current density $J^\alpha$ in general relativity. Then:
	\begin{equation}
		T^{\alpha\beta} = T^{\beta\alpha}, \p \nabla_\beta T^{\alpha\beta} = \fc{F}^{\alpha\beta} J_\beta, \p \nabla_\alpha J^\alpha = 0.
	\end{equation}
	
	The worldtube $\bb{W}$ is the convex hull of the union of the supports of $T^{\alpha\beta}$ and $J^\alpha$. The worldline $z^\mu(\lambda)$ is an arbitrary smooth future orientated timelike curve contained entirely within the worldtube parameterized by the arbitrary worldline time $\lambda$. The ``worldline time direction'' $u^\mu(\lambda)$ is an arbitrary differentiable future orientated timelike unit vector on the worldline at each worldline time. $\Sigma(\lambda)$ represents space at fixed time as determined geodesically by an observer following the worldline assigning $u^\mu$ to always be the time direction: $\Sigma(\lambda) = \{ x\ :\ \sigma^\mu(x, z(\lambda)) u_\mu(\lambda) = 0\}$ . The worldline clock $\tau(x)$ is the scalar field giving what time that observer assigns to the event $x$: $\tau(x) = \lambda \iff x \in \Sigma(\lambda)$. Let $\ec(\lambda) = -u\cdot \dot z$ be the einbein and let $w^\alpha(x)$ be any vector field satisfying $\ec w^\alpha\nabla_\alpha\tau = 1$. $d\Sigma_\alpha$ represents the directed volume measure on $\Sigma$ and $d\Sigma_w = w^\alpha d\Sigma_\alpha$. $\gamma_{\mu\nu} = g_{\mu\nu} + u_\mu u_\nu$ is the induced spatial metric on the worldline. $\doot{f}{}^\Omega = \frac{1}{\ec} \frac{D f^\Omega}{D\lambda}$ is the reparameterization invariant covariant time derivative along the worldline. 
	
	Suppose that the object is physically reasonable in the following ways:
	\setlist{nolistsep}
	\begin{enumerate}[label=(\Roman*)]
		\item For all $V^\alpha$ which are timelike, $T^{\alpha\beta} V_\beta$ is timelike everywhere that $T^{\alpha\beta} \neq 0$,
		\item For all $V^\alpha$ which are timelike, $T^{\alpha\beta} V_\alpha V_\beta \ge 0$ and at least somewhere this is positive,
		\item At each worldline time there is precisely one geodesic connecting any two points within $\bb{W}\cap \Sigma(\lambda)$,
		\item There is a bound $\ell$ independent of $\lambda$ so that the geodesic distance between any pair of points in $\bb{W}\cap \Sigma(\lambda)$ is less than or equal to $\ell$,
		\item $p_\mu$ as defined in \eqref{eq:1131} is timelike,
		\item $\eps_u$ as defined in \eqref{eq:e16} is less than 1,
		\item $\eps_s$ as defined in the text leading to \eqref{eq:d16} is positive.
	\end{enumerate} Given these conditions, we have all of the following results. 

	\subsection{Definition of Momentum and Spin}
	
	The total linear momentum and total spin about the worldline are given by:
	\begin{align}
		p_\mu(\lambda) = \int_{\Sigma(\lambda)}\( T^{\alpha\beta} K_{\beta\mu} + \Phi_\mu J^\alpha\)d\Sigma_\alpha, \ps \ps S^{\mu\nu}(\lambda) = 2\int_{\Sigma(\lambda)}\( v^{[\mu} H_\beta{}^{\nu]} T^{\alpha\beta} + v^{[\mu} \Psi^{\nu]} J^\alpha\) d\Sigma_\alpha \label{eq:1131}
	\end{align}
	where:
	\begin{equation}
		\Phi_\mu(x,z) = \int_0^1 \fc{F}_{\alpha\beta}(\zeta) \sigma^\alpha(\zeta, z) K^\beta{}_\mu(\zeta, z) \frac{ds}{s}, \ps \Psi_\mu(x,z) = \int_0^1 \fc{F}_{\alpha\beta}(\zeta)\sigma^\alpha(\zeta,z) H^\beta{}_\mu(\zeta,z) ds. 
	\end{equation}
	These definitions are unique in simultaneously being gauge invariant and also such that if the spacetime and electromagnetic field possess a Killing vector $\xi^\alpha(x)$ then the combination:
	\begin{equation}
		G(\lambda) = p_\mu(\lambda) \xi^\mu(z(\lambda)) + \frac{1}{2} S^{\mu\nu}(\lambda) \nabla_{[\mu}\xi_{\nu]}(z(\lambda)).
	\end{equation}
	is extensive and conserved. Consequently, by Noether's theorem these are the proper generators of translations of the worldline and Lorentz transformations about the worldline. 
	
	\subsection{Charge Multipole Moments}
	
	For $(z,u)$ given, there is a unique set of multipole moments for the electromagnetic current density so that (I) the different multipole moments are fully reduced, meaning (beyond the monopole moment) they are completely unconstrained by the charge continuity equation and so that (II) when the current density is integrated against a vector potential the result can be expressed as a series in these moments multiplying spatial tensor extensions of the electromagnetic field strength. They have the index symmetries:
	\begin{align}
		&Q_n^{\rho_1\dots\rho_n\mu\nu} = Q_n^{(\rho_1\dots\rho_n)[\mu\nu]}, \p u_{\rho_1} Q_n^{\rho_1\dots\rho_n\mu\nu} = 0, \nonumber \\
		&Q_n^{\rho_1\dots\rho_{n-1} \rho_n\mu\nu} + Q_n^{\rho_1\dots\rho_{n-1}\mu\nu\rho_n} + Q_n^{\rho_1\dots\rho_{n-1}\nu\rho_n\mu} = 0.
	\end{align}
	The result of integrating the current density against a vector potential in terms of these moments is:
	\begin{equation}
		\int J^\alpha A_\alpha D^4 x = \int\( q\doot z{}^\mu A_\mu(z) + \sum_{n=0}^\8 \frac{n+1}{(n+2)!} Q_n^{\rho_1\dots\rho_n\mu\nu} {\ee}_{\rho_1\dots\rho_n} \fc{F}_{\mu\nu}\) D\lambda. \label{eq:jasimpexpand}
	\end{equation}
	They are given precisely by spatial integrals of the current density by:
	\begin{equation}
		Q_n^{\rho_1\dots\rho_n\mu\nu} = \mc{j}^{\rho_1\dots\rho_n[\mu\nu]}_{n+1} + \frac{1}{n+1} \mc{r}_{n+1}^{\rho_1\dots\rho_n[\mu\nu]\sigma}\doot z{}_\sigma
	\end{equation}
	where:
	\begin{gather}
		\mc{j}_n^{\rho_1\dots\rho_n\mu} = \int_\Sigma v^{\rho_1}\dots v^{\rho_n} H^{-\mu}{}_\alpha J^\alpha d\Sigma_w, \p \mc{r}_n^{\rho_1\dots\rho_n\mu\nu} = \int_\Sigma v^{\rho_1}\dots v^{\rho_n} \Theta^{\mu\nu}_{n-1} J^\alpha d\Sigma_\alpha, \nonumber \\
		\Theta^{\mu\nu}_n(x,z) = (n+1)\int_0^1 \sigma^\mu{}_\alpha(\zeta, z) \sigma^{\alpha\nu} (\zeta, z) s^n ds. 
	\end{gather}
	The charge continuity equation $\nabla_\alpha J^\alpha = 0$ places absolutely no constraints on the time evolution of the $Q_n^{\cdots}$ moments. Given the form \eqref{eq:jasimpexpand} the charge continuity equation is exactly equivalent to the conservation of the body's total charge:
	\begin{equation}
		q = \int_{\Sigma} J^\alpha d\Sigma_\alpha.
	\end{equation}
	
	\subsection{Mass Multipole Moments}

	For $(z, u)$ given, there is a unique set of multipole moments for the energy-momentum tensor so that (I) the different multipole moments are fully reduced, meaning (beyond the dipole moment) they are completely unconstrained by the covariant conservation of energy and momentum and so that (II) when the energy-momentum tensor is integrated against a metric perturbation the result can be expressed as a series in these moments multiplying a curved spacetime generalized of the spatial derivatives of the Riemann tensor. This generalization is constructed to ensure that Bianchi identities are exactly maintained. It is given by the multipolar Riemann tensors:
	\begin{equation}
		R^{(n)}_{\kappa_1\dots\kappa_n\rho\mu\sigma\nu}(z) = -2\ee_{\kappa_1\dots\kappa_n [\rho['\sigma}g_{\mu]\nu]'}. 
	\end{equation}
	$R^{(0)}_{\rho\mu\sigma\nu}$ is just the ordinary Riemann tensor. The reduced moments have the index symmetries:
	\begin{align}
		&\fc{J}_n^{\kappa_1\dots\kappa_n\rho\mu\sigma\nu} = \fc{J}_n^{(\kappa_1\dots\kappa_n)[\rho\mu][\sigma\nu]} = \fc{J}_n^{\kappa_1\dots\kappa_n\sigma\nu\rho\mu}, \p u_{\kappa_1} \fc{J}_n^{\kappa_1\dots\kappa_n\rho\mu\sigma\nu} = 0, \nonumber \\
		&\fc{J}_n^{\kappa_1\dots\kappa_n\rho\mu\sigma\nu} + \fc{J}_n^{\kappa_1\dots\kappa_n\rho\sigma\nu\mu} + \fc{J}_n^{\kappa_1\dots\kappa_n\rho\nu\mu\sigma} = 0,\nonumber \\
		&\fc{J}_n^{\kappa_1\dots\kappa_n\rho\mu\sigma\nu} + \fc{J}_n^{\kappa_1\dots\kappa_{n-1}\rho\mu\kappa_n\sigma\nu} + \fc{J}_n^{\kappa_1\dots\kappa_{n-1}\mu\kappa_n\rho\sigma\nu} = 0.
	\end{align}
	The result of integrating the energy-momentum tensor against a metric perturbation in terms of these moments is:
		\begin{align}
		\frac{1}{2}\!\int\!\! T^{\alpha\beta} \delta g_{\alpha\beta} D^4 x =&\ \frac{1}{2}\!\int\(p^\mu \doot z{}^\nu \delta g_{\mu\nu} + S^{\rho\mu} \doot z{}^\nu \nabla_\rho \delta g_{\mu\nu}\) D\lambda \nonumber \\
		&+\! \int\! \sum_{n=0}^\8 \!\frac{n+1}{(n+2)!} Q_n^{\rho_1\dots\rho_n\mu\nu}\(\delta \ee_{\rho_1\dots\rho_n}\fc{F}_{\mu\nu} + \frac{1}{2}\delta g_{\tau\omega} \fc{G}^{\tau\omega} \ee_{\rho_1\dots\rho_n}\fc{F}_{\mu\nu}\) D\lambda \nonumber \\
		&-\! \int\! \sum_{n=0}^\8 \!\frac{n+1}{(n+3)!} \! \fc{J}_n^{\kappa_1\dots\kappa_n\rho\mu\sigma\nu}\! \(\delta R^{(n)}_{\kappa_1\dots\kappa_n\rho\mu\sigma\nu} + \frac{1}{2} \delta g_{\tau\omega}\fc{G}^{\tau\omega} R^{(n)}_{\kappa_1\dots\kappa_n\rho\mu\sigma\nu}\!\) D\lambda \label{eq:thsimpexpand}
	\end{align}
	where $\delta \ee_{\rho_1\dots\rho_n}\fc{F}_{\mu\nu}$ and $\delta R^{(n)}_{\kappa_1\dots\kappa_n\rho\mu\sigma\nu}$ represent the variations of $\ee_{\rho_1\dots\rho_n}\fc{F}_{\mu\nu}$ and $R^{(n)}_{\kappa_1\dots\kappa_n\rho\mu\sigma\nu}$ under metric perturbation. 

	The moments are given as spatial integrals of the energy-momentum tensor by:
	\begin{equation}
		\fc{J}_n^{\kappa_1\dots\kappa_n \rho\mu\sigma\nu} = \mc{t}_{n+2}^{\kappa_1\dots\kappa_n[\rho['\sigma\mu]\nu]'} + \frac{1}{n+1}\mc{p}_{n+2}^{\kappa_1\dots\kappa_n[\rho['\sigma\mu]\nu]'\iota}\doot z{}_\iota
	\end{equation}
	where:
	\begin{align}
		\mc{t}_n^{\kappa_1\dots\kappa_n\mu\nu} &= \int_\Sigma v^{\kappa_1}\dots v^{\kappa_n} H^{-\mu}{}_\alpha H^{-\nu}{}_\beta T^{\alpha\beta} d\Sigma_w	\nonumber	\\
		\mc{p}_n^{\kappa_1\dots\kappa_n\mu\nu\sigma} &= \int_{\Sigma} 2v^{\kappa_1}\dots v^{\kappa_n} \(T^{\alpha\beta} H_{\beta\rho} \text{V}_n^{\rho\sigma\mu\nu} + \frac{1}{n+1} J^\alpha v_\rho \text{W}_n^{\rho\sigma\mu\nu}\) d\Sigma_\alpha	\nonumber 	\\
		\text{V}_n^{\rho\sigma\mu\nu} &= (n-1)\int_0^1 \sigma^\rho{}_\alpha(\zeta,z)\sigma^{\alpha(\mu}(\zeta,z) \sigma^{\nu)}{}_\beta(\zeta, z) \sigma^{\beta\sigma}(\zeta,z) s^{n-2} ds	\nonumber	\\
		\text{W}_n^{\rho\sigma\mu\nu}(x,z) &= (n+1)\int_0^1 \fc{F}_{\alpha\beta}(\zeta) H^{\alpha\rho}(\zeta,z)H^\beta{}_\kappa(\zeta,z) \text{V}^{\kappa\sigma\mu\nu}_n(\zeta,z) s^n ds.
	\end{align}
	The covariant conservation of energy-momentum $\nabla_\beta T^{\alpha\beta} = \fc{F}^{\alpha\beta} J_\beta$ places absolutely no constraints on the time evolution of the $\fc{J}_n^{\cdots}$ moments. Given the form \eqref{eq:thsimpexpand}, the conservation of energy-momentum is exactly equivalent to the MPD equations of motion: 
	\begin{equation}
		\doot p_\mu = - \frac{1}{2}R_{\mu\nu\rho\sigma}\doot z{}^\nu S^{\rho\sigma} + q\fc{F}_{\mu\nu} \doot z{}^\nu + \text{\textbf{F}}_\mu, \p \doot S{}^{\mu\nu} = p^\mu\doot z{}^\nu - \doot z{}^\mu p^\nu + \text{\textbf{T}}^{\mu\nu}
	\end{equation}
	where the MPD force and torque are given by:
	\begin{align}
		\text{\textbf{F}}_\mu =&\  - \sum_{n=0}^\8\frac{n+1}{(n+3)!} \fc{J}^{\kappa_1\dots\kappa_n\rho\tau\sigma\omega}_n \nabla_\mu R^{(n)}_{\kappa_1\dots\kappa_n\rho\tau\sigma\omega} + \sum_{n=0}^\8 \frac{n+1}{(n+2)!}Q_n^{\kappa_1\dots\kappa_n\rho\sigma}\nabla_\mu \ee_{\kappa_1\dots\kappa_n} \fc{F}_{\rho\sigma} \nonumber \\
		\text{\textbf{T}}_{\mu\nu} =&\ 2\sum_{n=0}^\8\frac{n+1}{(n+3)!} \fc{J}^{\kappa_1\dots\kappa_n\rho\tau\sigma\omega}_n\fc{G}_{[\mu\nu]} R^{(n)}_{\kappa_1\dots\kappa_n\rho\tau\sigma\omega} -2\sum_{n=0}^\8 \frac{n+1}{(n+2)!}Q_n^{\kappa_1\dots\kappa_n\rho\sigma}\fc{G}_{[\mu\nu]} \ee_{\kappa_1\dots\kappa_n}\fc{F}_{\rho\sigma}.
	\end{align}
	
	\subsection{Specification of the worldline and time direction}

	It is possible to choose the time direction $u^\mu$ so that the resulting $p^\mu$ is parallel to it, i.e. $p^\mu = \fc{M} u^\mu$ with $\fc{M} = |p|$. This means the body's time direction is identified in its center of momentum frame. It is possible to choose the worldline $z^\mu$ so that the spin satisfies the covariant spin supplementary condition $S^{\mu\nu} p_\nu = 0$. This means the body's spin is defined in its center of momentum frame. With these choices, the unique such worldline of the body satisfies:
	\begin{equation}
		\fc{M}\doot z{}^\mu = p^\mu - \text{\textbf{T}}^{\mu\nu} u_\nu - S^{\mu\nu} \doot u_\nu. 
	\end{equation}
	In 4 spacetime dimensions, we can define the spin vector $S^\mu = -\frac{1}{2}\epsilon^{\mu\nu\rho\sigma} u_\nu S_{\rho\sigma}$ (from which the spin tensor can be recovered by $S^{\mu\nu} = \epsilon^{\mu\nu\rho\sigma} u_\rho S_\sigma)$. In terms of the spin vector, the dual field strength ${}^\star\fc{F}_{\mu\nu} = \frac{1}{2}\epsilon_{\mu\nu\rho\sigma}\fc{F}^{\rho\sigma}$, and the left, right, and double dual Riemann tensor (defined by dualizing its leftmost pair of indices, rightmost pair, or both), the MPD equations of motion become:
	\begin{align}
		(\fc{M}^2 + {}^\star R^\star_{uSuS} - q{}^\star \fc{F}_{uS})\, \doot z{}^\mu = &\ \fc{M}^2 u^\mu - {}^\star R^{\star\mu}{}_{SuS} + q{}^\star \fc{F}^{\mu}{}_S - \epsilon^{\mu\nu\rho\sigma} u_\nu S_\rho \text{\textbf{F}}_\sigma\nonumber \\
		& - \fc{M} \text{\textbf{T}}^{\mu\nu} u_\nu + \frac{\text{\textbf{T}}^{\rho\sigma} S_\rho u_\sigma}{\fc{M}}\({}^\star R^{\star\mu}{}_{uuS} - q{}^\star \fc{F}^\mu{}_S\), \nonumber \\
		\doot p{}_\mu =&\ -R^\star_{\mu\nu\rho\sigma} \doot z{}^\nu u^\rho S^\sigma + q \fc{F}_{\mu\nu} \doot z{}^\nu + \text{\textbf{F}}_\mu, \nonumber \\
		\doot S{}^\mu =&\ u^\mu \doot u{}_\nu S^\nu - {}^\star \text{\textbf{T}}^{\mu\nu} u_\nu.  \label{eq:mpdfinal}
	\end{align}
	
	\subsection{Variational Results}
	
	The entire MPD formalism can be packaged into a single variational principle with action $\fc{S}[z, p, \Lambda, S, \scri, \scr{P}, \ec, \beta, \chi]$. Here $\Lambda^\mu{}_I(\lambda)$ is a tetrad tracking the frame of the body (and so always satisfies $\Lambda^\mu{}_I \Lambda^\nu{}_J \eta^{IJ} = g^{\mu\nu}$) with angular velocity $\Omega^{\mu\nu} = \eta^{IJ} \Lambda^\mu{}_I \frac{D\Lambda^\nu{}_J}{D\lambda}$, $\scri^\Omega$ are the internal degrees of freedom of the body in the body frame with conjugate momenta $\scr{P}_\Omega$, and $\ec, \beta, \chi$ are Lagrange multipliers. The action is:
	\begin{equation}
		\fc{S} = \int_{-\8}^\8\(p_\mu\dot z^\mu + \frac{1}{2}S^{\mu\nu}\Omega_{\mu\nu} + q A_\mu \dot z^\mu + \scr{P}_\Omega\dot\scri^\Omega + \ec(|{p}| - \fc{M}) + \beta_{\mu} S^{\mu\nu} p_\nu + \chi_{\mu}(\Lambda^\mu{}_0 - u^\mu)\)d\lambda
	\end{equation}
	where $\fc{M}(z, u, \Lambda, S, \scri, \scr{P})$, the dynamical mass function, is a scalar (change-of-coordinates invariant) function that depends on $z$ only through $g_{\mu\nu}$, the multipolar Riemann tensors, and tensor extensions of the electromagnetic field strength all evaluated on the worldline. In order for the worldline, momentum, and spin of this action to coincide with Dixon's definitions, the dynamical mass function must be organized so that:
	\begin{equation}
		u_{\rho_1} \frac{\d \fc{M}}{\d \ee_{\rho_1\dots\rho_n}\fc{F}_{\mu\nu}} = 0, \p u_{\kappa_1} \frac{\d\fc{M}}{\d R^{(n)}_{\kappa_1\dots\kappa_n\rho\mu\sigma\nu}} = 0. 
	\end{equation}
	In addition reproducing the equations of motion \eqref{eq:mpdfinal} for $z, p, S$, this action produces the equations of motion for the body degrees of freedom $\Lambda, \scri, \scr{P}$:
	\begin{equation}
		\doot \Lambda{}^\mu{}_i = \(u^\mu \doot u{}^\nu + 2\gamma^{\mu\rho}\gamma^{\nu\sigma}\frac{\d\fc{M}}{\d S^{\rho\sigma}}\) \Lambda_{\nu i}, \p \doot{\scri}{}^\Omega = \frac{\d\fc{M}}{\d \scr{P}_{\Omega}}, \p \doot{\scr{P}}{}_\Omega = -\frac{\d\fc{M}}{\d \scri^\Omega}.
	\end{equation}
	The reduced moments are contained within the dynamical mass function as:
	\begin{equation}
		Q_n^{\rho_1\dots\rho_n\mu\nu} = -(n+2) n! \frac{\d\fc{M}}{\d \ee_{\rho_1\dots\rho_n}\fc{F}_{\mu\nu}}, \p \fc{J}_n^{\kappa_1\dots\kappa_n \rho\mu\sigma\nu} = (n+2)(n+3) n! \frac{\d\fc{M}}{\d R^{(n)}_{\kappa_1\dots\kappa_n\rho\mu\sigma\nu}}.
	\end{equation}

	\section{Conclusion}
	
	The MPD formalism recasts the energy-momentum tensor and the electric charge current density vector of a compact object in curved spacetime into a center of mass worldline, momentum, spin, and multipole moments for that object. While this formalism is entirely general and equivalent to the conservation laws \eqref{eq:tcons} and \eqref{eq:jcons}, it allows for the extraction of meaning about the body's overall macroscopic properties (position, motion, rotation) and determines how these quantities evolve in time.
	
	In reviewing the foundations of this theory we explored the relevant concepts in Newtonian gravity which serve as inspiration of the many generalizations presented in this work. We further discussed the mathematical foundations of Synge's worldfunction and bitensors which underpin the construction of the MPD theory. We then showed how the properties of Synge's worldfunction naturally led us to understand the general properties of symmetries in curved spacetime. Using this, we extracted the accompanying generators of translations and rotations which gave us the definitions of momentum and spin in curved spacetime. We also showed explicit conditions under which the worldline and time direction can be fixed uniquely such that the covariant spin supplementary condition is fulfilled. The succeeding sections iteratively generalized the concept of multipole moments.  We began with electromagnetism in flat spacetime, which gave us the necessary intuition for the reduction of the multipole moments into an independent form. We continued with a similar treatment of the multipoles for the energy-momentum tensor in the linearized theory of gravity near Minkowski space. We then showed how the multipoles of electromagnetism can be generalized to the case of curved spacetime. Extending this approach we combined both the gravitational and electromagnetic sectors in general relativity which culminated in a general action in terms of the momentum, spin, and fully reduced and independent mass and charge multipoles. These multipole moments were paired with the multipolar Riemann tensors to which they are conjugate in the variational principle. We showed how this action produces the MPD equations of motion for the momentum and spin in their full glory. Finally, we showed how this formalism can be developed via a worldline effective field theory variational principle, finding the relationship between the worldline dynamical mass function and the reduced mass and charge multipoles.
	
	Approaching the MPD formalism, and in particular the multipole moments of the energy-momentum tensor, from the perspective of an action in which the metric is varied, allowed us to naturally motivate and significantly simplify many of the results obtained in the original works by Dixon. This not only provides a more approachable context for the technical details of the theory, but also makes it directly useful for applications in effective worldline descriptions of relativistic extended objects. It is in tandem with suitably chosen approximation schemes (such as the post-Minkowskian or the gravitational self-force expansions) that the MPD formalism displays its power. The fact that the degrees of freedom in the MPD formalism are fully reduced and independent has the consequence that their effects also produce independent contributions to observables. Beyond that, they come with clear transformation properties and physical interpretations. 
	
	The generality and universality of the theory cannot be overstated. When brought into the form of a variational principle, objects treated with this formalism differ only in their specific dynamical mass function. Using the tools of effective field theory, this function can be constructed in a bottom-up way, with the rest of the framework directly providing the physical interpretations of the relevant features of the object.
	
	\section*{Acknowledgements}
	
	The authors thank Francisco Blanco, Richard Myers, Raj Patil, M.V.S. Saketh, Jan Steinhoff, and Justin Vines for valuable discussions regarding this work. TS would in particular like to further thank Justin Vines for introducing him to Dixon's formalism and to thank Jan Steinhoff for his encouragement that a review of this kind would be valuable. No LLMs or other generative AI tools were used in the research or writing of this work. 
	
	\paragraph{Funding information}

	T.S. is supported by ERC grant GWSky / 101167314, funded by the European Union. Views and opinions expressed are however those of the authors only and do not necessarily reflect those of the European Union or the European Research Council Executive Agency. Neither the European Union nor the granting authority can be held responsible for them.

	\begin{appendix}
		\numberwithin{equation}{section}
		
		\section{Computation of Tensor Extensions} \label{sec:tensext}
		
		In order to compute tensor extensions, we repeatedly think of the relationship between generic $x^\alpha$ coordinates and normal $v^\mu$ coordinates as a coordinate transformation. Correspondingly, in this section only, we change our notation slightly to reflect this thinking. Whereas previously ${\nabla}_\mu$ meant the covariant derivative with respect to $z$ at fixed $x$, in this section $z$ is \textit{always} fixed and $\nabla_\mu$ represents the covariant derivative expressed in the $v$ coordinate system. In this section we will have no use for the $x$ coordinate system; all calculations will be done in Riemann normal coordinates. Tensor extensions are just partial $v$ derivatives evaluated at $v = 0$ (which is just the point $z$). 
		
		Let $\zeta^\mu(s)$ be a geodesic in the $v^\mu$ coordinate system so that $\zeta^\mu(0) = z^\mu$ and with initial tangent $\dot\zeta^\mu(0) = u^\mu$ (in this section $u^\mu$ is just a generic vector in $T_z(\bb{T})$, it has no relationship to any time-slicing). In this coordinate system, we can explicitly solve the geodesic equation, because the point reached by such a geodesic in unit affine time has coordinate $u^\mu$ in the $v^\mu$ coordinate system by definition of the exponential map. Let $\zeta'^\mu(s') = \zeta^\mu(s s')$ be a geodesic, now with initial tangent $s u^\mu$ and reaching $\zeta'^\mu(1) = \zeta^\mu(s)$ in unit affine time. Again being such a geodesic, this final point matches the initial tangent and so:
		\begin{equation}
			\zeta^\mu(s) = su^\mu \label{eq:normalgeod}
		\end{equation}
		for all $s$. Returning this to the geodesic equation gives the fact that in normal coordinates:
		\begin{equation}
			\Gamma^\rho{}_{\mu\nu}(sv) v^\mu v^\nu = 0 \label{eq:normalchrist}
		\end{equation}
		for all $v$ and $s$. Differentiating $n$ times with respect to $s$ gives:
		\begin{equation}
			v^{\kappa_1}\dots v^{\kappa_n} \d^n_{\kappa_1\dots\kappa_n} \Gamma^\rho{}_{\mu\nu}(sv) v^\mu v^\nu = 0. 
		\end{equation}
		Now evaluating at $s=0$ and using that this result holds for all $v$, we have that at $v=0$:
		\begin{equation}
			\d^n_{(\kappa_1\dots\kappa_n}\Gamma^\rho{}_{\mu\nu)}(z) = 0.  \label{eq:conn_van}
		\end{equation}
		Tensor extensions can be computed recursively in terms of lower extensions and so-called ``normal tensors'' which are defined by:
		\begin{equation}
			\fc{N}^\rho{}_{\sigma\mu_1\dots\mu_n}(z) = \left.\d^{n-1}_{(\mu_1\dots\mu_{n-1}} \Gamma^\rho{}_{\mu_n)\sigma}(v)\right|_{v=0}. 
		\end{equation}
		The first normal tensor $\fc{N}^\rho{}_{\sigma\mu} = 0$ due to \eqref{eq:conn_van}. We will now find a formula for how to calculate all tensor extensions in terms of these normal tensors. Then we will derive a recursive formula for how to compute normal tensors in terms of derivatives of the Riemann tensor. 
		
		Let's start with the definition of the $n^\text{th}$ tensor extension of an arbitrary tensor $\phi^\Omega$ in \eqref{eq:tensexdef}: 
		\begin{equation}
			\ee_{\mu_1\dots\mu_n}\phi^{\Omega}(z) = \partial^n_{(\mu_1\dots\mu_n)}\phi^\Omega(v)\Big\lvert_{v=0}.
		\end{equation} 
		Now convert one of the partial derivatives into a covariant one by adding and subtracting the difference:
		\begin{equation}
			\partial^{n}_{(\mu_1\dots\mu_n)}\phi^\Omega(v) = \partial^{n-1}_{(\mu_1\dots\mu_{n-1}}\Big[\delta_{\mu_n)}^\nu\(\nabla_{\nu}\phi^\Omega - \Gamma^{\sigma}{}_{\rho\nu}\mathcal{G}^{\rho}{}_{\sigma} \phi^{\Omega}\)\Big]
		\end{equation}
		Using the product rule to expand the derivatives and evaluating at $v = 0$ (so that we arrive at a genuine tensor equation at $z$), this becomes:
		\begin{equation}
			\label{eq:any_tensor_extension}
			\ee_{\mu_1\dots\mu_n}\phi^{\Omega} = \ee_{(\mu_1\dots\mu_{n-1}}\nabla_{\mu_n)}\phi^{\Omega} - \sum_{k=1}^{n-1} \binom{n-1}{k} \mathcal{N}^{\rho}{}_{\sigma(\mu_1\dots\mu_{k+1}}\ee_{\mu_{k+2}\dots\mu_n)}(\mathcal{G}^{\sigma}{}_{\rho} \phi^\Omega).
		\end{equation}
		This expresses the $n^\text{th}$ extension of an arbitrary tensor in terms of extensions of degree up through $n-1$ and the first $n$ normal tensors. Thus, if we know the first $n$ normal tensors, through recursively applying this formula downward we can compute tensor extensions up to degree $n$. 
		
		Next let's derive a recursive formula for the normal tensors. To do so, we strip off one of the indices in the full symmetrization in \eqref{eq:conn_van} to get:
		\begin{equation}
			0 = (n+1)\d^{n-1}_{(\mu_1\dots\mu_{n-1}} \Gamma^\rho{}_{\mu_n\sigma)}(z) = 2 \fc{N}^\rho{}_{\sigma\mu_1\dots\mu_n} + (n-1)\d^{n-1}_{\sigma(\mu_n\dots\mu_{3}} \Gamma^\rho{}_{\mu_{2}\mu_1)}(z). 
		\end{equation}
		Using the component definition of the Riemann tensor \eqref{eq:3112}, we can write the term on the far right so that:
		\begin{equation}
			0 = 2 \fc{N}^\rho{}_{\sigma\mu_1\dots\mu_n} + (n-1) \delta^\tau_{(\mu_1}\delta^\omega_{\mu_2} \d^{n-2}_{\mu_3\dots \mu_n)}\(R^\rho{}_{\tau\sigma\omega} + \d_\omega \Gamma^\rho{}_{\tau\sigma} - \Gamma^\rho{}_{\sigma\lambda}\Gamma^\lambda{}_{\tau\omega} + \Gamma^\rho{}_{\omega\lambda}\Gamma^\lambda{}_{\tau\sigma}\).  \label{eq:norm_riem_part}
		\end{equation}
		Each of these terms after the $(n-1)$ factor can simplified into a tensor. The Riemann term, since we are evaluating at $z$, just turns into a Riemann tensor extension:
		\begin{equation}
			\delta^\tau_{(\mu_1}\delta^\omega_{\mu_2} \d^{n-2}_{\mu_3\dots\mu_n)} R^\rho{}_{\tau\sigma\omega} (z) = -\ee_{(\mu_n\dots \mu_3} R^\rho{}_{\mu_2\mu_1)\sigma}.
		\end{equation}
		The second term gives us more copies of the normal tensor:
		\begin{equation}
			\delta^\tau_{(\mu_1}\delta^\omega_{\mu_2}\d^{n-2}_{\mu_3\dots\mu_n)} \d_\omega \Gamma^\rho{}_{\tau\sigma}(z) = \fc{N}^\rho{}_{\sigma\mu_1\dots\mu_n}. 
		\end{equation}
		The third term vanishes on account of \eqref{eq:conn_van} after using the product rule:
		\begin{equation}
			\delta^\tau_{(\mu_1}\delta^\omega_{\mu_2}\d^{n-2}_{\mu_3\dots\mu_n)} (\Gamma^\rho{}_{\sigma\lambda}\Gamma^\lambda{}_{\tau\omega})(z) = \sum_{k=0}^{n-2}\binom{n-2}{k} \d^k_{(\mu_3\dots\mu_{k+2}} \Gamma^\lambda{}_{\mu_1\mu_2} \d^{n-2-k}_{\mu_{k+3}\dots\mu_n)}\Gamma^\rho{}_{\sigma\lambda} = 0. 
		\end{equation}
		The fourth term can be turned into a sum of products of normal tensors using the product rule (and recalling that $\fc{N}^\rho{}_{\sigma\mu} = 0$):
		\begin{align}
			\delta^\tau_{(\mu_1}\delta^\omega_{\mu_2}\d^{n-2}_{\mu_3\dots\mu_n)}(\Gamma^\rho{}_{\omega\lambda}\Gamma^\lambda{}_{\tau\sigma})(z) =&\ \sum_{k=0}^{n-2}\binom{n-2}{k} \d^k_{(\mu_3\dots\mu_{k+2}} \Gamma^\rho{}_{\mu_2|\lambda|} \d^{n-2-k}_{\mu_{k+3}\dots\mu_n} \Gamma^\lambda{}_{\mu_1)\sigma} \\
			=&\ \sum_{k=2}^{n-2}\binom{n-2}{k-1} \fc{N}^\rho{}_{\lambda(\mu_1\dots \mu_k} \fc{N}^\lambda{}_{|\sigma|\mu_{k+1}\dots\mu_n)}. 
		\end{align}
		Returning all of these ingredients to \eqref{eq:norm_riem_part}, we arrive at:
		\begin{equation}
			\label{eq:normal_tensor_recursion}
			{\mathcal{N}^\rho}{}_{\sigma\mu_1\dots\mu_n} = \frac{n-1}{n+1} \ee_{(\mu_n\dots\mu_3}R^\rho{}_{\mu_{2}\mu_1)\sigma} - \frac{n-1}{n+1}  \sum_{k=2}^{n-2} \binom{n-2}{k-1}\, {\mathcal{N}^\rho}_{\lambda(\mu_1\dots\mu_{k}}{\mathcal{N}^\lambda}_{\lvert\sigma\rvert\mu_{k+1}\dots\mu_n)}.
		\end{equation}
		This relation gives the $n^\text{th}$ normal tensor in terms of the $(n-1)^\text{th}$ tensor extension of the Riemann tensor (which itself, through \eqref{eq:any_tensor_extension} depends on the first $n-1$ normal tensors) and a sum of products of the first $n-2$ normal tensors. Thus, through iteratively using \eqref{eq:any_tensor_extension} and \eqref{eq:normal_tensor_recursion} as recursion relations, the normal tensors and tensor extensions of arbitrary degree can be calculated. The first few normal tensors are simply:
		\begin{gather}
			\fc{N}^\rho{}_{\sigma\mu} = 0, \p \fc{N}^\rho{}_{\sigma\mu\nu} = \frac{1}{3}R^\rho{}_{(\mu\nu)\sigma}, \p \fc{N}^\rho{}_{\sigma\mu_1\mu_2\mu_3} = \frac{1}{2}\nabla_{(\mu_1} R^\rho{}_{\mu_2\mu_3)\sigma}, \nonumber \\ \fc{N}^\rho{}_{\sigma\mu_1\mu_2\mu_3\mu_4} = \frac{3}{5}\nabla_{(\mu_1}\nabla_{\mu_2} R^\rho{}_{\mu_3\mu_4)\sigma} - \frac{2}{15}R^\rho{}_{(\mu_1\mu_2}{}^\lambda R_{|\lambda|\mu_3\mu_4)\sigma}.
		\end{gather}
		The resulting low order extensions of the metric are:
		\begin{gather}
			\ee_\rho g_{\mu\nu} = 0, \p \ee_{\rho\sigma} g_{\mu\nu} = \frac{2}{3} R_{\mu(\rho\sigma)\nu}, \p \ee_{\kappa \rho\sigma} g_{\mu\nu} = \nabla_{(\kappa} R_{|\mu|\rho\sigma)\nu}, \nonumber \\
			\ee_{\kappa\lambda\rho\sigma} g_{\mu\nu} = \frac{6}{5} \nabla^2_{(\kappa\lambda} R_{|\mu|\rho\sigma)\nu} + \frac{16}{15} R^\tau{}_{(\rho\sigma|\mu} R_{\tau|\kappa\lambda)\nu}.
		\end{gather}
		The resulting low order extensions of the Riemann tensor are:
		\begin{equation}
			\ee_\kappa R^\rho{}_{\mu\sigma\nu} = \nabla_\kappa R^\rho{}_{\mu\sigma\nu}, \p	\ee_{\kappa\lambda}R^\rho{}_{\mu\sigma\nu} = \nabla^2_{(\kappa\lambda)} R^\rho{}_{\mu\sigma\nu} - \frac{1}{3} R^\tau{}_{(\kappa\lambda)\omega} \fc{G}^\omega{}_\tau R^\rho{}_{\mu\sigma\nu}. 
		\end{equation}
		A Mathematica file \texttt{TensorExtensions.m} is included in the supplementary material which implements the calculation of normal tensors and tensor extensions algorithmically. It is able to comfortably produce the seventh normal tensor and seventh order tensor extensions on a laptop, and could be significantly optimized further. The main bottleneck at high orders is the efficient use of the Bianchi identity for simplification. 
		
		\section{Poincaré gauge for the Vector Potential} \label{sec:poinca}
		
		Given a bitensor satisfying \eqref{eq:bitensacond}, we immediately have that as a tangent tensor:
		\begin{equation}
			v^\mu \bar{\mc{A}}_\mu(z, v) = 0.  \label{eq:poingauge1}
		\end{equation}
		The $v^\mu$ derivative of this yields:
		\begin{equation}
			\bar{\mc{A}}_\mu(z, v) + v^\nu \bar{\d}_\mu \bar{\mc{A}}_\nu(z, v) = 0.  \label{eq:b2}
		\end{equation}
		Let $F_{\mu\nu}(z, v)$ be given by:
		\begin{equation}
			F_{\mu\nu}(z, v) = \bar{\d}_\mu \bar{\mc{A}}_\mu(z, v) - \bar{\d}_\nu \bar{\mc{A}}_\nu(z, v). 
		\end{equation}
		With this, \eqref{eq:b2} becomes:
		\begin{equation}
			\bar{\mc{A}}_\mu(z, v) + v^\nu \bar{\d}_\nu \bar{\mc{A}}_\mu(z, v) + F_{\mu\nu}(z, v) v^\nu = 0. 
		\end{equation}
		Evaluating this with $v$ replaced by $sv$ with $s$ a real number, we find:
		\begin{gather}
			\implies \frac{d}{ds}(s\bar{\mc{A}}_\mu(z, sv)) = -v^\nu F_{\mu\nu}(z, sv) s. 
		\end{gather}
		Integrating from $s = 0$ to $s = 1$, we find:
		\begin{equation}
			\bar{\mc{A}}_\mu(z, v) = -\bar{\mc{F}}_{\mu\nu}(z, v) v^\nu, \p \bar{\mc{F}}_{\mu\nu}(z, v) = \int_0^1 F_{\mu\nu}(z, sv) s ds.  \label{eq:poingaugesol1}
		\end{equation}
		Detangentizing the first of these yields the form \eqref{eq:bitensa}. Detangentizing the second, one can check by direct calculation that $\mc{F}_{\alpha\beta}$ satisfies the electromagnetic Bianchi identity.
		
		All of this analysis is performed at fixed $z$, and so the fact that \eqref{eq:poingauge1} allows $\bar{\mc{A}}_\mu$ to be put into the form \eqref{eq:poingaugesol1} is equivalent to the fact that given a vector potential $A_\mu(x)$ satisfying the ``Poincaré gauge'' condition:
		\begin{equation}
			x^\mu A_\mu(x) = 0
		\end{equation}
		one can recover that vector potential from its field strength $\fc{F}_{\mu\nu} = \d_\mu A_\nu - \d_\nu A_\mu$ by:
		\begin{equation}
			A_\mu(x) = -x^\nu \int_0^1 \fc{F}_{\mu\nu}(sx) s ds. 
		\end{equation}
		
		\section{Poincaré gauge for a Metric Perturbation} \label{sec:poincah}
		
		Given a bitensor metric perturbation that satisfies \eqref{eq:bitenshcond}, we immediately have that as a tangent tensor: 
		\begin{equation}
			\bar{{\mc{H}}}_{\mu\nu} v^\nu = 0.  \label{eq:poingauge2}
		\end{equation}
		The $v^\nu$ derivative of this gives:
		\begin{equation}
			\bar{{\mc{H}}}_{\mu\nu} + v^\rho \bar{\d}_\nu \bar{{\mc{H}}}_{\mu\rho} = 0. \label{eq:poingauge3}
		\end{equation}
		Taking another $v^\rho$ derivative:
		\begin{equation}
			\bar{\d}_\rho \bar{{\mc{H}}}_{\mu\nu} + \bar{\d}_\nu\bar{{\mc{H}}}_{\mu\rho} + v^\sigma \bar{\d}^2_{\rho\nu} \bar{{\mc{H}}}_{\mu\sigma} = 0. \label{eq:poingauge4}
		\end{equation}
		Symmetrizing \eqref{eq:poingauge3} and using \eqref{eq:poingauge4} then gives:
		\begin{equation}
			2\bar{{\mc{H}}}_{\mu\nu} - v^\rho v^\sigma \bar{\d}^2_{\mu\nu}\bar{{\mc{H}}}_{\rho\sigma} = 0. \label{eq:poingauge5}
		\end{equation}
		Let $\bar{\text{R}}_{\rho\mu\sigma\nu}(z,v)$ be the analog of the linearized tangentized Riemann tensor corresponding to $\bar{{\mc{H}}}_{\mu\nu}$:
		\begin{equation}
			\bar{\text{R}}_{\rho\mu\sigma\nu}(z,v) = \frac{1}{2}(\bar{\d}^2_{\mu\sigma} \bar{{\mc{H}}}_{\nu\rho} - \bar{\d}^2_{\mu\nu} \bar{{\mc{H}}}_{\rho\sigma} - \bar{\d}^2_{\rho\sigma} \bar{{\mc{H}}}_{\mu\nu} + \bar{\d}^2_{\rho\nu} \bar{{\mc{H}}}_{\sigma\mu} ). 
		\end{equation}
		With this, we can rewrite \eqref{eq:poingauge5} as:
		\begin{equation}
			2\bar{{\mc{H}}}_{\mu\nu} + v^\rho v^\sigma\bar{\d}^2_{\rho\sigma}\bar{{\mc{H}}}_{\mu\nu} - 2 v^\rho v^\sigma \bar{\d}^2_{\rho (\mu} \bar{{\mc{H}}}{}^{\phantom{2}}_{\nu)\sigma} + 2\bar{\text{R}}_{\rho\mu\sigma\nu} v^\rho v^\sigma = 0.  \label{eq:poingauge6}
		\end{equation}
		Using \eqref{eq:poingauge3} and \eqref{eq:poingauge4}, the third term becomes::
		\begin{equation}
			v^\rho v^\sigma \bar{\d}^2_{\rho (\mu} \bar{{\mc{H}}}{}^{\phantom{2}}_{\nu)\sigma} = \bar{{\mc{H}}}_{\mu\nu} - v^\rho\bar{\d}_\rho \bar{{\mc{H}}}_{\mu\nu}. 
		\end{equation}
		Returning this to \eqref{eq:poingauge6}, we have:
		\begin{equation}
			2 v^\rho\bar{\d}_\rho\bar{{\mc{H}}}_{\mu\nu} + v^\rho v^\sigma \bar{\d}^2_{\rho\sigma} \bar{{\mc{H}}}_{\mu\nu} + 2\bar{\text{R}}_{\rho\mu\sigma\nu} v^\rho v^\sigma = 0. 
		\end{equation}
		Evaluating this with $v$ replaced by $sv$ with $s$ a real number, we find:
		\begin{gather}
			\frac{d}{ds}(s^2 v^\rho \bar{\d}_\rho \bar{{\mc{H}}}_{\mu\nu}(z, sv)) = - 2s^2 v^\rho v^\sigma \bar{\text{R}}_{\rho\mu\sigma\nu}(z, sv). 
		\end{gather}
		Integrating from $0$ to $s$ we have:
		\begin{gather}
			s^2 v^\rho\bar{\d}_\rho\bar{{\mc{H}}}_{\mu\nu}(z, sv) = -2v^\rho v^\sigma\int_0^s \bar{\text{R}}_{\rho\mu\sigma\nu}(z, s'v) s'^2 ds' \\
			\implies \frac{d}{ds}\bar{{\mc{H}}}_{\mu\nu}(z, sv) = -2v^\rho v^\sigma \int_0^s\bar{\text{R}}_{\rho\mu\sigma\nu}(z, s'v) \frac{s'^2}{s^2} ds'.
		\end{gather}
		Again integrating from $0$ to $1$ and using that $\bar{{\mc{H}}}_{\mu\nu}(z, 0) = 0$ due to \eqref{eq:poingauge3}, and then integrating by parts with respect to $s$:
		\begin{equation}
			\bar{{\mc{H}}}_{\mu\nu}(z,v) = -\bar{\mc{R}}_{\rho\mu\sigma\nu} v^\rho v^\sigma, \p \bar{\mc{R}}_{\rho\mu\sigma\nu} = 2\int_0^1 s(1-s) \bar{\text{R}}_{\rho\mu\sigma\nu}(z, sv) ds.  \label{eq:poingaugesol2}
		\end{equation}
		Detangentizing the first of these yields the form \eqref{eq:bitensh}. Detangentizing the second, one can check by direct calculation that $\mc{R}_{\gamma\alpha\delta\beta}$ satisfies the algebraic and differential Bianchi identities. 
		
		All of this analysis is performed at fixed $z$, and so the fact that \eqref{eq:poingauge2} allows $\bar{\mc{H}}_{\mu\nu}$ to be put into the form \eqref{eq:poingaugesol2} is equivalent to the fact that given a metric perturbation $h_{\mu\nu}(x)$ about flat spacetime satisfying the ``Poincaré gauge'' condition:
		\begin{equation}
			x^\nu h_{\mu\nu}(x) = 0
		\end{equation}
		one can recover that metric perturbation from its linearized Riemann tensor $\delta R_{\rho\mu\sigma\nu}$ by:
		\begin{equation}
			h_{\mu\nu}(x) = -2 x^\rho x^\sigma\int_0^1 s(1-s) \delta R_{\rho\mu\sigma\nu}(sx)  ds. 
		\end{equation}
		
		\section{Time Direction Integral Bounds} \label{sec:uint}
		
		The angle of approach for this appendix takes significant inspiration from and builds on \cite{Beiglbock:1965fun} and the appendix of \cite{Costa:2014nta}. Recall $F^\mu(V)$ as defined in \eqref{eq:5612} as the function that takes a unit time direction to the unit vector in the direction of the momentum gotten by using that time direction to perform time-slicing. We are going to prove \eqref{eq:banachdist}, i.e. that two time directions are always brought proportionally closer to each other when this map is applied to each of them. Let $\xi^\mu$ be an arbitrary spacelike unit vector at $z$. Let $V_1$ and $V_2$ be two future oriented timelike unit vectors at $z$. Let $\Sigma_1$ and $\Sigma_2$ be the resulting Cauchy slices from taking $u = V_1$ and $u = V_2$. Then, let $p_{1}$ and $p_{2}$ be the resulting momenta from these slicings:
		\begin{gather}
			p_{1\mu} = \int_{\Sigma_1}\(T^{\alpha\beta}K_{\beta\mu} + J^\alpha \Phi_\mu\) d\Sigma_\alpha, \p p_{2\mu} = \int_{\Sigma_2}\(T^{\alpha\beta}K_{\beta\mu} + J^\alpha\Phi_\mu\) d\Sigma_\alpha. 
		\end{gather}
		The difference in these, contracted with $\xi^\mu$, gives:
		\begin{gather}
			\Delta_\xi p = \xi^\mu p_{2\mu} - \xi^\mu p_{1\mu} = \int_{\bb{V}} \nabla_\alpha(T^{\alpha\beta}K_{\beta\mu} \xi^\mu + J^\alpha \Phi_\mu \xi^\mu) d\bb{T}
		\end{gather}
		using Stokes' theorem, where $\bb{V}$ is the spacetime volume formed by the intersection of the worldtube with the region bounded by $\Sigma_1$ and $\Sigma_2$. Because $T^{\alpha\beta}$ and $J^\alpha$ are spatially compactly supported and $\Sigma_1$ and $\Sigma_2$ intersect at $z$, $\bb{V}$ is compact and has finite total volume	. Using \eqref{eq:tcons} and \eqref{eq:jcons}:
		\begin{gather}
			\Delta_\xi p = \int_{\bb{V}} \(T^{\alpha\beta} \nabla_\alpha K_{\beta\mu} \xi^\mu + J^\alpha\nabla_\alpha\Phi_\mu \xi^\mu - \fc{F}_{\alpha\beta} J^\alpha K^\beta{}_\mu \xi^\mu\) d\bb{T}.  \label{eq:e3}
		\end{gather}
		We will now systematically bound the right hand side of this equation. 
		
		Since by \eqref{eq:causalchar} $|v|$ is the geodesic distance between $x$ and $z$, $|v| \le \ell$ where $\ell$ retains its meaning from the text following \eqref{eq:posenergy}. Thus, we can count powers of $v$ as powers in the characteristic body size $\ell$. At coincidence, $\mlb  \nabla_\alpha K_{\beta\mu} \mrb = 0$, using \eqref{eq:coincide}. Because of this, and because $\nabla_\alpha K_{\beta\mu}$ depends smoothly on $x$ in the neighborhood of $z$, there exists a smooth bitensor field $\scr{K}_{\rho\sigma\mu\nu}(x,z)$ so that:
		\begin{gather}
			\nabla_{\alpha} K_{\beta\mu}(x, z) = \sigma^\rho{}_\alpha(x,z) \sigma^\sigma{}_\beta(x,z)\scr{K}_{\rho\sigma\mu\nu} (x, z) v^\nu(x, z) 
		\end{gather}
		with $v^\mu = -\sigma^\mu$ as always. And, due to the coincidence results, $\scr{K}_{\rho\sigma\mu\nu}(x,z)$ has a series expansion in powers of $v^\mu$ as:
		\begin{gather}
			\scr{K}_{\rho\sigma\mu\nu}(x,z) =R_{\rho\sigma\mu\nu}(z) + \(\frac{1}{6}\nabla_\mu R_{\rho(\nu\lambda)\sigma} - \frac{1}{2}\nabla_{(\nu} R_{\lambda)\mu\rho\sigma}\) v^\lambda(x,z) + \OO(v^2). 
		\end{gather}
		Using this (and series expanding $\bar{T}^{\mu\nu}$ similarly), we have:
		\begin{gather}
			T^{\alpha\beta}(x)\nabla_\alpha K_{\beta\mu}(x,z) \xi^\mu = \frac{1}{6} \xi^\mu \nabla_\mu R_{\rho \nu \lambda \sigma}(z) T^{\rho\sigma}(z) v^\nu(x,z) v^\lambda(x,z) + \ell^3 \scr{R}^{(g)}_\mu(x, z)\xi^\mu  \label{eq:e6}
		\end{gather}
		where $\scr{R}^{(g)}_\mu(x, z)$ is some remainder that is finite in the $\ell\to 0^+$ limit and depends on at least one power of the Riemann tensor and its derivatives. Let $|T\nabla R|$ be the supremum value attained by $\xi^\mu\nabla_\mu R_{\rho\nu\lambda\sigma}(z) T^{\rho\sigma}(z) n^\nu n^\lambda$ when it is ranged over all spacelike unit vectors $n^\mu$ at fixed $z$ and $\xi$, then over all spacelike unit vectors $\xi^\mu$ at fixed $z$, then over all $z \in \bb{V}$. This supremum is finite because the domain of values for each of these is compact ($n^\mu$ and $\xi^\mu$ are valued in the unit sphere). Let $\scr{R}^{(g)}$ be the supremum value attained by $|\scr{R}^{(g)}_\mu(x, z)\xi^\mu|$ when it is ranged over all spacelike unit vectors $\xi^\mu$, then over all $x \in \bb{V}$. Assume $|T\nabla R| \neq 0$. Then, \eqref{eq:e6} implies:
		\begin{gather}
			|T^{\alpha\beta}(x)\nabla_\alpha K_{\beta\mu}(x,z) \xi^\mu| \le \frac{\ell^2}{6} |T\nabla R| + \ell^3 \scr{R}^{(g)} = \ell^2 C_g |T \nabla R|  \label{eq:e7}
		\end{gather}
		where this last equation defines $C_g$ as some quantity which has a finite $\ell \to 0^+$ limit and a finite flat spacetime limit. If instead $|T\nabla R| = 0$, then \eqref{eq:e6} is identically 0 and \eqref{eq:e7} holds trivially with any finite choice of $C_g$. 
		
		We can perform a similar analysis for the electromagnetic terms in \eqref{eq:e3}. Using \eqref{eq:phipsidef} to explicitly calculate, we find:
		\begin{gather}
			J^\alpha\nabla_\alpha \Phi_\mu \xi^\mu - \fc{F}_{\alpha\beta} J^\alpha K^\beta{}_\mu \xi^\mu = \frac{1}{2}\xi^\mu v^\rho(x,z) J^\sigma(z) \nabla_\mu\fc{F}_{\rho\sigma}(z) + \ell^2\scr{R}^{(e)}_\mu(x, z)\xi^\mu \label{eq:e8}
		\end{gather}
		where $\scr{R}^{(e)}_\mu(x,z)$ is some remainder that is finite in the $\ell\to 0^+$ limit and which depends linearly on the electromagnetic field strength and its derivatives. Let $|J \nabla \fc{F}|$ be the supremum value attained by $\xi^\mu n^\rho J^\sigma \nabla_\mu \fc{F}_{\rho\sigma}$ when it is ranged over all spacelike unit vectors $n^\mu$ at fixed $z$ and $\xi$, then over all spacelike unit vectors $\xi^\mu$ at fixed $z$, then over all $z \in \bb{V}$. This supremum is finite for the same reason as in the gravitational case. Let $\scr{R}^{(e)}$ be the supremum value attained by $|\scr{R}^{(e)}_\mu(x,z)\xi^\mu|$ when it is ranged over all spacelike unit vectors $\xi^\mu$, then over all $x \in \bb{V}$. Assume $|J\nabla \fc{F}| \neq 0$. Then, \eqref{eq:e8} implies:
		\begin{gather}
			|J^\alpha(x)\nabla_\alpha \Phi_\mu(x,z) \xi^\mu - \fc{F}_{\alpha\beta}(x) J^\alpha(x) K^\beta{}_\mu(x,z) \xi^\mu| \le \frac{\ell}{2}|J\nabla\fc{F}| + \ell^2 \scr{R}^{(e)} = \ell C_e |J \nabla \fc{F}|  \label{eq:e9}
		\end{gather}
		where this last equation defines $C_e$ as some quantity which has a finite $\ell \to 0^+$ limit and a finite no electromagnetism limit. If instead $|J\nabla\fc{F}| = 0$, then \eqref{eq:e8} is identically 0 and \eqref{eq:e9} holds trivially with any finite choice of $C_e$. 
		
		The final quantity in \eqref{eq:e3} requiring a bound is the volume $\bb{V}$ itself. Viewed in the time-slicing defined by $V_1$, $\bb{V}$ at any fixed time has a spatial volume of $\OO(\ell^3)$. Because $\bb{V}$ is compact, the range of times covered by $\bb{V}$ is compact and so there is a maximum spatial volume achieved in that range, allowing us to bound this $\ell^3$ with a finite prefactor. The maximum time that can have elapsed in the $V_1$ time-slicing when moving from $z$ to any point in $\bb{V}$ would occur when one leaves $z$ at the speed of light and moves all the way to the edge of $\bb{V}$. Doing so on a constant time-slice (as defined by $V_2$) for a distance $\ell$ gives an elapsed time $v_\text{rel} \ell$ where $v_\text{rel}$ is the relative velocity of the $V_2$ frame to the $V_1$ frame at $z$. $v_\text{rel}$ is of order $|V_2-V_1|$ (to see this, go to the $V_1$ frame in normal coordinates at $z$ so that $V_1^\mu = (1, 0, 0, 0)^\mu$ and $V_2^\mu = (\gamma_\text{rel}, 0, 0, \gamma_\text{rel} v_\text{rel})^\mu$). Thus, the spacetime volume $\bb{V}$, for some finite coefficient $C_\bb{V}$ with finite $\ell \to 0^+$ limit is bounded by:
		\begin{gather}
			\bb{V} \le C_\bb{V} \ell^4 |V_2 - V_1|. 
		\end{gather}
		Combining this with \eqref{eq:e7} and \eqref{eq:e9}, \eqref{eq:e3} implies:
		\begin{gather}
			|\Delta_\xi p| \le C_\bb{V} \ell^5 \(C_g |T\nabla R| \ell + C_e |J\nabla \fc{F}|\) |V_2- V_1|. 
		\end{gather}
	
		$(p_2-p_1)^\mu$ can be decomposed into a component purely along $V_1$ and a remainder which is purely orthogonal to $V_1$ as $(p_2-p_1)^\mu = A V_1^\mu + B^\mu$ where $A = -V_{1\mu}(p_2-p_1)^\mu$ and $B^\mu V_{1\mu} = 0$. Then, $|p_2 - p_1| = \sqrt{-A^2 + |B|^2} \le |B|$. As well, $|B|$ can be written as just an example of $\Delta_\xi p$ with $\xi^\mu = B^\mu$ by noting that $|B| = \sqrt{(p_2-p_1)^\mu B_\mu}$. Thus, $|p_2 - p_1|$ is bounded as:
		\begin{gather}
			|p_2 - p_1| \le C_\bb{V} \ell^5 \(C_g |T\nabla R| \ell + C_e |J\nabla \fc{F}|\) |V_2- V_1|.  \label{eq:e12}
		\end{gather}
		We can now properly turn the left hand side into an approximation of $d(F(V_1), F(V_2))$ as in \eqref{eq:banachdist}. Let $\fc{M}_1 = |p_1|$ and $\fc{M}_2 = |p_2|$ be the masses of the two momenta as in \eqref{eq:massdefi}. Then:
		\begin{align}
			\fc{M}_1|F(V_2) - F(V_1)| &= |\fc{M}_1 F(V_2) - \fc{M}_1 F(V_1)| \nonumber \\
			&= |p_2 - p_1 - (\fc{M}_2 - \fc{M}_1) F(V_2)| \le |p_2 - p_1| + |\fc{M}_2-\fc{M}_1|. \label{eq:d13}
		\end{align}
		We can estimate $|\fc{M}_2 - \fc{M}_1|$ as:
		\begin{align}
			|\fc{M}_2 - \fc{M}_1| &= \frac{|\fc{M}_2^2 - \fc{M}_1^2|}{\fc{M}_2 + \fc{M}_1} = \frac{|p_2^2 - p_1^2|}{\fc{M}_2+\fc{M}_1} = \frac{|(p_2-p_1)\cdot(p_2+p_1)|}{\fc{M}_2+\fc{M}_1} \nonumber \\
			& \le \frac{|p_2+p_1|}{\fc{M}_2+\fc{M}_1} |p_2 - p_1| \le |p_2 - p_1|. 
		\end{align}
		Combining this with \eqref{eq:d13}
		\begin{gather}
			|F(V_2) - F(V_1)| \le \frac{2}{\fc{M}_1}|p_2 - p_1|.  \label{eq:d15}
		\end{gather}
		As explained in the text leading to \eqref{eq:udef}, $p_\mu$ is timelike and future oriented for any valid $V$. Thus $\fc{M} = |p| > 0$  for all $V$. We require further that there is a greatest lower bound $m > 0$ so that $\fc{M} \ge m$ for all $V$. As in, there is no limit of time-slicings so that in that limit the body's apparent mass converges to $0$. For any choice of $V$, only points within the lightcone around $z$ which intersect the worldtube are possibly accessible in the integrand of $p_\mu$ in \eqref{eq:psdef}. This intersection is a compact subset of spacetime, and so the existence of $m$ can be justified by requiring sufficient positivity of the time component of this integrand in this compact region. Given the existence of $m$, \eqref{eq:d15} becomes:
		\begin{gather}
			|F(V_2) - F(V_1)| \le \frac{2}{m}|p_2 - p_1|. 
		\end{gather}
		Finally then, using this and \eqref{eq:e12}, we find that \eqref{eq:banachdist} is true with:
		\begin{gather}
			\eps_u = \(\frac{2 C_\bb{V} \ell^5}{m}\(C_g |T\nabla R| \ell + C_e |J\nabla \fc{F}|\) \)^2 < 1. \label{eq:e16}
		\end{gather}
		We need $\eps_u < 1$ in order to apply the Banach fixed point theorem to \eqref{eq:banachdist}. If the gravitational and electromagnetic fields the body is placed into are not too large and do not vary too quickly on the scale of the body (so that $|T\nabla R|$ and $|J \nabla \fc{F}|$ are sufficiently small), then this will hold, guaranteeing that a $u^\mu$ satisfying \eqref{eq:udef} exists. For any fixed fields this holds for sufficiently small bodies due to the overall factor of $\ell^5$. Requiring $\eps_u < 1$ is the technical specification on what we meant in Subsection \ref{sec:bpa} when we demanded that the fields the body experiences not be overwhelmingly strong. If one would like to include the self-field of the body, then use of the full fields in \eqref{eq:e16} gives an $\eps_u$ that is of order 1, and thus the inequality cannot be guaranteed. However, results in self-force theory~\cite{Harte:2025tmd, Harte:2014wya}  suggest it is possible to use the self-field to renormalize the body variables (mass, momentum, spin) and if one does so, then \eqref{eq:e16} would be replaced by the same expression with the renormalized values and the regularized remainder fields. In this case we can again believe $\eps_u < 1$. 
		
		\end{appendix}
		
		\section{SSC Integral Bounds} \label{sec:sscint}
		
		Let us properly prove that it is possible to choose the worldline so that \eqref{eq:ssc} is satisfied. We will accomplish this by showing that when an initial choice of $z$ is continuously pushed in the direction of $\text{K}$ as defined in \eqref{eq:kpush}, it converges to a unique fixed point at which \eqref{eq:ssc} holds. For any choice of $z$, we have some choice of future oriented unit time direction $u^\mu(z)$ whose details we are agnostic to for now. From $u^\mu$ we construct $\Sigma(z)$ by shooting out all geodesics orthogonal to it just as in \eqref{eq:sigdef} (but here we are considering only a single worldline time $\lambda$ and multiple possible choices of the worldline location at that time). ${\mc{K}}^\mu$ as defined in \eqref{eq:skdef} is a function of the chosen $z$. Let $z_0(0)$ and $z_1$(0) be two spacelike separated points within the worldtube, thus representing two candidate choices for the worldline position at some time. For each of them, we define a one-parameter family of valid such choices, $z_0(r)$ and $z_1(r)$, by the solutions to the differential equation:
		\begin{gather}
			\frac{dz^\mu}{dr} = -{\mc{K}}^\mu(z) \label{eq:e1}
		\end{gather}
		with the corresponding initial conditions. ${\mc{K}}^\mu(z)$ is a spatial vector for each $z$ since ${\mc{K}}_\mu u^\mu = 0$ by \eqref{eq:skdef}. As such, this flow stays entirely within the set of points $\bb{V}$ which are both within the worldtube and spacelike separated from the initial conditions. $\bb{V}$ is automatically compact by the spatial compactness of the worldtube. 
		
		Let $\text{dist}(x,z)$ be the geodesic distance between two points $x$ and $z$. By \eqref{eq:causalchar}:
		\begin{gather}
			\text{dist}(z_1(r),z_0(r)) = \sqrt{2 \sigma(z_1(r), z_0(r))}. 
		\end{gather}
		Differentiating with respect to $r$ and using \eqref{eq:syngder}:
		\begin{gather}
			\frac{d}{dr}\text{dist}(z_1(r), z_0(r)) = \frac{1}{\text{dist}(z_1(r), z_0(r))}\(-\dot\zeta_\alpha {\mc{K}}^\alpha(z_1) + \dot \zeta_\mu {\mc{K}}^\mu(z_0)\)
		\end{gather}
		where $\zeta^\alpha(r,s)$ is the geodesic with affine parameter $s$ and boundary conditions $\zeta^\mu(r, 0) = z_0^\mu(r)$ and $\zeta^\alpha(r, 1) = z_1^\alpha(r)$ and where dots indicate $s$ derivatives. The right hand side can be rewritten using the fundamental theorem of calculus as an $s$ integral and then simplified using the geodesic equation to find:
		\begin{gather}
			\frac{d}{dr}\text{dist}(z_1(r), z_0(r)) = -\(\int_0^1  \xi^\alpha \xi^\beta\nabla_\alpha {\mc{K}}_\beta ds\)\, \text{dist}(z_1(r), z_0(r)) \label{eq:e4}
		\end{gather}
		where $\xi^\alpha(r,s)$ is the unit vector in the direction of the geodesic tangent $\dot\zeta^\alpha(r,s)$. 
		
		If the integral in \eqref{eq:e4} is always positive, then $z_1$ and $z_0$ always get closer under the flow \eqref{eq:e1} and converge to a unique fixed point of such by the Banach fixed point theorem. We will next estimate the value of this integral to understand how positivity can be guaranteed by appropriate conditions on the body and the fields it experiences. We start by simplifying $\xi^\alpha \xi^\beta \nabla_\alpha {\mc{K}}_\beta$ using the integral definition of ${\mc{K}}^\mu(z)$ in \eqref{eq:558}. Under a change in $z$, both the integrand of ${\mc{K}}^\mu$ and the integration region $\Sigma(z)$ will change. The explicit integrand changing is easy; we just take the covariant $z$-derivative. The change in the integration region is slightly more complicated, but we actually already worked out how this works in \eqref{eq:volintder}. Consider a continuous one-parameter family $z_\eps$ of points (becoming $z$ at $\eps = 0$) and $\Sigma_\eps = \Sigma(z_\eps)$. We want to think about the change from $\Sigma$ to $\Sigma_\eps$ as a flow of each point $x$ in $\Sigma$ to a new point $x_\eps$. Each point in $\Sigma_\eps$ takes the form $x_\eps = \exp_{z_\eps} v_\eps$ with $v_\eps^\mu u_\mu(z_\eps) = 0$. Since we want $z$ to change in the $\xi^\mu$ direction, we take:
		\begin{gather}
			\left.\frac{dz^\mu_\eps}{d\eps}\right|_{\eps = 0} = \xi^\mu(z). 
		\end{gather}
		We define $v_\eps$ by taking $v^\mu$, parallel transporting it from $z$ to $z_\eps$, then projecting out the component along $u_\eps$ so that the result is orthogonal to $u_\eps$ (as it must be for $x_\eps$ to be in $\Sigma_\eps$):
		\begin{gather}
			v_\eps^\mu = (\delta^\mu_\nu + u_\eps^\mu u_{\eps \nu}) \text{PT}_{z\to z_\eps}(v)^\nu. 
		\end{gather}
		Then, with $\lambda$ replaced by $\eps$ in \eqref{eq:volintder}, the change in $x_\eps^\alpha$ at $\eps = 0$ gives the analog of the vector field $w^\alpha$, and we have:
		\begin{gather}
			\left.\frac{d}{d\eps}\int_{\Sigma_\eps} f^\alpha(x) d\Sigma_\alpha\right|_{\eps = 0} = \int_\Sigma \nabla_\alpha f^\alpha \left.\frac{dx^\beta_\eps}{d\eps}\right|_{\eps = 0} d\Sigma_\beta \label{eq:e7a}
		\end{gather}
		for a generic smooth $z$-independent vector field of compact support $f^\alpha(x)$ and:
		\begin{gather}
			\left.\frac{dx^\alpha_\eps}{d\eps}\right|_{\eps=0} = K^\alpha{}_\mu \xi^\mu + H^\alpha{}_\mu u^\mu v^\nu \nabla_\xi u_\nu \label{eq:e8a}
		\end{gather} 
		by \eqref{eq:desicvari} where $\nabla_\xi = \xi^\mu \nabla_\mu$. 
		
		With \eqref{eq:e7a} and \eqref{eq:e8a}, we can differentiate the integral form of ${\mc{K}}^\mu$ in \eqref{eq:558} as:
		\begin{align}
			\xi^\mu \xi^\nu\nabla_\mu {\mc{K}}_\nu =&\ \int_\Sigma \xi_\mu \nabla_\xi\(v^\mu\(T^\alpha{}_\beta H^\beta{}_\nu + J^\alpha \Psi_\nu\) u^\nu\) d\Sigma_\alpha \nonumber \\ 
			&\ +\int_\Sigma \xi_\mu \nabla_\alpha\(v^\mu\(T^\alpha{}_\beta H^\beta{}_\nu + J^\alpha\Psi_\nu\) u^\nu\) \(K^\gamma{}_\rho \xi^\rho + H^\gamma{}_\rho u^\rho v^\sigma \nabla_\xi u_\sigma\) d\Sigma_\gamma
		\end{align}
		where the first line accounts for the explicit $z$ dependence of ${\mc{K}}^\mu$'s integrand, while the second accounts for the change in integration region. Fully expanding the derivatives and simplifying yields:
		\begin{gather}
			\xi^\mu \xi^\nu \nabla_\mu {\mc{K}}_\nu = \ee + \xi_\mu \nabla_\xi u_\nu S^{\mu\nu} + \Delta^I_\xi{\mc{K}}_\text{EM} + \Delta^I_\xi{\mc{K}}_\text{GR} \label{eq:e12a}
		\end{gather}
		where $\Delta^I_\xi{\mc{K}}_\text{EM}$, the intrinsic change in ${\mc{K}}^\mu$ due to the electromagnetic field, is:
		\begin{gather}
			\Delta^I_\xi{\mc{K}}_\text{EM} = \frac{1}{2}\fc{F}_{u \rho}\(\delta^\rho_\sigma - \xi^\rho \xi_\sigma\) Q^{\sigma u}_0 + \OO_\text{EM}(\ell^2)
		\end{gather}
		($Q^{\mu\nu}_0$ is the reduced field dipole moment of the current density) and where $\Delta^I_\xi{\mc{K}}_\text{GR}$, the intrinsic change in ${\mc{K}}^\mu$ due to the gravitational field, is:
		\begin{align}
			\Delta^I_\xi{\mc{K}}_\text{GR} =&\ \frac{1}{3}R_{\rho\mu\sigma u} \fc{J}^{\rho\mu\sigma u}_0 + \frac{2}{9} R_{\rho u \sigma u} \fc{J}^{\rho u \sigma u}_0 - \frac{1}{2} R_{u \xi \rho \sigma} \fc{J}^{u\xi\rho\sigma}_0 \nonumber \\
			&\ - \frac{2}{9} R_{u\xi\rho\sigma}\fc{J}^{u\xi\rho u}_0 - \frac{4}{9} R_{\xi\rho \xi\sigma} \fc{J}^{\rho u \sigma u}_0 + \frac{2}{3} R_{\xi(\rho\sigma) u} \fc{J}^{\xi(\rho\sigma) u}_0 + \OO_\text{GR}(\ell^3)
		\end{align}
		($\fc{J}^{\rho\mu\sigma\nu}_0$ is the reduced field quadupole moment of the energy-momentum tensor). $\ell$ is the maximal body size as always. The notation $\OO_\text{EM}(\ell^2)$ indicates higher order corrections that all linear in the electromagnetic field strength and its derivatives while $\OO_\text{GR}(\ell^3)$ indicates higher order corrections that all depend on at least one power of the Riemann tensor and its derivatives but not on the electromagnetic field strength. 
		
		So far we have been agnostic to the selection of $u^\mu$, but practically we always choose $u^\mu$ as in Appendix \ref{sec:uint} so that $p^\mu = \fc{M} u^\mu$. With this choice, we can compute $\nabla_\xi u_\mu$ via applying the calculation strategy used on $\nabla_\xi {\mc{K}}^\mu$ to $\nabla_\xi p_\mu$ from \eqref{eq:psdef}. Doing so yields:
		\begin{gather}
			\nabla_\xi p_\mu = -\frac{1}{2}R_{\mu \xi \rho \sigma} S^{\rho\sigma} + q \fc{F}_{\mu \xi} + \OO_\text{EM}(\ell) + \OO_\text{GR}(\ell^2), \p \nabla_\xi u_\mu = \frac{\delta^\nu_\mu + u^\nu u_\mu}{\fc{M}} \nabla_\xi p_\nu.  \label{eq:e13}
		\end{gather}
		Returning these to \eqref{eq:e12a}, we find that:
		\begin{gather}
			\xi^\mu \xi^\nu \nabla_\mu \mc{K}_\nu = \ee + \Delta_\xi \mc{K}_{\text{EM}} + \Delta_\xi \mc{K}_{\text{GR}} \label{eq:e14}
		\end{gather}
		where $ \Delta_\xi \mc{K}_{\text{EM}}$, the total change in $\mc{K}_\mu$ due to electromagnetism is:
		\begin{gather}
			\Delta_\xi \mc{K}_{\text{EM}} = \Delta^I_\xi{\mc{K}}_\text{EM} + \frac{q}{\fc{M}} \xi_\mu S^{\mu\nu}(\delta^\rho_\nu + u^\rho u_\nu) \fc{F}_{\rho \xi} + \OO_\text{EM}(\ell^2) \label{eq:e16a}
		\end{gather}
		and where $\Delta_\xi\mc{K}_{\text{GR}}$, the total change in $\mc{K}_\mu$ due to gravity is:
		\begin{gather}
			\Delta_\xi \mc{K}_{\text{GR}} = \Delta^I_\xi{\mc{K}}_\text{GR} - \frac{1}{2\fc{M}} \xi_\mu S^{\mu\nu}(\delta^\rho_\nu + u^\rho u_\nu) R_{\rho \xi\kappa\lambda} S^{\kappa\lambda} + \OO_\text{GR}(\ell^3).  \label{eq:e17a}
		\end{gather}
		
		It is important to note that multipoles scale with the body size (the monopole by default scaling like $\ell^0$, the dipole like $\ell^1$, and so on) due to the appearance of explicit factors of $v^\mu$ in their integral definitions. Let $|\Delta\mc{K}_{\text{EM}}|(z)$ be the supremum of $|\Delta_\xi\mc{K}_{\text{EM}}|$ when ranged over all spacelike unit vectors $\xi^\mu$ at fixed $z$. Because the set of such vectors is compact, this supremum exists and is finite. By \eqref{eq:e16a}, that value necessarily scales at least as $\ell$ with the body size. Since $z(r)$ stays in the compact set $\bb{V}$, there is also finite a supremum of $|\Delta\mc{K}_{\text{EM}}|(z)$ when ranged over all $z$; call it $\ell |\hat Q \fc{F}|$. This result is named to indicate that the remainder after stripping off the leading $\ell$ scaling depends on the electromagnetic field strength in the worldtube, is 0 in the absence of such a field, and that at leading order in the body size it is determined by the multipole moments of the body up to dipolar order. Thus:
		\begin{gather}
			|\Delta_\xi\mc{K}_{\text{EM}}| \le \ell |\hat Q \fc{F}|. \label{eq:e17b}
		\end{gather}
		
		We can bound the gravitational contributions to $\xi^\mu \xi^\nu \nabla_\mu \mc{K}_\nu$ in precisely the same way. Let $|\Delta\mc{K}_{\text{GR}}|(z)$ be the supremum of $|\Delta_\xi\mc{K}_{\text{GR}}|$ when ranged over all spacelike unit vectors $\xi^\mu$ at fixed $z$. By \eqref{eq:e17a}, that value necessarily scales at least as $\ell^2$ with the body size. Let $\ell^2 |\hat{\fc{J}} R|$ be the supremum of $|\Delta\mc{K}_{\text{GR}}|(z)$ when ranged over all $z \in \bb{V}$. Again the result is named to indicate that the remainder after stripping off the leading $\ell^2$ scaling depends on the Riemann tensor in the worldtube, is 0 in flat spacetime, and that at leading order in the body size it is determined by the multipole moments of the body up to quadrupolar order. Thus:
		\begin{gather}
			|\Delta_\xi\mc{K}_{\text{GR}}| \le \ell^2 |\hat{\fc{J}} \fc{R}|. \label{eq:e18b}
		\end{gather}
		
		Lastly we let $\ee_0$ be the minimum value taken by the body energy $\ee$ when ranged over all $z\in\bb{V}$. Altogether \eqref{eq:e14}, \eqref{eq:e17b}, and \eqref{eq:e18b} imply:
		\begin{gather}
			\xi^\mu \xi^\nu \nabla_\mu \mc{K}_\nu \ge \ee_0 - \ell |\hat Q \fc{F}| - \ell^2 |\hat{\fc{J}} \fc{R}|. 
		\end{gather}
		We can now properly state our condition on the physics of the body and its environment. We require the existence of $\eps_s$ such that:
		\begin{gather}
			\eps_s = \frac{\ell |\hat Q \fc{F}| + \ell^2 |\hat{\fc{J}} \fc{R}|}{\ee_0} < 1. \label{eq:d16}
		\end{gather}
		Given that $\eps_s < 1$, we have by \eqref{eq:e4} that:
		\begin{gather}
			\frac{d}{dr}\text{dist}(z_1(r), z_0(r)) \le -\ee_0(1-\eps_s)\text{dist}(z_1(r), z_0(r)) .
		\end{gather}
		By the monotonicity of the integral, this implies:
		\begin{gather}
			\text{dist}(z_1(r), z_0(r)) \le e^{-\ee_0(1-\eps_s)r}\text{dist}(z_1(0), z_0(0)) .
		\end{gather}
		Thus, by the Banach fixed point theorem this flow is a contraction mapping and all initial candidate points converge to the same unique fixed point as $r\to\8$. That fixed point is precisely the worldline position at the given time of consideration which satisfies the covariant SSC. The condition $\eps_s < 1$ is automatic for a sufficiently small body, given the $\ell$ scaling in \eqref{eq:d16}. It is also valid for finite size objects, provided the fields they experience are not too strong. Physically, \eqref{eq:d16} requires that the electromagnetic and gravitational potential energies of the body in the field it experiences are not greater in magnitude than the body's inertia. 
		
		Similarly to the selection of the body frame, there are subtleties in guaranteeing \eqref{eq:d16} in consideration of the self-field of the body. Use of the full fields again gives an $\eps_s$ that is of order 1. self-force results~\cite{Harte:2025tmd, Harte:2014wya} suggest that using the renormalized body variables and regularized fields permit an effective $\eps_s < 1$.

\end{document}